\documentclass[english,preprintnumbers,nofootinbib,twocolumn,superscriptaddress]{revtex4-2}

\usepackage{graphicx}
\usepackage{scrextend}
\usepackage{mathtools}
\usepackage{amssymb}
\usepackage{accents}
\usepackage{verbatim}
\usepackage{hyperref}
\usepackage{xcolor}
\usepackage{stmaryrd}
\usepackage{tabulary}
\usepackage{multirow}
\usepackage[outline]{contour}

\hypersetup{
    colorlinks,
    linkcolor={red!50!black},
    citecolor={blue!50!black},
    urlcolor={blue!80!black}
}

\newcommand{\Utm} {U_{\text{TM}_1}\!({\scriptstyle \frac{\pi}{12},\, \text{-}\frac{\pi}{2} })}
\newcommand{\T} {{\scriptscriptstyle T}}

\newcommand{\om} {\omega}
\newcommand{\ob} {\bar{\omega}}
\newcommand{\taub} {\bar{\tau}}
\newcommand{\eig} {\varsigma}
\newcommand{\eib} {\bar{\varsigma}}
\newcommand{\be} {\begin{equation}}
\newcommand{\ee} {\end{equation}}

\newcommand{\xia} {\xi_{\mathrm a}}
\newcommand{\xib} {\xi_{\mathrm b}}
\newcommand{\xiq} {\xi_{\mathrm q}}
\newcommand{\xibq} {\xi_{\mathrm {bq}}}
\newcommand{\xic} {\xi_{\mathrm c}}
\newcommand{\xiab} {\xi_{\mathrm {ab}}}
\newcommand{\xiabc} {\xi_{\mathrm {abc}}}

\newcommand{\phia} {\phi_{\mathrm a}}
\newcommand{\phib} {\phi_{\mathrm b}}
\newcommand{\phiq} {\phi_{\mathrm q}}
\newcommand{\phibq} {\phi_{\mathrm {bq}}}
\newcommand{\phic} {\phi_{\mathrm c}}
\newcommand{\phibc} {\phi_{\mathrm {bc}}}
\newcommand{\phiabc} {\phi_{\mathrm {abc}}}

\newcommand{\Ga} {\acute{G}}
\newcommand{\Ht} {\tilde{H}}
\newcommand{\Hb} {\breve{H}}
\newcommand{\Ha} {\acute{H}}

\newcommand{\Sa} {\acute{S}}
\newcommand{\Tt} {\tilde{T}}
\newcommand{\Tb} {\breve{T}}

\newcommand{\mup} {\mu_{\mathrm p}}
\newcommand{\muq} {\mu_{\mathrm q}}
\newcommand{\mupq} {\mu_{\mathrm {pq}}}

\newcommand{\nup} {\nu_{\mathrm p}}
\newcommand{\nuq} {\nu_{\mathrm q}}
\newcommand{\nur} {\nu_{\mathrm r}}
\newcommand{\nupq} {\nu_{\mathrm {pq}}}
\newcommand{\nupr} {\nu_{\mathrm {pr}}}
\newcommand{\nuqr} {\nu_{\mathrm{qr}}}

\newcommand{\lamp} {\lambda_{\mathrm p}}
\newcommand{\lamq} {\lambda_{\mathrm q}}
\newcommand{\lamr} {\lambda_{\mathrm r}}
\newcommand{\lampq} {\lambda_{\mathrm {pq}}}
\newcommand{\lampr} {\lambda_{\mathrm {pr}}}
\newcommand{\lamqr} {\lambda_{\mathrm {qr}}}

\newcommand{\phiarho} {\phi_\mathrm {a\rho}}
\newcommand{\xicrho} {\xi_{\mathrm {c\rho}}}
\newcommand{\rhop} {\rho_{\mathrm p}}
\newcommand{\rhoq} {\rho_{\mathrm q}}
\newcommand{\rhor} {\rho_{\mathrm r}}
\newcommand{\rhos} {\rho_{\mathrm s}}
\newcommand{\rhoqr} {\rho_{\mathrm {qr}}}
\newcommand{\rhopqr} {\rho_{\mathrm {pqr}}}

\newcommand{\phiceta} {\phi_{\mathrm {c\eta}}}
\newcommand{\xiceta} {\xi_{\mathrm {c\eta}}}
\newcommand{\etap} {\eta_{\mathrm p}}
\newcommand{\etaq} {\eta_{\mathrm q}}
\newcommand{\etar} {\eta_{\mathrm r}}
\newcommand{\etas} {\eta_{\mathrm s}}
\newcommand{\etaqr} {\eta_{\mathrm {qr}}}
\newcommand{\etapqr} {\eta_{\mathrm {pqr}}}

\newcommand{\mati}{{\scriptstyle{\mathbf I}}}
\newcommand{\matt}{\mathbf{t}}
\newcommand{\matr}{\mathbf{r}}

\newcommand{{\matv}}{\mathbf{x}}
\newcommand{\matw}{\boldsymbol{\omega}}
\newcommand{\mattau}{\boldsymbol{\tau}}
\newcommand{\sigx}{\boldsymbol{\sigma}_{\!x}}
\newcommand{\sigy}{\boldsymbol{\sigma}_{\!y}}
\newcommand{\sigz}{\boldsymbol{\sigma}_{\!z}}

\newcommand{\reps} {\boldsymbol{1}}
\newcommand{\repsp} {\boldsymbol{1}'}
\newcommand{\repd} {\boldsymbol{2}}
\newcommand{\repdd} {\boldsymbol{\underline{2}}}
\newcommand{\repddp} {\boldsymbol{\underline{2}}'}
\newcommand{\rept} {\boldsymbol{3}}
\newcommand{\reptp} {\boldsymbol{3}'}
\newcommand{\repqq} {\boldsymbol{\underline{4}}}
\newcommand{\repq} {\boldsymbol{4}}
\newcommand{\repx} {\boldsymbol{6}}

\newcommand{\gencp}{\mathbf{C}_{\sigma}}
\newcommand{\gent}{\mathbf{T}}
\newcommand{\genr}{\mathbf{R}}
\newcommand{\gens}{\mathbf{S}}
\newcommand{\genu}{\mathbf{U}}
\newcommand{\gene}{\mathbf{E}}

\newcommand{\genzl}{\mathbf{Z}_l}
\newcommand{\genzn}{\mathbf{Z}_\nu}
\newcommand{\genx}{\mathbf{X}}

\newcommand{\genxmu}{\mathbf{X}_{\mathrm \mu}}
\newcommand{\genxnu}{\mathbf{X}_{\mathrm \nu}}
\newcommand{\genxlam}{\mathbf{X}_{\mathrm \lambda}}
\newcommand{\genxrho}{\mathbf{X}_{\mathrm \rho}}
\newcommand{\genxeta}{\mathbf{X}_{\mathrm \eta}}

\begin{document}
	
\title{The framework of the auxiliary group: two birds with one stone}
\author{R.~Krishnan}
\email{krisphysics@gmail.com}
\homepage{https://orcid.org/0000-0002-0707-3267}
\affiliation{Saha Institute of Nuclear Physics, 1/AF Bidhannagar, Kolkata 700064, India}  
\begin{abstract}
Flavon models in the literature assume constraints on the components of the vacuum expectation values (vevs) of flavons, and typically, these constraints are not fully determined by the residual symmetry group of the set of vevs. This poses a problem because the general potential of the flavons cannot have a minimum that leads to such constraints unless additional mechanisms involving supersymmetry, extra dimensions etc., are invoked. In this paper, we show that the framework of the auxiliary group naturally results in vevs satisfying the required constraints, and using it, we construct a type-1 seesaw model with two right-handed neutrinos, which predicts the ratio of the light neutrino masses $m_2/m_3=(\sqrt{2}-1)/(\sqrt{2}+1)$ and $\text{TM}_1$ mixing with $\sin\theta_{13}=\frac{1}{\sqrt{3}}\sin\frac{\pi}{12}$ and $\sin\delta_\text{CP}=-1$. Our framework posits auxiliary group transformations which act on the flavons but not on the fermions. We construct the general renormalizable potential without invoking additional mechanisms and show that it has a minimum that leads to the required constraints.\footnote{A video presentation of this paper is available \href{https://www.youtube.com/playlist?list=PLjRYJtC1E1HfEbur89bH3CT5BvMtO4FWK}{here}.}
\end{abstract}
\maketitle

Almost every discrete flavour symmetry group in the literature is constructed as the direct product of a finite subgroup of $U(3)$ and one or more Abelian groups. It was believed that there was no reason to go beyond $U(3)$ to model flavour symmetries because there are only three families of fermions. However, \cite{1006.0203,1111.1730,1211.5143} went beyond this $U(3)$-subgroup paradigm by utilizing group extensions. By doing so, the authors succeeded in avoiding dangerous cross terms between the flavons in the charged-lepton and neutrino sectors. These cross terms, if present, would spoil the required constraints on the vevs by spoiling the residual symmetry groups obtained separately for the two sectors. Other papers in the literature assume mechanisms\cite{hep-ph/0504165,hep-ph/0512103,hep-ph/0601001,1004.0321,1205.3617} involving supersymmetry, extra dimensions etc.,~to forbid the problematic cross terms.

In \cite{1111.1730, 1211.5143}, the flavons in the neutrino sector transform as a quartet representation of the flavour group $Q_8\rtimes A_4$ while the fermions transform as multiplets isomorphic to the representions of $A_4$, i.e.,~the fermions form an unfaithful representation of the flavour group. Two quartet flavons are combined to obtain a multiplet isomorphic to the $A_4$ triplet so that it can couple with the fermions in the neutrino sector. The authors showed that this construction generates accidental symmetries in the renormalizable potential and avoids dangerous cross terms. This approach is very appealing since it is minimal: it utilizes discrete symmetries and scalar fields only without requiring additional mechanims. 

Refs.~\cite{1901.01205, 1912.02451} put forward a framework that also went beyond the $U(3)$-subgroup paradigm. The author proposed that the {\it elementary} flavons transform as irreducible multiplets (irreps) of a group that is an extension of a $U(3)$-subgroup by an auxiliary group whereas the fermions transform as multiplets that are isomorphic to the irreps of the $U(3)$-subgroup. An {\it effective} irrep that can couple with the fermions was constructed by combining several elementary flavons. The author assigned vevs with specific residual symmetries to each elementary flavon, which led to constraints on the components of the vev of the effective irrep. These constraints did not originate from the residual symmetries of the vev under the $U(3)$-subgroup. Such constraints are similar to the ones proposed in the so-called indirect models\cite{1510.02091} and lead to highly predictive phenomenology.

Refs.~\cite{1006.0203, 1111.1730, 1211.5143} and \cite{1901.01205, 1912.02451} used group extensions by auxiliary groups for different purposes. In the former, the purpose was to prevent dangerous cross terms among the various irreps such that they acquire vevs with specific residual symmetries. The later simply assigned residual symmetries to the vevs of the elementary irreps without addressing the problem of cross terms. In the later, the purpose was to obtain constraints in the vev of the effective irrep that are not attributable to its residual symmetries. In the current paper, the author kills both these birds using one stone, i.e.,~the author prevents the dangerous cross terms and obtains the aforementioned constraints using group extensions.

\subsection{$2O$ - the binary octahedral group}
\label{sec:2O}

We construct the faithful $2$-dimensional irrep of the binary octahedral group, $2O$, using the generators
\begin{equation}\label{eq:bogen}
\matt_2=\frac{1}{\sqrt{2}}\left(\begin{matrix} \taub & \taub\\
-\tau & \tau
\end{matrix}\right)\!,\,\quad \matr_2=\frac{1}{\sqrt{2}}\left(\begin{matrix} 1 & 1\\
-1 & 1
\end{matrix}\right)\!,\!
\end{equation}
where $\tau=\frac{1+i}{\sqrt{2}}$, $\taub=\frac{1-i}{\sqrt{2}}$ are the complex eighth roots of unity. Let us call this irrep $\repdd$. Let $\gent$ and $\genr$ be the abstract elements of $2O$ corresponding to $\matt_2$ and $\matr_2$, respectively. In this paper, we use lowercase bold letters to denote square matrices, e.g.,~$\matt_2$, $\matr_2$, and uppercase bold letters to denote abstract group elements, e.g.,~$\gent$, $\genr$. $2O$ is a group with $48$ elements. By taking the tensor product expansion of two $\repdd$s of $2O$, we can understand its connection with the chiral octahedral group $S_4$,
\begin{equation}\label{eq:dtimesd}
\repdd\otimes \repdd = \rept \oplus \reps.
\end{equation}
The resulting $\rept$ is isomorphic to the triplet irrep of $S_4$. Its generators are given by
\begin{equation}\label{eq:ogen}
\matt_{3}= \left(\begin{matrix} 0 & 1 & 0\\
0 & 0 & 1\\
1 & 0 & 0
\end{matrix}\right)\!, \,\, \matr_{3}= \left(\begin{matrix} 1 & 0 & 0\\
0 & 0 & 1\\
0 & -1 & 0
\end{matrix}\right)\!.
\end{equation}
$\rept$ and $\reps$ in (\ref{eq:dtimesd}) correspond to the symmetric and the antisymmetric parts of the tensor product. Given  $x=(x_1, x_2)^\T$ and $y=(y_1, y_2)^\T$ transforming as $\repdd$, we construct $\rept$ as
\be\label{eq:triptrace}
(x,y)_{\rept}=\!\big(x_1 y_1+ x_2 y_2,  -i(x_1 y_2+x_2 y_1) , i(x_1 y_1-x_2 y_2)\big)^{\!\T}\!\!\!.
\ee

$S_4$ is widely utilized for model building in the flavour physics literature\cite{Pakvasa:1978tx, Brown:1984dk, hep-ph/9403201, hep-ph/0301234, hep-ph/0508231, hep-ph/0602244, hep-ph/0611078, hep-ph/0612214, 0705.2275, 0811.0345, 1211.2000}. It has five irreps given by $\reps$, $\repsp$, $\repd$, $\rept$ and $\reptp$. These irreps of $S_4$ form all the unfaithful irreps of $2O$. The singlet $\repsp$ remains invariant under $\gent$ and changes sign under $\genr$. We may obtain the irrep $\reptp$ as $\reptp=\rept\otimes\repsp$. The representation matrices of $\repd$ corresponding to $\gent$, $\genr$ are given by
\be\label{eq:ddgen2}
\matw_2=\left(\begin{matrix} \ob & 0\\
0 & \om
\end{matrix}\right), \quad \sigx=\left(\begin{matrix} 0 & 1\\
1 & 0
\end{matrix}\right),
\ee
respectively, where $\om=\frac{-1+i\sqrt{3}}{2}$, $\ob=\frac{-1-i\sqrt{3}}{2}$ are the complex cube roots of unity. 

Besides $\repdd$, $2O$ has two more faithful irreps, which we denote by $\repddp$ and $\repqq$. We may obtain them as
\be\label{eq:dtimesdd}
\repddp=\repdd\otimes\repsp, \quad \repqq=\repd\otimes\repdd.
\ee
By taking the Kronecker products of matrices representing $\repd$ (\ref{eq:ddgen2}) and $\repdd$ (\ref{eq:bogen}), we obtain the generators of $\repqq$ as
\be\label{eq:qgen}
\matw_2\otimes\matt_2=\left(\begin{matrix} \ob\matt_2 & 0\\
0 & \om\matt_2
\end{matrix}\right)\!,\quad\sigx\otimes\matr_2= \left(\begin{matrix} 0 & \matr_2\\
\matr_2 & 0
\end{matrix}\right)\!,
\ee
representing $\gent$ and $\genr$, respectively. With $x=(x_1, x_2)^\T$ and $y=(y_1, y_2)^\T$ transforming as $\repd$ and $\repdd$, respectively, we obtain
\be\label{eq:quarttrace}
(x,y)_{\repqq}=\left(x_1 y_1, x_1 y_2, x_2 y_1, x_2 y_2\right)^\T.
\ee

In our model, we also utilize the expansion,
\begin{equation}\label{eq:tptimesd}
\reptp\otimes\repdd=\repddp\oplus\repqq.
\end{equation}
With $\reptp\equiv x=(x_1, x_2, x_3)^\T$ and $\repdd\equiv y=(y_1, y_2)^\T$, the above expansion is given by
\begin{align}
(x,y)_{\repddp}&\!=\!( -x_1 y_2 + i x_2 y_1 + i x_3 y_2, x_1 y_1- i x_2 y_2 + i x_3 y_1)^\T\!\!,\notag\\
(x,y)_{\repqq}&\!=\!(-i x_1 y_2\!-\!\om x_2 y_1\!-\!\ob x_3 y_2,\,i x_1 y_1\!+\!\om x_2 y_2\!-\!\ob x_3y_1, \notag\\
&\quad i x_1 y_2\!+\!\ob x_2 y_1\!+\!\om x_3 y_2, -i x_1 y_1\!-\!\ob x_2 y_2\!+\!\om x_3 y_1)^{\T}\!\!.\label{eq:4from3p2}
\end{align}

We give more details about $2O$, such as its presentation and character table, in Appendix~\ref{sec:appone}. We also define the generalized CP\cite{Ecker:1987qp, Neufeld:1987wa, 1211.5560,1211.6953,1303.6180,1402.0507} (gCP) for $2O$. We denote the generator of gCP by $\gencp$. In Appendix~\ref{sec:appone}, we obtain the action of $\gencp$ on the irreducible multiplets of $2O$ as
\be\label{eq:gcpdefs}
\begin{split}
&\reps\rightarrow\reps^*\!,\,\,\,\reps'\rightarrow{\reps'}^*\!,\,\,\,\repd\rightarrow\sigx\repd^*\!,\,\,\,\rept\rightarrow\rept^*\!,\,\,\,\reptp\rightarrow\reptp^*\!,\!\!\!\\
&\repdd\rightarrow i\sigy\repdd^*,\quad\repddp\rightarrow i\sigy{\repddp}^*,\quad\repqq\rightarrow(\sigx\otimes i\sigy)\repqq^*,
\end{split}
\ee
where $^*$ denotes complex conjugation, and $\sigx$ and $\sigy$ are the first and second Pauli matrices, respectively.
\vspace{-2mm}

\subsection{The model - with unbroken symmetries}
\label{sec:themodel}
We construct our model in the type-1 seesaw framework after incorporporating discrete flavour symmetries and gCP, which act on the Standard model leptons, a doublet of right-handed neutrinos and a set of flavons. We introduce seven flavon multiplets: $\chi\!=\!(\chi_1, \chi_2, \chi_3)^\T$\!, $\xia$,\, $\xib\!=\!({\xib}_1, {\xib}_2)^\T$\!, $\xic\!=\!({\xic}_1, {\xic}_2)^\T$\!, $\phia\!=\!({\phia}_1, {\phia}_2)^\T$\!, $\phib\!=\!({\phib}_1, {\phib}_2)^\T$ and $\phic\!=\!({\phic}_1, {\phic}_2)^\T$\!. Their components are complex numbers. In Section~\ref{sec:lagrangian}, we will show that the leading order flavon contribution in the construction of the Dirac mass term of the neutrinos is a $\repqq$ of $2O$ obtained from the tensor product $\xia\otimes\xib\otimes\xic$. We denote this multiplet by $\xi=(\xi_1, \xi_2, \xi_3,\xi_4)^\T$. Similarly, we will show that a $\rept$ of $2O$ obtained from $\phia\otimes\phib\otimes\phic$ and denoted by $\phi=(\phi_1, \phi_2, \phi_3)^\T$ is the leading order contribution in the Majorana mass term. We call the multiplets $\xi$ and $\phi$ effective flavons as opposed to the earlier-mentioned seven flavons, which we call elementary.

{\renewcommand{\arraystretch}{1.0}
	\setlength{\tabcolsep}{5pt}
	\begin{table}[]
	\begin{center}
	\begin{tabular}{|c|c c c c c c c c c|}
\hline
&&&&&&&&&\\[-1em]
	&$L$ & $\tau_R$	&$\mu_R$&$e_R$&$N$&$H$&$\chi$&$\xi$&$\phi$\\
\hline
&&&&&&&&&\\[-1em]
	$2O$ &$\reptp$ & $\reps$&$\reps$&$\reps$&$ \repd$&$\reps$&$ \reptp$&$\repqq$&$\rept$\\
$\genzl$ &$1$ & $\ob$ &$\om$&$1$&$1$&$1$&$\om$&$1$&$1$\\
$\genzn$ &$\taub$ & $\taub$ &$\taub$&$\taub$&$\tau$&$1$&$1$&$i$&$i$\\
				\hline
			\end{tabular}
		\end{center}
		\caption{The fermions, the elementary flavon $\chi$ and the effective flavons $\xi$ and $\phi$ transforming as the real or pseudo-real irreps of $2O$. $\genzl$ and $\genzn$ complexify these irreps such that we can impose gCP generated by $\gencp$. Under $\gencp$, the flavons transform as $\chi^*$\!, $\phi^*$ and $(\sigx\otimes i\sigy)\xi^*$. For the spinors, $\mathbf{CP}$ is defined as $\mathbf{CP}\psi_L=-i\sigy\psi_L^*$ and $\mathbf{CP}\psi_R=i\sigy\psi_R^*$. Since $N$ is a $\repdd$ of $2O$, we have $\gencp N = i\sigy \mathbf{CP} N= i\sigy i\sigy N^*$ where the first and the second $i\sigy$ are acting on the $2O$ and the spinor indices, respectively. $\gencp$ is equal to $\mathbf{CP}$ for $L$, $\tau_R$, $\mu_R$ and~$e_R$. For the Higgs also, $\gencp=\mathbf{CP}$, i.e.,~$\gencp H=H^*$. }
		\label{tab:flavourcontent}
\end{table}}

TABLE~\ref{tab:flavourcontent} shows how the leptons and the scalar fields transform under $2O$. We have the three families of left-handed weak isospin doublets of leptons forming a triplet under $2O$, denoted by $L=(L_e,L_\mu,L_\tau)^\T$. The right-handed charged-lepton fields $e_R,$ $\mu_R$ and $\tau_R$ are singlets. The right-handed neutrinos form a doublet under $2O$, denoted by $N=(N_1,N_2)^\T$. We introduce $\genzl$ and $\genzn$, which generate the Abelian groups $Z_3$ and $Z_8$, respectively, along with the generalized CP, under which the various fields transform. We call the flavour-gCP group generated by $\gent$, $\genr$, $\genzl$, $\genzn$ and $\gencp$, as defined by their actions on the fermions, TABLE~\ref{tab:flavourcontent}, $G_r$.

\subsubsection{The auxiliary generator} 
\label{sec:aux}
\vspace{-0.2cm}
Let us introduce an auxiliary generator $\genx$ under which the elementary flavons $\xia$, $\xib$, $\xic$, $\phia$, $\phib$ and $\phic$ transform while the fermions, the elementary flavon $\chi$ and the effective flavons $\xi$ and $\phi$ remain invariant. Its introduction is a group extension of $G_r$. To explain how $\xia$, $\xib$, $\xic$, $\phia$, $\phib$ and $\phic$ transform under $\genx$ and to achieve our group extension, we place the components of these flavons as well as the related effective flavons, $\xi$ and $\phi$, inside matrices as given below,
\begin{align}\notag
\llbracket \xia\rrbracket&=\!\frac{1}{2}\!\left(\begin{matrix} \xia & 0 & \xia^* & 0\\
0 & \xia & 0 & \xia^*
\end{matrix}\right)\!,&\llbracket \xib\rrbracket&=\!\frac{1}{\sqrt{2}}\!\left(\begin{matrix} {\xib}_1 & 0\\
0 & {\xib}_2\\
0 & {\xib}^*_2\\
{\xib}^*_1 & 0
\end{matrix}\right)\!,\\
\llbracket\xic\rrbracket&=\!\frac{1}{\sqrt{2}}\!\left(\begin{matrix} {\xic}_1 & {\xic}_2\\
{\xic}^*_2 & -{\xic}^*_1
\end{matrix}\right)\!,&\llbracket \xi\rrbracket&=\!\left(\begin{matrix} \xi_1 & \xi_2\\
\xi_3 & \xi_4
\end{matrix}\right)\!,\qquad\qquad\label{eq:direlem}
\end{align}
\vspace{-5mm}
\begin{align}\notag
\llbracket\phia\rrbracket&=\!\frac{1}{\sqrt{2}}\!\left(\begin{matrix} {\phia}_1 & {\phia}^*_2\\
{\phia}_2 & -{\phia}^*_1
\end{matrix}\right)\!\!,\!\!\!\!&\llbracket\phib\rrbracket\!&=\!\frac{1}{\sqrt{2}}\!\left(\begin{matrix} {\phib}_1 & {\phib}^*_2\\
{\phib}_2 & {\phib}^*_1
\end{matrix}\right)\!,\\
\llbracket\phic\rrbracket&=\!\frac{1}{\sqrt{2}}\!\left(\begin{matrix} {\phic}_1 & {\phic}_2\\
{\phic}^*_2 & -{\phic}^*_1
\end{matrix}\right)\!\!,\!\!\!\!&\llbracket\phi\rrbracket\!&=\!\frac{1}{\sqrt{2}}\!\left(\begin{matrix}  \phi_1-i\phi_3 & i\phi_2\\
i\phi_2 & \phi_1+i\phi_3
\end{matrix}\right)\!.\label{eq:majelem}
\end{align}
We call this kind of placement the {\it placeholder notation}\footnote{We use the factors $\frac{1}{2}$ and $\frac{1}{\sqrt{2}}$ in (\ref{eq:direlem}) and (\ref{eq:majelem}) for convenience so that the norms of the various multiplets are equal to the Frobenius norms of the corresponding matrices, i.e.,~$|\xia|^2=\text{tr}(\llbracket \xia\rrbracket^\dagger\llbracket \xia\rrbracket)$, $|\xib|^2=\text{tr}(\llbracket \xib\rrbracket^\dagger\llbracket \xib\rrbracket)$, etc.}. Next, we make the following assignment,
\be\label{eq:assignment}
\llbracket \xi\rrbracket=\llbracket\xia\rrbracket\llbracket\xib\rrbracket\llbracket\xic\rrbracket\,, \quad \llbracket\phi\rrbracket=(\llbracket\phia\rrbracket\llbracket\phib\rrbracket\llbracket \phic\rrbracket)_\text{sym}\,,
\ee
where $()_\text{sym}$ denotes the symmetric part, i.e.,~$(M)_\text{sym}=(M+M^\T)/2$. Given this assignment, we can obtain the components of $\xi$ and $\phi$ in terms of the components of the respective elementary flavons. 

We define the actions of the generators $\gent$ and $\genr$ on the elementary flavons $\xia$, $\xib$, $\xic$, $\phia$, $\phib$ and $\phic$ by defining their actions on the the products $\llbracket\xia\rrbracket\llbracket\xib\rrbracket\llbracket\xic\rrbracket$ and $\llbracket\phia\rrbracket\llbracket\phib\rrbracket\llbracket\phic\rrbracket$ in the following way,
\begin{align}
\gent:\,\, \begin{split}\label{eq:taction}
&\matw_2\llbracket\xia\rrbracket(\mati_2\otimes\matw_2)^\dagger\,\,\,(\mati_2\otimes\matw_2)\llbracket\xib\rrbracket \,\,\,\llbracket\xic\rrbracket\matt_2^\T\\
&\matt_2\llbracket\phia\rrbracket\,\,\,\llbracket\phib\rrbracket\,\,\,\llbracket\phic\rrbracket\matt_2^\T
\end{split}\,,\\
\genr:\,\, \begin{split}\label{eq:raction}
&\sigx\llbracket\xia\rrbracket(\sigx\otimes\sigx)^\dagger\,\,\,(\sigx\otimes\sigx)\llbracket\xib\rrbracket \,\,\,\llbracket\xic\rrbracket\matr_2^\T\\
&\matr_2\llbracket\phia\rrbracket\,\,\,\llbracket\phib\rrbracket \,\,\,\llbracket\phic\rrbracket\matr_2^\T
\end{split}\,,
\end{align}
where $\mati_2$ is the $2\times 2$ identity matrix. As an example, let us evaluate the action of $\genr$ on $\xia$. The first line of (\ref{eq:raction}) shows that $\llbracket\xia\rrbracket\rightarrow\sigx\llbracket\xia\rrbracket(\sigx\otimes\sigx)^\dagger$ under $\genr$. This action keeps the positions of the zeros in $\llbracket\xia\rrbracket$ (\ref{eq:direlem}) unchanged while exchanging $\xia$ and $\xia^*$. Therefore, (\ref{eq:raction}) keeps the form of the matrix $\llbracket\xia\rrbracket$ unchanged and results in $\xia\rightarrow\xia^*$, i.e.,~the action of $\genr$ on $\xia$ is complex conjugation. We can verify that the placeholder notations of the elementary flavons $\llbracket\xia\rrbracket$, $\llbracket\xib\rrbracket$, etc., remain form-invariant under the group actions (\ref{eq:taction}) and (\ref{eq:raction}). We evaluate the corresponding actions on $\xia$, $\xib$ etc., and list them in TABLE~\ref{tab:neutrinosector}.

We can also see that the matrices acting between the flavons, e.g.,~$(\mati_2\otimes\matw_2)^\dagger$ and $(\mati_2\otimes\matw_2)$ between $\llbracket\phia\rrbracket$ and $\llbracket\phib\rrbracket$ in the first line of (\ref{eq:taction}), are inverses of each other. Since these {\it internal} multiplications cancel each other, the overall group actions on the products $\llbracket\xia\rrbracket\llbracket\xib\rrbracket\llbracket\xic\rrbracket$ and $\llbracket\phia\rrbracket\llbracket\phib\rrbracket\llbracket\phic\rrbracket$ are {\it external}, i.e.,~multiplying on the left and the right sides of these products. For $\llbracket\xia\rrbracket\llbracket\xib\rrbracket\llbracket\xic\rrbracket$, the left and the right multiplications are by $\matw_2$ and $\matt_2^\T$, respectively, in (\ref{eq:taction}), and by $\sigx$ and $\matr_2^\T$, respectively, in (\ref{eq:raction}). These are the representation matrices corresponding to $\repd$ (\ref{eq:ddgen2}) and $\repdd$ (\ref{eq:bogen}), respectively. For $\llbracket\phia\rrbracket\llbracket\phib\rrbracket\llbracket\phic\rrbracket$, on the other hand, the actions on both sides correspond to $\repdd$ (\ref{eq:bogen}). Given the tensor products (\ref{eq:dtimesdd}) and (\ref{eq:dtimesd}), we can conclude that $\llbracket\xia\rrbracket\llbracket\xib\rrbracket\llbracket\xic\rrbracket$ and the symmetric part of $\llbracket\phia\rrbracket\llbracket\phib\rrbracket\llbracket\phic\rrbracket$ transform as $\repqq$ and $\rept$ of $2O$, respectively, i.e.,~$\llbracket\xia\rrbracket\llbracket\xib\rrbracket\llbracket\xic\rrbracket\equiv(\xia,\xib,\xic)_{\repqq}$ and $\llbracket\phia\rrbracket\llbracket\phib\rrbracket\llbracket\phic\rrbracket\equiv(\phia,\phib,\phic)_{\rept}$. Therefore, our assignement (\ref{eq:assignment}) is in agreement with the effective flavons $\xi$ and $\phi$ transforming as $\repq$ and $\rept$ of $2O$, TABLE~\ref{tab:flavourcontent}.

{\renewcommand{\arraystretch}{1.0}
	\setlength{\tabcolsep}{5pt}
	\begin{table}[tbp]
	\begin{center}
	\begin{tabular}{|c|c c c c c |c|}
\hline
&&&&&&\\[-1em]
	&$\gent$&$\genr$&$\genzn$&$\gencp$&$\genx$&$d$\\
\hline
&&&&&&\\[-1em]
$\xia$ &$1$&$()^*$&$()^*$&$()^*$&$\om()^*$&$2$\\
$\xib$ &$\matw_2$&$()^*$&$\sigx()^*$&$i\sigy ()^*$& $\ob i\sigy ()^*$&$4$\\
$\xic$ &$\matt_2$&$\matr_2$&$-\sigy()^*$&$-1$&$i\sigy ()^*$&$4$\\
$\phia$ &$\matt_2$&$\matr_2$&$1$&$i\sigy ()^*$&$\tau\om i\sigy ()^*$&$4$\\
$\phib$ &$1$&$1$&$\sigx()^*$&$()^*$&$\tau\om\mattau_2\matw_2\sigz()^*$&$4$\\
$\phic$&$\matt_2$&$\matr_2$&$-\sigy()^*$&$i\sigy()^*$&$\taub\ob i\sigy()^*$&$4$\\
				\hline
			\end{tabular}
		\end{center}
		\caption{The group actions on the elementary flavons in the neutrino sector. Even though we express these flavons as complex multiplets, they transform as real irreps because the group actions on them involve complex conjugation. See Appendix~\ref{sec:appone} for a detailed discussion. The last column lists the dimensions of the flavons as real irreps. $\sigx$, $\sigy$ and $\sigz$ are the Pauli matrices.\vspace{-1.5mm}}
		\label{tab:neutrinosector}
\end{table}} 

We define the actions of $\genzn$ and $\gencp$ on the elementary flavons in the following way,
\begin{align}
\genzn:\begin{split}\label{eq:zaction}
&\llbracket\xia\rrbracket(\sigx\otimes\mati_2)^\dagger\,\,\,(\sigx\otimes\mati_2)\llbracket \xib\rrbracket\sigx^\dagger \,\,\,\sigx\llbracket \xic\rrbracket i\\
&\llbracket\phia\rrbracket\,\,\,\llbracket \phib\rrbracket\sigx^\dagger \,\,\,\sigx\llbracket \phic\rrbracket i
\end{split}\,,\\
\gencp:\begin{split}\label{eq:caction}
&\sigx\llbracket\xia\rrbracket^*(\mati_2\!\otimes\!\sigx)^{\!\dagger}\,(\mati_2\!\otimes\!\sigx)\llbracket \xib\rrbracket^*(i\sigy)^{\!\dagger}\,(i\sigy)\llbracket \xic\rrbracket^* (i\sigy)^{\!\T}\!\!\\
&i\sigy\llbracket\phia\rrbracket^*\,\,\llbracket \phib\rrbracket^*\,\,\llbracket \phic\rrbracket^*(i\sigy)^{\!\T}\!\!\!
\end{split}.
\end{align}
(\ref{eq:zaction}) and (\ref{eq:caction}) lead to the transformations $\llbracket\xia\rrbracket\llbracket\xib\rrbracket\llbracket\xic\rrbracket\rightarrow\llbracket\xia\rrbracket\llbracket\xib\rrbracket\llbracket\xic\rrbracket i$ and $\llbracket\phia\rrbracket\llbracket\phib\rrbracket\llbracket\phic\rrbracket\rightarrow\llbracket\phia\rrbracket\llbracket\phib\rrbracket\llbracket\phic\rrbracket i$ under $\genzn$ and $\llbracket\xia\rrbracket\llbracket\xib\rrbracket\llbracket\xic\rrbracket\rightarrow\sigx(\llbracket\xia\rrbracket\llbracket\xib\rrbracket\llbracket\xic\rrbracket)^*(i\sigy)^\T$ and $\llbracket\phia\rrbracket\llbracket\phib\rrbracket\llbracket\phic\rrbracket\rightarrow i\sigy(\llbracket\phia\rrbracket\llbracket\phib\rrbracket\llbracket\phic\rrbracket)^*(i\sigy)^{\!\T}$ under $\gencp$. These are in agreement with the transformation rules for the effective flavons given in TABLE~\ref{tab:flavourcontent}, i.e.,~$\xi\rightarrow i \xi$ and $\phi\rightarrow i \phi$ under $\genzn$ and $\xi\rightarrow (\sigx\otimes i\sigy)\xi^*$ and $\phi\rightarrow \phi^*$ under $\gencp$.

Besides the generators (\ref{eq:taction})-(\ref{eq:caction}), we introduce the auxiliary generator $\genx$, whose action is defined by
\begin{align}\label{eq:xaction}
\genx\!:\!\begin{split}
&\llbracket\xia\rrbracket(\matw_2\sigx\otimes\mati_2)^{\dagger}\,\,(\matw_2\sigx\otimes\mati_2)\llbracket \xib\rrbracket(i\sigy)^{\dagger}\,\,(i\sigy)\llbracket \xic\rrbracket\!\!\!\\
&\llbracket\phia\rrbracket(\mattau_2\matw_2i\sigy)^{\!\dagger}(\mattau_2\matw_2i\sigy)\llbracket \phib\rrbracket(\mattau_2\matw_2i\sigy)^{\!\dagger}(\mattau_2\matw_2i\sigy)\llbracket \phic\rrbracket\!\!\!\!
\end{split}\,\,\,\,\,\,,\\
& \!\!\!\!\!\!\!\!\text{where} \qquad\qquad\qquad\,\,\, \mattau_2=\text{diag}(\taub, \tau).
\end{align}
The action of $\genx$ on the products $\llbracket\xia\rrbracket\llbracket\xib\rrbracket\llbracket\xic\rrbracket$ and $\llbracket\phia\rrbracket\llbracket\phib\rrbracket\llbracket\phic\rrbracket$ is only internal, which cancels. Therefore, $\genx$ acts trivially on the products and thus act trivially on $\xi$ and $\phi$. We can verify that the group actions as defined in (\ref{eq:zaction}), (\ref{eq:caction}) and (\ref{eq:xaction}) keep the placeholder notations of the elementary flavons $\llbracket\xia\rrbracket$, $\llbracket\xib\rrbracket$, etc., form-invariant. We evaluate the respective group actions on $\xia$, $\xib$, etc., and list them in TABLE~\ref{tab:neutrinosector}. 

We call the flavour-gCP group generated by $\gent$, $\genr$, $\genzl$, $\genzn$, $\gencp$ and $\genx$, as defined by their actions on all the fermions and the flavons, TABLE~\ref{tab:flavourcontent}, \ref{tab:neutrinosector}, $G_{\!f}$. We verify that every multiplet in TABLE~\ref{tab:neutrinosector} is irreducible by studying the corresponding quotient group, i.e.,~the group generated by $\gent$, $\genr$, $\genzn$, $\gencp$ and $\genx$, as defined by their actions on the given multiplet. We perform this verification using the computational discrete algebra system GAP\cite{GAP4} without providing the details in the paper. The elements of $G_{\!f}$ that act trivially on the fermions constitute a normal subgroup of $G_{\!f}$. We call it the auxiliary group $G_x$. The group $G_r$, which we defined at the beginning of Section~\ref{sec:themodel}, is the quotient of $G_{\!f}$ by $G_x$, i.e.,~using our technique that invoves the placeholder notation, we extended $G_r$ by $G_x$ to obtain $G_{\!f}$\footnote{We have $G_{\!f}=G_x\cdot G_r$, which denotes the extension of $G_r$ by $G_x$. The group extension can be split or non-split. Split extension is a special case that corresponds to the semidirect product $G_x\rtimes G_r$. The direct product $G_x\times G_r$ is a special case of $G_x\rtimes G_r$.}. In this paper, we study the action of the various generators on the irreps individually and then analyze the cross terms among the irreps. This approach does not require the characterization of the groups $G_r$, $G_{\!f}$ and $G_x$ as a whole.


\subsubsection{Construction of the Lagrangian} 
\label{sec:lagrangian}
\vspace{-0.3cm}
In this section, we determine the leading-order flavon contributions to the various mass terms in the Lagrangian. The fermion fields $\tau_R$ and $\mu_R$ transform as $\ob$ and $\om$, respectively, under $\genzl$. The only elementary flavon that transforms nontrivially under $\genzl$ is $\chi$. Therefore, $\chi$ and $\chi^*$ are the only leading order multiplets that can couple with $\tau_R$ and $\mu_R$, respectively. The corresponding mass terms are $L^\dagger \chi \tau_R H + h.c.$ and $L^\dagger \chi^* \mu_R H + h.c.$. Under $\genzl$, $e_R$ is invariant, hence we need a multiplet that is invariant under $\genzl$ and transforms as $\reptp$ to couple with $L^\dagger$ and $e_R$. Since such an elementary flavon does not exist in our model, we search for second-order flavon contributions. To aid us in finding the suitable pairs of flavons, we define the following group elements: $\gene_{\xia}=(\genr\genx)^8$, $\gene_{\xib}=\genx^2(\genzn\gencp)^2$, $\gene_{\xic}\!=\!\gencp^2\genx^2$, $\gene_{\phia}=\genzn^2(\genzn\gencp)^2$, $\gene_{\phib}\!=\!\genzn^2\gencp^2\genx^2(\genzn\genx)^{12}$ and $\gene_{\phic}=\genzn^2\gencp^2\genx^2$. How they act on the various multiplets in our model is shown in TABLE~\ref{tab:eelems}. 

{\renewcommand{\arraystretch}{0.85}
	\setlength{\tabcolsep}{5pt}
	\begin{table}[tbp]
	\begin{center}
	\begin{tabular}{|c|c c c c c c c|}
\hline
&&&&&&&\\[-.8em]
	&$\genzl$&$\gene_{\xia}$&$\gene_{\xib}$&$\gene_{\xic}$&$\gene_{\phia}$&$\gene_{\phib}$&$\gene_{\phic}$\\
&&&&&&&\\[-.9em]
\hline
&&&&&&&\\[-.8em]
$L, e_R$&$1$&$1$&$1$&$1$&$-i$&$i$&$-i$\\
$\mu_R$&$\om$&$1$&$1$&$1$&$-i$&$i$&$-i$\\
$\tau_R$&$\ob$&$1$&$1$&$1$&$-i$&$i$&$-i$\\
$N$&$1$&$1$&$-1$&$-1$&$-i$&$i$&$-i$\\
\hline
&&&&&&&\\[-.8em]
$\chi$&$\om$&$1$&$1$&$1$&$1$&$1$&$1$\\
$\xia$&$1$&$\om$&$1$&$1$&$1$&$1$&$1$\\
$\xib$&$1$&$\ob$&$-1$&$1$&$1$&$1$&$1$\\
$\xic$&$1$&$1$&$1$&$-1$&$1$&$1$&$1$\\
$\phia$&$1$&$1$&$1$&$1$&$-1$&$1$&$1$\\
$\phib$&$1$&$1$&$1$&$1$&$1$&$-1$&$1$\\
$\phic$&$1$&$1$&$1$&$1$&$1$&$1$&$-1$\\
\hline
			\end{tabular}
		\end{center}
		\caption{The action of $\genzl$, $\gene_{\xia}$, ..., $\gene_{\phic}$.\vspace{-1.5mm}}
		\label{tab:eelems}
\end{table}} 


$L^\dagger e_R$ is invariant under $\genzl$ as well as under $\gene_{\xia},..,\gene_{\phic}$. As a consequence and by examining TABLE~\ref{tab:eelems}, we can rule out every pair made with two different flavons, i.e.,~the only suitable pairs are the ones that involve the same flavon. For further analysis, we evaluate the tensor product of each flavon with itself,
\begin{align}
\chi\otimes\chi&=|\chi|^2\oplus\chi^{2\alpha}\!\oplus\!\chi^{2\beta}\!\oplus\!\chi^{2\gamma}\!\oplus\!\chi^{2\delta}\!\oplus\!\chi^{2\epsilon}\!\oplus\!\chi^{2\zeta}\!,\!\label{eq:prodchi}\\
\xia\otimes\xia&=|\xia|^2\oplus{\xia}^{\!\!\!2},\\
\xib\otimes\xib&=|\xib|^2\oplus{\xib}^{\!\!\!2\alpha}\oplus{\xib}^{\!\!\!2\beta}\oplus{\xib}^{\!\!\!2\gamma}\oplus{\xib}^{\!\!\!2\delta},\\
\xic\otimes\xic&=|\xic|^2\oplus{\xic}^{\!\!2\alpha}\oplus{\xic}^{\!\!2\beta}\oplus{\xic}^{\!\!2\gamma},\\
\phia\otimes\phia&=|\phia|^2\oplus{\phia}^{\!\!\!2\alpha}\oplus{\phia}^{\!\!\!2\beta},\\
\phib\otimes\phib\!&=|\phib|^2\oplus{\phib}^{\!\!\!2\alpha}\oplus{\phib}^{\!\!\!2\beta}\oplus{\phib}^{\!\!\!2\gamma}\!\oplus{\phib}^{\!\!\!2\delta},\\
\phic\!\otimes\phic\!&=|\phic|^2\oplus{\phic}^{\!\!\!2\alpha}\oplus{\phic}^{\!\!\!2\beta},\label{eq:prodphic}
\end{align}
where the expressions of the multiplets in the RHS are given in TABLE~\ref{tab:prodexp}. Throughout this paper, when a complex flavon multiplet appears in the LHS of a tensor product, it denotes the corresponding real irrep of twice the dimension, e.g.,~in (\ref{eq:prodchi}), $\chi\otimes\chi$ denotes $(\chi,\chi^*)^\T\otimes(\chi,\chi^*)^\T$. This is a consequence of using complex conjugation to represent group transformations. See Appendix~\ref{sec:appone} for a detailed discussion. In (\ref{eq:prodchi}), $\chi\otimes\chi$ produces  $6\times6=36$ components, of which $21$ form the symmetric part. In the RHS of (\ref{eq:prodchi})--(\ref{eq:prodphic}), we have omitted the antisymmetric parts since they vanish. How the multiplets in the RHS transform is given in TABLE~\ref{tab:prodtrof}. By examining it, we can see that the only multiplet that can couple with $L^\dagger e_R$ is $\chi^{2\beta}$.  Thus, we obtain the leading order mass term that involves $e_R$ as $L^\dagger \chi^{2\beta} e_R H + h.c.$. 

To construct the Dirac mass term, we consider the tensor product expansion $L^\dagger\!\otimes\!N=(L^\dagger\!, N)_{\repddp}\oplus(L^\dagger\!, N)_{\repqq}$, vis-a-vis (\ref{eq:tptimesd}). The expressions of $(L^\dagger\!, N)_{\repddp}$ and $(L^\dagger\!, N)_{\repqq}$ can be evaluated using (\ref{eq:4from3p2}) with the substitution $x=L^\dagger$ and $y=N$. The product $L^\dagger\otimes N$ changes sign under $\gene_{\xib}$ and $\gene_{\xic}$ and remains invariant under the other group elements listed in TABLE~\ref{tab:eelems}. No elementary flavon has this transformation propertly. At the second order, $\xib\otimes\xic$ is the only combination that transforms in this way. However, under $\gene_{\xia}$, $L^\dagger \otimes N$ and $\xic$ remain invariant while $\xia$ and $\xib$ transform nontrivially. Therefore, the only way to obtain a multiplet that can couple with $L^\dagger\otimes N$ is to consider the third order and utilize the product $\xia\otimes\xib\otimes\xic$. We analyze the tensor product expansion of $\xia\otimes\xib\otimes\xic$ in Appendix~\ref{sec:apptwo}. None of the resulting multiplets can couple with $(L^\dagger\!, N)_{\repddp}$. On the other hand, we find that only one multiplet exists that can couple with $(L^\dagger\!, N)_{\repqq}$, which is nothing but the effective flavon $\xi$ defined in (\ref{eq:assignment}). Thus, we obtain the invariant term in the neutrino Dirac sector, $(L^\dagger\!, N)_{\repqq}^\T \xi^*i\sigy H^*+h.c.$. 

{\renewcommand{\arraystretch}{1.2}
	\setlength{\tabcolsep}{5pt}
\begin{table}[tbp]
\begin{center}
	\begin{tabular}{|c|c|}
\hline
&\\[-1em]
	&Expressions of the multiplets\\
\hline
\!\!$\chi^{2\alpha}$\!\!&$\chi^*_1\chi_1\!+\!\om\chi^*_2\chi_2\!+\!\ob\chi^*_3\chi_3$\\
\!\!$\chi^{2\beta}$\!\!&$(\chi_2^*\chi_3+\chi_3^*\chi_2,\,  \chi_3^*\chi_1+\chi_1^*\chi_3,\, \chi_1^*\chi_2+\chi_2^*\chi_1)^\T\!\!$\\
\!\!$\chi^{2\gamma}$\!\!&$i(\chi_2^*\chi_3-\chi_3^*\chi_2,\,  \chi_3^*\chi_1-\chi_1^*\chi_3,\, \chi_1^*\chi_2-\chi_2^*\chi_1)^\T\!\!$\\
\!\!$\chi^{2\delta}$\!\!&$\chi_1^2+\chi^2_2+\chi^2_3$\\
\!\!$\chi^{2\epsilon}$&$(\chi^2_1+\om\chi^2_2+\ob\chi^2_3,\,\chi^2_1+\ob\chi^2_2+\om\chi^2_3)^\T\!\!$\\
\!\!$\chi^{2\zeta}$\!\!&$(\chi_2\chi_3,\, \chi_3\chi_1,\, \chi_1\chi_2)^\T\!\!\!$\\
\hline
\!\!${\xib}^{\!\!\!2\alpha}$\!\!&${\xib^*}_1{\xib}_1-{\xib^*}_2{\xib}_2,$\\
\!\!${\xib}^{\!\!\!2\beta}$\!\!&${\xib^*}_1{\xib}_2,$\\
\!\!${\xib}^{\!\!\!2\gamma}$\!\!&${\xib}_1{\xib}_2,$\\
\!\!${\xib}^{\!\!\!2\delta}$\!\!&$({\xib^2}_1,\, {\xib^2}_2)^\T\!\!,$\\
\hline
\!\!${\xic}^{\!\!2\alpha}$\!\!&$(i({\xic^*}_1{\xic}_2-{\xic^*}_2{\xic}_1),\,{\xic^*}_1{\xic}_1\!-\!{\xic^*}_2{\xic}_2,{\xic^*}_1{\xic}_2+{\xic^*}_2{\xic}_1)^\T\!\!$\\
\!\!${\xic}^{\!\!2\beta}$\!\!&$\text{re}({\xic^2}_1+{\xic^2}_2, -2i{\xic}_1{\xic}_2, i({\xic^2}_1-{\xic^2}_2))^\T\!\!$\\
\!\!${\xic}^{\!\!2\gamma}$\!\!&$\text{im}({\xic^2}_1+{\xic^2}_2, -2i{\xic}_1{\xic}_2, i({\xic^2}_1-{\xic^2}_2))^\T\!\!$\\
\hline
\!\!${\phia}^{\!\!\!2\alpha}$\!\!&$\!\!\!(i({\phia^*}_1{\phia}_2\!-\!{\phia^*}_2{\phia}_1),{\phia^*}_1{\phia}_1\!-\!{\phia^*}_2{\phia}_2,{\phia^*}_1{\phia}_2\!+\!{\phia^*}_2{\phia}_1)^{\!\T}\!\!$\!\!\!\\
\!\!${\phia}^{\!\!\!2\beta}$\!\!&$({\phia^2}_1+{\phia^2}_2, -2i{\phia}_1{\phia}_2, i({\phia^2}_1-{\phia^2}_2))^{\!\T}$\\
\hline
\!\!${\phib}^{\!\!\!2\alpha}$\!\!&${\phib^*}_1{\phib}_1-{\phib^*}_2{\phib}_2$\\
\!\!${\phib}^{\!\!\!2\beta}$\!\!&${\phib^*}_1{\phib}_2$\\
\!\!${\phib}^{\!\!\!2\gamma}$\!\!&${\phib}_1{\phib}_2$\\
\!\!${\phib}^{\!\!\!2\delta}$\!\!&$({\phib^2}_1,\, {\phib^2}_2)^\T\!\!$\\
\hline
\!\!${\phic}^{\!\!\!2\alpha}$\!\!&$\!\!\!(i({\phic^*}_1{\phic}_2\!-\!{\phic^*}_2{\phic}_1),\,{\phic^*}_1{\phic}_1\!-\!{\phic^*}_2{\phic}_2,{\phic^*}_1{\phic}_2\!+\!{\phic^*}_2{\phic}_1)^\T\!\!$\\
\!\!${\phic}^{\!\!\!2\beta}$\!\!&$({\phic^2}_1+{\phic^2}_2, -2i{\phic}_1{\phic}_2, i({\phic^2}_1-{\phic^2}_2))^\T\!\!$\!\!\!\\
\hline
\end{tabular}
\end{center}
\caption{The expressions of the multiplets in the RHS of (\ref{eq:prodchi})--(\ref{eq:prodphic}). Invariants such as $|\chi|^2=\chi^\dagger\chi$ are not listed here.}
\label{tab:prodexp}
\begin{center}
\begin{tabular}{|c|c c c c c c| c |}
\hline
&&&&&&&\\[-1em]
	&$\gent$&$\genr$&$\genzl$&$\genzn$&$\gencp$&$\genx$&$d$\\
\hline
$\chi^{2\alpha}$&$\ob$&$()^*$&$1$&$1$&$1$&$1$&$2$\\
$\chi^{2\beta}$&$\matt_3$&$-\matr_3$&$1$&$1$&$1$&$1$&$3$\\
$\chi^{2\gamma}$&$\matt_3$&$\matr_3$&$1$&$1$&$-1$&$1$&$3$\\
$\chi^{2\delta}$&$1$&$1$&$\ob$&$1$&$()^*$&$1$&$2$\\
$\chi^{2\epsilon}$&$\matw_2$&$\sigx$&$\ob$&$1$&$\sigx()^*$&$1$&$4$\\
$\chi^{2\zeta}$&$\matt_3$&$-\matr_3$&$\ob$&$1$&$()^*$&$1$&$6$\\
\hline
${\xia}^{\!\!\!2}$&$1$&$()^*$&$1$&$()^*$&$()^*$&$\ob()^*$&$2$\\
\hline
${\xib}^{\!\!\!\!2\alpha}$&$1$&$1$&$1$&$-1$&$-1$&$-1$&$1$\\
${\xib}^{\!\!\!\!2\beta}$&$\ob$&$()^*$&$1$&$1$&$-1$&$-1$&$2$\\
${\xib}^{\!\!\!\!2\gamma}$&$1$&$()^*$&$1$&$()^*$&$-()^*$&$-\om()^*$&$2$\\
${\xib}^{\!\!\!\!2\delta}$&$\matw_2^*$&$()^*$&$1$&$\sigx()^*$&$\sigx()^*$&$\om\sigx()^*$&$4$\\
\hline
${\xic}^{\!\!\!2\alpha}$&$\matt_3$&$\matr_3$&$1$&$-1$&$1$&$-1$&$3$\\
${\xic}^{\!\!\!2\beta}$&$\matt_3$&$\matr_3$&$1$&$-1$&$1$&$1$&$3$\\
${\xic}^{\!\!\!2\gamma}$&$\matt_3$&$\matr_3$&$1$&$1$&$1$&$-1$&$3$\\
\hline
${\phia}^{\!\!\!2\alpha}$&$\matt_3$&$\matr_3$&$1$&$1$&$-1$&$-1$&$3$\\
${\phia}^{\!\!\!2\beta}$&$\matt_3$&$\matr_3$&$1$&$1$&$()^*$&$i\ob()^*$&$6$\\
\hline
${\phib}^{\!\!\!\!2\alpha}$&$1$&$1$&$1$&$-1$&$1$&$1$&$1$\\
${\phib}^{\!\!\!\!2\beta}$&$1$&$1$&$1$&$1$&$()^*$&$-i\ob()^*$&$2$\\
${\phib}^{\!\!\!\!2\gamma}$&$1$&$1$&$1$&$()^*$&$()^*$&$-i\ob()^*$&$2$\\
${\phib}^{\!\!\!\!2\delta}$&$1$&$1$&$1$&$\sigx()^*$&$()^*$&$\ob\matw^*_2\sigz()^*$&$4$\\
\hline
${\phic}^{\!\!\!2\alpha}$&$\matt_3$&$\matr_3$&$1$&$-1$&$-1$&$-1$&$3$\\
${\phic}^{\!\!\!2\beta}$&$\matt_3$&$\matr_3$&$1$&$-()^*$&$()^*$&$-i\om()^*$&$6$\\
\hline
\end{tabular}
\end{center}
\caption{The transformation rules for the multiplets in the RHS of (\ref{eq:prodchi})--(\ref{eq:prodphic}). The last column lists the dimensions of these multiplets as real irreps. The invariants are not listed.}
\label{tab:prodtrof}
\end{table}}

The Majorana mass term consists of the symmetric part of the product $i\sigy N\!\otimes\!N$, where $i\sigy$ is acting on the spinor index. According to (\ref{eq:dtimesd}), the symmetric part transforms as $\rept$ of $2O$, and we denote it by $(i\sigy N, N)_{\rept}$. We can evaluate $(i\sigy N, N)_{\rept}$ using (\ref{eq:triptrace}) with the substitution $x=i\sigy N$ and $y=N$. The product $i\sigy N\!\otimes\!N$ changes sign under $\gene_{\phia}$,  $\gene_{\phib}$ and  $\gene_{\phic}$, while remaining invariant under the other group elements in TABLE~\ref{tab:eelems}. We can show that $\phia\otimes\phib\otimes\phic$ is the only leading-order combination of flavons that transforms in this way. We analyze this tensor product in Appendix~\ref{sec:apptwo}. Among the multiplets obtained, the only one that can couple with $(i\sigy N, N)_{\rept}$ is the effective flavon $\phi$ defined in (\ref{eq:assignment}). This results in the invariant term in the Majorana sector, $(i\sigy N, N)_{\rept}^\T \phi^*+h.c.$.

We summarize by writing down the Lagrangian,
\begin{align}\label{eq:lagr}
\begin{split}
-{\mathcal L}=&y_\tau L^\dagger \frac{\chi}{\Lambda} \tau_R H+{y_\mu L^\dagger \frac{\chi^*}{\Lambda} \mu_R H + y_e L^\dagger \frac{\chi^{2\beta}}{\Lambda^2} e_R H }+\\
&y_\xi (L^\dagger\!, N)_{\repqq}^\T \frac{\xi^*}{\Lambda^3}i\sigy H^*+ \frac{y_\phi}{2}(i\sigy N, N)_{\rept}^\T \,\frac{\phi^*}{\Lambda^2}\,+ h.c.,
\end{split}
\end{align}
where $\Lambda$ is the cut-off scale. Invariance under gCP ensures that the Yukawa couplings $y_\tau$, $y_\mu$, $y_e$, $y_\xi$ and $y_\phi$ are real. We  assume that all these couplings are of the order of one. The charged-lepton mass term, the neutrino Dirac mass term and the Majorana mass term in (\ref{eq:lagr}) can be written in matrix form as
\begin{align}
&\left(\begin{matrix} L_e\\
L_\mu\\
L_\tau
\end{matrix}\right)^{\!\!\!\dagger}\!\!\left(\begin{matrix} \frac{y_e}{\Lambda}(\chi^*_2\chi_3+\chi^*_3\chi_2) & y_\mu\chi^*_1 & y_\tau\chi_1\\
\frac{y_e}{\Lambda} (\chi^*_3\chi_1+\chi^*_1\chi_3)& y_\mu\chi^*_2 & y_\tau\chi_2\\
\frac{y_e}{\Lambda} (\chi^*_1\chi_2+\chi^*_2\chi_1)& y_\mu\chi^*_3 & y_\tau\chi_3
\end{matrix}\right)\!\frac{H}{\Lambda}\!\!\left(\begin{matrix} e_R\\
\mu_R\\
\tau_R
\end{matrix}\right)\!,\label{eq:lagrcl}\\
&\left(\begin{matrix} L_e\\
L_\mu\\
L_\tau
\end{matrix}\right)^{\!\!\!\dagger}\!\!\left(\begin{matrix}  -i \xi^*_4+i\xi^*_2 & i\xi^*_3-i\xi^*_1\\
 \ob\xi^*_3-\om\xi^*_1 & -\ob\xi^*_4+\om\xi^*_2\\
 \om\xi^*_4-\ob\xi^*_2 & \om\xi^*_3-\ob\xi^*_1
\end{matrix}\right)\!\!\frac{y_\xi i\sigy H^*}{\Lambda^3}\!\!\left(\begin{matrix} N_1\\
N_2
\end{matrix}\right)\!,\label{eq:lagrdr}\\
&\text{and}\,\,\,\left(\begin{matrix} i\sigy N_1\\
i\sigy N_2
\end{matrix}\right)^{\!\!\T}\!\left(\begin{matrix}  \phi^*_1+i\phi^*_3 & -i\phi^*_2\\
-i\phi^*_2 & \phi^*_1-i\phi^*_3
\end{matrix}\right)\frac{y_\phi}{2\Lambda^2}\left(\begin{matrix} N_1\\
N_2
\end{matrix}\right)\!,\label{eq:lagrmj}
\end{align}
respectively. For the explicit expressions of $\xi$ and $\phi$ in terms of the components of the elementary flavons, please see (\ref{eq:xiexpr2}) and (\ref{eq:phiexpr1}), respectively, in Appendix~\ref{sec:apptwo}.

\subsection{Spontaneous symmetry breaking}
\label{sec:ssb} 

The Higgs field acquires the vev $\langle H\rangle =(0, v)^\T$, leading to the spontaneous breaking of electroweak symmetry. The flavour and gCP symmetries are broken at a higher energy scale when the elementary flavons acquire their vevs,
\begin{align}\label{eq:elemvevs}
\langle\chi\rangle&=\frac{r_{\!\chi}}{\sqrt{3}}(-1, \om, \ob)^\T\!\!,\notag\\
\langle\xia\rangle\!&=\!r_{\!\xia}\ob,\qquad\quad\, \langle\xib\rangle=\frac{r_{\!\xib}}{\sqrt{2}}(\om, \ob)^\T\!\!,\,\,\,\,\, \langle\xic\rangle\!=\!\frac{r_{\!\xic}}{\sqrt{2}}\tau(\eib, \eig)^\T\!\!,\,\notag\\
\langle\phia\rangle\!&=\!\frac{r_{\!\phia}}{\sqrt{2}}(1, \tau)^\T\!\!,\quad \langle\phib\rangle\!=\!\frac{r_{\!\phib}}{\sqrt{2}}(\om, 1)^\T\!\!,\quad \langle\phic\rangle\!=\!\frac{r_{\!\phic}}{\sqrt{2}}(\taub, 1)^\T\!\!\!,
\end{align}
where $\eig=e^{\frac{\pi}{8}}$ and $\eib=e^{-\frac{\pi}{8}}$, and $r_{\!\chi}$, $r_{\!\xia}$, etc., are the norms of the vevs.  We will obtain these vevs as the minimum of the general renormalizable flavon potential by following the procedure developed by the author in a companion paper\cite{2306.07325}. First, we consider the flavons individually. We use arguments based on symmetries to show that the potential of each flavon is guaranteed to have a stationary point that corresponds to its vev (\ref{eq:elemvevs}). In the second step, we show that the renormalizable cross terms among the flavons do not spoil these stationary points. The final step is to introduce driving flavons\footnote{If a flavon does not appear in the fermion mass terms at the leading order, we call it a driving flavon. This definition differs from that of the driving fields with $R$-charge equal to two proposed in supersymmetric flavour models, e.g.,~\cite{hep-ph/0504165}.} to break accidental continuous symmetries that may be present in the renormalizable potentials of the individual flavons.

Ref.~\cite{2306.07325} used theorems by Michel\cite{MICHEL1971,MICHEL200111}, Golubitsky and Stewart\cite{GOLUBITSKY1988} to prove the existence of specific stationary points of the potential of an irrep without the need to explicitly extremize the potential. An alignment of an irrep is guaranteed to be a stationary point of the potential of the irrep if its stabilizer (residual symmetry group) under the symmetry group of the potential fully determines it up to its norm. The symmetry group of the potential of each flavon is the corresponding quotient group of $G_{\!f}$. We obtain the stabilizers of the vevs (\ref{eq:elemvevs}) under these quotient groups. We list them along with their generators in TABLE~\ref{tab:elemstabilizers}. We can see that these generators fully determine the vevs up to their norms. Hence, the vevs (\ref{eq:elemvevs}) are guaranteed to be the stationary points of the corresponding potentials.

Our second step involves analyzing the cross terms among the flavons. TABLE~\ref{tab:eelems} shows that $\chi$ transforms as $\chi\rightarrow\om\chi$ under $\genzl$. Since every other flavon is invariant under $\genzl$, the contribution of $\chi$ to cross terms cannot be linear. Rather, $\chi$ may contribute quadratically from $\chi\otimes\chi$ or cubically from $\chi\otimes\chi\otimes\chi$. Every flavon except $\chi$ transforms nontrivially under $\genx$. This rules out cubic contributions from $\chi\otimes\chi\otimes\chi$ to cross terms. Under $\gene_{\xib}$, $\gene_{\xic}$, $\gene_{\phia}$, $\gene_{\phib}$ and $\gene_{\phic}$, the flavons $\xib$, $\xic$, $\phia$, $\phib$ and $\phic$, respectively, change sign, while every other flavon remains invariant. Therefore, these flavons cannot contribute linearly or cubically to cross terms. The flavon $\xia$ cannot contribute cubically since we have ruled out a linear contribution from every other flavon. Under $\gene_{\xia}$, only $\xia$ and $\xib$ transform nontrivially. Therefore, $\xia$ may appear linearly only in the combination $\xia\otimes\xib\otimes\xib$.  None of the multiplets obtained from $\xib\otimes\xib$, TABLE~\ref{tab:prodtrof}, transforms in the same way as $\xia$, which rules out the combination $\xia\otimes\xib\otimes\xib$. Hence, we conclude that no flavon can appear linearly or cubically in cross terms.

\setlength{\tabcolsep}{5pt}
\begin{table}[tbp]
\begin{center}
\begin{tabular}{|c|c|c|c|}
\hline
&\multicolumn{2}{c|}{}&\\[-1em]
\multirow{2}{*}{\!vev\!}&\multicolumn{2}{c|}{Generators of the stabilizer}&\!\!\!Stabi-\!\!\!\\
\cline{2-3}
&&&\\[-1em]
&Representation&Abstract form&\!\!lizer\!\!\\
\hline
&&&\\[-1em]
$\!\!\!\langle\chi\rangle\!\!\!$&\!\!$\om\matt_3\matr_3^2\matt_3, -\om\matr_3\matt_3()^*$\!\!\! & {\footnotesize$\gent\genr^2\gent\genzl,\,\, \genr\gent\genzl\gencp$}&\!$S_{3\chi}$\!\\
$\!\!\!\langle\xia\rangle\!\!\!$ & $\om()^*$ &{\footnotesize$\genx$}&\!\!$Z_{2\xia}$\!\!\\
$\!\!\!\langle\xib\rangle\!\!\!$ & $\matw_2\sigx,\,\, \matw_2()^*$ &{\footnotesize$\gent\genr\genzn,\,\, \gent\genr$}&\!\!$K_{4\xib}$\!\!\\
$\!\!\!\langle\xic\rangle\!\!\!$ & \!\!$\mattau_2\sigx,\,\,i\mattau_2()^*$\!\! & \!\!{\footnotesize$\genr\gent^2\genr^2\genzn^3\genx,\,\,\gent\genr\genzn\gencp$}\!\!&\!\!$K_{4\xic}$\!\!\\
$\!\!\!\langle\phia\rangle\!\!\!$ & \!\!$\mattau_2\sigx,\,\,\tau\mattau_2()^*$\!\!&\!\!{\footnotesize$\genr\gent^2\genr^2(\gencp\genx)^6,\,\gent\genr(\gencp\genx)^3\gencp^3$}\!\!&\!\!$K_{4\phia}$\!\!\\
$\!\!\!\langle\phib\rangle\!\!\!$ &$\matw_2^*\sigx,\,\,\om\matw_2^*()^*$ &\!\!\!\footnotesize$(\gencp\genx)^5(\genzn\genx)^7$\!, $\genzn\genx(\gencp\genx)^7\genzn$\!\!\!&\!\!$K_{4\phib}$\!\!\\
$\!\!\!\langle\phic\rangle\!\!\!$ & \!\!\!\!$\mattau_2\sigx,\,\,\taub\mattau_2()^*$\!\!\!\! & \!\!{\footnotesize$\genr\gent^2\genr^2\gencp\genzn,\,\,\gent\genr(\gencp\genx)^3\gencp^3$}\!\!&\!\!$K_{4\phic}$\!\!\\
\hline
\end{tabular}
\end{center}
\caption{Stabilizers of the vevs of the elementary flavons. They are isomorphic to $S_3$, $Z_2$ and the Klein four group $K_4$.}
\label{tab:elemstabilizers}
\end{table}

TABLE~\ref{tab:prodtrof} lists the multiplets obtained from the quadratic products of the flavons. No two rows of the table match, implying that no cross term can be constructed among these flavons, with the exception of the cross terms in the form of products of norms such as $|\chi|^2|\xia|^2$. Ref.~\cite{2306.07325} derived the {\it compatibility condition} which determines if the symmetries of a stationary point are preserved under the inclusion of a new term in the potential. The products of norms were shown to satisfy this condition. Therefore, the renormalizable cross terms that we obtained do not spoil the vevs (\ref{eq:elemvevs}), i.e.,~they do not destroy the constraints among the components of these vevs. 

We defer the final step to Appendix~\ref{sec:appfour}. We find that the renormalizable potentials of several flavons contain accidental continuous symmetries. To break them, we introduce driving flavons. We need to ensure that they do not produce dangerous cross terms. To this end, the driving flavons are accompanied by group extensions implemented using auxiliary generators beyond $\genx$. We construct the general renormalizable potential of all the flavons, including the driving flavons, and show that it has a minimum that corresponds to the vevs (\ref{eq:elemvevs}).

\subsection{Phenomenology} 
\label{sec:phen}

Substituting $\langle\chi\rangle$ (\ref{eq:elemvevs}) in (\ref{eq:lagrcl}), we obtain the charged-lepton mass matrix,
\begin{equation}
M_l=\left(\begin{matrix} \,-y_e \frac{r_{\!\chi}}{\,\sqrt{3}\Lambda} & -y_\mu & -y_\tau\\
y_e\frac{r_{\!\chi}}{\sqrt{3}\Lambda}& \ob y_\mu & \om y_\tau\\
y_e\frac{r_{\!\chi}}{\sqrt{3}\Lambda}& \om y_\mu & \ob y_\tau
\end{matrix}\right) v\frac{r_{\!\chi}}{\sqrt{3}\Lambda}.
\end{equation}
We diagonalize $M_l$ as 
\begin{equation}\label{eq:diagcl}
U_l M_l = \text{diag}(m_e, m_\mu, m_\tau),
\end{equation}
\be\label{eq:ulept}
\!\!\!\!\!\!\!\text{where}\qquad\quad\qquad U_l = \frac{1}{\sqrt{3}}\left(\begin{matrix} -1 & 1 & 1\\
-1 & \om & \ob\\
-1 & \ob & \om
\end{matrix}\right),\qquad\qquad
\ee
\vspace{-4mm}
\be\label{eq:leptmass}
\!\!\!\!\!\!\!\text{and}\quad m_e=y_e v\frac{r_{\!\chi}^2}{\sqrt{3}\Lambda^2}\,,\,\, m_\mu=y_\mu v\frac{r_{\!\chi}}{\Lambda}\,,\,\, m_\tau=y_\tau v\frac{r_{\!\chi}}{\Lambda}\,.\!\!\!\!
\ee
We define the ratio of the flavon vev and the cut-off scale as the factor $f$, i.e.,~$f\approx\frac{r_{\!x}}{\Lambda}$ where $r_{\!x}$ denotes $r_{\!\chi}$, $r_{\!\xia}$, etc. According to (\ref{eq:leptmass}), the masses of the muon and the tau are suppressed by $f$ in relation to the Higgs vev, while the electron mass is suppressed by $f^2$. Knowing these masses, we can estimate that $f\approx0.002$.

Substituting $\langle\xia\rangle$, $\langle\xib\rangle$, $\langle\xic\rangle$ and $\langle\phia\rangle$, $\langle\phib\rangle$, $\langle\phic\rangle$ (\ref{eq:elemvevs}) in (\ref{eq:xiexpr2}) and (\ref{eq:phiexpr1}), we obtain the vevs of the effective flavons,
\begin{align}
\begin{split}
\langle\xi\rangle&=\frac{r_{\!\xia}r_{\!\xib}r_{\!\xic}}{8\sqrt{2}}\big((-1+\sqrt{3})\eib \om,\,(-1-\sqrt{3})i\eig\om,\\[-5pt]
&\qquad\qquad\qquad\,\,\,(-1-\sqrt{3})\eib\ob,\,(-1+\sqrt{3})i\eig\ob\big)^\T\!\!,\!
\end{split}\label{eq:xivev}\\
\langle\phi\rangle&=\frac{r_{\!\phia}r_{\!\phib}r_{\!\phic}}{8}(-1-i\sqrt{2},\, \sqrt{6},\, -1+i\sqrt{2})^\T.\label{eq:phivev}
\end{align}
The constraints among the components of $\langle\xi\rangle$ and $\langle\phi\rangle$ are highly nontrivial. These constraints do not originate from the stabilizers of $\langle\xi\rangle$ and $\langle\phi\rangle$ under the group $G_r$. In this sense, these vevs are similar to those found in the indirect models in the literature. However, unlike in the previous works, we obtain these constraints using discrete symmetries alone without the help of mechanisms involving supersymmetry, extradimensions etc. The constraints on the vevs of the effective flavons (\ref{eq:xivev}), (\ref{eq:phivev}), originate indirectly from the stabilizers of the vevs of the elementary flavons (\ref{eq:elemvevs}), TABLE~\ref{tab:elemstabilizers}.

Substituting (\ref{eq:xivev}) and (\ref{eq:phivev}) in (\ref{eq:lagrdr}) and (\ref{eq:lagrmj}), we obtain the neutrino Dirac and the Majorana mass matrices,
\begin{align}
M_D &= \left(\begin{matrix}  1 & \tau\\
 -\sqrt{2} & \sqrt{2}\tau\\
1 & \tau
\end{matrix}\right)\sqrt{3}\,\eig y_\xi v\frac{r_{\!\xia}r_{\!\xib}r_{\!\xic}}{8\Lambda^3},\label{eq:mdir}\\
M_M&= \left(\begin{matrix}  \taub (\sqrt{2}-1) & -\sqrt{3}\\
-\sqrt{3} &\tau (\sqrt{2}+1)
\end{matrix}\right)i y_\phi \frac{r_{\!\phia} r_{\!\phib} r_{\!\phic}}{4\sqrt{2}\Lambda^2}.\label{eq:mmaj}
\end{align}
Using (\ref{eq:mdir}) and (\ref{eq:mmaj}), we obtain the seesaw mass matrix,
\be\label{eq:seesaw}
M_{ss}=-M_D M_M^{-1}M_D^\T.
\ee
Since the mass terms in the neutrino Dirac sector and the Majorana sector are constructed with a single irrep each, i.e.,~$\xi$ and $\phi$, respectively, their vevs (\ref{eq:xivev}) and (\ref{eq:phivev}) fully constrain the neutrino mass matrices (\ref{eq:mdir}), (\ref{eq:mmaj}) and (\ref{eq:seesaw}) up to proportionality constants. Fully constrained mass matrices result in highly predictive models, e.g.,~\cite{1801.10197}.

We diagonalize $M_{ss}$ in the following way:
\be
U_\nu^\dagger M_{ss} U_\nu^* = \text{diag} (m_1,m_2,m_3),
\ee
\be\label{eq:uneutri}
\text{where}\,\,\,\, U_\nu=\left(\begin{matrix} \frac{1}{\sqrt{2}} & 0 & \frac{1}{\sqrt{2}}\\
0 & 1 & 0\\
\frac{-1}{\sqrt{2}} & 0 & \frac{1}{\sqrt{2}}
\end{matrix}\right)\left(\begin{matrix} 1 & 0 & 0\\
0 & i\cos\frac{\pi}{12} & -\sin\frac{\pi}{12}\\
0 & i\sin\frac{\pi}{12} & \cos\frac{\pi}{12}
\end{matrix}\right)\!\!,\!\!
\ee
\vspace{-4mm}
\be\label{eq:numasses}
m_1=0,\quad m_2=(\sqrt{2}-1)m_\nu,\quad m_3 = (\sqrt{2}+1)m_\nu,
\ee
\vspace{-5.5mm}
\be\label{eq:nufactor}
\text{with}\qquad m_\nu=\frac{3}{4}\frac{y_\xi^2}{y_\phi} \frac{v^2r_{\!\xia}^2r_{\!\xib}^2r_{\!\xic}^2}{r_{\!\phia}r_{\!\phib}r_{\!\phic}\Lambda^4}\,.\qquad
\ee
Higher-order terms in the Lagrangian ensure a non-zero mass for the lightest neutrino. Even though we do not discuss such terms in this paper, we can infer that $m_1$ is suppressed by at least a factor of $f\approx0.002$ in relation to the other neutrinos. Therefore, we use $m_1\approx0$ for our calculations in this section. 

The PMNS neutrino mixing matrix is given by
\be\label{eq:pmnsform}
U=U_l U_\nu.
\ee
Substituting (\ref{eq:ulept})and (\ref{eq:uneutri}) in the above equation and simplifying, we obtain
\be\label{eq:upmns}
U = -\,\text{diag}(1,\om,\ob)\,\,\Utm\,\,\text{diag}(1,-i,1),
\ee
where the expression of $\Utm$ is provided in Appendix~\ref{sec:appthree}. It denotes the $\text{TM}_1$ mixing matrix\cite{hep-ph/0607302,0812.0436,1004.2798,1108.4278,1212.3247, 1304.6264,1306.2358,1312.4401,1509.06915, 1802.00425, 1709.02136,1705.02027,1305.4846, 1512.07531,1607.05276,1802.00425, 2003.00506} with its free parameters having values $\frac{\pi}{12}$ and $-\frac{\pi}{2}$. The author recently showed\cite{1912.02451} that $\Utm$ is consistent with the neutrino oscillation data. Please see Appendix~\ref{sec:appthree} for a comparison of the predicted mixing angles and the CP phase with the data. The phase factors in the left side of $\Utm$ in (\ref{eq:upmns}) are unobservable, while $-i$ in the right side is a Majorana phase.

Global fit\cite{2006.11237} of the oscillation data gives the neutrino mass-squared differences, $\Delta m^2_{21}=75.0^{+2.2}_{-2.0}~\text{meV}^2$ and $\Delta m^2_{31}=2550^{+20}_{-30}~\text{meV}^2$. From (\ref{eq:numasses}), we predict the ratio of the masses $m_2$ and $m_3$,
\be
\frac{m_2}{m_3}=\frac{\sqrt{2}-1}{\sqrt{2}+1}=\tan^2\frac{\pi}{8}.
\ee
This ratio is in excellent agreement with the data (within $1\sigma$ errors) assuming $m_1\approx 0$. By fitting the data with (\ref{eq:numasses}), we obtain 
\be\label{eq:propmass}
m_\nu=20.92^{+0.08}_{-0.12}~\text{meV},
\ee
and also $m_2=8.66^{+0.13}_{-0.12}~\text{meV}$ and $m_3=50.50^{+0.20}_{-0.30}~\text{meV}$. 

Neutrinoless double-beta decay experiments\cite{1902.04097} seek to determine the effective neutrino mass,
\be\label{eq:dbetaformula}
\langle m_{\beta\beta}\rangle=|U_{e1}^2 m_1 + U_{e2}^2 m_2 +U_{e3}^2 m_3|,
\ee
where $U_{e1}$, $U_{e2}$ and $U_{e3}$ are the elements of the first row of the PMNS matrix. Substituting the values of these elements from (\ref{eq:tma}) with $\theta=\frac{\pi}{12}$ and using (\ref{eq:numasses}) and (\ref{eq:propmass}), we obtain
\be
\langle m_{\beta\beta}\rangle\,\,= \,\,\frac{1}{3}\left(\sqrt{\frac{3}{2}}-1\right)\!m_\nu\,\,=\,\,1.57^{+0.01}_{-0.01}\text{ meV}.
\ee
This value is well below the upper bounds set by the recent $0\nu\beta\beta$ experiments\cite{1605.02889,1906.02723,2104.06906}, which are around $100~\text{meV}$. The sum of the neutrino masses, $\Sigma m_i$, is constrained by cosmological observations. The strongest upper bound provided by these observations is $90~\text{meV}$~\cite{2106.15267,1811.02578}. Since $m_1\approx 0$, we obtain
\be\label{eq:sum}
\Sigma m_i = \sqrt{\Delta m^2_{21}}+ \sqrt{\Delta m^2_{31}}= 59.16^{+0.32}_{-0.41}\text{ meV},
\ee
which is the lowest value consistent with the neutrino oscillation experiments. 

Using (\ref{eq:nufactor}), (\ref{eq:propmass}) and $f\approx 0.002$, we make the following order of magnitude estimates: $r_{\!x}=\mathcal{O}(10^{4}~\text{GeV})$, $\Lambda=\mathcal{O}(10^{7}~\text{GeV})$, $M_D=\mathcal{O}(1~\text{keV})$ and $M_M=\mathcal{O}(100~\text{MeV})$. As $M_D$ (\ref{eq:mdir}) contains three flavon insertions, it is heavily suppressed. The seesaw scale,  i.e.,~$M_M$, is correspondingly suppressed such that we obtain the correct masses for the light neutrinos. Low-scale seesaw models, such as ours, have interesting phenomenological consequences, e.g., they may result in heavy neutrinos detectable in the current and planned experiments\cite{2203.08039}. Since our model breaks the flavour-gCP symmetries at a relatively low scale compared to, say, the grand unification scale, we can make rather precise predictions without taking into account their running under the renormalization group.


\subsection{Summary} 

In this paper, we construct a type-1 seesaw model with two right-handed neutrinos and several flavons in the framework of the auxiliary group. According to this framework, the fermions transform under a finite group $G_r$ while the elementary flavons transform under its group extension $G_{\!f}$. We have $G_r=G_{\!f}/G_x$, where $G_x$ is the auxiliary group. In our model, the fermions transform as multiplets under the binary octahedral group generated by $\gent$ and $\genr$. We also assign the Abelian charges $\genzl$ and $\genzn$ and the generalized CP transformation $\gencp$ to the fermions. The actions of $\gent$, $\genr$, $\genzl$, $\genzn$ and $\gencp$ on the fermions generate the flavour-gCP group $G_r$. We introduce the flavons $\chi$, $\xi$ and $\phi$ that transform as irreps under $G_r$ and couple them with the fermions to form the charged-lepton, neutrino Dirac and Majorana mass terms, respectively. 

We construct $\xi$ and $\phi$ as the effective flavons using $\xia$, $\xib$ and $\xic$, and $\phia$, $\phib$ and $\phic$, respectively, which are the elementary flavons transforming as irreps of the extended flavour-gCP group $G_{\!f}$. To implement this group extension, we invent a technique that involves expressing the flavons in the {\it placeholder notation}. Using it, we define how $\gent$, $\genr$, $\genzl$, $\genzn$ and $\gencp$ act on the elementary flavons $\xia$, $\xib$, $\xic$, $\phia$, $\phib$ and $\phic$, while being consistent with how they act on the effective flavons $\xi$ and $\phi$. Then, we introduce the auxiliary generator $\genx$ that acts nontrivially on these elementary flavons but trivially on $\xi$ and $\phi$. We define $G_{\!f}$ as the group generated by $\gent$, $\genr$, $\genzl$, $\genzn$, $\gencp$ and $\genx$. In this way, our technique enables us to build $G_{\!f}$ as a group extension of $G_r$. To simplify our analysis of the extended flavour group, we develop further techniques such as constructing specific group elements, TABLE~\ref{tab:eelems}, under which the fermions and the flavons transform in a simple way, utilizing complex conjugation to represent group actions beyond gCP and combining a complex multiplet with its conjugate as a real irrep to evaluate tensor products. In Section~\ref{sec:lagrangian} and Appendix~\ref{sec:apptwo}, we verify that the leading-order contributions of the elementary flavons in the construction of the Dirac and Majorana mass terms are the effective irreps $\xi$ and $\phi$ obtained from $\xia\otimes\xib\otimes\xic$ and $\phia\otimes\phib\otimes\phic$, respectively. 

We obtain the vacuum expectation values (vevs) of the flavons with the help of the mathematical procedure developed in \cite{2306.07325}. To each elementary flavon, we assign a vev fully determined by its stabilizer, up to its norm. Such a vev is guaranteed to be a stationary point of the potential of the flavon. We analyze the renormalizable cross terms among the flavons and find that all of them are products of norms, which implies they satisfy the {\it compatibility condition}\cite{2306.07325}. Therefore, the cross terms would not spoil our assigned vevs.

We explicitly construct the flavon potential in Appendix~\ref{sec:appfour}. Renormalizable potentials of the various flavons contain accidental continuous symmetries. To break them, we introduce driving flavons after ensuring that they do not produce dangerous cross terms among themselves or with the other flavons appearing in the model. This is achieved by assigning specific auxiliary generators to the driving flavons, which results in further group extensions enlarging $G_x$ and $G_{\!f}$. We show that the general renormalizable potential involving all the flavons has a minimum that corresponds to our assigned vevs. In Appendix~\ref{sec:appfive}, we show that the driving flavons cannot provide any leading-order contribution to the mass terms.

We obtain the vevs of the effective flavons $\langle\xi\rangle$ and  $\langle\phi\rangle$ using the vevs of the elementary flavons. The constraints in $\langle\xi\rangle$ and $\langle\phi\rangle$ are similar to those in indirect models and lead to highly predictive mass matrices. We obtain the mixing matrix to be in $\text{TM}_1$ form with $\sin^2 \theta_{13}=\frac{1}{3}\sin^2 \frac{\pi}{12}$, $\sin^2 \theta_{12}=1-\frac{2}{3\cos^2 \theta_{13}}$, $\sin^2 \theta_{23}=\frac{1}{2}$ and $\sin\delta_{\text{CP}}=-1$. We estimate the lightest neutrino mass $m_1$ to be sub-meV and predict the ratio of the other light neutrino masses as $\frac{m_2}{m_3}=\frac{\sqrt{2}-1}{\sqrt{2}+1}$. Using the experimental values of the neutrino mass-squared differences, we obtain $m_2=8.66^{+0.13}_{-0.12}~\text{meV}$, $m_3=50.50^{+0.20}_{-0.30}~\text{meV}$, $\langle m_{\beta\beta}\rangle = 1.57^{+0.01}_{-0.01}\text{ meV}$ and $\Sigma m_i = 59.16^{+0.32}_{-0.41}\text{ meV}$. Since the Dirac mass matrix is suppressed by three flavon insertions, we have a relatively low-scale seesaw. 

In this paper and in the companion paper\cite{2306.07325}, we set up the tools to construct flavon models in the framework of the auxiliary group, which does not assume mechanisms beyond the first principles of discrete symmetries and scalar fields. We construct such a model, predicting the neutrino mixing angles and the masses consistent with the current experimental data. 

\vspace{5mm}
\centerline{\bf \small Acknowledgments} 
\vspace{1.5mm}
I am grateful to Paul Harrison, who persuaded me to seek ways to construct flavon models using discrete symmetries alone. Without his encouragement and support, I would not have been able to persevere in my attempts. I thank my parents, K Venugopal and J Saraswathi Amma, for always reassuring me that it is worthwhile to follow my pursuits. I owe them everything for keeping a watchful eye on my physical and mental health these past few years, and I dedicate this work to them.

\onecolumngrid
\vspace{6mm}
\noindent\makebox[\linewidth]{\resizebox{0.5\linewidth}{1pt}{$\bullet$}}\bigskip
\vspace{6mm}
\twocolumngrid


\appendix

\section{The group $2O$}
\label{sec:appone}

Abstractly, the binary octahedral group, $2O$, can be defined using the presentation,
\begin{equation}\label{eq:2Opres}  
\langle \gent, \genr |(\gent\genr)^2=\gent^3 = \genr^4, \gent^6 = 1\rangle.
\end{equation}
The unfaithful irrep $\rept$ of $2O$ is isomorphic to the group $S_4$. Comparing (\ref{eq:2Opres}) with the presentation of $S_4$,
\begin{equation}  
\langle \mathbb{T}, \mathbb{R} |(\mathbb{T}\mathbb{R})^2=\mathbb{T}^3 = \mathbb{R}^4, \mathbb{T}^3 = 1\rangle,
\end{equation}
we can see that the kernel of the irrep $\rept$ of $2O$ is its $Z_2$ subgroup generated by its element $\gent^3$. In terms of the representation matrices of $\repdd$ (\ref{eq:bogen}) and $\rept$ (\ref{eq:ogen}), we have $\matt_2^3=-1$ and $\matt_3^3=1$, as expected. The $Z_2$ group generated by $\gent^3$ is a normal subgroup of $2O$. The quotient of $2O$ by $Z_2$ is nothing but $S_4$. Therefore, $2O$ is a double cover of $S_4$\footnote{$2O$ is the group extension of $S_4$ by $Z_2$, i.e.,~~$2O=Z_2\cdot S_4$. This is not a split extension; therefore, $2O$ cannot be expressed as the semidirect product, i.e.,~$2O\neq Z_2\rtimes S_4$. In the recent literature\cite{2006.03058, 2006.10722, 2211.04546}, a double cover of $S_4$, namely $S_4'$, has been studied in the context of modular symmetries. We note that $2O$ and $S_4'$ are different groups with quite different structures even though both are double covers of $S_4$.}. Using the computational discrete algebra system GAP\cite{GAP4}, we can obtain $2O$ as SmallGroup(48,28). 

We define the following group elements of $2O$,
\begin{equation}
\gens=\genr^2,\quad \genu=\gent\genr^2\gent^2\genr,
\end{equation}
and obtain their matrix representations under $\repdd$ and $\rept$ using (\ref{eq:bogen}) and (\ref{eq:ogen}),
\begin{align}
&\gens(\repdd)=\left(\begin{matrix} 0 & 1\\
-1 & 0
\end{matrix}\right),\qquad\quad\,\genu(\repdd)= \frac{i}{\sqrt{2}}\left(\begin{matrix} -1 & 1\\
1 & 1
\end{matrix}\right),\\
&\gens(\rept)=\left(\begin{matrix} 1 & 0 & 0\\
0 & -1 & 0\\
0 & 0 & -1
\end{matrix}\right),\quad\genu(\rept)=-\left(\begin{matrix} 1 & 0 & 0\\
0 & 0 & 1\\
0 & 1 & 0
\end{matrix}\right).
\end{align}
The matrices $\gens(\rept)$, $\matt_3$ (\ref{eq:ogen}) and $\genu(\rept)$ (or those isomorphic to them) are widely used to generate and study $S_4$ in the flavour physics literature. We may obtain the abstract generator $\genr$ of $2O$ from $\gens$, $\gent$ and $\genu$ as $\genr=\gent\gens\gent^5\genu$. 

We provide the character table of $2O$ in TABLE~\ref{tab:char}. The elements $\gent$, $\genr$, $\gens$ and $\genu$ belong to the conjugacy classes $(123)$, $(1234)$, $(12)(34)$ and $(12)$, respectively. Also, $\gent^3$, $\gent^3 \gent$ and $\gent^3\genr$ belong to the conjugacy classes $()^-$, $(123)^-$ and $(1234)^-$, respectively. The first five irreps, $\reps$, $\repsp$, $\repd$, $\rept$ and $\reptp$, are isomorphic to the irreps of $S_4$. The trivial, the $3$-cycle and the $4$-cycle conjugacy classes of $S_4$ correspond to the pairs of classes of $2O$ given by $()$,~$()^-$; $(123)$, $(123)^-$; and $(1234)$, $(1234)^-$, respectively. These pairs of classes are identical for the first five irreps. The representation matrices of $\repdd$, $\repddp$ and $\repqq$ in these pairs of classes differ by a sign with respect to the normal subgroup $Z_2$. 

\setlength{\tabcolsep}{3pt}
\begin{table}[tbp]
		\begin{center}
			\begin{tabular}{|c|c c c c c c c c|}
				\hline
&&&&&&&&\\[-1em]
				&$()$ &$()^-$  & $(123)$ & $(123)^-$ & $(1234)$ & $(1234)^-$ & $(12)(34)$	& $(12)$\\
				\hline
&&&&&&&&\\[-1em]
				$\reps$ & $1$ & $1$ & $1$ & $1$  & $1$ & $1$ & $1$ & $1$ \\
				$\repsp$ & $1$ & $1$ & $1$ & $1$ & $-1$ & $-1$ & $1$ & $-1$  \\
				$\repd$ & $2$ & $2$ & $-1$& $-1$ & $0$ & $0$ & $2$ & $0$ \\
				$\rept$ & $3$ & $3$ & $0$ & $0$ & $1$ & $1$ & $-1$ & $-1$ \\
				$\reptp$ & $3$ & $3$ & $0$ & $0$ & $-1$ & $-1$ & $-1$ & $1$\\
				$\repdd$ & $2$ & $-2$ & $1$ & $-1$ & $\sqrt{2}$ & $-\sqrt{2}$ & $0$ & $0$\\
				$\repddp$ & $2$ & $-2$ & $1$ & $-1$ & $-\sqrt{2}$ & $\sqrt{2}$ & $0$ & $0$\\
				$\repqq$ & $4$ & $-4$ & $-1$ & $1$ & $0$ & $0$ & $0$ & $0$\\
				\hline
			\end{tabular}
		\end{center}
		\caption{The character table of $2O$.}
		\label{tab:char}
\end{table}

$\repdd$,  $\repddp$ and  $\repqq$ are pseudo-real representations, also known as quaternionic representations. We require complex numbers to represent them, but their characters are real. The complex conjugation of pseudo-real representations are isomorphic to themselves. The representation matrices of $\repdd$ (\ref{eq:bogen}),  $\repddp$ (\ref{eq:dtimesdd}) and $\repqq$ (\ref{eq:qgen}) are related to their complex conjugates through the similarity transformations,
\begin{align}
\repdd&=(i\sigy)\,\repdd^*\,(i\sigy)^\T,\label{eq:sim1}\\
\repddp&=(i\sigy)\,\repddp^*\,(i\sigy)^\T,\label{eq:sim2}\\
\repqq&=(\sigx\otimes i\sigy)\,\repqq^*\,(\sigx\otimes i\sigy)^\T.\label{eq:sim3}
\end{align}
$\repd$ is a real irrep even though, for convenience, we have used complex numbers  (\ref{eq:ddgen2}) to construct it. For $\repd$  (\ref{eq:ddgen2}), the corresponding similarity transformation is
\be\label{eq:sim4}
\repd=\sigx\,\repd^*\sigx^\T.
\ee

After introducing flavons transforming as irreps under a non-Abelian finite group, flavour models may impose additional Abelian transformations, such as $Z_3$, on the irreps. This complexifies the real and pseudo-real irreps, i.e.,~changes them to complex irreps, and enlarges the flavour group. Models may also impose complex conjugation on the flavons as a symmetry transformation, which usually corresponds to the CP transformation. For the complex conjugation to be consistent\cite{1211.6953} with the flavour group, a similarity transformation must exist such that the combined action of the conjugation and the similarity transformation is an automorphism on the representation matrices of the flavour group. This combined action is called the generalized CP (gCP)\cite{Ecker:1987qp, Neufeld:1987wa,1211.5560,1211.6953,1303.6180,1402.0507}. 

The irreps $\reps$, $\repsp$, $\rept$, $\reptp$ of $2O$ are real, and we have used real matrices to construct them. Therefore, complex conjugation maps every element of these irreps to itself. For $\repdd$, $\repddp$, $\repqq$ and $\repd$ also, conjugation combined with the corresponding similarity transformation (\ref{eq:sim1})-(\ref{eq:sim4}), maps every element to itself. Given these mappings, let us define the action of gCP on the irreducible multiplets as (\ref{eq:gcpdefs}). This definition implies that gCP maps every representation matrix to itself, i.e.,~gCP produces the trivial automorphism. This is expected because the irreps of $2O$ are real or pseudo-real. On the other hand, we can show that gCP (\ref{eq:gcpdefs}) produces a nontrivial automorphism if we complexify the irreps using the aforementioned Abelian transformations. Therefore, our definition of gCP is consistent and nontrivial for the complexified irreps of $2O$. 

There are two equivalent ways to approach the flavour and gCP symmetries: either construct the flavour group and study gCP separately as an automorphism or study them together by constructing the flavour-gCP group as the semidirect product\cite{1211.6953} of the flavour group (normal subgroup) and the gCP group (quotient). In this paper, we follow the second approach. We construct $2O$ using the generators $\gent$ and $\genr$. In Section~\ref{sec:themodel}, we complexify the irreps by introducing the generators $\genzl$ and $\genzn$, and then introduce gCP using the generator $\gencp$. Using these five generators, we construct the flavour-gCP group $G_r$. Complex irreps in the flavour group become real irreps of twice the dimension in the flavour-gCP group. For example, in TABLE~\ref{tab:flavourcontent}, we introduce $\chi$ as a $\reptp$ of $2O$, which gets complexified by $\genzl$. With the inclusion of gCP as a group transformation, $\chi$ becomes $(\chi,\chi^*)^\T$ transforming as a six-dimensional real irrep. 

We may use complex conjugation to represent group transformations other than gCP also. For example, in TABLE~\ref{tab:neutrinosector}, complex conjugation appears in relation to not only $\gencp$ but also $\genr$, $\genzn$ and $\genx$. Every flavon expressed as a complex multiplet in this paper undergoes complex conjugation as a part of one or more group generators. Hence, every one of them transforms as a real irrep of twice the dimension. We use complex multiplets to represent them because of the ease of construction and analysis. This usage is justified because all degrees of freedom in the real multiplet are present in the corresponding complex multiplet, and all group actions on the real multiplet are faithfully represented with the help of matrix multiplications and complex conjugation on the complex multiplet. However, evaluating tensor products using these complex multiplets results in difficulties, similar to those encountered when gCP is studied separately as an automorphism rather than studied together with the rest of the group\cite{1211.6953}. To avoid these difficulties, we use real irreps for evaluating the tensor products, e.g.,~$\chi\otimes\chi$ denotes $(\chi,{\chi}^*)^\T\!\otimes(\chi,{\chi}^*)^\T$.

\section{The expressions of $\xi$ and $\phi$}
\label{sec:apptwo}

In this section, we obtain the explicit expression of $\xi$ defined in (\ref{eq:assignment}) after showing that among the irreps constructed from the tensor product $\xia\otimes\xib\otimes\xic$, the only irrep that can couple in the Dirac mass term is $\xi$. Similarly, we obtain the expression of $\phi$ (\ref{eq:assignment}) after showing that it is the only irrep from $\phia\otimes\phib\otimes\phic$ that can couple in the Majorana mass term.

{\renewcommand{\arraystretch}{1.1}
	\setlength{\tabcolsep}{5pt}
	\begin{table}[tbp]
	\begin{center}
	\begin{tabular}{|c|c c c c c |c|}
\hline
	&$\gent$&$\genr$&$\genzn$&$\gencp$&$\genx$&$d$\\
\hline
&&&&&&\\[-1em]
\!\!${\xiab}^{\!\!\!\!\!\!2\alpha}$\!\! &$\matw_2$&$()^*$&$\sigx()^*$&$i\sigy ()^*$& $i\sigy ()^*$&$4$\\
\!\!${\xiab}^{\!\!\!\!\!\!2\beta}$\!\! &$\matw_2$&$()^*$&$\sigx()^*$&$i\sigy ()^*$& $\om i\sigy ()^*$&$4$\\
\!\!\!${\xiabc}^{\!\!\!\!\!\!\!\!\!3\alpha\,\,\,}$\!\!\! &\!\!$\matw_2\!\otimes\!\matt_2$\!\!&\!\!$\sigx\!\otimes\!\matr_2$\!\!&$i$&\!\!\!$\sigx\!\otimes\!i\sigy()^*$\!\!& $1$&$8$\\
\!\!\!${\xiabc}^{\!\!\!\!\!\!\!\!\!3\beta\,\,\,}$\!\!\! &\!\!$\matw_2\!\otimes\!\matt_2$\!\!&\!\!$\sigx\!\otimes\!\matr_2$\!\!&$i$&\!\!\!$-\sigx\!\otimes\!i\sigy()^*$\!\!& $-1$&$8$\\
\hline
&&&&&&\\[-1em]
\!\!${\phibc}^{\!\!\!\!\!\!\!2\alpha}$\!\! &\!\!$\mati_2\!\otimes\!\matt_2$\!\!&\!\!$\mati_2\!\otimes\!\matr_2$\!\!&\!\!\!$-\sigx\!\otimes\!\sigy()^*$\!\!&\!\!$\mati_2\!\otimes\!i\sigy()^*$\!\!&\!\!\!\!$\matt_2\matw_2\sigz\!\otimes\!i\sigy()^*$\!\!\!&$8$\\
\!\!${\phibc}^{\!\!\!\!\!\!\!2\beta}$\!\! &\!\!$\mati_2\!\otimes\!\matt_2$\!\!&\!\!$\mati_2\!\otimes\!\matr_2$\!\!&\!\!\!$-\sigx\!\otimes\!\sigy()^*$\!\!&\!\!$\mati_2\!\otimes\!i\sigy()^*$\!\!&\!\!\!\!$\om\matt_2\matw_2^*\!\otimes\!i\sigy()^*$\!\!\!&$8$\\
\!\!\!${\phiabc}^{\!\!\!\!\!\!\!\!\!\!3\alpha\,\,\,}$\!\!\! &\!\!$\matt_3$\!\!&\!\!$\matr_3$\!\!&$i$&\!\!\!$()^*$\!\!& $1$&$6$\\
\!\!\!${\phiabc}^{\!\!\!\!\!\!\!\!\!\!3\beta\,\,\,}$\!\!\! &\!\!$1$\!\!&\!\!$1$\!\!&$i$&\!\!\!$()^*$\!\!& $1$&$2$\\
\!\!\!${\phiabc}^{\!\!\!\!\!\!\!\!\!\!3\gamma\,\,\,}$\!\!\! &\!\!$\matt_3$\!\!&\!\!$\matr_3$\!\!&$i$&\!\!\!$()^*$\!\!& $-1$&$6$\\
\!\!\!${\phiabc}^{\!\!\!\!\!\!\!\!\!\!3\delta\,\,\,}$\!\!\! &\!\!$1$\!\!&\!\!$1$\!\!&$i$&\!\!\!$()^*$\!\!& $-1$&$2$\\
\!\!\!${\phiabc}^{\!\!\!\!\!\!\!\!\!\!3\epsilon\,\,\,}$\!\!\! &\!\!$\mati_2\!\otimes\!\matt_3$\!\!&\!\!$\mati_2\!\otimes\!\matr_3$\!\!&$i$&\!\!\!$()^*$\!\!& $\matw_2\sigy\!\otimes\!\mati_3$&$\!\!12\!\!$\\
\!\!\!${\phiabc}^{\!\!\!\!\!\!\!\!\!\!3\zeta\,\,\,}$\!\!\! &\!\!$1$\!\!&\!\!$1$\!\!&$i$&\!\!\!$()^*$\!\!& $\matw_2\sigy$&$4$\\
\hline
			\end{tabular}
		\end{center}
		\caption{The group action on the irreps constructed from $\xia$, $\xib$ and $\xic$ as well as from $\phia$, $\phib$ and $\phic$.}
		\label{tab:cubicprods}
\end{table}}

We proceed by first evaluating
\be
\text{\small$\xia\!\otimes\xib={\xiab}^{\!\!\!\!\!\!2\alpha}\oplus{\xiab}^{\!\!\!\!\!\!2\beta}$},
\ee
where
\be
\text{\small${\xiab}^{\!\!\!\!\!\!2\alpha}=(\xia{\xib}_1,\xia{\xib}_2)^\T,\quad {\xiab}^{\!\!\!\!\!\!2\beta}=(\xia^*{\xib}_1,\xia^*{\xib}_2)^\T$}.
\ee
How these irreps transform is given in TABLE~\ref{tab:cubicprods}. Since $L$ and $N$ are invariant under $\gene_{\xia}$, TABLE~\ref{tab:eelems}, the irrep that can couple in the Dirac mass term must also be invariant under it. We can show that ${\xiab}^{\!\!\!\!\!\!2\alpha}$ is invariant while ${\xiab}^{\!\!\!\!\!\!2\beta}$ transforms as ${\xiab}^{\!\!\!\!\!\!2\beta}\rightarrow \om{\xiab}^{\!\!\!\!\!\!2\beta}$ under $\gene_{\xia}$. Since $\xic$ also is invariant, we can rule out ${\xiab}^{\!\!\!\!\!\!2\beta}$ from our construction. Using ${\xiab}^{\!\!\!\!\!\!2\alpha}$ we obtain

\be
\text{\small${\xiab}^{\!\!\!\!\!\!2\alpha}\!\otimes\xic={\xiabc}^{\!\!\!\!\!\!\!\!\!3\alpha\,\,\,}\oplus{\xiabc}^{\!\!\!\!\!\!\!\!\!3\beta\,\,\,}$},
\ee
where 
\begin{align}
\begin{split}
\text{\small${\xiabc}^{\!\!\!\!\!\!\!\!\!3\alpha\,\,\,\,}$}&\text{\small$=\big((\llbracket \xia\rrbracket\llbracket \xib\rrbracket\llbracket \xic\rrbracket)_{11},(\llbracket \xia\rrbracket\llbracket \xib\rrbracket\llbracket \xic\rrbracket)_{12},$}\\
&\text{\small$\qquad(\llbracket \xia\rrbracket\llbracket \xib\rrbracket\llbracket \xic\rrbracket)_{21},(\llbracket \xia\rrbracket\llbracket \xib\rrbracket\llbracket \xic\rrbracket)_{22}\big)^\T\!\!,\!\!$}
\end{split}\label{eq:xiexpr1}\\
\begin{split}
\text{\small${\xiabc}^{\!\!\!\!\!\!\!\!\!3\alpha\,\,\,\,}$}&\text{\small$=\big((\llbracket \xia\rrbracket\llbracket \xib\rrbracket\sigx\llbracket \xic\rrbracket)_{11},(\llbracket \xia\rrbracket\llbracket \xib\rrbracket\sigx\llbracket \xic\rrbracket)_{12},$}\\
&\text{\small$\qquad(\llbracket \xia\rrbracket\llbracket \xib\rrbracket\sigx\llbracket \xic\rrbracket)_{21},(\llbracket \xia\rrbracket\llbracket \xib\rrbracket\sigx\llbracket \xic\rrbracket)_{22}\big)^\T\!\!.\!\!$}
\end{split}
\end{align}
Their transformation rules are given in TABLE~\ref{tab:cubicprods}. Since ${\xiabc}^{\!\!\!\!\!\!\!\!\!3\beta\,\,\,\,}$ transforms nontrivially under $\genx$, it cannot couple in the mass term. Hence, we are left with only ${\xiabc}^{\!\!\!\!\!\!\!\!\!3\alpha\,\,\,\,}$, which is nothing but $\xi$ (\ref{eq:assignment}). Its explicit expression is given by,
\begin{equation}
\begin{split}
\text{\small$\xi\,\,$}&\text{\small$=\frac{1}{4}\big(\xia{\xib}_1{\xic}_1+\xia^*{\xib}_2^{\!\!*}{\xic}_2^{\!\!*}, \xia{\xib}_1{\xic}_2-\xia^*{\xib}_2^{\!\!*}{\xic}_1^{\!\!*},$}\\
&\text{\small$\qquad\xia^*{\xib}_1^{\!\!*}{\xic}_1+\xia{\xib}_2{\xic}_2^{\!\!*},\xia^*{\xib}_1^{\!\!*}{\xic}_2-\xia{\xib}_2{\xic}_1^{\!\!*}\big)^\T\!\!.\!\!$}
\end{split}\label{eq:xiexpr2}
\end{equation} 

Let us analyze the tensor product $\phia\otimes\phib\otimes\phic$. First, we evaluate $\phib\otimes\phic$. We obtain
\be
\text{\small$\phib\!\otimes\phic={\phibc}^{\!\!\!\!\!\!2\alpha}\oplus{\phibc}^{\!\!\!\!\!\!2\beta},$}
\ee
where
\begin{align}
\text{\small${\phibc}^{\!\!\!\!\!\!2\alpha}$}&\text{\small$\,\,=({\phib}_1{\phic}_1,{\phib}_1{\phic}_2,{\phib}_2{\phic}_1,{\phib}_2{\phic}_2)^\T,$}\\ 
\text{\small${\phibc}^{\!\!\!\!\!\!2\beta}$}&\text{\small$\,\,=({\phib}_1^{\!*}{\phic}_1,{\phib}_1^{\!*}{\phic}_2,{\phib}_2^{\!*}{\phic}_1,{\phib}_2^{\!*}{\phic}_2)^\T.$}
\end{align}
How these irreps transform is given in TABLE~\ref{tab:cubicprods}. For further analysis, consider the abstract element $(\genzn\genx)^4$. $N$ is invariant under it, hence the irrep obtained from $\phia\otimes\phib\otimes\phic$ that can couple in the Majorana mass term must also be invariant under it. We can show that $\phia$ and ${\phibc}^{\!\!\!\!\!\!2\alpha}$ are invariant while ${\phibc}^{\!\!\!\!\!\!2\beta}$ transforms as ${\phibc}^{\!\!\!\!\!\!2\beta}\rightarrow\ob{\phibc}^{\!\!\!\!\!\!2\beta}$ under $(\genzn\genx)^4$. Therefore, we rule out ${\phibc}^{\!\!\!\!\!\!2\beta}$ from our construction. Using ${\phibc}^{\!\!\!\!\!\!2\alpha}$, we obtain
\be
\text{\small$\phia\otimes{\phibc}^{\!\!\!\!\!\!2\alpha}={\phiabc}^{\!\!\!\!\!\!\!\!\!3\alpha\,\,\,}\,\oplus{\phiabc}^{\!\!\!\!\!\!\!\!\!3\beta\,\,\,}\,\oplus{\phiabc}^{\!\!\!\!\!\!\!\!\!3\gamma\,\,\,}\,\oplus{\phiabc}^{\!\!\!\!\!\!\!\!\!3\delta\,\,\,}\,\,\oplus{\phiabc}^{\!\!\!\!\!\!\!\!\!3\epsilon\,\,\,}\,\,\oplus{\phiabc}^{\!\!\!\!\!\!\!\!\!3\zeta\,\,\,.}$}
\ee
where
\begin{align}
\text{\small${\phiabc}^{\!\!\!\!\!\!\!\!\!3\alpha\,\,\,}$}&\text{\small$\,=(\llbracket \phia\rrbracket\llbracket \phib\rrbracket\llbracket \phic\rrbracket)_\text{sym},$}\\
\text{\small${\phiabc}^{\!\!\!\!\!\!\!\!\!3\beta\,\,\,}$}&\text{\small$\,=(\llbracket \phia\rrbracket\llbracket \phib\rrbracket\llbracket \phic\rrbracket)_\text{asym},$}\\
\text{\small${\phiabc}^{\!\!\!\!\!\!\!\!\!3\gamma\,\,\,}$}&\text{\small$\,=(\llbracket \phia\rrbracket\sigz\llbracket \phib\rrbracket\llbracket \phic\rrbracket)_\text{sym},$}\\
\text{\small${\phiabc}^{\!\!\!\!\!\!\!\!\!3\delta\,\,\,}$}&\text{\small$\,=(\llbracket \phia\rrbracket\sigz\llbracket \phib\rrbracket\llbracket \phic\rrbracket)_\text{asym},$}\\
\begin{split}
\text{\small${\phiabc}^{\!\!\!\!\!\!\!\!\!3\epsilon\,\,\,}$}&\text{\small$\,=\big((\llbracket \phia\rrbracket(\sigx\!+\!i\sigy)\llbracket \phib\rrbracket\llbracket \phic\rrbracket)_\text{sym},$}\\
&\qquad\text{\small$(\llbracket \phia\rrbracket(\sigx\!-\!i\sigy)\llbracket \phib\rrbracket\llbracket \phic\rrbracket)_\text{sym}\big)^{\!\T}\!\!\!,$}
\end{split}\\
\begin{split}
\text{\small${\phiabc}^{\!\!\!\!\!\!\!\!\!3\zeta\,\,\,}$}&\text{\small$\,=\big((\llbracket \phia\rrbracket(\sigx\!+\!i\sigy)\llbracket \phib\rrbracket\llbracket \phic\rrbracket)_\text{asym},$}\\
&\qquad\text{\small$(\llbracket \phia\rrbracket(\sigx\!-\!i\sigy)\llbracket \phib\rrbracket\llbracket \phic\rrbracket)_\text{asym}\big)^{\!\T}\!\!\!,$}\!\!
\end{split}
\end{align}
with `sym' and `asym' denoting the symmetric and the antisymmetric parts of the matrices. How these irreps transform is given in TABLE~\ref{tab:cubicprods}. Every irrep except ${\phiabc}^{\!\!\!\!\!\!\!\!\!3\alpha\,\,\,}$ and ${\phiabc}^{\!\!\!\!\!\!\!\!\!3\beta\,\,\,}$ transforms nontrivially under $\genx$, hence they cannot couple in the Majorana mass term. Since the Majorana mass matrix is symmetric, we discard the antisymmetric contribution ${\phiabc}^{\!\!\!\!\!\!\!\!\!3\beta\,\,\,}$. Hence, we are left with only ${\phiabc}^{\!\!\!\!\!\!\!\!\!3\alpha\,\,\,}$, which is equal to $\phi$ (\ref{eq:assignment}). Its expression is given by
\begin{align}\label{eq:phiexpr1}
\begin{split}
\text{\small$\phi\,\,$}&\text{\small$=\frac{1}{4}\big({\phia}_1({\phib}_1{\phic}_1+{\phib}_2^{\!\!*}{\phic}_2^{\!\!*})+{\phia}_2({\phib}_1{\phic}_2-{\phib}_2^{\!\!*}{\phic}_1^{\!\!*})$}\\
&\text{\small$\qquad\,\,+{\phia}_2^{\!\!*}({\phib}_2{\phic}_1+{\phib}_1^{\!\!*}{\phic}_2^{\!\!*})-{\phia}_1^{\!\!*}({\phib}_2{\phic}_2-{\phib}_1^{\!\!*}{\phic}_1^{\!\!*}),$}\\
&\text{\small$\qquad-i{\phia}_1({\phib}_1{\phic}_2-{\phib}_2^{\!\!*}{\phic}_1^{\!\!*})-i{\phia}_2({\phib}_1{\phic}_1+{\phib}_2^{\!\!*}{\phic}_2^{\!\!*})$}\\
&\text{\small$\qquad\,\,-i{\phia}_2^{\!\!*}({\phib}_2{\phic}_2-{\phib}_1^{\!\!*}{\phic}_1^{\!\!*})+i{\phia}_1^{\!\!*}({\phib}_2{\phic}_1+{\phib}_1^{\!\!*}{\phic}_2^{\!\!*}),$}\\
&\text{\small$\qquad+i{\phia}_1({\phib}_1{\phic}_1+{\phib}_2^{\!\!*}{\phic}_2^{\!\!*})-i{\phia}_2({\phib}_1{\phic}_2-{\phib}_2^{\!\!*}{\phic}_1^{\!\!*})$}\\
&\text{\small$\qquad\,\,+i{\phia}_2^{\!\!*}({\phib}_2{\phic}_1+{\phib}_1^{\!\!*}{\phic}_2^{\!\!*})+i{\phia}_1^{\!\!*}({\phib}_2{\phic}_2-{\phib}_1^{\!\!*}{\phic}_1^{\!\!*}\big)^\T\!\!.$}
\end{split}
\end{align}

\section{$\text{TM}_1$ mixing and the oscillation data}
\label{sec:appthree}

$\text{TM}_1$ mixing\cite{hep-ph/0607302,0812.0436,1004.2798,1108.4278,1212.3247, 1304.6264,1306.2358,1312.4401,1509.06915, 1802.00425, 1709.02136,1705.02027,1305.4846, 1512.07531,1607.05276,1802.00425, 2003.00506} preserves the first column of the tribimaximal mixing\cite{hep-ph/0202074}. We use the $\text{TM}_1$ parametrization presented in \cite{1802.00425},
\begin{align}\label{eq:tma}
U_{\text{TM}_1} (\theta,\zeta) &
= \left(\begin{matrix} \frac{\sqrt{2}}{\sqrt{3}} &  \frac{\cos \theta}{\sqrt{3}} &  \frac{\sin \theta }{\sqrt{3}}\\[6pt]
\frac{-1}{\sqrt{6}}  &   \frac{\cos \theta}{\sqrt{3}} - \frac{ \sin \theta}{\sqrt{2}} e^{i\zeta} & \frac{\sin \theta }{\sqrt{3}}+ \frac{\cos \theta}{\sqrt{2}}e^{i\zeta}  \\[6pt]
\frac{-1}{\sqrt{6}}  & \frac{\cos \theta}{\sqrt{3}} + \frac{ \sin \theta}{\sqrt{2}} e^{i\zeta} & \frac{\sin \theta }{\sqrt{3}}- \frac{\cos \theta}{\sqrt{2}}e^{i\zeta} 
\end{matrix}\right).
\end{align}
We express the neutrino mixing angles and the Dirac CP phase in terms of the $\text{TM}_1$ parameters $\theta$ and $\zeta$ using the following equations,
\begin{align}
\sin^2 \theta_{13}& = \frac{\sin^2 \theta}{3}\,, \label{eq:theta13}\\
\sin^2 \theta_{12}& = 1-\frac{2}{3-\sin^2 \theta}\,, \label{eq:theta12}\\
\sin^2 \theta_{23}& = \frac{1}{2}\left(1+\frac{\sqrt{6}\sin 2\theta \cos \zeta}{3-\sin^2 \theta}\right)\,,\label{eq:theta23}\\
\begin{split}
J_\text{CP}&=\text{im}(U_{e2}U_{\mu3}U^*_{e3}U^*_{\mu2})\\
&=\frac{1}{8}\sin \delta_\text{CP} \sin 2\theta_{12}\sin 2\theta_{23}\sin 2\theta_{13}\cos \theta_{13}\\
& = \frac{\sin 2 \theta \sin \zeta}{6 \sqrt{6}} \label{eq:jcp},
\end{split}
\end{align}
where $J_\text{CP}$ is the Jarlkog's rephasing invariant\cite{Jarlskog:1985ht,Jarlskog:1985cw}, which parametrizes CP violation.

$\text{TM}_1$ mixing scheme with $\theta=\frac{\pi}{12}$ and $\zeta=\frac{\pi}{2}$ was proposed in \cite{1205.0761} to account for the nonzero reactor angle. Ref.~\cite{1912.02451} presented a model in the framework of the auxiliary group that resulted in $\Utm$. Substituting $\theta=\frac{\pi}{12}$ and $\zeta=-\frac{\pi}{2}$ in (\ref{eq:theta13})-(\ref{eq:jcp}), we obtain
\begin{align}
\sin^2 \theta_{13}& = \frac{1}{3}\sin^2 \frac{\pi}{12}=0.0223\,,\label{eq:13}\\
\sin^2 \theta_{12}& = 1-\frac{2}{3-\sin^2\frac{\pi}{12}}=0.318\,,\label{eq:12}\\
\sin^2 \theta_{23}& = \frac{1}{2}\,,\label{eq:23}\\
J_\text{CP}& = -\frac{1}{12 \sqrt{6}},\quad \delta_{\text{CP}}=-\frac{\pi}{2}\label{eq:cp}.
\end{align}
Our predictions (\ref{eq:13}) and (\ref{eq:12}) are within $1\sigma$ errors of the global fit\cite{2006.11237} values given by $\sin^2 \theta_{13}=0.022000^{+0.00069}_{-0.00062}$ and $\sin^2 \theta_{12}=0.318^{+16}_{-16}$. Since our model preserves $\mu\tau$-reflection symmetry\cite{hep-ph/0203209, hep-ph/0206292,hep-ph/0207352,hep-ph/0210197,hep-ph/0305309}, we obtain maximal atmospheric mixing ($\sin^2 \theta_{23}= \frac{1}{2}$) and maximal CP violation ($\delta_{\text{CP}}=-\frac{\pi}{2}$). These predictions are in tension with the global fit at $2\sigma$ level even though they are consistent at $3\sigma$ level\cite{2006.11237} given by $\sin^2 \theta_{23}=0.434-0.610$ and $\delta_{\text{CP}}=0.71\pi-1.99\pi$. 

\section{The renormalizable flavon potential} 
\label{sec:appfour}

We follow the procedure developed in \cite{2306.07325} to construct the general renormalizable flavon potential. In this procedure, we infer the existence of specific minima of the potential using symmetry arguments without the need for explicit extremization of the potential. These arguments utilize theorems by Michel\cite{MICHEL1971, MICHEL200111}, reduction lemma by Golubitsky and Stewart\cite{GOLUBITSKY1988} and the compatibility condition obtained by the author\cite{2306.07325}. 

We analyze the potentials of the elementary flavons one by one. We show that the potentials of $\chi$ and $\xia$ have the required minima. On the other hand, the potentials constructed with the rest of the flavons have accidental continuous symmetries. To break the continuous symmetries for each of these elementary flavons, we introduce driving flavons. We construct the potential of these driving flavons and obtain its minimum, and then construct the cross terms of the elementary flavon with the driving flavons. These cross terms break the continuous symmetries, resulting in the required discrete minimum for the flavon. Once this procedure is completed for all the elementary flavons, we comprehensively search for all possible cross terms among the flavons, including the driving flavons. We show that all cross terms are compatible\cite{2306.07325} with the stabilizers of the individual flavons; hence, their presence does not spoil the minimum obtained earlier. Thus, by the end of this section, we will have constructed the general renormalizable potential $\mathcal{V}$ involving all the flavons and shown that it has a minimum that corresponds to our vevs (\ref{eq:elemvevs}). 

The stabilizers associated with the elementary flavons as well as those associated with driving flavons, which we will identify in this section, remain unaffected by a general change in the coefficients of terms in $\mathcal{V}$. These stabilizers generate constraints among the components of the vevs. For example, $\Ht_\chi= S_{3\chi}$, TABLE~\ref{tab:elemstabilizers}, generates five constraints among the components of $\langle\chi\rangle$ (\ref{eq:elemvevs}). Since the stabilizers are coefficient-independent, the resulting constraints are also coefficient-independent even though the norms of the vevs change as a function of the coefficients. Such constraints were named homogeneous linear intrinsic constraints (HLICs) in \cite{2306.07325}. The predictions we make in Section~\ref{sec:phen} are the result of the HLICs in our vevs (\ref{eq:elemvevs}).

\subsection{The potential involving $\chi$}

Using the quadratic invariant $|\chi|^2$ and the quartic invariant  $|\chi|^4$, we construct 
\be\label{eq:chipot1}
\text{\small$\breve{\mathcal{V}}_\chi=c_1m^2 |\chi|^2+c_2|\chi|^4$}.
\ee
For $c_2>0$ and $c_1<0$, the minimum of this potential is a $5$-sphere. This can be made manifest by rewriting $\breve{\mathcal{V}}_\chi$ as
\be\label{eq:chipot2}
\text{\small$\breve{\mathcal{V}}_\chi=k_1(|\chi|^2-r_{\!\chi}^2)^2-k_1r_{\!\chi}^4$}\,,
\ee
with $r_{\!\chi}$ and $k_1$ being positive real numbers. The above potential has a minimum when $|\chi|=r_{\!\chi}$. Here, $r_{\!\chi}$ forms the radius of the $5$-sphere. By comparing (\ref{eq:chipot1}) and (\ref{eq:chipot2}), we can express $r_{\!\chi}$ and $k_1$ as functions of $c_1$ and $c_2$. We call $r_{\!\chi}$ and $k_1$ the arbitrary constants, and $c_1$ and $c_2$ the coefficients. In \cite{2306.07325}, we denoted the arbitrary constants as functions of the coefficients explicitly, e.g.,~$r_{\!\chi}(c_1,c_2)$ and $k_1(c_1,c_2)$. In this paper, we do not show their functional dependence on the coefficients for the sake of brevity, e.g.,~$r_{\!\chi}$ and $k_1$.

The symmetry group\footnote{In this Appendix, we use the notation introduced in \cite{2306.07325}. $\Tb$ denotes the continuous symmetry group of {\it Transitive} action on the stationary manifold (minimum). $\Ga$ denotes the group obtained by the breaking of $\Tb$ by the newly added terms. $\Ha$ and $\Sa$ denote the pointwise and setwise stabilizers, respectively, of the minimum of the newly obtained potential. When the multiplet is an irrep, the symmetry groups associated with the minimum are given by $\Ht=\Ha$ and $\Tt=\Sa/\Ha$. Please see Section~C and FIG.~3 in \cite{2306.07325} for more details.} of the minimum of $\breve{\mathcal{V}}_\chi$, i.e.,~the $5$-sphere, is $\Tb_\chi=O(6)_\chi$. We break the continous group $\Tb_\chi$ to the discrete group $\Ga_\chi=((S_4\times Z_3)\rtimes Z_2)_\chi$ by adding the cubic invariant $\text{re}(\chi^\T\chi^{2\zeta})$ and the quartic invariant $|\chi^{2\delta}|^2$ to $\breve{\mathcal{V}}_\chi$ (\ref{eq:chipot1}), 
\be\label{eq:chipot3}
\text{\small$\tilde{\mathcal{V}}_\chi=\breve{\mathcal{V}}_\chi+c_3 m\,\text{re}(\chi^\T\chi^{2\zeta})+c_4|\chi^{2\delta}|^2$}.
\ee
The alignment $\langle\chi\rangle$ (\ref{eq:elemvevs}) breaks $\Ga_\chi$ to its stabilizer $\Ha_\chi=S_{3\chi}$, TABLE~\ref{tab:elemstabilizers}. Every degree of freedom of $\langle\chi\rangle$ except its norm is determined by $\Ha_\chi$. In other words, the stabilizer $\Ha_\chi$ fully determines $\langle\chi\rangle$ in the $5$-sphere. Therefore, symmetry arguments\cite{MICHEL1971, MICHEL200111, GOLUBITSKY1988} guarantee that this alignment is a stationary point. To make this more apparent and also to show that the stationary point can be obtained as a minimum, we rewrite $\tilde{\mathcal{V}}_\chi$ (\ref{eq:chipot3}) in the form
\be\label{eq:chipot4}
\text{\small$\tilde{\mathcal{V}}_\chi=\breve{\mathcal{V}}_\chi+k_{2}|\chi^{2\zeta}\!+\frac{r_{\!\chi}}{\sqrt{3}}\chi^*|^2+k_3|\chi^{2\delta}|^2+\frac{k_2}{3}|\chi^{2\alpha}|^2$},
\ee
where $\breve{\mathcal{V}}_\chi$ has the form (\ref{eq:chipot2}). The above potential attains its minimum when $|\chi|=r_{\!\chi}$, $\chi^{2\zeta}=-\frac{r_{\!\chi}}{\sqrt{3}}\chi^*$, $\chi^{2\delta}=0$ and $\chi^{2\alpha}=0$. It can be shown that these equations have a discrete set of solutions, and $\langle\chi\rangle$ is one among them. (\ref{eq:chipot3}) contains two quartic invariants $|\chi|^4$ and $|\chi^{2\delta}|^2$ while (\ref{eq:chipot4}) contains four quartic invariants, $|\chi|^4$, $|\chi^{2\zeta}|^2$, $|\chi^{2\delta}|^2$ and $|\chi^{2\alpha}|^2$. We are able to equate (\ref{eq:chipot3}) and (\ref{eq:chipot4}) because we have constructed (\ref{eq:chipot4}) in such a way that its quartic part can be expressed as a linear combination of $|\chi|^4$ and $|\chi^{2\delta}|^2$ only. (\ref{eq:chipot3}) contains $4$ coefficients $c_1, .., c_4$. Correspondingly, (\ref{eq:chipot4}) contains $4$ arbitrary constants $r_{\!\chi}$, $k_1$, $k_2$, $k_3$. By comparing (\ref{eq:chipot3}) and (\ref{eq:chipot4}), we can obtain the arbitrary constants as functions of the coefficients. Finally\footnotemark[\value{footnote}], we have $\Ht_\chi=\Ha_\chi=S_{3\chi}$.

It can be shown that among the various quartic invariants that we can construct, only three are linearly independent. In (\ref{eq:chipot3}), we have utilized two. We construct the general renormalizable potential by adding $|\chi^{2\alpha}|^2$ as the third quartic invariant to (\ref{eq:chipot3}),
\be\label{eq:chipot5}
\text{\small$\mathcal{V}_\chi=\tilde{\mathcal{V}}_\chi+c_5|\chi^{2\alpha}|^2$}.
\ee
$|\chi^{2\alpha}|^2$ is compatible\cite{2306.07325} with $\Ht_\chi$ since it remains invariant it. Hence, even after the addition of $|\chi^{2\alpha}|^2$ the minimum of the potential continues to have $\Ht_\chi$ as its stabilizer, i.e., the constraints among the components of $\langle\chi\rangle$, which are generated by $\Ht_\chi$, remain unaffected by the addition of $|\chi^{2\alpha}|^2$. We make this manifest by rewriting $\mathcal{V}_\chi$ (\ref{eq:chipot5}) as
\be\label{eq:chipot6}
\text{\small$\mathcal{V}_\chi=\tilde{\mathcal{V}}_\chi+k_4|\chi^{2\alpha}|^2$},
\ee
where $\tilde{\mathcal{V}}_\chi$ has the form (\ref{eq:chipot4}). $\mathcal{V}_\chi$ (\ref{eq:chipot6}) has manifestly a discrete minimum at $\langle\chi\rangle$. $\mathcal{V}_\chi$ (\ref{eq:chipot6}) contains five arbitrary constants $r_{\!\chi}, k_1, .., k_4$, which can be expressed as functions of the five coefficients $c_1, .., c_5$ present in $\mathcal{V}_\chi$ (\ref{eq:chipot5}). Even though the norm of the vev, $r_{\!\chi}$, is a function of the coefficients, the constraints among the components of the vev remain unaffected by a general variation of the coefficients. 

We are able to rewrite the general potential (\ref{eq:chipot5}) in the form (\ref{eq:chipot6}) in terms of the arbitrary constants to make $\langle\chi\rangle$ manifest precisely because the symmetries of $\langle\chi\rangle$ guarantee its existence as a stationary point of the potential. In the rest of this Appendix, we provide the potentials constructed with the various flavons in two forms: the first form similar to (\ref{eq:chipot5}) containing the coefficients ($c_i$) and the second form similar to (\ref{eq:chipot6}) containing the arbitrary constants (the norms and $k_i$). The number of arbitrary constants will always be equal to the number of coefficients and we will always be able to express the arbitrary constants as functions of the coefficients.

\subsection{The potential involving $\xia$}

Using $\xia$, we construct
\begin{align}
\text{\small$\breve{\mathcal{V}}_{\xia}$}&\text{\small$=c_{6}m^2|\xia|^2+c_{7}|\xia|^4$}\label{eq:xiapot1}\\
&\text{\small$=k_{5}(|\xia|^2-r_{\!\xia}^2)^2-k_5 r_{\!\xia}^4$}\label{eq:xiapot2}
\end{align}
This potential has $O(2)$ symmetry and has a circle with radius $r_{\!\xia}$ as its minimum with $\Tb_{\xia}=O(2)_{\xia}$. Comparing the above equations, the arbitrary constants $r_{\!\xia}$ and $k_5$ can be obtained as functions of the coefficients $c_6$ and $c_7$.

The general potential of $\xia$ is obtained by adding the cubic invariant $\text{re}(\xia^3)$ to $\breve{\mathcal{V}}_{\xia}$ (\ref{eq:xiapot1}),
\be\label{eq:xiapot3}
\text{\small$\mathcal{V}_{\xia}=\breve{\mathcal{V}}_{\xia}+c_{8}\,m\,\text{re}(\xia^3)$}.
\ee
The cubic term breaks $\Tb_{\xia}$ to $\Ga_{\xia}=(Z_3\rtimes Z_2)_{\xia}$. The alignment $\langle\xia\rangle$ (\ref{eq:elemvevs}) breaks $\Ga_{\xia}$ to its stabilizer $\Ha_{\xia}=Z_{2\xia}$, TABLE~\ref{tab:elemstabilizers}. $\Ht_{\xia}$ fully determines $\langle\xia\rangle$ in the circle. Hence, symmetry arguments guarantee that $\langle\xia\rangle$ forms a stationary point of $\mathcal{V}_{\xia}$ (\ref{eq:xiapot3}). To make this apparent, we rewrite $\mathcal{V}_{\xia}$ (\ref{eq:xiapot3}) as
\be
\text{\small$\mathcal{V}_{\xia}=\breve{\mathcal{V}}_{\xia}+k_{6}|\xia^2-r_{\!\xia}\xia^*|^2$},
\ee
where $\breve{\mathcal{V}}_{\xia}$ has the form (\ref{eq:xiapot2}). At a discrete set of points, which includes $\langle\xia\rangle$, both $(|\xia|^2-r_{\!\xia}^2)$ and $|\xia^2-r_{\!\xia}\xia^*|$ vanish, which corresponds to the minima of the potential. Here also, we may obtain the arbitrary constants $r_{\!\xia}$, $k_{5}$ and $k_{6}$ as functions of $c_{6}$, $c_{7}$ and $c_{8}$. Finally, we have $\Ht_{\xia}=\Ha_{\xia}=Z_{2\xia}$.

\subsection{The potential involving $\xib$}

Using $\xib$, we construct
\begin{align}
\text{\small$\breve{\mathcal{V}}_{\xib}$}&\text{\small$=c_{9}m^2|\xib|^2+c_{10}|\xib|^4$}\label{eq:xibpot1}\\
&\text{\small$=k_{7}(|\xib|^2-r_{\!\xib}^2)^2-k_7 r_{\!\xib}^4$}\label{eq:xibpot2}
\end{align}
This potential has a $3$-sphere with radius $r_{\!\xib}$ as its minimum, having the continuous symmetry $\Tb_{\xib}=O(4)_{\xib}$.

The general renormalizable potential of $\xib$ is given by
\be\label{eq:xibpot3}
\text{\small$\mathcal{V}_{\xib}=\breve{\mathcal{V}}_{\xib}+c_{11}({\xib}^{\!\!\!2\alpha})^2,$}
\ee
where $\breve{\mathcal{V}}_{\xib}$ has the form (\ref{eq:xibpot1}). The quartic term $({\xib}^{\!\!\!2\alpha})^2$ breaks the $\Tb_{\xib}$ to $\Ga_{\xib}=(O(2)_1\times O(2)_2\rtimes Z_2)_{\xib}$ generated by the group actions ${\xib}_1\rightarrow e^{i\theta_1}{\xib}_1$, ${\xib}_1\rightarrow{\xib}^*_1$, ${\xib}_2\rightarrow e^{i\theta_2}{\xib}_2$, ${\xib}_2\rightarrow{\xib}^*_2$ and $\xib\rightarrow \sigx\xib$. The $3$-sphere gets stratified\footnote{Please see the Appendix of \cite{2306.07325} for a brief discussion about the theory of stratification of a manifold under a symmetry group.} under $\Ga_{\xib}$. It can be shown that the $2$-dimensional manifold $\frac{r_{\!\xib}}{\sqrt{2}}(e^{i\theta_1},e^{i\theta_2})^\T$ parametrized by $\theta_1$ and $\theta_2$, which contains the point $\langle\xib\rangle$ (\ref{eq:elemvevs}), is isolated in its stratum in the $3$-sphere. Hence, symmetry arguments guarantee that it is a stationary manifold. To make this apparent and to show that the stationary manifold forms minima, we rewrite $\mathcal{V}_{\xib}$ as
\be\label{eq:xibpot4}
\text{\small$\mathcal{V}_{\xib}=\breve{\mathcal{V}}_{\xib}+k_{8}({\xib}^{\!\!\!2\alpha})^2$},
\ee
where $\breve{\mathcal{V}}_{\xib}$ has the form (\ref{eq:xibpot2}). This potential attains minimum when $|\xib|=r_{\!\xib}$ and ${\xib}^{\!\!\!2\alpha}=0$. Every point in the manifold $\frac{r_{\!\xib}}{\sqrt{2}}(e^{i\theta_1},e^{i\theta_2})^\T$ satisfies these two equations. The arbitrary constants $r_{\!\xib}, k_7$ and $k_8$ in (\ref{eq:xibpot4}) can be obtained as functions of the coefficients $c_9, c_{10}$ and $c_{11}$ in (\ref{eq:xibpot3}). The pointwise stabilizer of the manifold $\frac{r_{\!\xib}}{\sqrt{2}}(e^{i\theta_1},e^{i\theta_2})^\T$ under $\Ga_{\xib}$ is $\Ha_{\xib}=1$\footnote{This pointwise stabilizer, being trivial, does not generate HLICs in the manifold. However, the manifold contains a non-linear constraint, i.e.,~the norms of its two components are equal.}, i.e., we have $\Ht_{\xib}=1$. The group of transitive action on the manifold is $\Tt_{\xib}=\Ga_{\xib}=(O(2)_1\times O(2)_2\rtimes Z_2)_{\xib}$.

{\renewcommand{\arraystretch}{1.15}
	\setlength{\tabcolsep}{5pt}
	\begin{table}[tbp]
	\begin{center}
	\begin{tabular}{|c|c c c c c c |c|}
\hline
&&&&&&&\\[-1em]
	&$\gent$&$\genr$&$\genzn$&$\gencp$&$\genx$&$\genxmu$&$d$\\
\hline
$\mup$ &$\matw_2$&$()^*$&$\sigx()^*$&$\sigx ()^*$& $\ob \sigx ()^*$&$\sigz$&$4$\\
$\muq$&$1$&$1$&$\sigx$&$\sigx$&$\sigx$&$\sigz$&$2$\\
\hline
${\mup}^{\!\!\!\!2\alpha}$&$1$&$1$&$-1$&$-1$&$-1$&$1$&$1$\\
${\mup}^{\!\!\!\!2\beta}$&$\ob$&$()^*$&$1$&$1$&$1$&$-1$&$2$\\
${\mup}^{\!\!\!\!2\gamma}$&$1$&$()^*$&$()^*$&$()^*$&$\om()^*$&$-1$&$2$\\
${\mup}^{\!\!\!\!2\delta}$&$\matw_2^*$&$()^*$&$\sigx()^*$&$\sigx()^*$&$\om\sigx()^*$&$1$&$4$\\
\hline
${\muq}^{\!\!\!2\alpha}$&$1$&$1$&$-1$&$-1$&$-1$&$1$&$1$\\
${\muq}^{\!\!\!2\beta}$&$1$&$1$&$1$&$1$&$1$&$-1$&$1$\\
\hline
\!\!${\mupq}^{\!\!\!\!\!\!\!2\alpha}$\!\!&$\matw_2^*$&$()^*$&$\sigx()^*$&$\sigx()^*$&$\om\sigx()^*$&$1$&$4$\\
\!\!${\mupq}^{\!\!\!\!\!\!\!2\beta}$\!\!&$\matw_2^*$&$()^*$&$\sigx()^*$&$\sigx()^*$&$\om\sigx()^*$&$-1$&$4$\\
\hline
			\end{tabular}
		\end{center}
		\caption{The transformation rules for the driving flavons $\mup$ and $\muq$ and the irreps obtained from them.}
		\label{tab:musector}
\end{table}} 

To break the continuous group $\Tt_{\xib}$, we introduce an auxiliary generator, namely $\genxmu$, and two diving flavons $\mup=({\mup}_1, {\mup}_2)^\T$ and $\muq=({\muq}_1, {\muq}_2)^\T$, which transform under it as well as the other generators introduced in our model. How these driving flavons transform is given in TABLE~\ref{tab:musector}. We define $\mup$ as a complex doublet, which is, in fact, a real irrep of dimension four, i.e.,~$\mup\equiv(\mup, \mup^*)^\T$. On the other hand, $\muq$ is defined as a real irrep of dimension two. The tensor products relevant to our analysis are given below
\begin{align}
\text{\small$\mup\otimes\mup$}&\text{\small$=|\mup|^2\oplus{\mup}^{\!\!\!\!2\alpha}\oplus{\mup}^{\!\!\!\!2\beta}\oplus{\mup}^{\!\!\!\!2\gamma}\oplus{\mup}^{\!\!\!\!2\delta},$}\label{eq:muquad1}\\
\text{\small$\muq\otimes\muq$}&\text{\small$=|\muq|^2\oplus{\muq}^{\!\!\!2\alpha}\oplus{\muq}^{\!\!\!\!2\beta},$}\label{eq:muquad2}\\
\text{\small$\mup\otimes\muq$}&\text{\small$={\mupq}^{\!\!\!\!\!\!\!2\alpha}\oplus{\mupq}^{\!\!\!\!\!\!\!2\beta}$}\,,\label{eq:muquad3}
\end{align}
where 
\begin{align}
\text{\small${\mup}^{\!\!\!\!2\alpha}$}&\text{\small$={\mup^*}_1{\mup}_1-{\mup^*}_2{\mup}_2$},&\text{\small${\mup}^{\!\!\!\!2\beta}$}&\text{\small$={\mup}_1^{\!\!*}{\mup}_2$},\\
\text{\small${\mup}^{\!\!\!\!2\gamma}$}&\text{\small$={\mup}_1{\mup}_2$},&\text{\small${\mup}^{\!\!\!\!2\delta}$}&\text{\small$=({\mup}_1^{\!\!2},{\mup}_2^{\!\!2})^\T$},\\
\text{\small${\muq}^{\!\!\!2\alpha}$}&\text{\small$={\muq}_1^{\!2}-{\muq}_2^{\!2}$},&\text{\small${\muq}^{\!\!\!2\beta}$}&\text{\small$={\muq}_1{\muq}_2$},\\
\text{\small${\mupq}^{\!\!\!\!\!\!\!2\alpha}$}&\text{\small$=({\mup}_1^{\!\!*}{\muq}_1,{\mup}_2^{\!\!*}{\muq}_2)^\T,$}&\text{\small${\mupq}^{\!\!\!\!\!\!\!2\beta}$}&\text{\small$=({\mup}_1^{\!\!*}{\muq}_2,{\mup}_2^{\!\!*}{\muq}_1)^\T$}.\!\!\!
\end{align}
How these irreps transform is also given in TABLE~\ref{tab:musector}. 

\setlength{\tabcolsep}{5pt}
\begin{table}[tbp]
\begin{center}
\begin{tabular}{|c|c|c|c|}
\hline
&\multicolumn{2}{c|}{}&\\[-1em]
\multirow{2}{*}{vev}&\multicolumn{2}{c|}{Generators of the stabilizer}&\!\!Stabi-\!\!\\
\cline{2-3}
&&&\\[-1em]
&Representation&Abstract form&lizer\\
\hline
&&&\\[-1em]
$\!\!\langle\mup\rangle\!\!$ & $\matw_2\sigx,\,\, \matw_2()^*$ &{\footnotesize$\gent\genr\genzn,\,\, \gent\genr$}&\!$K_{4\mup}$\!\\
$\!\!\langle\muq\rangle\!\!$&\!\!$\sigx$\!\!\! & {\footnotesize$\genzn$}&\!$Z_{2\muq}$\!\\
$\!\!\langle\nup\rangle\!\!$&\!\!$\matw_2^*\sigx,\,\, \om\matw_2^*()^*$\!\!\! &\!\!\!{\footnotesize$(\genzn\genx)^5\gencp\genx,\,(\genzn\genx)^4\gencp\genx\gencp$}\!\!\!&\!$K_{4\nup}$\!\\
$\!\!\langle\nuq\rangle\!\!$&\!\!$\sigx$\!\!\! & {\footnotesize$\genzn$}&\!$Z_{2\nuq}$\!\\
$\!\!\langle\nur\rangle\!\!$&\!\!$\sigx$\!\!\! & {\footnotesize$\genzn$}&\!$Z_{2\nur}$\!\\
$\!\!\langle\lamp\rangle\!\!$ & $\matt_4,\,\, \matr_4$ &{\footnotesize$\gent,\,\, \genr$}&\!$S_{4\lamp}$\!\\
$\!\!\langle\lamq\rangle\!\!$ & \!\!$-\matt_4{\matv}_\lambda,\,\,-{\matv}_\lambda\matr_4$\!\! & \!\!{\footnotesize$\gent\genxlam^3,\,\,\genxlam^3\genr$}\!\!&\!$S_{4\lamq}$\!\\
$\!\!\langle\lamr\rangle\!\!$ & \!\!$\matt_4\matr_4, \,{\matv}_\lambda\matt_4({\matv}_\lambda\matr_4)^3$\!\!&\!\!{\footnotesize$\gent\genr,\,\,\genxlam\gent(\genxlam\genr)^3$}\!\!&\!$D_{8\lamr}$\!\\
$\!\!\langle\rhop\rangle\!\!$ &\!\!$\mattau_2^*\sigx,\,\, \mattau_2^*()^*$\!\!\!&\!\!\!{\footnotesize$(\gencp\genx)^3\gencp,\,\gencp^3(\gencp\genx)^3\genxrho$}\!\!\!&$K_{4\rhop}$\\
$\!\!\langle\rhoq\rangle\!\!$ &$-i()^*$&\!\!\!{\footnotesize$(\gencp\genx)^3\gencp$}\!\!\!&\!$Z_{2\rhoq}$\!\\
$\!\!\langle\rhor\rangle\!\!$ &$i()^*$&\!\!\!{\footnotesize$(\gencp\genx)^3\gencp$}\!\!\!&\!$Z_{2\rhor}$\!\\
$\!\!\langle\rhos\rangle\!\!$ &$()^*$&\!\!\!{\footnotesize$\genzn$}\!\!\!&\!$Z_{2\rhos}$\!\\
$\!\!\langle\etap\rangle\!\!$ &$\mattau_2\sigx,\,\, \mattau_2()^*$&\!\!\!{\footnotesize$(\gencp\genx)^3\gencp,\,\gencp^3(\gencp\genx)^3\genxeta$}\!\!\!&$K_{4\etap}$\\
$\!\!\langle\etaq\rangle\!\!$ &$i()^*$ &\!\!\footnotesize$(\gencp\genx)^3\gencp$\!\!&\!$Z_{2\etaq}$\!\\
$\!\!\langle\etar\rangle\!\!$ & \!\!\!\!$-i()^*$\!\!\!\! & \!\!{\footnotesize$(\gencp\genx)^3\gencp$}\!\!&\!$Z_{2\etar}$\!\\
$\!\!\langle\etas\rangle\!\!$ & \!\!\!\!$-()^*$\!\!\!\! & \!\!{\footnotesize$\genxeta^2\genzn$}\!\!&\!$Z_{2\etas}$\!\\
\hline
\end{tabular}
\end{center}
\caption{Stabilizers of the vevs of the driving flavons. They are isomorphic to $K_4$, $Z_2$, $S_4$ and $D_8$.}
\label{tab:flavonstabilizers}
\end{table}

We can see that ${\mup}^{\!\!\!\!2\delta}$ transforms in the same way as ${\xib}^{\!\!\!\!2\delta}$, TABLE~\ref{tab:prodtrof}. By coupling these together using the cross term $|{\xib}^{\!\!\!\!2\delta}-\frac{r_{\!\xib}^2}{r_{\!\mup}^2}{\mup}^{\!\!\!\!2\delta}|^2$, we can break $\Tt_{\xib}$. For $\mup$, we assign the alignment 
\be\label{eq:mupvev}
\langle\mup\rangle=\frac{r_{\!\mup}}{\sqrt{2}}(\om,\ob)^\T.
\ee
Its stabilizer $K_{4\mup}$, TABLE~\ref{tab:flavonstabilizers}, fully determines it up to its norm. Hence, it is guaranteed by symmetry arguments to be a stationary point of the potential of $\mup$. At $\langle\xib\rangle$ (\ref{eq:elemvevs}) and $\langle\mup\rangle$ (\ref{eq:mupvev}), the cross term $|{\xib}^{\!\!\!\!2\delta}-\frac{r_{\!\xib}^2}{r_{\!\mup}^2}{\mup}^{\!\!\!\!2\delta}|^2$ vanishes. 

We construct the potentials of $\mup$ in the same way as that of $\xib$. Similar to $\breve{\mathcal{V}}_{\xib}$, (\ref{eq:xibpot1}), (\ref{eq:xibpot2}), we construct 
\begin{align}
\text{\small$\breve{\mathcal{V}}_{\mup}$}&\text{\small$=c_{12}m^2|\mup|^2+c_{13}|\mup|^4$}\label{eq:muppot1}\\
&\text{\small$=k_{9}(|\mup|^2-r_{\!\mup}^2)^2-k_9 r_{\!\mup}^4$}.\label{eq:muppot2}
\end{align}
This potential has a $3$-sphere with radius $r_{\!\mup}$ as its minimum with $\Tb_{\mup}=O(4)_{\mup}$. Similar to $\breve{\mathcal{V}}_{\xib}$, (\ref{eq:xibpot3}), (\ref{eq:xibpot4}), we construct the general potential of $\mup$,
\begin{align}
\text{\small$\mathcal{V}_{\mup}$}&\text{\small$=\breve{\mathcal{V}}_{\mup}+c_{14}({\mup}^{\!\!\!\!2\alpha})^2$}\label{eq:muppot3}\\
&\text{\small$=\breve{\mathcal{V}}_{\mup}+k_{10}({\mup}^{\!\!\!\!2\alpha})^2$}.\label{eq:muppot4}
\end{align}
The newly added term breaks the $\Tb_{\mup}$ to $\Ga_{\mup}=(O(2)_1\times O(2)_2\rtimes Z_2)_{\mup}$ akin to $\Ga_{\xib}$ that we obtained earlier in relation to $\xib$. We obtain the manifold $\frac{r_{\!\mup}}{\sqrt{2}}(e^{i\theta_1},e^{i\theta_2})^\T$ as the minimum of $\mathcal{V}_{\mup}$ along with the symmetry groups $\Ht_{\mup}=\Ha_{\mup}=1$ and $\Tt_{\mup}=\Ga_{\mup}=(O(2)_1\times O(2)_2\rtimes Z_2)_{\mup}$.

For breaking the continuous symmetry $\Tt_{\mup}$, we introduce $\muq$. The irreps ${\mup}^{\!\!\!\!2\delta}$ and ${\mupq}^{\!\!\!\!\!\!\!2\alpha}$ transform in the same way, TABLE~\ref{tab:musector}. Hence, we can utilize the cross term $|{\mup}^{\!\!\!\!2\delta}-\frac{r_{\!\mup}}{r_{\!\muq}}{\mupq}^{\!\!\!\!\!\!\!2\alpha}|^2$ to break $\Tt_{\mup}$. For $\muq$, we assign the vev
\be\label{eq:muqvev}
\langle\muq\rangle=\frac{r_{\!\muq}}{\sqrt{2}}(1,1)^\T,
\ee
whose stabilizer $Z_{2\muq}$, TABLE~\ref{tab:flavonstabilizers}, fully determines it up to its norm. Hence, it is guaranteed to be a stationary point of the potential of $\muq$.  At $\langle\mup\rangle$ (\ref{eq:mupvev}) and $\langle\muq\rangle$ (\ref{eq:muqvev}), the cross term $|{\mup}^{\!\!\!\!2\delta}-\frac{r_{\!\mup}}{r_{\!\muq}}{\mupq}^{\!\!\!\!\!\!\!2\alpha}|^2$ vanishes.

{\renewcommand{\arraystretch}{1.0}
	\setlength{\tabcolsep}{5pt}
	\begin{table}[tbp]
	\begin{center}
	\begin{tabular}{|c|c c c c c c c| c c c c c|}
\hline
&&&&&&&&&&&&\\[-1em]
	&\!$\genzl$\!&\!\!\!$\gene_{\xia}$\!\!\!&\!\!\!$\gene_{\xib}$\!\!\!&\!\!\!$\gene_{\xic}$\!\!\!&\!\!\!$\gene_{\phia}$\!\!\!&\!\!\!$\gene_{\phib}$\!\!\!&\!\!\!$\gene_{\phic}$\!\!\!&\!\!\!$\gene_{\mu}$\!\!\!&\!\!\!$\gene_{\nu}$\!\!\!&\!\!\!$\gene_{\lambda}$\!\!\!&\!\!\!$\gene_{\rho}$\!\!&\!\!\!$\gene_{\eta}$\!\!\\
\hline
$\chi$&$\om$&$1$&$1$&$1$&$1$&$1$&$1$&$1$&$1$&$1$&$1$&$1$\\
$\xia$&$1$&$\om$&$1$&$1$&$1$&$1$&$1$&$1$&$1$&$1$&$1$&$1$\\
$\xib$&$1$&$\ob$&\!$-1$\!&$1$&$1$&$1$&$1$&$1$&$1$&$1$&$1$&$1$\\
$\xic$&$1$&$1$&$1$&\!$-1$\!&$1$&$1$&$1$&$1$&$1$&$1$&$1$&$1$\\
$\phia$&$1$&$1$&$1$&$1$&\!$-1$\!&$1$&$1$&$1$&$1$&$1$&$1$&$1$\\
$\phib$&$1$&$1$&$1$&$1$&$1$&\!$-1$\!&$1$&$1$&$1$&$1$&$1$&$1$\\
$\phic$&$1$&$1$&$1$&$1$&$1$&$1$&\!$-1$\!&$1$&$1$&$1$&$1$&$1$\\
\hline
$\mup$&$1$&$\ob$&$1$&$1$&$1$&$1$&$1$&\!$-1$\!&$1$&$1$&$1$&$1$\\
$\muq$&$1$&$1$&$1$&$1$&$1$&$1$&$1$&\!$-1$\!&$1$&$1$&$1$&$1$\\
\!\!\!$\nup$,$\nuq$,$\nur$\!\!\!&$1$&$1$&$1$&$1$&$1$&$1$&$1$&$1$&\!$-1$\!&$1$&$1$&$1$\\
\!\!\!$\lamp$,$\lamq$,$\lamr$\!\!\!&$1$&$1$&$1$&$1$&$1$&$1$&$1$&$1$&$1$&\!$-1$\!&$1$&$1$\\
$\rhop$&$1$&$1$&$1$&\!$-1$\!&\!$-1$\!&$1$&$1$&$1$&$1$&$1$&\!$-1$\!&$1$\\
\!\!\!$\rhoq$, $\rhor$\!\!\!&$1$&$1$&$1$&$1$&$1$&$1$&$1$&$1$&$1$&$1$&$1$&$1$\\
$\rhos$&$1$&$1$&$1$&$1$&$1$&$1$&$1$&$1$&$1$&$1$&\!$-1$\!&$1$\\
$\etap$&$1$&$1$&$1$&\!$-1$\!&$1$&$1$&\!$-1$\!&$1$&$1$&$1$&$1$&\!$-1$\!\\
\!\!\!$\etaq$, $\etar$\!\!\!&$1$&$1$&$1$&$1$&$1$&$1$&$1$&$1$&$1$&$1$&$1$&$1$\\
$\etas$&$1$&$1$&$1$&$1$&$1$&$1$&$1$&$1$&$1$&$1$&$1$&\!$-1$\!\\
\hline
			\end{tabular}
		\end{center}
		\caption{The action of $\genzl$, $\gene_{\xia}$, ..., $\gene_{\eta}$ on the flavons.}
		\label{tab:eelemflavons}
\end{table}} 

To make these arguments clear, we construct the potentials explicitly. To aid our construction, we define the group element $\gene_\mu=\genzn^2(\genzn\genxmu)^2$. We can show that $\mup$ and $\muq$ change sign under $\gene_\mu$ while every other flavon remains invariant under it, TABLE~\ref{tab:eelemflavons}. Therefore, $\mup$ and $\muq$ can couple with other flavons via quadratic products only, which we obtained in (\ref{eq:muquad1})-(\ref{eq:muquad3}). Using $\muq$, we construct
\begin{align}
\text{\small$\breve{\mathcal{V}}_{\muq}$}&\text{\small$=c_{15}m^2|\muq|^2+c_{16}|\muq|^4$}\label{eq:muqpot1}\\
&\text{\small$=k_{11}(|\muq|^2-r_{\!\muq}^2)^2-k_{11} r_{\!\muq}^4$}.\label{eq:muqpot2}
\end{align}
This potential has a circle with radius $r_{\!\muq}$ as its minimum, having the continuous symmetry $\Tb_{\muq}=O(2)_{\muq}$. The general potential of $\muq$ is
\begin{align}
\text{\small$\mathcal{V}_{\muq}$}&\text{\small$=\breve{\mathcal{V}}_{\muq}+c_{17}({\muq}^{\!\!\!\!2\alpha})^2$}\label{eq:muqpot3}\\
&\text{\small$=\breve{\mathcal{V}}_{\muq}+k_{12}({\muq}^{\!\!\!\!2\alpha})^2$}.\label{eq:muqpot4}
\end{align}
The newly added term breaks the $\Tb_{\muq}$ to $\Ga_{\muq}=D_{8\muq}$, the dihedral group of eight elements generated by the group actions $\muq\rightarrow\sigx\muq$ and $\muq\rightarrow \sigz\muq$. The vev $\langle\muq\rangle$ (\ref{eq:muqvev}) breaks $\Ga_{\muq}$ to $\Ha_{\muq}=Z_{2\muq}$, TABLE~\ref{tab:flavonstabilizers}. The stabilizer $\Ha_{\muq}$ fully determines $\langle\muq\rangle$ in the circle; hence, $\langle\muq\rangle$ is guaranteed to be a stationary point of $\mathcal{V}_{\muq}$ (\ref{eq:muqpot3}). The rewritten potential (\ref{eq:muqpot4}) shows that it can be obtained as a minimum. At the minimum of (\ref{eq:muqpot4}), we have $|\muq|=r_{\!\muq}$ and ${\mup}^{\!\!\!\!2\alpha}=0$, which corresponds to $\langle\muq\rangle$. The arbitrary constants $r_{\!\muq}, k_{11}$ and $k_{12}$ in (\ref{eq:muqpot4}) are functions of the coefficients $c_{15}, c_{16}$ and $c_{17}$ (\ref{eq:muqpot3}). We have $\Ht_{\muq}=\Ha_{\muq}=Z_{2\muq}$. We also have $\Tt_{\muq}=1$ since the minimum is a single point rather than a manifold. 

Utilizing both $\mup$ and $\muq$, we construct
\be\label{eq:mupot1}
\text{\small$\breve{\mathcal{V}}_{\mu}=\mathcal{V}_{\mup}+\mathcal{V}_{\muq}$}.
\ee
Here, $\mu$ denotes the irreps $\mup$ and $\muq$ combined, i.e.,~$\mu=(\mup,\muq)^\T$. This potential has a minimum consisting of the manifold $\frac{r_{\!\mup}}{\sqrt{2}}(e^{i\theta_1},e^{i\theta_2})^\T$ for $\mup$ and the point $\frac{r_{\!\muq}}{\sqrt{2}}(1,1)^\T$ for $\muq$. The corresponding symmetry groups are $\Hb_\mu=\Ht_{\mup}\times\Ht_{\muq}=\Ht_{\muq}=Z_{2\muq}$ and $\Tb_\mu=\Tt_{\mup}\times\Tt_{\muq}=\Tt_{\mup}=(O(2)_1\times O(2)_2\rtimes Z_2)_{\mup}$. The continuous group $\Tb_{\mu}$ is broken by the cross term between $\mup$ and $\muq$,
\be\label{eq:mupot2}
\text{\small$\tilde{\mathcal{V}}_{\mu}=\breve{\mathcal{V}}_{\mu}+c_{18}\,\text{re}({\mup}^{\!\!\!\!2\delta\dagger}{\mupq}^{\!\!\!\!\!\!\!2\alpha}),$}
\ee
to the discrete group $\Ga_\mu=(D_{6_1}\times D_{6_2}\rtimes Z_2)_{\mup}$ generated by the group actions ${\mup}_1\rightarrow \om{\mup}_1$, ${\mup}_1\rightarrow{\mup}^*_1$, ${\mup}_2\rightarrow \om{\mup}_2$, ${\mup}_2\rightarrow{\mup}^*_2$ and $\mup\rightarrow \sigx\mup$. The manifold $\frac{r_{\!\mup}}{\sqrt{2}}(e^{i\theta_1},e^{i\theta_2})^\T$ gets stratified under $\Ga_\mu$. Our vev $\langle\mup\rangle=\frac{r_{\!\mup}}{\sqrt{2}}(\om,\ob)^\T$ (\ref{eq:mupvev}) is a point in the manifold, and it breaks $\Ga_\mu$ to its stabilizer $\Ha_{\mu}=(K_4\rtimes Z_2)_{\mup}$, generated by ${\mup}_1\rightarrow \ob{\mup}_1^*$, ${\mup}_2\rightarrow \om{\mup}_2^*$ and $\mup\rightarrow \matw_2\sigx\mup$. Since the stabilizer $\Ha_{\mu}$ fully determines $\langle\mup\rangle$ in the manifold, symmetry arguments guarantee that it is a stationary point of the potential. To show that it can be obtained as a minimum, we rewrite $\tilde{\mathcal{V}}_{\mu}$ (\ref{eq:mupot2}) as
\be\label{eq:mupot3}
\setlength{\jot}{0pt}
\begin{split}
\text{\small$\tilde{\mathcal{V}}_{\mu}$}&\text{\small$=\breve{\mathcal{V}}_{\mu}+k_{13}|{\mup}^{\!\!\!\!2\delta}-\frac{r_{\!\mup}}{r_{\!\muq}}{\mupq}^{\!\!\!\!\!\!\!2\alpha}|^2$}\\
&\quad\text{\small$+\frac{k_{13}}{4}({\mup}^{\!\!\!\!2\alpha}-\frac{r_{\!\mup}^2}{r_{\!\muq}^2}{\muq}^{\!\!\!2\alpha})^2+\frac{k_{13}}{4}(|\mup|^2\!-\frac{r_{\!\mup}^2}{r_{\!\muq}^2}|\muq|^2)^2.\!\!\!$}
\end{split}
\ee
The above potential contains seven arbitrary constants $r_{\!\mup}, r_{\!\muq}, k_9, .., k_{13}$, which can be expressed as functions of the seven coefficients $c_{12}, .., c_{18}$. Since the arbitrary constants are all positive, the potential attains its minimum when the various terms in it vanish. It can be shown that they do vanish at our vevs $\langle\mup\rangle$ (\ref{eq:mupvev}) and $\langle \muq\rangle$ (\ref{eq:muqvev}). For the combined vev $\langle\mu\rangle=(\langle\mup\rangle, \langle\muq\rangle)^\T$, we have $\Ht_\mu=\Ha_\mu\times\Hb_\mu=(K_4\rtimes Z_2)_{\mup}\times Z_{2\muq}$. We also have $\Tt_\mu=1$. 

The stabilizer associated with $\muq$ that we obtained here, i.e.,~$(K_4\rtimes Z_2)_{\mup}$, is different from that we provided in TABLE~\ref{tab:flavonstabilizers}, i.e.,~$K_{4\mup}$. In TABLE~\ref{tab:flavonstabilizers}, we listed the stabilizers corresponding to the various flavons under the assumption that the potentials of these flavons do not contain accidental symmetries. In reality, for many flavons, such as $\mup$, the renormalizable potentials do contain accidental continuous symmetries; hence, we introduce other flavons, such as $\muq$, to break them. In such a scenario, we may obtain a stabilizer different from what we expect in relation to the potential of the original flavon alone, e.g.,~$(K_4\rtimes Z_2)_{\mup})\neq K_{4\mup}$. However, the fixed point subspace\cite{2306.07325} remains unique, e.g.,~$\text{fix}((K_4\rtimes Z_2)_{\mup})=\text{fix}(K_{4\mup})$ which is nothing but the space of all points of the form $\mup=a(\om,\ob)^\T$ where $a$ is a real number.

The general potential of $\mup$ and $\muq$ is given by
\be\label{eq:mupot4}
\text{\small$\mathcal{V}_{\mu}=\tilde{\mathcal{V}}_{\mu}+c_{19}{\mup}^{\!\!\!\!2\alpha}{\muq}^{\!\!\!\!2\alpha}\!+c_{20}|\mup|^2|\muq|^2$}
\ee
Since ${\mup}^{\!\!\!\!2\alpha}$ and ${\muq}^{\!\!\!\!2\alpha}$ vanish at our vev $\langle \mu\rangle$, i.e.,~at (\ref{eq:mupvev}) and (\ref{eq:muqvev}), the term ${\mup}^{\!\!\!\!2\alpha}{\muq}^{\!\!\!\!2\alpha}$ is compatible with $\Ht_\mu$, according to corollary~B in \cite{2306.07325}. Note that corollary~B in \cite{2306.07325} was the main result in \cite{2011.11653}. The term $|\mup|^2|\muq|^2$ is compatible since it is a product of norms. Therefore, the addition of the above two terms does not spoil $\langle \mu\rangle$, i.e.,~$\Ht_\mu$ remains unaffected. To make this manifest, we may rewrite (\ref{eq:mupot4}) as
\be\label{eq:mupot5}
\text{\small$\mathcal{V}_{\mu}=\tilde{\mathcal{V}}_{\mu}+k_{14}({\mup}^{\!\!\!\!2\alpha}-\frac{r_{\!\mup}^2}{r_{\!\muq}^2}{\muq}^{\!\!\!2\alpha})^2+k_{15}(|\mup|^2-\frac{r_{\!\mup}^2}{r_{\!\muq}^2}|\muq|^2)^2$},
\ee
where $\tilde{\mathcal{V}}_{\mu}$ has the form (\ref{eq:mupot3}). $\mathcal{V}_{\mu}$ (\ref{eq:mupot5}) contains nine arbitrary constants $r_{\!\mup}, r_{\!\muq}, k_9, .., k_{15}$, which are functions of the nine coeffiecients $c_{12}, .., c_{20}$ present in (\ref{eq:mupot4}).

We combine the potentials $\mathcal{V}_{\xib}$ and $\mathcal{V}_\mu$ to obtain
\be\label{eq:xibmupot1}
\text{\small$\breve{\mathcal{V}}_{\xib\mu}=\mathcal{V}_{\xib}+\mathcal{V}_{\mu}$},
\ee
where $\xib\mu$ denotes the combined multiplet of $\xib$, $\mup$ and $\muq$, i.e.,~$\xib\mu=(\xib, \mup, \muq)^\T$. This potential contains $12$ arbitrary constants and $12$ coefficients. It has a minimum consisting of the manifold $\frac{r_{\!\xib}}{\sqrt{2}}(e^{i\theta_1},e^{i\theta_2})^\T$ for $\xib$, the point $\frac{r_{\!\mup}}{\sqrt{2}}(\om,\ob)^\T$ for $\mup$ and the point $\frac{r_{\!\muq}}{\sqrt{2}}(1,1)^\T$ for $\muq$. The symmetry groups of this minimum are $\Hb_{\xib\mu}\!=\!\Ht_{\xib}\!\times\!\Ht_\mu\!=\!\Ht_\mu\!=\!(K_4\rtimes Z_2)_{\mup}\!\times\!Z_{2\muq}$ and $\Tb_{\xib\mu}\!=\!\Tt_{\xib}\!\times\!\Tt_{\mu}\!=\!\Tt_{\xib}\!=\!(O(2)_1\times O(2)_2\rtimes Z_2)_{\xib}$. To break the continuous group $\Tb_{\xib\mu}$, we couple $\xib$ with $\mup$ as given below,
\begin{align}
\text{\small$\tilde{\mathcal{V}}_{\xib\mu}$}&\text{\small$=\breve{\mathcal{V}}_{\xib\mu}+c_{21}\text{re}({\xib}^{\!\!\!\!2\delta\dagger}{\mup}^{\!\!\!\!2\delta})$}\label{eq:xibmupot2}\\
&\text{\small$=\breve{\mathcal{V}}_{\xib\mu}+k_{16}|{\xib}^{\!\!\!\!2\delta}-\frac{r_{\!\xib}^2}{r_{\!\mup}^2}{\mup}^{\!\!\!\!2\delta}|^2\!\!$}.\label{eq:xibmupot3}
\end{align}
The newly added term breaks $\Tb_{\xib\mu}$ to $\Ga_{\xib\mu}=(K_{4_1}\times K_{4_2}\rtimes Z_2)_{\xib}$ generated by the group actions ${\xib}_1\rightarrow -{\xib}_1$, ${\xib}_1\rightarrow\ob{\xib}^*_1$, ${\xib}_2\rightarrow -{\xib}_2$, ${\xib}_2\rightarrow\om{\xib}^*_2$ and $\xib\rightarrow \matw_2\sigx\xib$. The manifold $\frac{r_{\!\xib}}{\sqrt{2}}(e^{i\theta_1},e^{i\theta_2})^\T$ gets stratified under $\Ga_{\xib\mu}$. Our vev $\langle\xib\rangle$ (\ref{eq:elemvevs}) breaks $\Ga_{\xib\mu}$ to its stabilizer $\Ha_{\xib\mu}=(K_4\rtimes Z_2)_{\xib}$ generated by the group actions ${\xib}_1\rightarrow \ob{\xib}_1^*$, ${\xib}_2\rightarrow \om{\xib}_2^*$ and $\xib\rightarrow \matw_2\sigx\xib$. The stabilizer $(K_4\rtimes Z_2)_{\xib}$ fully determines $\langle\xib\rangle$ in the manifold; hence, $\langle\xib\rangle$ is guaranteed to be a stationary point of $\tilde{\mathcal{V}}_{\xib\mu}$. The rewritten form of $\tilde{\mathcal{V}}_{\xib\mu}$ (\ref{eq:xibmupot3}) shows that $\langle\xib\rangle$ can be obtained as its minimum. The potentials (\ref{eq:xibmupot2}) and (\ref{eq:xibmupot3}) contain $13$ coefficients and $13$ arbitrary constants, respectively. We also obtain $\Ht_{\xib\mu}=\Ha_{\xib\mu}\times\Hb_{\xib\mu}=(K_4\rtimes Z_2)_{\xib}\times (K_4\rtimes Z_2)_{\mup}\times Z_{2\muq}$. In this analysis, we obtain the stabilizer associated with $\xib$ as $(K_4\rtimes Z_2)_{\xib}$. This group is different from $K_{4\xib}$, TABLE~\ref{tab:elemstabilizers}. However, we have $\text{fix}((K_4\rtimes Z_2)_{\xib})=\text{fix}(K_{4\xib})$.

There are some more cross terms between $\xib$ and the driving flavons that we have not yet included in the potential. By including those terms also, we construct the general potential containing $\xib$, $\mup$ and $\muq$,
\begin{align}\label{eq:xibmupot4}
\begin{split}
\text{\small${\mathcal{V}}_{\xib\mu}$}&\text{\small$=\tilde{\mathcal{V}}_{\xib\mu}+c_{22}\,\text{re}({\xib}^{\!\!\!\!2\delta\dagger}{\mupq}^{\!\!\!\!\!\!\!2\alpha})+c_{23}{\xib}^{\!\!\!\!2\alpha}{\mup}^{\!\!\!\!2\alpha}$}\\
&\text{\small$\quad\,\,+c_{24}{\xib}^{\!\!\!\!2\alpha}{\muq}^{\!\!\!\!2\alpha}+c_{25}|\xib|^2|\mup|^2\!+c_{26}|\xib|^2|\muq|^2$}.\!\!\!
\end{split}
\end{align}
Let us verify if the newly added cross terms are compatible with $\Ht_{\xib\mu}$. The $25^\text{th}$ and $26^\text{th}$ terms are compatible since they are products of norms. It is straightforward to show that ${\xib}^{\!\!\!\!2\alpha}$, ${\mup}^{\!\!\!\!2\alpha}$ and ${\muq}^{\!\!\!\!2\alpha}$ vanish at $\langle\xib\rangle$ (\ref{eq:elemvevs}), $\langle\mup\rangle$ (\ref{eq:mupvev}) and $\langle\muq\rangle$ (\ref{eq:muqvev}). Hence, by corollary~B\cite{2306.07325}, the $23^\text{rd}$ and $24^\text{th}$ terms are compatible. To verify if the $22^\text{nd}$ term is compatible with $\Ht_{\xib\mu}$, we need to utilize the general result involving the gradient and the fixed-point subspace given in equation~(15) in \cite{2306.07325}. We may denote a point in the space of $\xib$, $\mup$ and $\muq$ as
\be\label{eq:pointcomp}
\text{\small$\big(({\xib}_{1r}, {\xib}_{1i}, {\xib}_{2r}, {\xib}_{2i}), ({\mup}_{1r}, {\mup}_{1i}, {\mup}_{2r}, {\mup}_{2i}), ({\muq}_1, {\muq}_2)\big)^\T$},
\ee
where ${\xib}_{1r}=\text{re}({\xib}_1)$, ${\xib}_{1i}=\text{im}({\xib}_1)$ and so on. In the above basis, a general point in $\text{fix}(\Ht_{\xib\mu})$ is given by
\be\label{eq:xibmufix}
\begin{split}
&\text{\small$\big(\frac{a}{\sqrt{2}}(\frac{-1}{2}, \frac{\sqrt{3}}{2}, \frac{-1}{2}, \frac{-\sqrt{3}}{2}),$}\\
&\qquad\qquad\qquad\text{\small$\frac{b}{\sqrt{2}}(\frac{-1}{2}, \frac{\sqrt{3}}{2}, \frac{-1}{2}, \frac{-\sqrt{3}}{2}),\! \frac{c}{\sqrt{2}}(1,1)\big)^{\!\T}$}
\end{split}
\ee
where $a$, $b$ and $c$ are real numbers. The expression of the $22^\text{nd}$ term is given by
\be\label{eq:crossexp}
\begin{split}
\text{\small$\text{re}({\xib}^{\!\!\!\!2\delta\dagger}{\mupq}^{\!\!\!\!\!\!\!2\alpha})$}&\text{\small$=({\xib}^{\!\!2}_{1r}-{\xib}^{\!\!2}_{1i}){\mup}_{1r}{\muq}_1-2{\xib}_{1r}{\xib}_{1i}{\mup}_{1i}{\muq}_1$}\\
&\quad\text{\small$-2{\xib}_{2r}{\xib}_{2i}{\mup}_{2i}{\muq}_2+({\xib}^{\!\!2}_{2r}-{\xib}^{\!\!2}_{2i}){\mup}_{2r}{\muq}_2$}.
\end{split}
\ee
Differentiating (\ref{eq:crossexp}) with the components of (\ref{eq:pointcomp}) and substituting the values from (\ref{eq:xibmufix}), we obtain the gradient of the cross term $\text{re}({\xib}^{\!\!\!\!2\delta\dagger}{\mupq}^{\!\!\!\!\!\!\!2\alpha})$ at $\text{fix}(\Ht_{\xib\mu})$ as
\be\label{eq:eq:xibmupartial}
\begin{split}
&\text{\small$\big(\frac{abc}{\sqrt{2}}(\frac{-1}{2}, \frac{\sqrt{3}}{2}, \frac{-1}{2}, \frac{-\sqrt{3}}{2}),$}\\
&\qquad\qquad\quad\text{\small$\frac{a^2c}{2\sqrt{2}}(\frac{-1}{2}, \frac{\sqrt{3}}{2}, \frac{-1}{2}, \frac{-\sqrt{3}}{2}),\frac{a^2b}{2\sqrt{2}}(1,1)\big)^\T$}\!\!\!.
\end{split}
\ee
The above expression can be obtained from (\ref{eq:xibmufix}) with the substitution $a\rightarrow abc$, $b\rightarrow\frac{a^2c}{2}$ and $c\rightarrow\frac{a^2b}{2}$. In other words, the gradient of the term $\text{re}({\xib}^{\!\!\!\!2\delta\dagger}{\mupq}^{\!\!\!\!\!\!\!2\alpha})$ calculated at every point in $\text{fix}(\Ht_{\xib\mu})$ lies in $\text{fix}(\Ht_{\xib\mu})$. Hence, $\text{re}({\xib}^{\!\!\!\!2\delta\dagger}{\mupq}^{\!\!\!\!\!\!\!2\alpha})$ is compatible with $\Ht_{\xib\mu}$. 

Since all the added terms are compatible with $\Ht_{\xib\mu}$, the newly obtained potential ${\mathcal{V}}_{\xib\mu}$ (\ref{eq:xibmupot4}) continues to have a minimum of the form $\langle\xib\rangle$ (\ref{eq:elemvevs}), $\langle\mup\rangle$ (\ref{eq:mupvev}) and $\langle\muq\rangle$ (\ref{eq:muqvev}) with the symmetry group $\Ht_{\xib\mu}$. This is made manifest by rewriting ${\mathcal{V}}_{\xib\mu}$ as 
\begin{align}\label{eq:xibmupot5}
\begin{split}
\text{\small${\mathcal{V}}_{\xib\mu}$}&=\text{\small$\tilde{\mathcal{V}}_{\xib\mu}+k_{17}|{\xib}^{\!\!\!\!2\delta}-\frac{r_{\!\xib}^2}{r_{\!\mup}r_{\!\muq}}{\mupq}^{\!\!\!\!\!\!\!2\alpha}|^2\!\!$}\\
&\quad\,\text{\small$+k_{18} ({\xib}^{\!\!\!\!2\alpha}-\frac{r_{\!\xib}^2}{r_{\!\mup}^2}{\mup}^{\!\!\!\!2\alpha})^2+k_{19} ({\xib}^{\!\!\!\!2\alpha}-\frac{r_{\!\xib}^2}{r_{\!\muq}^2}{\muq}^{\!\!\!\!2\alpha})^2$}\\
&\quad\,\text{\small$+k_{20}(|\xib|^2\!-\!\frac{r_{\!\xib}^2}{r_{\!\mup}^2}|\mup|^2)^2\!+k_{21}(|\xib|^2\!-\!\frac{r_{\!\xib}^2}{r_{\!\muq}^2}|\muq|^2)^2$},\!\!\!
\end{split}
\end{align}
where $\tilde{\mathcal{V}}_{\xib\mu}$ has the form (\ref{eq:xibmupot3}). ${\mathcal{V}}_{\xib\mu}$ (\ref{eq:xibmupot4}) contains $18$ coefficients $c_9,..,c_{26}$. This matches with the number of arbitrary constants $r_{\!\xib}, r_{\!\mup}, r_{\!\muq}, k_7,..,k_{21}$ in (\ref{eq:xibmupot5}), which are functions of these coefficients.

\subsection{The potential involving $\phib$}

Using $\phib$, we construct
\begin{align}
\text{\small$\breve{\mathcal{V}}_{\phib}$}&\text{\small$=c_{27}m^2|\phib|^2+c_{28}|\phib|^4$}\label{eq:phibpot1}\\
&\text{\small$=k_{22}(|\phib|^2-r_{\!\phib}^2)^2-k_{22} r_{\!\phib}^4$},\label{eq:phibpot2}
\end{align}
which has a $3$-sphere with radius $r_{\!\phib}$ as its minimum. We have $\Tb_{\phib}=O(4)_{\phib}$. The general renormalizable potential of $\phib$ is given by
\begin{align}
\text{\small$\mathcal{V}_{\phib}$}&\text{\small$=\breve{\mathcal{V}}_{\phib}+c_{29}({\phib}^{\!\!\!2\alpha})^2$}\label{eq:phibpot3}\\
&\text{\small$=\breve{\mathcal{V}}_{\phib}+k_{23}({\phib}^{\!\!\!2\alpha})^2$}.\label{eq:phibpot4}
\end{align}
Similar to the case of $\mathcal{V}_{\xib}$ (\ref{eq:xibpot3}), (\ref{eq:xibpot4}), the added term breaks $\Tb_{\phib}$ to $\Ga_{\phib}=(O(2)_1\times O(2)_2\rtimes Z_2)_{\phib}$. $\mathcal{V}_{\phib}$ has the manifold $\frac{r_{\!\phib}}{\sqrt{2}}(e^{i\theta_1},e^{i\theta_2})^\T$ as its minimum with the symmetry groups $\Ht_{\phib}=\Ha_{\phib}=1$ and $\Tt_{\phib}=\Ga_{\phib}$.

To break the continuous group $\Tt_{\phib}$, we introduce an auxiliary generator $\genxnu$ and three driving flavons: a complex multiplet $\nup=({\nup}_1, {\nup}_2)^\T$ and two real doublets $\nuq=({\nuq}_1, {\nuq}_2)^\T$ and $\nur=({\nur}_1, {\nur}_2)^\T$ that transform under $\genxnu$ and other generators in the model. The multiplet $\nup$ is a four-component real irrep $\nup\equiv(\nup,\nup^*)^\T$. How the driving flavons transform is given in TABLE~\ref{tab:nusector}. To help the construction of the potentials, we define a group element $\gene_\nu=\genzn^2(\genzn\genxnu)^2$. Since $\nup$, $\nuq$ and $\nur$ are the only flavons that transform nontrivially (change of sign) under $\gene_\nu$, TABLE~\ref{tab:eelemflavons}, they can appear only quadratically in cross terms. The quadratic tensor products involving $\nup$, $\nuq$ and $\nur$ are given below
\begin{align}
\text{\small$\nup\otimes\nup$}&\text{\small$=|\nup|^2\!\oplus\!{\nup}^{\!\!\!2\alpha}\!\oplus\!{\nup}^{\!\!\!2\beta}\oplus{\nup}^{\!\!\!2\gamma}\oplus{\nup}^{\!\!\!2\delta},$}\\
\text{\small$\nuq\otimes\nuq$}&\text{\small$=|\nuq|^2\oplus{\nuq}^{\!\!\!2\alpha}\oplus{\nuq}^{\!\!\!2\beta},$}\\
\text{\small$\nur\otimes\nur$}&\text{\small$=|\nur|^2\oplus{\nur}^{\!\!2\alpha}\oplus{\nur}^{\!\!2\beta},$}\\
\text{\small$\nup\otimes\nuq$}&\text{\small$={\nupq}^{\!\!\!\!\!\!\!2\alpha}\oplus{\nupq}^{\!\!\!\!\!\!\!2\beta},$}\\
\text{\small$\nup\otimes\nur$}&\text{\small$={\nupr}^{\!\!\!\!\!\!2\alpha}\oplus{\nupr}^{\!\!\!\!\!\!2\beta},$}\\
\text{\small$\nuq\otimes\nur$}&\text{\small$={\nuqr}^{\!\!\!\!\!2\alpha}\oplus{\nuqr}^{\!\!\!\!\!2\beta},$}
\end{align}
where 
\begin{align}
\text{\small${\nup}^{\!\!\!\!2\alpha}$}&\text{\small$={\nup^*}_1{\nup}_1-{\nup^*}_2{\nup}_2$},&\text{\small${\nup}^{\!\!\!\!2\beta}$}&\text{\small$={\nup}_1^{\!\!*}{\nup}_2$},\\
\text{\small${\nup}^{\!\!\!\!2\gamma}$}&\text{\small$={\nup}_1{\nup}_2$},&\text{\small${\nup}^{\!\!\!\!2\delta}$}&\text{\small$=({\nup}_1^{\!\!2},{\nup}_2^{\!\!2})^\T$},\\
\text{\small${\nuq}^{\!\!\!2\alpha}$}&\text{\small$={\nuq}_1^{\!2}-{\nuq}_2^{\!2},$}&\text{\small${\nuq}^{\!\!\!2\beta}$}&\text{\small$={\nuq}_1{\nuq}_2$},\\
\text{\small${\nur}^{\!\!2\alpha}$}&\text{\small$={\nur}_1^{\!2}-{\nur}_2^{\!2}$},&\text{\small${\nur}^{\!\!2\beta}$}&\text{\small$={\nur}_1{\nur}_2$},\\
\text{\small${\nupq}^{\!\!\!\!\!\!\!2\alpha}$}&\text{\small$=({\nup}_1^{\!\!*}{\nuq}_1,{\nup}_2^{\!\!*}{\nuq}_2)^\T,$}&\text{\small${\nupq}^{\!\!\!\!\!\!\!2\beta}$}&\text{\small$=({\nup}_1^{\!\!*}{\nuq}_2,{\nup}_2^{\!\!*}{\nuq}_1)^\T$},\!\!\!\\
\text{\small${\nupr}^{\!\!\!\!\!\!2\alpha}$}&\text{\small$=({\nup}_1^{\!\!*}{\nur}_1,{\nup}_2^{\!\!*}{\nur}_2)^\T,$}&\text{\small${\nupr}^{\!\!\!\!\!\!2\beta}$}&\text{\small$=({\nup}_1^{\!\!*}{\nur}_2,{\nup}_2^{\!\!*}{\nur}_1)^\T$},\!\!\!\\
\text{\small${\nuqr}^{\!\!\!\!\!2\alpha}$}&\text{\small$=({\nuq}_1{\nur}_1,{\nuq}_2{\nur}_2)^\T,$}&\text{\small${\nuqr}^{\!\!\!\!\!2\beta}$}&\text{\small$=({\nuq}_1{\nur}_2,{\nuq}_2{\nur}_1)^\T$}.\!\!\!
\end{align}
How these irreps transform is given in TABLE~\ref{tab:nusector}. We can see that ${\nupr}^{\!\!\!\!\!\!2\alpha}$ transform in the same way as ${\phib}^{\!\!\!\!2\delta}$, TABLE~\ref{tab:prodtrof}. The cross term $|{\phib}^{\!\!\!\!2\delta}-\frac{r_{\!\phib}^2}{r_{\!\nup}r_{\!\nuq}}{\nupr}^{\!\!\!\!\!\!2\alpha}|^2$ can be used to break the continuous symmetries in $\mathcal{V}_{\phib}$ (\ref{eq:phibpot3}), (\ref{eq:phibpot4}). To achieve our alignment $\langle\phib\rangle$ (\ref{eq:elemvevs}), we assign 
\be\label{eq:nuprvevs}
\langle\nup\rangle=\frac{r_{\!\nup}}{\sqrt{2}}(\om,1)^\T,\quad\langle\nur\rangle=\frac{r_{\!\nur}}{\sqrt{2}}(1,1)^\T,
\ee
so that the above-mentioned cross term vanishes. The stabilizers of $\langle\nup\rangle$ and $\langle\nur\rangle$, TABLE~\ref{tab:flavonstabilizers}, fully determine them up to their norms. Hence, symmetry arguments guarantee that these vevs are the stationary points of the potentials of $\nup$ and $\nur$, respectively.

{\renewcommand{\arraystretch}{1.15}
	\setlength{\tabcolsep}{5pt}
	\begin{table}[tbp]
	\begin{center}
	\begin{tabular}{|c|c c c c c c |c|}
\hline
&&&&&&&\\[-1em]
	&$\gent$&$\genr$&$\genzn$&$\gencp$&$\genx$&$\genxnu$&$d$\\
\hline
$\nup$ &$1$&$1$&$\sigx()^*$&$()^*$&$\om\matw_2()^*$&$\sigz$&$4$\\
$\nuq$ &$1$&$1$&$\sigx$&$1$&$1$&$\sigz$&$2$\\
$\nur$ &$1$&$1$&$\sigx$&$1$&$\sigz$&$\sigz$&$2$\\
\hline
${\nup}^{\!\!\!\!2\alpha}$&$1$&$1$&$-1$&$1$&$1$&$1$&$1$\\
${\nup}^{\!\!\!\!2\beta}$&$1$&$1$&$1$&$()^*$&$\ob()^*$&$-1$&$2$\\
${\nup}^{\!\!\!\!2\gamma}$&$1$&$1$&$()^*$&$()^*$&$\ob()^*$&$-1$&$2$\\
${\nup}^{\!\!\!\!2\delta}$&$1$&$1$&$\sigx()^*$&$()^*$&$\ob\matw^*_2()^*$&$1$&$4$\\
\hline
${\nuq}^{\!\!2}_\alpha$&$1$&$1$&$-1$&$1$&$1$&$1$&$1$\\
${\nuq}^{\!\!2}_\beta$&$1$&$1$&$1$&$1$&$1$&$-1$&$1$\\
\hline
${\nur}^{\!\!2}_\alpha$&$1$&$1$&$-1$&$1$&$1$&$1$&$1$\\
${\nur}^{\!\!2}_\beta$&$1$&$1$&$1$&$1$&$-1$&$-1$&$1$\\
\hline
${\nupq}^{\!\!\!\!\!\!2\alpha}$&$1$&$1$&$\sigx()^*$&$()^*$&$\ob\matw_2^*()^*$&$1$&$4$\\
${\nupq}^{\!\!\!\!\!\!2\beta}$&$1$&$1$&$\sigx()^*$&$()^*$&$\ob\matw_2^*()^*$&$-1$&$4$\\
${\nupr}^{\!\!\!\!\!\!\!2\alpha}$&$1$&$1$&$\sigx()^*$&$()^*$&$\ob\matw_2^*\sigz()^*$&$1$&$4$\\
${\nupr}^{\!\!\!\!\!\!\!2\beta}$&$1$&$1$&$\sigx()^*$&$()^*$&$-\ob\matw_2^*\sigz()^*$&$-1$&$4$\\
${\nuqr}^{\!\!\!\!\!2\alpha}$&$1$&$1$&$\sigx$&$1$&$\sigz$&$1$&$2$\\
${\nuqr}^{\!\!\!\!\!2\beta}$&$1$&$1$&$\sigx$&$1$&$-\sigz$&$-1$&$2$\\
\hline
			\end{tabular}
		\end{center}
		\caption{The transformation rules for the driving flavons $\nup$, $\nuq$ and $\nur$ and the irreps obtained from them.}
		\label{tab:nusector}
\end{table}} 

Using $\nup$, we construct
\begin{align}
\text{\small$\breve{\mathcal{V}}_{\nup}$}&\text{\small$=c_{30}m^2|\nup|^2+c_{31}|\nup|^4$}\\
&\text{\small$=k_{24}(|\nup|^2-r_{\!\nup}^2)^2-k_{24} r_{\!\nup}^4$},
\end{align}
which has a $3$-sphere with radius $r_{\!\nup}$ as its minimum with $\Tb_{\nup}=O(4)_{\nup}$. The general renormalizable potential of $\nup$ is given by
\begin{align}
\text{\small$\mathcal{V}_{\nup}$}&\text{\small$=\breve{\mathcal{V}}_{\nup}+c_{32}({\nup}^{\!\!\!\!2\alpha})^2$}\\
&\text{\small$=\breve{\mathcal{V}}_{\nup}+k_{25}({\nup}^{\!\!\!\!2\alpha})^2$}.
\end{align}
The term $({\nup}^{\!\!\!\!2\alpha})^2$ breaks $\Tb_{\nup}=O(4)_{\nup}$ to $\Ga_{\nup}=(O(2)_1\times O(2)_2\rtimes Z_2)_{\nup}$. $\mathcal{V}_{\nup}$ has the manifold $\frac{r_{\!\nup}}{\sqrt{2}}(e^{i\theta_1},e^{i\theta_2})^\T$ as its minimum with $\Ht_{\nup}=\Ha_{\nup}=1$ and $\Tt_{\nup}=\Ga_{\nup}=(O(2)_1\times O(2)_2\rtimes Z_2)_{\nup}$. To break $\Tt_{\nup}$, we utilize $\nuq$. From TABLE~\ref{tab:nusector}, we see that ${\nup}^{\!\!\!\!2\delta}$ and ${\nupq}^{\!\!\!\!\!\!\!2\alpha}$ transform in the same way. Therefore, we use the cross term $|{\nup}^{\!\!\!\!2\delta}-\frac{r_{\!\nup}}{r_{\!\nur}}{\nupq}^{\!\!\!\!\!\!\!2\alpha}|^2$  to break $\Tt_{\nup}$. We assign the vev
\be\label{eq:nuqvev}
\langle\nuq\rangle=\frac{r_{\!\nuq}}{\sqrt{2}}(1,1)^\T,
\ee
so that $|{\nup}^{\!\!\!\!2\delta}-\frac{r_{\!\nup}}{r_{\!\nur}}{\nupq}^{\!\!\!\!\!\!\!2\alpha}|^2$ vanishes at $\langle\nup\rangle$ (\ref{eq:nuprvevs}) and $\langle\nuq\rangle$ (\ref{eq:nuqvev}). $\langle\nuq\rangle$ is guaranteed to be a stationary point of the potential of $\nuq$ since its stabilizer $Z_{2\nuq}$, TABLE~\ref{tab:flavonstabilizers}, fully determines it up to its norm. We construct the potential of $\nuq$ as follows:
\begin{align}
\text{\small$\breve{\mathcal{V}}_{\nuq}$}&\text{\small$=c_{33}m^2|\nuq|^2+c_{34}|\nuq|^4$}\label{eq:nuqpot1}\\
&\text{\small$=k_{26}(|\nuq|^2-r_{\!\nuq}^2)^2-k_{26} r_{\!\nuq}^4$},\label{eq:nuqpot2}\\
\text{\small$\mathcal{V}_{\nuq}$}&\text{\small$=\breve{\mathcal{V}}_{\nuq}+c_{35}({\nuq}^{\!\!\!2\alpha})^2$}\label{eq:nuqpot3}\\
&\text{\small$=\breve{\mathcal{V}}_{\nuq}+k_{27}({\nuq}^{\!\!\!2\alpha})^2$}.\label{eq:nuqpot4}
\end{align}
$\breve{\mathcal{V}}_{\nuq}$ has a circle with radius $r_{\!\nuq}$ as its minimum with $\Tb_{\nuq}=O(2)_{\nuq}$. The term 
$({\nuq}^{\!\!\!2\alpha})^2$ (\ref{eq:nuqpot3}) breaks $\Tb_{\nuq}$ to $\Ga_{\nuq}=D_{8\nup}$. The vev $\langle\nuq\rangle$ (\ref{eq:nuqvev}) forms a minimum of $\mathcal{V}_{\nuq}$ with the symmetry groups $\Ht_{\nuq}=\Ha_{\nuq}=Z_{2\nuq}$, TABLE~\ref{tab:flavonstabilizers}, and $\Tt_{\nuq}=1$.

We construct the potentials containing both $\mathcal{V}_{\nup}$ and $\mathcal{V}_{\nuq}$ in the following way:
\begin{align}
\text{\small$\breve{\mathcal{V}}_{\nupq}$}&\text{\small$=\mathcal{V}_{\nup}+\mathcal{V}_{\nuq}$},\label{eq:nupqpot1}\\
\text{\small$\tilde{\mathcal{V}}_{\nupq}$}&\text{\small$=\breve{\mathcal{V}}_{\nu_\mathrm{pq}}+c_{36}\, \text{re}({\nup}^{\!\!\!\!2\delta\dagger}{\nupq}^{\!\!\!\!\!\!\!2\alpha})$}\label{eq:nupqpot2}\\
\begin{split}
&\text{\small$=\breve{\mathcal{V}}_{\nupq}+k_{28}|{\nup}^{\!\!\!\!2\delta}-\frac{r_{\!\nup}}{r_{\!\nuq}}{\nupq}^{\!\!\!\!\!\!\!2\alpha}|^2$}\\
&\text{\small$\quad\,\,+\frac{k_{28}}{4}({\nup}^{\!\!\!\!2\alpha}\!-\frac{r_{\!\nup}^2}{r_{\!\nuq}^2}{\nuq}^{\!\!\!2\alpha})^2\!+\frac{k_{28}}{4}(|\nup|^2\!-\frac{r_{\!\nup}^2}{r_{\!\nuq}^2}|\nuq|^2)^2\!\!,\!\!$}
\end{split}\label{eq:nupqpot3}\\
\text{\small$\mathcal{V}_{\nupq}$}&\text{\small$=\tilde{\mathcal{V}}_{\nu_\mathrm{pq}}+\,c_{37}{\nup}^{\!\!\!\!2\alpha}{\nuq}^{\!\!\!2\alpha}+c_{38}|\nup|^2|\nuq|^2\!$}\label{eq:nupqpot4}\\
&\text{\small$=\tilde{\mathcal{V}}_{\nupq}\!+k_{29}({\nup}^{\!\!\!\!2\alpha}-\frac{r_{\!\nup}^2}{r_{\!\nuq}^2}{\nuq}^{\!\!\!2\alpha})^2\!+k_{30}(|\nup|^2-\frac{r_{\!\nup}^2}{r_{\!\nuq}^2}|\nuq|^2)^2\!\!.\!\!\!\!$}\label{eq:nupqpot5}
\end{align}
We use $\nupq$ to denote the combined multiplet $(\nup,\nuq)^\T$. The minimum of $\mathcal{V}_{\nup}$, i.e.,~$\frac{r_{\!\nup}}{\sqrt{2}}(e^{i\theta_1},e^{i\theta_2})^\T$, and $\mathcal{V}_{\nuq}$, i.e.,~$\frac{r_{\!\nuq}}{\sqrt{2}}(1,1)^\T$, constitute the minimum of $\breve{\mathcal{V}}_{\nu_\mathrm{pq}}$. Correspondingly, we obtain $\Hb_{\nupq}=\Ht_{\nup}\times\Ht_{\nuq}=\Ht_{\nuq}=Z_{2\nuq}$ and $\Tb_{\nupq}=\Tt_{\nup}\times\Tt_{\nuq}=\Tt_{\nup}=(O(2)_1\times O(2)_2\rtimes Z_2)_{\nup}$. The term $\text{re}({\nup}^{\!\!\!\!2\delta\dagger}{\nupq}^{\!\!\!\!\!\!\!2\alpha})$ in $\tilde{\mathcal{V}}_{\nupq}$ (\ref{eq:nupqpot2}) breaks $\Tb_{\nupq}$ to the discrete group $\Ga_{\nupq}=(D_{6_1}\times D_{6_2}\rtimes Z_2)_{\nup}$ generated by the group actions ${\nup}_1\rightarrow \om{\nup}_1$, ${\nup}_1\rightarrow{\nup}^*_1$, ${\nup}_2\rightarrow \om{\nup}_2$, ${\nup}_2\rightarrow{\nup}^*_2$ and $\nup\rightarrow \sigx\nup$. The alignment $\langle\nup\rangle$ (\ref{eq:nuprvevs}) breaks $\Ga_{\nupq}$ to $\Ha_{\nupq}\!=\!(K_4\rtimes Z_2)_{\nup}$ generated by ${\nup}_1\rightarrow\ob{\nup}^*_1$, ${\nup}_2\rightarrow{\nup}^*_2$ and $\nup\rightarrow\matw_2^*\sigx\nup$. For the combined vev $\langle\nupq\rangle\!=\!(\langle\nup\rangle, \langle\nuq\rangle)^\T$, we obtain $\Ht_{\nupq}\!\!=\Ha_{\nupq}\times\Hb_{\nupq}\!=\!(K_4\rtimes Z_2)_{\nup}\times Z_{2\nuq}$. We also have $\Tt_{\nupq}=1$. $\mathcal{V}_{\nu_\mathrm{pq}}$ (\ref{eq:nupqpot4}), (\ref{eq:nupqpot5}), is the general potential constructed with  $\mathcal{V}_{\nup}$ and $\mathcal{V}_{\nuq}$. The term ${\nup}^{\!\!\!\!2\alpha}{\nuq}^{\!\!\!2\alpha}$ is compatible since ${\nup}^{\!\!\!\!2\alpha}$ and ${\nuq}^{\!\!\!2\alpha}$ vanish at $\langle\nup\rangle$ and $\langle\nuq\rangle$. The term $|\nup|^2|\nuq|^2$ is also compatible. Hence, $\Ht_{\nupq}$ is unaffected by the addition of these terms, and we obtain $\langle\nup\rangle$ and $\langle\nuq\rangle$ as the minimum of the general potential $\mathcal{V}_{\nu_\mathrm{pq}}$. $\mathcal{V}_{\nu_\mathrm{pq}}$ contains $9$ coefficients $c_{30},..,c_{38}$ and correspondingly $9$ arbitrary constants $r_{\!\nup},r_{\!\nuq},k_{24},..,k_{30}$. 

Using $\nur$, we construct
\begin{align}
\text{\small$\breve{\mathcal{V}}_{\nur}$}&\text{\small$=c_{39}m^2|\nur|^2+c_{40}|\nur|^4$}\label{eq:nurpot1}\\
&\text{\small$=k_{31}(|\nur|^2-r_{\!\nur}^2)^2-k_{31} r_{\!\nur}^4$},\label{eq:nurpot2}\\
\text{\small$\mathcal{V}_{\nur}$}&\text{\small$=\breve{\mathcal{V}}_{\nur}+c_{41}({\nur}^{\!\!2\alpha})^2$}\label{eq:nurpot3}\\
&\text{\small$=\breve{\mathcal{V}}_{\nur}+k_{32}({\nur}^{\!\!2\alpha})^2$}.\label{eq:nurpot4}
\end{align}
$\breve{\mathcal{V}}_{\nur}$ has a circle with radius $r_{\!\nur}$ as its minimum with $\Tb_{\nur}=O(2)_{\nur}$.  The term 
$({\nur}^{\!\!\!2\alpha})^2$ (\ref{eq:nurpot3}) breaks $\Tb_{\nur}$ to $\Ga_{\nur}=D_{8\nur}$. The vev $\langle\nur\rangle$ (\ref{eq:nuprvevs}) forms a minimum of $\mathcal{V}_{\nur}$ with the symmetry groups $\Ht_{\nur}=\Ha_{\nur}=Z_{2\nur}$, TABLE~\ref{tab:flavonstabilizers}, and $\Tt_{\nur}=1$.

Using $\nup$, $\nuq$ and $\nur$, we obtain
\be\label{eq:nupot1}
\text{\small$\tilde{\mathcal{V}}_{\nu}=\mathcal{V}_{\nu_\mathrm{pq}}+\mathcal{V}_{\nur}$},
\ee
where $\nu$ denotes the combined multiplet $(\nup,\nuq,\nur)^\T$. It contains $12$ coefficients and $12$ arbitrary constants. It has a minimum corresponding to $\langle\nup\rangle$ (\ref{eq:nuprvevs}), $\langle\nuq\rangle$ (\ref{eq:nuqvev}) and $\langle\nur\rangle$ (\ref{eq:nuprvevs}), with the symmetry groups $\Ht_\nu=\Ht_{\nupq}\times\Ht_{\nur}=(K_4\rtimes Z_2)_{\nup}\times Z_{2\nuq}\times Z_{2\nur}$ and $\Tt_\nu=\Tt_{\nupq}\times\Tt_{\nur}=1$. We construct the general potential containing $\nup$, $\nuq$ and $\nur$ by adding to $\tilde{\mathcal{V}}_{\nu}$ the cross terms of $\nup$ and $\nuq$ with $\nur$,
\begin{align}
\begin{split}
\text{\small$\mathcal{V}_{\nu}$}&\text{\small$=\tilde{\mathcal{V}}_{\nu}+c_{42} {\nup}^{\!\!\!\!2\alpha}{\nur}^{\!\!\!2\alpha}+ c_{43} |\nup|^2|\nur|^2$}\\
&\text{\small$\qquad\qquad+c_{44} {\nuq}^{\!\!\!\!2\alpha}{\nur}^{\!\!\!2\alpha}+c_{45} |\nuq|^2|\nur|^2$},
\end{split}\label{eq:nupot2}\\
\begin{split}
\text{\small$=\tilde{\mathcal{V}}_{\nu}+k_{33} ({\nup}^{\!\!\!\!2\alpha}-\frac{r_{\!\nup}^2}{r_{\!\nur}^2}{\nur}^{\!\!\!2\alpha})^2+ k_{34} (|\nup|^2-\frac{r_{\!\nup}^2}{r_{\!\nur}^2}|\nur|^2)^2\!\!$}\\
\text{\small$\quad\, +k_{35} ({\nuq}^{\!\!\!\!2\alpha}-\frac{r_{\!\nuq}^2}{r_{\!\nur}^2}{\nur}^{\!\!\!2\alpha})^2+k_{36} (|\nuq|^2-\frac{r_{\!\nuq}^2}{r_{\!\nur}^2}|\nur|^2)^2.\!\!\!\!\!$}\label{eq:nupot3}
\end{split}
\end{align}
Since ${\nup}^{\!\!\!2\alpha}$, ${\nuq}^{\!\!\!2\alpha}$ and ${\nur}^{\!\!\!2\alpha}$ vanish at our vevs, the cross terms ${\nup}^{\!\!\!2\alpha}{\nur}^{\!\!\!2\alpha}$ and ${\nuq}^{\!\!\!2\alpha}{\nur}^{\!\!\!2\alpha}$ are compatible. The other two cross terms, which are products of norms, are also compatible. Therefore, our vevs and the group $\Ht_\nu$ remain unaffected by the addition of these terms. $\mathcal{V}_{\nu}$ (\ref{eq:nupot2}), (\ref{eq:nupot3}), contains $16$ coefficients $c_{30},..,c_{45}$ and $16$ arbitrary constants $r_{\!\nup},r_{\!\nuq},r_{\!\nur},k_{24},..,k_{36}$.

Using $\phib$ and the driving flavons, we construct
\begin{align}
\text{\small$\breve{\mathcal{V}}_{\phib\nu}$}&\text{\small$=\mathcal{V}_{\phib}+\mathcal{V}_{\nu}$},\label{eq:phibnupot1}\\
\text{\small$\tilde{\mathcal{V}}_{\phib\nu}$}&\text{\small$=\breve{\mathcal{V}}_{\phib\nu}+c_{46}\text{re}({\phib}^{\!\!\!\!2\delta\dagger}{\nupr}^{\!\!\!\!\!\!2\alpha})$}\label{eq:phibnupot2}\\
&\text{\small$=\breve{\mathcal{V}}_{\phib\nu}+k_{37}|{\phib}^{\!\!\!\!2\delta}-\frac{r_{\!\phib}^2}{r_{\!\nup}r_{\!\nur}}{\nupr}^{\!\!\!\!\!\!2\alpha}|^2$},\label{eq:phibnupot3}
\end{align}
where $\phib\nu$ denotes the combined multiplet $(\phib,\nup,\nuq,\nur)^\T$. The respective minima of $\mathcal{V}_{\phib}$ and $\mathcal{V}_{\nu}$, i.e.,~$\frac{r_{\!\phib}}{\sqrt{2}}(e^{i\theta_1},e^{i\theta_2})^\T$, $\langle\nup\rangle$ (\ref{eq:nuprvevs}), $\langle\nuq\rangle$ (\ref{eq:nuqvev}) and $\langle\nur\rangle$ (\ref{eq:nuprvevs}), constitute a minimum of $\breve{\mathcal{V}}_{\phib\nu}$ (\ref{eq:phibnupot1}). We have the symmetry groups $\Hb_{\phib\nu}=\Ht_{\phib}\times\Ht_{\nu}=\Ht_{\nu}=(K_4\rtimes Z_2)_{\nup}\times Z_{2\nuq}\times Z_{2\nur}$ and $\Tb_{\phib\nu}=\Tt_{\phib}\times\Tt_{\nu}=\Tt_{\phib}=(O(2)_1\times O(2)_2\rtimes Z_2)_{\phib}$. The cross term $\text{re}({\phib}^{\!\!\!\!2\delta\dagger}{\nupr}^{\!\!\!\!\!\!2\alpha})$ in (\ref{eq:phibnupot2}) breaks $\Tb_{\phib\nu}$ to $\Ga_{\phib\nu}=(K_{4_1}\times K_{4_2}\rtimes Z_2)_{\phib}$ generated by ${\phib}_1\!\rightarrow\!-{\phib}_1$, ${\phib}_1\!\rightarrow\!\ob{\phib}_1^*$, ${\phib}_2\!\rightarrow\!-{\phib}_2$, ${\phib}_1\!\rightarrow\!{\phib}_2^*$ and $\phib\!\rightarrow\!\matw_2^*\sigx\phib$. The alignment $\langle\phib\rangle$ (\ref{eq:elemvevs}) breaks $\Ga_{\phib\nu}$ to its the stabilizer $\Ha_{\phib\nu}=(K_4\rtimes Z_2)_{\phib}$ generated by ${\phib}_1\!\rightarrow\!\ob{\phib}_1^*$, ${\phib}_2\!\rightarrow\!{\phib}_2^*$ and $\phib\!\rightarrow\!\matw_2^*\sigx\phib$. $\langle\phib\rangle$ forms a guaranteed stationary point of (\ref{eq:phibnupot2}) and a minimum of (\ref{eq:phibnupot3}). We obtain $\Ht_{\phib\nu}\!=\!\Ha_{\phib\nu}\times\Hb_{\phib\nu}\!=\!(K_4\rtimes Z_2)_{\phib}\times (K_4\rtimes Z_2)_{\nup}\times Z_{2\nuq}\times Z_{2\nur}$ and $\Tt_{\phib\nu}=1$. $\tilde{\mathcal{V}}_{\phib\nu}$ contains $20$ coefficients $c_{27},..,c_{46}$ and correspondingly $20$ arbitrary constants $r_{\!\phib},r_{\!\nup},r_{\!\nuq},r_{\!\nur}, k_{22},..,k_{37}$.

To construct the general potential involving $\phib$ and the driving flavons $\nup$, $\nuq$ and $\nur$, we need to include all the cross terms among them. Considering how they transform under $\gene_{\phib}$ and $\gene_\nu$, TABLE~\ref{tab:eelemflavons}, we can see that the possible cross terms can only involve the irreps obtained from the quadratic product $\phib\otimes\phib$, TABLE~\ref{tab:prodtrof}, and the irreps obtained from the quadratic products of $\nup$, $\nuq$ and $\nur$, TABLE~\ref{tab:nusector}. Adding all such cross terms, we construct the general potential,
\begin{align}
\begin{split}
\text{\small$\mathcal{V}_{\phib\nu}$}&\text{\small$=\tilde{\mathcal{V}}_{\phib\nu}+c_{47}{\phib}^{\!\!\!\!2\alpha}{\nup}^{\!\!\!\!2\alpha}+c_{48}{\phib}^{\!\!\!\!2\alpha}{\nuq}^{\!\!\!2\alpha}+c_{49}{\phib}^{\!\!\!\!2\alpha}{\nur}^{\!\!2\alpha}$}\\
&\text{\small$\quad\,+c_{50}|\phib|^2|\nup|^2+c_{51}|\phib|^2|\nuq|^2+c_{52}|\phib|^2|\nur|^2$}
\end{split}\\
\begin{split}
&\text{\small$=\tilde{\mathcal{V}}_{\phib\nu}+k_{38}({\phib}^{\!\!\!\!2\alpha}-\frac{r_{\!\phib}^2}{r_{\!\nup}^2}{\nup}^{\!\!\!\!2\alpha})^2+k_{39}({\phib}^{\!\!\!\!2\alpha}-\frac{r_{\!\phib}^2}{r_{\!\nuq}^2}{\nuq}^{\!\!\!2\alpha})^2$}\\
&\text{\small$\quad\,\,+k_{40}({\phib}^{\!\!\!\!2\alpha}-\frac{r_{\!\phib}^2}{r_{\!\nur}^2}{\nur}^{\!\!2\alpha})^2+k_{41}(|\phib|^2-\frac{r_{\!\phib}^2}{r_{\!\nup}^2}|\nup|^2)^2$}\\
&\text{\small$\quad\,\,+k_{42}(|\phib|^2\!-\!\frac{r_{\!\phib}^2}{r_{\!\nuq}^2}|\nuq|^2)^2\!+k_{43}(|\phib|^2\!-\!\frac{r_{\!\phib}^2}{r_{\!\nur}^2}|\nur|^2)^2$}.\!\!\!\!
\end{split}
\end{align}
Using arguments similar to those provided earlier, we can show that all these cross terms are compatible. Therefore, the general renormalizable potential $\mathcal{V}_{\phib\nu}$ has a minimum that corresponds to $\langle\phib\rangle$ (\ref{eq:elemvevs}), $\langle\nup\rangle$ (\ref{eq:nuprvevs}), $\langle\nuq\rangle$ (\ref{eq:nuqvev}) and $\langle\nur\rangle$ (\ref{eq:nuprvevs}), along with the symmetry group $\Ht_{\phib\nu}$. $\mathcal{V}_{\phib\nu}$ has $26$ coefficients $c_{27},..,c_{52}$ and $26$ arbitrary constants $r_{\!\phib},r_{\!\nup},r_{\!\nuq},r_{\!\nur}, k_{22},..,k_{43}$.

\subsection{The potential involving $\xic$}

The general potential of $\xic$ is given by
\begin{align}
\text{\small$\mathcal{V}_{\xic}$}&\text{\small$=c_{53}m^2|\xic|^2+c_{54}|\xic|^4$}\\
&\text{\small$=k_{44}(|\xic|^2-r_{\!\xic}^2)^2-k_{44}r_{\!\xic}^4$}.
\end{align}
This potential has a $3$-sphere as its minimum, with $\Tt_{\xic}=O(4)_{\xic}$. To break $\Tt_{\xic}$, we can utilize driving flavons transforming as triplets of $S_4$ that couple with the $S_4$ triplets obtained from $\xic\otimes\xic$. However, such driving flavons will produce dangerous cross terms involving the $S_4$ triplet $\chi$, which would spoil our vevs. The solution to avoid this problem is to introduce driving flavons that are not $S_4$ triplets but whose tensor products with themselves generate the required $S_4$ triplets. We propose that these flavons transform as quartets of $Q_8\rtimes S_4$, generated by $\gent$, $\genr$ and the auxiliary generator $\genxlam$. The representation matrices $\matt_4$, $\matr_4$ and ${\matv}_\lambda$, corresponding to the abstract elements $\gent$, $\genr$ and $\genxlam$, respectively, that act on these flavons are given by  
\begin{equation}\label{}
\matt_4=\left(\begin{matrix} 1 & 0\\
0 & \matt_3
\end{matrix}\right)\!, \,\,\, \matr_4= \left(\begin{matrix} 1 & 0\\
0 & \matr_3
\end{matrix}\right)\!,\,\,\, {\matv}_\lambda= \left(\begin{matrix} -i\sigy & 0\\
0 & i\sigy
\end{matrix}\right)\!.\!\!
\end{equation}
These matrices generate the faithful quartet real representation of the group $Q_8\rtimes S_4$. The aforementioned solution to prevent dangerous cross terms was introduced in \cite{1111.1730} in relation to an $A_4$-based model with the help of quartets of $Q_8\rtimes A_4$. Table~2 in \cite{1111.1730} lists $Q_8\rtimes S_4$ as one of the groups relevant to $S_4$-based models.

The tensor product of $x=(x_1, x_2, x_3, x_4)^\T$ and $y=(y_1, y_2, y_3, y_4)^\T$ transforming as quartets of $Q_8\rtimes S_4$ is given by
\begin{equation}
x\otimes y=(xy)_{\reps}\oplus(xy)_{\rept_1}\oplus(xy)_{\repx}\oplus(xy)_{\rept}\oplus(xy)_{\rept_2},
\end{equation}
where $(xy)_{\reps}=x^\T y$ is the invariant singlet and
\begin{align}
(xy)_{\rept_1}&=(x^\T(\sigz\otimes\mati_2)y, x^\T(\mati_2\otimes\sigz)y, x^\T(\sigz\otimes\sigz)y)^\T,\notag\\
\begin{split}
(xy)_{\repx}&=(x^\T(\sigz\otimes \sigx)y, x^\T(\mati_2\otimes\sigx)y, x^\T(\sigx\otimes \sigz)y,\notag\\
&\quad\,\,\,\,\,\, x^\T(\sigx\otimes\mati_2)y, x^\T(i\sigy\otimes i\sigy)y, x^\T(\sigx\otimes \sigx)y)^\T,
\end{split}\notag\\
(xy)_{\rept}&=(x^\T(\mati_2\otimes i\sigy)y, x^\T(i\sigy\otimes\sigz)y, x^\T(i\sigy\otimes\sigx)y)^\T,\notag\\
(xy)_{\rept_2}&=(x^\T(\sigz\otimes i\sigy)y, x^\T(i\sigy\otimes\mati_2)y, x^\T(\sigx\otimes i\sigy)y)^\T.\notag
\end{align}
These irreps transform as given in TABLE~\ref{tab:q8prods}. The triplet $(xy)_{\rept}$ obtained from the antisymmetric part of the tensor product transforms as $\matt_3$ and $\matr_3$ under $\gent$ and $\genr$, respectively, and remains invariant under $\genxlam$. Hence, it is well suited for breaking $\Tt_{\xic}=O(4)_{\xic}$.

{\renewcommand{\arraystretch}{1.0}
	\setlength{\tabcolsep}{5pt}
	\begin{table}[tbp]
	\begin{center}
	\begin{tabular}{|c|c c c |c|}
\hline
&&&&\\[-1em]
	&$\gent$&$\genr$&$\genxlam$&$d$\\
\hline
&&&&\\[-0.9em]
$(xy)_{\rept_1}$&$\matt_3$&$-\matt_3\matr_3^2\matt_3^2\matr_3$&$\matr_3^2$&$3$\\
$(xy)_{\repx}$&$\matt_3\otimes\mati_2$&$\matr_3\otimes\sigx$&$\mati_2\otimes(\matt_3\matr_3^2\matt_3^2)$&$6$\\
$(xy)_{\rept}$&$\matt_3$&$\matr_3$&$1$&$3$\\
$(xy)_{\rept_2}$&$\matt_3$&$\matr_3$&$\matr_3^2$&$3$\\
\hline
			\end{tabular}
		\end{center}
		\caption{The transformation rules for the irreps obtained from the tensor product of two quartets of $Q_8\rtimes A_4$.}
		\label{tab:q8prods}
\end{table}} 

{\renewcommand{\arraystretch}{1.0}
	\setlength{\tabcolsep}{5pt}
	\begin{table}[tbp]
	\begin{center}
	\begin{tabular}{|c|c c c c c c |c|}
\hline
&&&&&&&\\[-1em]
	&\!$\gent$\!&\!$\genr$\!&$\genzn$&$\gencp$&$\genx$&$\genxlam$&$d$\\
\hline
$\lamp$ &\!$\matt_4$\!&\!$\matr_4$\!&\!\!$1$\!\!&$1$&\!\!$1$\!\!&${\matv}_\lambda$&$4$\\
$\lamq$ &\!$\matt_4$\!&\!$\matr_4$\!&\!\!$-1$\!\!&$1$&\!\!$1$\!\!&${\matv}_\lambda$&$4$\\
$\lamr$ &\!$\matt_4$\!&\!$\matr_4$\!&\!\!$1$\!\!&$1$&\!\!$-1$\!\!&${\matv}_\lambda$&$4$\\
\hline
\!\!${\lamp}^{\!\!\!\T}\!\lamq$\!\!&$1$&$1$&$-1$&$1$&$1$&$1$&$1$\\
\!\!${\lamp}^{\!\!\!\T}\!\lamr$\!\!&$1$&$1$&$1$&$1$&$-1$&$1$&$1$\\
\!\!${\lamq}^{\!\!\!\T}\!\lamr$\!\!&$1$&$1$&$-1$&$1$&$-1$&$1$&$1$\\
\!\!${\lampq}^{\!\!\!\!\!\!\!2\gamma}$\!\!&$\matt_3$&$\matr_3$&$-1$&$1$&$1$&$1$&$3$\\
\!\!${\lampr}^{\!\!\!\!\!\!2\gamma}$\!\!&$\matt_3$&$\matr_3$&$1$&$1$&$-1$&$1$&$3$\\
\!\!${\lamqr}^{\!\!\!\!\!\!2\gamma}$\!\!&$\matt_3$&$\matr_3$&$-1$&$1$&$-1$&$1$&$3$\\
\hline
			\end{tabular}
		\end{center}
		\caption{The transformation rules for the driving flavons $\lamp$, $\lamq$ and $\lamr$ and the irreps obtained from them.}
		\label{tab:lamsector}
\end{table}} 

We introduce the flavons $\lamp$, $\lamq$ and $\lamr$ that transform as quartets of $Q_8\rtimes S_4$, TABLE~\ref{tab:lamsector}. The flavons $\lamq$ and $\lamr$ change sign under $\genzn$ and $\genx$, respectively. In preparation to constructing the potentials involving $\lamp$, $\lamq$ and $\lamr$, we define the group element $\gene_\lambda=\genxlam^2$. As shown in TABLE~\ref{tab:eelemflavons}, $\lamp$, $\lamq$ and $\lamr$ change sign under $\gene_\lambda$ while every other flavon remains invariant under it. Hence, $\lamp$, $\lamq$ and $\lamr$ can appear in cross terms as quadratic products only. We define ${\lambda_{\mathrm x}}^{\!\!\!\!2\alpha}=(\lambda_{\mathrm x}\lambda_{\mathrm x})_{\rept_1}$, ${\lambda_{\mathrm {x y}}}^{\!\!\!\!\!\!\!2\alpha}=(\lambda_{\mathrm x}\lambda_{\mathrm y})_{\rept_1}$, ${\lambda_{\mathrm x}}^{\!\!\!\!2\beta}=(\lambda_{\mathrm x}\lambda_{\mathrm x})_{\repx}$ and ${\lambda_{\mathrm {x y}}}^{\!\!\!\!\!\!\!2\gamma}=(\lambda_{\mathrm x}\lambda_{\mathrm y})_{\rept}$ with $\mathrm x$, $\mathrm y$ being equal to $\mathrm p$, $\mathrm q$ or $\mathrm r$. Besides the norms, i.e.,~$|\lambda_{\mathrm {x}}|^2$, the only quadratic-order irreps that remain invariant under $\genxlam$ are $\lambda_{\mathrm {x}}^\T\lambda_{\mathrm {y}}$ and ${\lambda_{\mathrm {x y}}}^{\!\!\!\!\!\!\!2\gamma}$. We list them in TABLE~\ref{tab:lamsector}. The irreps ${\xic}^{\!\!\!2\beta}$ and ${\xic}^{\!\!\!2\gamma}$, TABLE~\ref{tab:prodtrof}, transform in the same way as ${\lambda_{\mathrm {pq}}}^{\!\!\!\!\!\!\!2\gamma}$ and ${\lambda_{\mathrm {pr}}}^{\!\!\!\!\!\!2\gamma}$, respectively. We can break $\Tt_{\xic}$ by coupling these together using two cross terms $|{\xic}^{\!\!\!2\beta}-\frac{r_{\!\xic}^2}{r_{\lamp}r_{\lamq}}{\lambda_{\mathrm {pq}}}^{\!\!\!\!\!\!\!2\gamma}|^2$ and $|{\xic}^{\!\!\!2\gamma}-\frac{r_{\!\xic}^2}{r_{\lamp}r_{\lamr}}{\lambda_{\mathrm {pr}}}^{\!\!\!\!\!\!2\gamma}|^2$. 

We make the following assignments of vevs:
\begin{align}
\langle\lamp\rangle&=r_{\lamp}(1,0,0,0)^\T,\label{eq:lampvev}\\
\langle\lamq\rangle&=r_{\lamq}(0,0,1,0)^\T,\label{eq:lamqvev}\\
\langle\lamr\rangle&=\frac{r_{\lamr}}{\sqrt{2}}(0,1,0,1)^\T.\label{eq:lamrvev}
\end{align}
These are fully defined, up to their norms, by their stabilizers, TABLE~\ref{tab:flavonstabilizers}; hence, they are guaranteed to be stationary points of the corresponding potentials. It can be shown that the two cross terms mentioned earlier vanish at the above vevs and $\langle\xic\rangle$ (\ref{eq:elemvevs}). We now proceed to explicitly construct the potentials of $\lamp$, $\lamq$ and $\lamr$ to show that the above vevs are obtained as a minimum.

Using $\lamp$, we construct
\begin{align}
\text{\small$\breve{\mathcal{V}}_{\lamp}$}&\text{\small$=c_{55}m^2|\lamp|^2+c_{56}|\lamp|^4$}\label{eq:lamppot1}\\
&\text{\small$=k_{45}(|\lamp|^2-r_{\lamp}^2)^2-k_{45}r_{\lamp}^4$}.\label{eq:lamppot2}
\end{align}
$\breve{\mathcal{V}}_{\lamp}$ has a $3$-sphere as its minimum, with the symmetry group $\Tb_{\lamp}=O(4)_{\lamp}$. The general potential of $\lamp$ is given by
\begin{align}
\text{\small$\mathcal{V}_{\lamp}$}&\text{\small$=\breve{\mathcal{V}}_{\lamp}+c_{57}|{\lamp}^{\!\!\!\!2\beta}|^2$}\label{eq:lamppot3}\\
&\text{\small$=\breve{\mathcal{V}}_{\lamp}+k_{46}|{\lamp}^{\!\!\!\!2\beta}|^2$}.\label{eq:lamppot4}
\end{align}
The symmetry group of $\mathcal{V}_{\lamp}$ is $((Z_{2_1}\times Z_{2_2}\times Z_{2_3}\times Z_{2_4})\rtimes S_4)_{\lamp}$ generated by the sign changes as well as the permutations of the four components of $\lamp$. In other words, the term $|{\lamp}^{\!\!\!\!2\beta}|^2$ (\ref{eq:lamppot3}) breaks $\Tb_{\lamp}$ to $\Ga_{\lamp}=((Z_{2_1}\times Z_{2_2}\times Z_{2_3}\times Z_{2_4})\rtimes S_4)_{\lamp}$. Note that $\Ga_{\lamp}$ is larger than $Q_8\rtimes S_4$ because of accidental symmetries. 

The alignment $\langle\lamp\rangle$ (\ref{eq:lampvev}) remains invariant under the sign changes as well as the permutations of the last three components of $\lamp$, which generates the group $(S_4\times Z_2)_{\lamp}$, i.e.,~$\Ha_{\lamp}=(S_4\times Z_2)_{\lamp}$ forms the stabilizer of $\langle\lamp\rangle$ under $\Ga_{\lamp}$. Since $\langle\lamp\rangle$ is fully determined by its stabilizer in the $3$-sphere, symmetry arguments guarantee that it forms a stationary point. The rewritten potential (\ref{eq:lamppot4}) shows that we can obtain it as a minimum. The solution of $|\lamp|=r_{\lamp}$ and ${\lamp}^{\!\!\!\!2\beta}=0$ is a discrete set of points, with $\langle\lamp\rangle$ being one among them. We have $\Ht_{\lamp}=\Ha_{\lamp}=(S_4\times Z_2)_{\lamp}$.

The construction of the potential of $\lamq$ follows the same lines as that of $\lamp$. We construct
\begin{align}
\text{\small$\breve{\mathcal{V}}_{\lamq}$}&\text{\small$=c_{58}m^2|\lamq|^2+c_{59}|\lamq|^4$}\\
&\text{\small$=k_{47}(|\lamq|^2-r_{\lamq}^2)^2-k_{47}r_{\lamq}^4$},\\
\text{\small$\mathcal{V}_{\lamq}$}&\text{\small$=\breve{\mathcal{V}}_{\lamq}+c_{60}|{\lamq}^{\!\!\!\!2\beta}|^2$}\\
&\text{\small$=\breve{\mathcal{V}}_{\lamq}+k_{48}|{\lamq}^{\!\!\!\!2\beta}|^2$}.
\end{align}
$\mathcal{V}_{\lamq}$ has a discrete minimum at $\langle\lamq\rangle$ (\ref{eq:lamqvev}) with $\Ht_{\lamq}=(S_4\times Z_2)_{\lamq}$, generated by the sign changes and permutations of the first, second and fourth elements of $\lamq$.

Combining $\mathcal{V}_{\lamp}$ and $\mathcal{V}_{\lamq}$, we construct 
\be
\text{\small$\tilde{\mathcal{V}}_{\lampq}=\mathcal{V}_{\lamp}+\mathcal{V}_{\lamq}$},
\ee
where $\lampq$ denotes the combined multiplet $(\lamp,\lamq)^\T$. $\tilde{\mathcal{V}}_{\lampq}$ has a minimum at $\lampq=(\langle\lamp\rangle,\langle\lamq\rangle)^\T$ with $\Ht_{\lampq}=\Ht_{\lamp}\times\Ht_{\lamq}=(S_4\times Z_2)_{\lamp}\times (S_4\times Z_2)_{\lamq}$. To obtain the general potential involving $\lamp$ and $\lamq$, we need to include their cross terms. There are three independent cross terms between $\lamp$ and $\lamq$. By including them, we construct
\begin{align}
\text{\small$\mathcal{V}_{\lampq}$}&\text{\small$=\tilde{\mathcal{V}}_{\lampq}\!+c_{61}{\lamp}^{\!\!\!\!2\alpha\T}{\lamq}^{\!\!\!\!2\alpha}+c_{62}{\lamp}^{\!\!\!\!2\beta\T}{\lamq}^{\!\!\!\!2\beta}\!+c_{63}|\lamp|^2|\lamq|^2\!\!\!\!$}\label{eq:lamppot5}\\
&\text{\small$=\tilde{\mathcal{V}}_{\lampq}\!+k_{49}|{\lambda_{\mathrm {pq}}}^{\!\!\!\!\!\!\!2\alpha}|^2\!+k_{50}(\lamp^\T\lamq)^2\!+k_{51}(|\lamp|^2\!-\frac{r_{\lamp}^2}{r_{\lamq}^2}|\lamq|^2)^2$}.\label{eq:lamppot6}
\end{align}
Let us determine if the three cross terms in (\ref{eq:lamppot5}) are compatible with $\Ht_{\lampq}$. First, we consider the cross term ${\lamp}^{\!\!\!\!2\alpha\T}{\lamq}^{\!\!\!\!2\alpha}$. With $\lampq$ being the eight component multiplet $(({\lamp}_1, {\lamp}_2, {\lamp}_3, {\lamp}_4), ({\lamq}_1, {\lamq}_2, {\lamq}_3, {\lamq}_4))^\T$, a point in $\text{fix}(\Ht_{\lampq})$ is given by $(a(1, 0, 0, 0), b(0, 0, 1, 0))^\T$ where $a$ and $b$ are real numbers. It can be shown that the gradient of ${\lamp}^{\!\!\!\!2\alpha\T}{\lamq}^{\!\!\!\!2\alpha}$ calculated at the above-mentioned point is $(-2ab^2(1, 0, 0, 0), -2a^2b(0, 0, 1, 0))^\T$. Since the gradient also lies in $\text{fix}(\Ht_{\lampq})$, the term ${\lamp}^{\!\!\!\!2\alpha\T}{\lamq}^{\!\!\!\!2\alpha}$ is compatible with $\Ht_{\lampq}$. For the case of ${\lamp}^{\!\!\!\!2\beta\T}{\lamq}^{\!\!\!\!2\beta}$, we can show that both ${\lamp}^{\!\!\!\!2\beta}$ and ${\lamq}^{\!\!\!\!2\beta}$ vanish at $(a(1, 0, 0, 0), b(0, 0, 1, 0))^\T$. Hence, according to corollary~B in \cite{2306.07325}, the term ${\lamp}^{\!\!\!\!2\beta\T}{\lamq}^{\!\!\!\!2\beta}$ is compatible with $\Ht_{\lampq}$. The term $|\lamp|^2|\lamq|^2$, which is the product of norms, is also compatible. Therefore, $\Ht_{\lampq}$ is unaffected by the addition of the cross terms, i.e.,~$\langle\lamp\rangle$ and $\langle\lamq\rangle$ remain unspoiled. This is made apparent in (\ref{eq:lamppot6}). Both ${\lambda_{\mathrm {pq}}}^{\!\!\!\!\!\!\!2\alpha}$ and $(\lamp^\T\lamq)$ vanish at our vevs. $|{\lambda_{\mathrm {pq}}}^{\!\!\!\!\!\!\!2\alpha}|^2$ and $(\lamp^\T\lamq)^2\!$ can be expressed as linear combinations of the three earlier-mentioned cross terms. $\mathcal{V}_{\lampq}$ (\ref{eq:lamppot6}) has $9$ arbitrary constants $r_{\lamp}, r_{\lamq}, k_{45}, .., k_{51}$ which are functions of the $9$ coefficients $c_{55}, .., c_{63}$ in (\ref{eq:lamppot5}).

Using $\lamr$, we construct
\begin{align}
\text{\small$\mathcal{V}_{\lamr}$}&\text{\small$=c_{64}m^2|\lamr|^2+c_{65}|\lamr|^4,$}\label{eq:lamrpot1}\\
&\text{\small$=k_{52}(|\lamr|^2-r_{\lamr}^2)^2-k_{52}r_{\lamr}^4$}\label{eq:lamrpot2}
\end{align}
$\mathcal{V}_{\lamr}$ has a $3$-sphere as its minimum with the symmetry group $\Tt_{\lamr}=O(4)_{\lamr}$. We may add $|{\lamr}^{\!\!\!\!2\beta}|^2$ to $\mathcal{V}_{\lamr}$ to break $O(4)_{\lamr}$ to the discrete group $((Z_{2_1}\times Z_{2_2}\times Z_{2_3}\times Z_{2_4})\rtimes S_4)_{\lamr}$. Our alignment $\langle\lamr\rangle$ (\ref{eq:lamrvev}) is fully determined by its stabilizer under this discrete group. Therefore, in the $3$-sphere, $\langle\lamr\rangle$ is guaranteed to appear as a stationary point of the newly obtained potential. However, it so happens that this stationary point can never be a minimum. Rather, it is a saddle point for every choice of the coefficients. In the discussion below, we utilize the cross terms of $\lamr$ with $\lamp$ and $\lamq$ along with the term $|{\lamr}^{\!\!\!\!2\beta}|^2$ to break $\Tt_{\lamr}=O(4)_{\lamr}$ and obtain a minimum for $\lamr$. 

First, we combine $\mathcal{V}_{\lampq}$ and $\mathcal{V}_{\lamr}$ to obtain,
\be\label{eq:lampot1}
\text{\small$\breve{\mathcal{V}}_{\lambda}=\mathcal{V}_{\lampq}+\mathcal{V}_{\lamr}$},
\ee
where $\lambda=(\lampq,\lamr)^\T=(\lamp,\lamq,\lamr)^\T$. $\breve{\mathcal{V}}_{\lambda}$ has a minimum consisting of $\langle\lamp\rangle$, $\langle\lamq\rangle$ and a $3$-sphere for $\lamr$. For this minimum, we have the symmetry groups $\Hb_\lambda=\Ht_{\lampq}=(S_4\times Z_2)_{\lamp}\times(S_4\times Z_2)_{\lamq}$ and $\Tb_\lambda=\Tt_{\lamr}=O(4)_{\lamr}$. We add two cross terms in addition to the term $|{\lamr}^{\!\!\!\!2\beta}|^2$ to $\breve{\mathcal{V}}_{\lambda}$, resulting in
\be\label{eq:lampot2}
\text{\small$\tilde{\mathcal{V}}_{\lambda}=\breve{\mathcal{V}}_{\lambda}+c_{66}|{\lamr}^{\!\!\!\!2\beta}|^2\!+c_{67}{\lamp}^{\!\!\!\!2\alpha\T}{\lamr}^{\!\!\!\!2\alpha}+c_{68}{\lamq}^{\!\!\!\!2\alpha\T}{\lamr}^{\!\!\!\!2\alpha}$}.
\ee
These terms break $\Tb_{\lambda}=O(4)_{\lamr}$ to $\Ga_\lambda=((Z_{2_1}\times Z_{2_2}\times Z_{2_3}\times Z_{2_4})\rtimes Z_2)_{\lamr}$ generated by the sign changes of the four components of $\lamr$ and the permutation of its second and fourth components. The alignment $\langle\lamr\rangle$ (\ref{eq:lamrvev}) breaks $\Ga_\lambda$ to $\Ha_\lambda=((Z_{2_1}\times Z_{2_3})\rtimes Z_2)_{\lamr}$. Since $\Ha_\lambda$ fully determines $\langle\lamr\rangle$ in the $3$-sphere, we are guaranteed to obtain it as a stationary point. We obtain $\Ht_\lambda=\Hb_\lambda\times\Ha_\lambda=(S_4\times Z_2)_{\lamp}\times(S_4\times Z_2)_{\lamq}\times ((Z_{2_1}\times Z_{2_3})\rtimes Z_2)_{\lamr}$. We can show that the stationary point forms a minimum by rewriting $\tilde{\mathcal{V}}_{\lambda}$ (\ref{eq:lampot2}) as
\be\label{eq:lampot3}
\begin{split}
\text{\small$\tilde{\mathcal{V}}_{\lambda}$}&\text{\small$=\breve{\mathcal{V}}_{\lambda}+k_{53}|2{\lamr}^{\!\!\!\!2\alpha}+\frac{r_{\lamr}^2}{r_{\lamp}^2}{\lamp}^{\!\!\!\!2\alpha}+\frac{r_{\lamr}^2}{r_{\lamq}^2}{\lamq}^{\!\!\!\!2\alpha}|^2$}\\
&\quad\text{\small$+k_{54}|{\lambda_{\mathrm {pr}}}^{\!\!\!\!\!\!2\alpha}|^2+k_{54}(\lamp^\T\lamr)^2+\frac{k_{54}}{2}\frac{r_{\lamp}^2}{r_{\lamr}^2}(|\lamr|^2-\frac{r_{\lamr}^2}{r_{\lamp}^2}|\lamp|^2)^2$}\\
&\quad\text{\small$+k_{55}|{\lambda_{\mathrm {qr}}}^{\!\!\!\!\!\!2\alpha}|^2+k_{55}(\lamq^\T\lamr)^2+\frac{k_{55}}{2}\frac{r_{\lamq}^2}{r_{\lamr}^2}(|\lamr|^2-\frac{r_{\lamr}^2}{r_{\lamq}^2}|\lamq|^2)^2$}.
\end{split}
\ee
The $14$ arbitrary constants $r_{\lamp},r_{\lamq},r_{\lamr}, k_{45},..,k_{55}$ in (\ref{eq:lampot3}) can be obtained as functions of the $14$ coefficients $c_{55}, .., c_{68}$ in (\ref{eq:lampot2}). $\tilde{\mathcal{V}}_{\lambda}$ attains its minimum when each term in (\ref{eq:lampot3}) vanishes. Our vev $\langle\lambda\rangle=(\langle\lamp\rangle,\langle\lamq\rangle,\langle\lamr\rangle)^\T$ (\ref{eq:lampvev})-(\ref{eq:lamrvev}) constitutes one of the points (among a discrete set) where this happens.

The general potential involving these driving flavons is obtained by including four more cross terms of $\lamr$ with $\lamp$ and $\lamq$,
\begin{align}\label{eq:lampot4}
\begin{split}
\text{\small$\mathcal{V}_{\lambda}$}&\text{\small$=\tilde{\mathcal{V}}_{\lambda}+c_{69}{\lamp}^{\!\!\!\!2\beta\T}{\lamr}^{\!\!\!\!2\beta}+c_{70}{\lamq}^{\!\!\!\!2\beta\T}{\lamr}^{\!\!\!\!2\beta}$}\\
&\text{\small$\qquad\qquad+c_{71}|\lamp|^2|\lamr|^2+c_{72}|\lamq|^2|\lamr|^2$}.
\end{split}
\end{align}
A point in $\text{fix}(\Ht_\lambda)$ is given by $\lambda=(\lamp,\lamq,\lamr)^\T=(a(1, 0, 0, 0), b(0, 0, 1, 0), \frac{c}{\sqrt{2}}(0, 1, 0, 1))^\T$ where $a$, $b$ and $c$ are real numbers. Let us analyze the cross term ${\lamp}^{\!\!\!\!2\beta\T}{\lamr}^{\!\!\!\!2\beta}$. Since it involves only $\lamp$ and $\lamr$, we may consider the point 
\be\label{eq:fixspace}
(\lamp,\lamr)^\T=(a(1, 0, 0, 0), \frac{c}{\sqrt{2}}(0, 1, 0, 1))^\T. 
\ee
We can show that at this point, the gradient of ${\lamp}^{\!\!\!\!2\beta\T}{\lamr}^{\!\!\!\!2\beta}$ is $((0, 0, 0, 0), (0, 0, 0, 0))^\T$, which corresponds to (\ref{eq:fixspace}) with $a\rightarrow 0$, $c\rightarrow 0$, i.e.,~the gradient calculated at $\text{fix}(\Ht_\lambda)$ lies in $\text{fix}(\Ht_\lambda)$. Therefore, the cross term ${\lamp}^{\!\!\!\!2\beta\T}{\lamr}^{\!\!\!\!2\beta}$ is compatible with $\Ht_\lambda$. Similarly, the gradient of ${\lamq}^{\!\!\!\!2\beta\T}{\lamr}^{\!\!\!\!2\beta}$ at $(\lamq,\lamr)^\T=(b(0, 0, 1, 0), \frac{c}{\sqrt{2}}(0, 1, 0, 1))^\T$ vanishes, proving that it is compatible $\Ht_\lambda$. The cross terms $|\lamp|^2|\lamr|^2$ and $|\lamq|^2|\lamr|^2$, which are products of norms, are also compatible. Hence, $\Ht_\lambda$ is preserved, and the vevs (\ref{eq:lampvev})-(\ref{eq:lamrvev}) form a minimum of $\mathcal{V}_{\lambda}$ (\ref{eq:lampot4}). To make this apparent, we rewrite $\mathcal{V}_{\lambda}$ as 
\begin{align}\label{eq:lampot5}
\begin{split}
\text{\small$\mathcal{V}_{\lambda}$}&\text{\small$=\tilde{\mathcal{V}}_{\lambda}+k_{56}(\lamp^\T\lamr)^2+k_{57}(|\lamr|^2-\frac{r_{\lamr}^2}{r_{\lamp}^2}|\lamp|^2)^2$}\\
&\text{\small$\qquad\qquad+k_{58}(\lamq^\T\lamr)^2+k_{59}(|\lamr|^2-\frac{r_{\lamr}^2}{r_{\lamq}^2}|\lamq|^2)^2$}.
\end{split}
\end{align}
$\mathcal{V}_{\lambda}$ (\ref{eq:lampot4}) contains $18$ coefficients, and correspondingly $\mathcal{V}_{\lambda}$ (\ref{eq:lampot5}) contains $18$ arbitrary constants.

We combine the potential of the flavon $\xic$ with that of its driving flavons $\lamp$, $\lamq$ and $\lamr$ to obtain
\be\label{eq:clampot1}
\text{\small$\breve{\mathcal{V}}_{\xic\lambda}=\mathcal{V}_{\xic}+\mathcal{V}_{\lambda}$},
\ee
where $\xic\lambda$ denotes the combined multiplet. A minimum of this potential corresponds to a $3$-sphere for $\xic$ and (\ref{eq:lampvev})-(\ref{eq:lamrvev}) for $\lambda$. We have $\Hb_{\xic\lambda}=\Ht_\lambda$ and $\Tb_{\xic\lambda}=O(4)_{\xic}$. We break $\Tb_{\xic\lambda}$ by coupling $\xic$ with the driving flavons, 
\begin{align}
\text{\small$\tilde{\mathcal{V}}_{\xic\lambda}$}&\text{\small$=\breve{\mathcal{V}}_{\xic\lambda}+c_{73}\,{\xic}^{\!\!\!2\beta\,\T}{\lambda_{\mathrm {pq}}}^{\!\!\!\!\!\!\!2\gamma}+c_{74}\,{\xic}^{\!\!\!2\gamma\,\T}{\lambda_{\mathrm {pr}}}^{\!\!\!\!\!\!2\gamma}$}\label{eq:clampot2}\\
&\text{\small$=\breve{\mathcal{V}}_{\xic\lambda}+k_{60}|{\xic}^{\!\!\!2\beta}\!-\frac{r_{\!\xic}^2}{r_{\lamp}r_{\lamq}}{\lambda_{\mathrm {pq}}}^{\!\!\!\!\!\!\!2\gamma}|^2+k_{61}|{\xic}^{\!\!\!2\gamma}\!-\frac{r_{\!\xic}^2}{r_{\lamp}r_{\lamr}}{\lambda_{\mathrm {pr}}}^{\!\!\!\!\!\!2\gamma}|^2$}.\label{eq:clampot3}
\end{align}
The cross terms in (\ref{eq:clampot2}) break $\Tb_{\xic\lambda}$ to $\Ga_{\xic\lambda}=(K_4\times Z_2)_{\xic}$ generated by $\xic\rightarrow\mattau_2\sigx\xic$, $\xic\rightarrow i\mattau_2\xic^*$ and $\xic\rightarrow-\xic$. The stabilizer of $\langle\xic\rangle$ (\ref{eq:elemvevs}) under $\Ga_{\xic\lambda}$ is given by $\Ha_{\xic\lambda}=K_{4\xic}$, TABLE~\ref{tab:elemstabilizers}, generated by $\xic\rightarrow\mattau_2\sigx\xic$ and $\xic\rightarrow i\mattau_2\xic^*$. This stabilizer fully determines $\langle\xic\rangle$ in the $3$-sphere; therefore, symmetry arguments guarantee that it forms a stationary point. We obtain $\Ht_{\xic\lambda}=\Ha_{\xic\lambda}\times \Hb_{\xic\lambda}=K_{4\xic}\times(S_4\times Z_2)_{\lamp}\times(S_4\times Z_2)_{\lamq}\times ((Z_{2_1}\times Z_{2_3})\rtimes Z_2)_{\lamr}$. The existence of a minimum as the stationary point is made apparent in (\ref{eq:clampot3}). At $\langle\lamp\rangle$, $\langle\lamq\rangle$, $\langle\lamr\rangle$ and $\langle\xic\rangle$, each term in (\ref{eq:clampot3}) vanishes. $\tilde{\mathcal{V}}_{\xic\lambda}$ has $22$ coefficients and correspondingly $22$ arbitrary coefficients.

The general potential constructed with $\xic$ and its driving flavons is given by
\begin{align}
\begin{split}\label{eq:clampot4}
\text{\small$\mathcal{V}_{\xic\lambda}$}&\text{\small$=\tilde{\mathcal{V}}_{\xic\lambda}+c_{75}\,{\xic}^{\!\!\!2\alpha\,\T}{\lambda_{\mathrm {rq}}}^{\!\!\!\!\!\!2\gamma}+c_{76}|\xic|^2|\lamp|^2$}\\
&\text{\small$\qquad\qquad\qquad+c_{77}|\xic|^2|\lamq|^2+c_{78}|\xic|^2|\lamr|^2$}
\end{split}\\
\begin{split}\label{eq:clampot5}
&\text{\small$=\tilde{\mathcal{V}}_{\xic\lambda}+k_{62}|{\xic}^{\!\!\!2\alpha}\!-\frac{r_{\!\xic}^2}{r_{\lamr}r_{\lamq}}{\lambda_{\mathrm {rq}}}^{\!\!\!\!\!\!\!2\gamma}|^2\!+k_{63}(|\xic|^2\!-\frac{r_{\!\xic}^2}{r_{\lamp}^2}|\lamp|^2)^2$}\\
&\text{\small$\qquad\quad\,+k_{64}(|\xic|^2\!-\frac{r_{\!\xic}^2}{r_{\lamq}^2}|\lamq|^2)^2+k_{65}(|\xic|^2\!-\frac{r_{\!\xic}^2}{r_{\lamr}^2}|\lamr|^2)^2$}
\end{split}
\end{align}
To analyze the cross term ${\xic}^{\!\!\!2\alpha\,\T}{\lambda_{\mathrm {rq}}}^{\!\!\!\!\!\!2\gamma}$, we consider the multiplet $(\xic,\lamq, \lamr)^\T=(({\xic}_{1r},{\xic}_{1i},{\xic}_{2r},{\xic}_{2i}),$ $({\lamq}_1,{\lamq}_2,{\lamq}_3,{\lamq}_4),({\lamr}_1,{\lamr}_2,{\lamr}_3,{\lamr}_4))^\T$. In this basis, a general point in $\text{fix}(\Ht_{\xic\lambda})$ is given by 
\be\label{eq:fixspace1}
\text{\small$(\frac{a}{\sqrt{2}}(\cos\frac{\pi}{8},\sin\frac{\pi}{8},\cos\frac{3\pi}{8},\sin\frac{3\pi}{8}),b(0,0,1,0),\frac{c}{\sqrt{2}}(0,1,0,1))^\T$}.
\ee
At this point, we obtain the gradient of the cross term ${\xic}^{\!\!\!2\alpha\,\T}{\lambda_{\mathrm {rq}}}^{\!\!\!\!\!\!2\gamma}$ as $(\sqrt{2}abc(\cos\frac{\pi}{8},\sin\frac{\pi}{8},\cos\frac{3\pi}{8},\sin\frac{3\pi}{8}),$  $a^2c(0,0,1,0),\frac{a^2b}{\sqrt{2}}(0,1,0,1))^\T$, which lies in $\text{fix}(\Ht_{\xic\lambda})$. Therefore, the cross term is compatible with $\Ht_{\xic\lambda}$. The other cross terms in (\ref{eq:clampot4}) are products of norms; hence, they also are compatible. Therefore, $\Ht_{\xic\lambda}$ is preserved, and we obtain our assigned vevs as one of the minima of $\mathcal{V}_{\xic\lambda}$ (\ref{eq:clampot4}). This is made explicit in (\ref{eq:clampot5}). (\ref{eq:clampot4}) and (\ref{eq:clampot5}) contain $26$ coefficients $c_{53},..,c_{78}$ and $26$ arbitrary constants $r_{\!\xic}, r_{\lamp}, r_{\lamq}, r_{\lamr}, k_{44},..,k_{65}$, respectively.

\subsection{The potential involving $\phia$}
\label{sec:phiapot}

The general potential for $\phia$ is given by
\begin{align}
\text{\small$\mathcal{V}_{\phia}$}&\text{\small$=c_{79}\,m^2|\phia|^2+c_{80}|\phia|^4$}\\
&\text{\small$=k_{66}(|\phia|^2-r_{\!\phia}^2)^2-k_{66}r_{\!\phia}^4.$}
\end{align}
It has a $3$-sphere as its minimum with the symmetry group $\Tt_{\phia}=O(4)_{\phia}$. To break $\Tt_{\phia}$, we introduce an auxiliary generator $\genxrho$ and four driving flavons that transform under it: a complex doublet $\rhop=({\rhop}_1, {\rhop}_2)^\T$, which is a real irrep of dimension four, and three complex singlets $\rhoq$, $\rhor$ and $\rhos$, which are real irreps of dimension two each. How they transform is given in TABLE~\ref{tab:rhosector}. The tensor products of the $2O$ doublets $\phia$ and $\xic$ with $\rhop$ and $\rhos$ are given by
\begin{align}
\phia\!\otimes\rhop&={\phiarho}^{\!\!\!\!\!\!2\alpha}\oplus{\phiarho}^{\!\!\!\!\!\!2\beta}\oplus{\phiarho}^{\!\!\!\!\!\!2\gamma},\label{eq:phiarhop}\\
\xic\!\otimes\rhos&={\xicrho}^{\!\!\!\!\!\!2\alpha}\oplus{\xicrho}^{\!\!\!\!\!\!2\beta},\label{eq:xicrhos}
\end{align}
where 
\begin{align}
{\phiarho}^{\!\!\!\!\!\!2\alpha}&=\llbracket \phia\rrbracket\rhop,\,\,\,\,\,\qquad{\phiarho}^{\!\!\!\!\!\!2\beta}=\llbracket \phia\rrbracket\sigz\rhop,\\
{\phiarho}^{\!\!\!\!\!\!2\gamma}&=({\phia}_1{\rhop}_2,{\phia}_2{\rhop}_2,{\phia}_2^{\!\!*}{\rhop}_1,-{\phia}_1^{\!\!*}{\rhop}_1)^\T,\\
{\xicrho}^{\!\!\!\!\!\!2\alpha}&=\llbracket \xic\rrbracket^\T\llbracket \rhos\rrbracket,\qquad{\xicrho}^{\!\!\!\!\!\!2\beta}=\llbracket \xic\rrbracket^\T\llbracket \rhos\rrbracket^*,
\end{align}
with $\llbracket \rhos\rrbracket=\frac{1}{\sqrt{2}}(\rhos,\rhos^*)^\T$, which is the flavon $\rhos$ expressed in the placeholder notation. How these objects transform is given in TABLE~\ref{tab:rhosector}.

{\renewcommand{\arraystretch}{1.0}
	\setlength{\tabcolsep}{5pt}
	\begin{table}[tbp]
	\begin{center}
	\begin{tabular}{|c|c c c c c c |c|}
\hline
&&&&&&&\\[-1em]
	&\!$\gent$\!&\!$\genr$\!&$\genzn$&$\gencp$&$\genx$&$\genxrho$&$d$\\
\hline
$\rhop$&$1$&$1$&$i$&$-i\sigy$&$\mattau_2\matw_2\sigy$&$\sigy()^*$&$4$\\
$\rhoq$&$1$&$1$&$-1$&$()^*$&$-i\ob()^*$&$-1$&$2$\\
$\rhor$&$1$&$1$&$-1$&$()^*$&$i()^*$&$-1$&$2$\\
$\rhos$ &$1$&$1$&$()^*$&$1$&$-i()^*$&$-i$&$2$\\
\hline
&&&&&&&\\[-1em]
\!\!\!${\phiarho}^{\!\!\!\!\!\!2\alpha}$\!\!\! &$\matt_2$&$\matr_2$&$i$&$-1$&$-i$& $-\sigy()^*$&$4$\\
\!\!\!${\phiarho}^{\!\!\!\!\!\!2\beta}$\!\!\! &$\matt_2$&$\matr_2$&$i$&$1$&$i$& $\sigy()^*$&$4$\\
\!\!\!${\phiarho}^{\!\!\!\!\!\!2\gamma}$\!\!\!&\!\!\!$\mati_2\!\otimes\!\matt_2$\!\!\!&\!\!\!$\mati_2\!\otimes\!\matr_2$\!\!\!&$i$&\!\!\!\!\!$\sigx\!\otimes \!\mati_2$\!\!\!&\!\!\!$-\matw_2i\sigy\!\otimes\!\mati_2$\!\!&\!\!$\sigx\!\otimes\!\sigy()^*$\!\!\!&$8$\\
\hline
&&&&&&&\\[-1em]
\!\!\!${\xicrho}^{\!\!\!\!\!\!2\alpha}$\!\!\! &$\matt_2$&$\matr_2$&$i$&$-1 $&$-i$& $-\sigy()^*$&$4$\\
\!\!\!${\xicrho}^{\!\!\!\!\!\!2\beta}$\!\!\! &$\matt_2$&$\matr_2$&$i$&$-1$&$i$& $\sigy()^*$&$4$\\
\hline
&&&&&&&\\[-1em]
${\rhop}^{\!\!\!\!2\alpha}$&$1$&$1$&$1$&$-1$&$-1$&$-1$&$1$\\
${\rhop}^{\!\!\!\!2\beta}$&$1$&$1$&$1$&$-()^*$&$-i\ob()^*$&$-1$&$2$\\
${\rhop}^{\!\!\!\!2\gamma}$&$1$&$1$&$-1$&$-1$&$1$&$1$&$1$\\
${\rhop}^{\!\!\!\!2\delta}$&$1$&$1$&$-1$&$-1$&$1$&$-1$&$1$\\
${\rhop}^{\!\!\!\!2\epsilon}$&$1$&$1$&$-1$&$()^*$&$-i\ob()^*$&$-1$&$2$\\
${\rhop}^{\!\!\!\!2\zeta}$&$1$&$1$&$-1$&$-()^*$&$i\ob()^*$&$1$&$2$\\
\hline
&&&&&&&\\[-1em]
${\rhoq}^{\!\!\!2}$&$1$&$1$&$1$&$()^*$&$-\om()^*$&$1$&$2$\\
${\rhor}^{\!\!\!2\alpha}$&$1$&$1$&$1$&$1$&$-1$&$1$&$1$\\
${\rhor}^{\!\!\!2\beta}$&$1$&$1$&$1$&$-1$&$1$&$1$&$1$\\
\!\!\!$\rhoq^*\rhor$\!\!\!&$1$&$1$&$1$&$()^*$&$-\om()^*$&$1$&$2$\\
\!\!\!$\rhoq\rhor$\!\!\!&$1$&$1$&$1$&$()^*$&$\ob()^*$&$1$&$2$\\
\hline
&&&&&&&\\[-1em]
${\rhos}^{\!\!\!2\alpha}$ &$1$&$1$&$1$&$1$&$-1$&$-1$&$1$\\
${\rhos}^{\!\!\!2\beta}$ &$1$&$1$&$-1$&$1$&$1$&$-1$&$1$\\
\hline
			\end{tabular}
		\end{center}
		\caption{The transformation rules for the driving flavons $\rhop$, $\rhoq$, $\rhor$ and $\rhos$ and the irreps obtained from them.}
		\label{tab:rhosector}
\end{table}} 

From the table, we see that the objects ${\phiarho}^{\!\!\!\!\!\!2\alpha}$ and ${\xicrho}^{\!\!\!\!\!\!2\alpha}$ transform in the same way. By coupling them together using the cross term $|{\phiarho}^{\!\!\!\!\!\!2\alpha}-\frac{r_{\!\phia}r_{\!\rhop}}{r_{\!\xic}r_{\!\rhos}}{\xicrho}^{\!\!\!\!\!\!2\alpha}|^2$, we can break $\Tt_{\phia}$. This term vanishes at the vevs $\langle\phia\rangle$ and $\langle\xic\rangle$ (\ref{eq:elemvevs}) if we assign
\be\label{eq:rhopsvevs}
\langle\rhop\rangle=\frac{r_{\!\rhop}}{\sqrt{2}}(\eig,\eib)^\T,\qquad\langle\rhos\rangle=r_{\!\rhos}.
\ee
The stabilizers of $\langle\rhop\rangle$ and $\langle\rhos\rangle$ (\ref{eq:rhopsvevs}) are given in TABLE~\ref{tab:flavonstabilizers}. These stabilizers fully determine $\langle\rhop\rangle$ and $\langle\rhos\rangle$ up to their norms. Hence, symmetry arguments guarantee that they appear as stationary points of the potentials of $\rhop$ and $\rhos$. 

Let us construct the general renormalizable potential involving $\rhop$, $\rhoq$, $\rhor$ and $\rhos$. Towards this end, we define the group element $\gene_\rho=\genx_\rho^2$. As shown in TABLE~\ref{tab:eelemflavons}, $\rhop$ and $\rhos$ change sign under $\gene_\rho$ while all other flavons remain invariant. Also, under both $\gene_{\xic}$ and $\gene_{\phia}$, $\rhop$ changes sign while $\rhoq$, $\rhor$ and $\rhos$ remain invariant. This implies that $\rhop$ and $\rhos$ can appear only quadratically in the cross terms in the potential of $\rhop$, $\rhoq$, $\rhor$ and $\rhos$. These quadratic products are given by
\begin{align}
\rhop\!\otimes\!\rhop\!&=\!|\rhop|^2\!\oplus{\rhop}^{\!\!\!\!2\alpha}\!\oplus\!{\rhop}^{\!\!\!\!2\beta}\!\oplus{\rhop}^{\!\!\!\!2\gamma}\!\oplus{\rhop}^{\!\!\!\!2\delta}\!\oplus\!{\rhop}^{\!\!\!\!2\epsilon}\!\oplus{\rhop}^{\!\!\!\!2\zeta},\!\!\!\\
\rhos\!\otimes\rhos\!&=\!|\rhos|^2\!\oplus{\rhos}^{\!\!\!2\alpha}\!\oplus{\rhos}^{\!\!\!2\beta},
\end{align}
where
\begin{align}
{\rhop}^{\!\!\!\!2\alpha}&={\rhop}_1^{\!\!*}{\rhop}_1-{\rhop}_2^{\!\!*}{\rhop}_2,&{\rhop}^{\!\!\!\!2\beta}&={\rhop}_1^{\!\!*}{\rhop}_2,\\
{\rhop}^{\!\!\!\!2\gamma}&=\text{re}({\rhop}_1{\rhop}_2),&{\rhop}^{\!\!\!\!2\delta}&=\text{im}({\rhop}_1{\rhop}_2),\\
{\rhop}^{\!\!\!\!2\epsilon}&={\rhop}_1^{\!\!*2}+{\rhop}_2^2,&{\rhop}^{\!\!\!\!2\zeta}&={\rhop}_1^{\!\!*2}-{\rhop}_2^2,\\
{\rhos}^{\!\!\!2\alpha}&=\text{re}({\rhos}^{\!\!2}), &{\rhos}^{\!\!\!2\beta}&=\text{im}({\rhos}^{\!\!2}).
\end{align}
How these objects transform is given in TABLE~\ref{tab:rhosector}. 

Using $\rhop$, we construct
\begin{align}
\text{\small$\breve{\mathcal{V}}_{\rhop}$}&\text{\small$=c_{81}\,m^2|\rhop|^2+c_{82}|\rhop|^4$}\label{eq:rhoppot1}\\
&\text{\small$=k_{67}(|\rhop|^2-r_{\!\rhop}^2)^2-k_{67}r_{\!\rhop}^4,$}\label{eq:rhoppot2}\\
\text{\small$\mathcal{V}_{\rhop}$}&\text{\small$=\breve{\mathcal{V}}_{\rhop}+c_{83} ({\rhop}^{\!\!\!\!2\alpha})^2+c_{84} ({\rhop}^{\!\!\!\!2\delta})^2$}\label{eq:rhoppot3}\\
&\text{\small$=\breve{\mathcal{V}}_{\rhop}+k_{68} ({\rhop}^{\!\!\!\!2\alpha})^2+k_{69} ({\rhop}^{\!\!\!\!2\delta})^2$}.\label{eq:rhoppot4}
\end{align}
$\breve{\mathcal{V}}_{\rhop}$ has a $3$-sphere with radius $r_{\!\rhop}$ as its minimum with $\Tb_{\rhop}\!=\!O(4)_{\rhop}$. We obtained $\mathcal{V}_{\rhop}$ (\ref{eq:rhoppot3}), which is the general renormalizable potential of $\rhop$, by adding the terms $({\rhop}^{\!\!\!\!2\alpha})^2$ and $({\rhop}^{\!\!\!\!2\delta})^2$ to $\breve{\mathcal{V}}_{\rhop}$ (\ref{eq:rhoppot1}). The newly added terms break $\Tb_{\rhop}$ to $\Ga_{\rhop}\!=\!((U(1)\!\times\!Z_2)\rtimes K_4)_{\rhop}$ generated by the group actions $({\rhop}_1,{\rhop}_2)^\T\rightarrow (e^{i\theta}{\rhop}_1,e^{-i\theta}{\rhop}_2)^\T$, $({\rhop}_1,{\rhop}_2)^\T\rightarrow (-{\rhop}_1,{\rhop}_2)^\T$, $({\rhop}_1,{\rhop}_2)^\T\rightarrow ({\rhop}_2,{\rhop}_1)^\T$ and $({\rhop}_1,{\rhop}_2)^\T\rightarrow ({\rhop}_1^*,{\rhop}_2^*)^\T$. The $3$-sphere gets stratified under this group. Let us consider the one-dimensional manifold $\frac{r_{\!\rhop}}{\sqrt{2}}(e^{i\theta},e^{-i\theta})^\T$ in the $3$-sphere. We can show that this manifold is isolated in its stratum; hence, we are guaranteed to obtain it as a stationary manifold of $\mathcal{V}_{\rhop}$ (\ref{eq:rhoppot3}). The rewritten potential (\ref{eq:rhoppot4}) shows that it can be obtained as a minimum since it corresponds to $|\rhop|=r_{\!\rhop}$, ${\rhop}^{\!\!\!\!2\alpha}=0$ and ${\rhop}^{\!\!\!\!2\delta}=0$. The symmetry groups associated with the minimum $\frac{r_{\!\rhop}}{\sqrt{2}}(e^{i\theta},e^{-i\theta})^\T$ are $\Ht_{\rhop}\!=\!\Ha_{\rhop}\!=\!Z_{2\rhop}$ generated by $({\rhop}_1,{\rhop}_2)^\T\rightarrow ({\rhop}_2^*,{\rhop}_1^*)^\T$ and $\Tt_{\rhop}=(U(1)\rtimes Z_2)_{\rhop}=O(2)_{\rhop}$ generated by $({\rhop}_1,{\rhop}_2)^\T\rightarrow (e^{i\theta}{\rhop}_1,e^{-i\theta}{\rhop}_2)^\T$ and $({\rhop}_1,{\rhop}_2)^\T\rightarrow ({\rhop}_1^*,{\rhop}_2^*)^\T$.

To break the continuous group $\Tt_{\rhop}=O(2)_{\rhop}$, we introduce the driving flavon $\rhoq$, which transforms in the same way as ${\rhop}^{\!\!\!\!2\epsilon}$, TABLE~\ref{tab:rhosector}. We utilize the cross term $|{\rhop}^{\!\!\!\!2\epsilon}-\frac{r_{\!\rhop}^2}{r_{\!\rhoq}}\rhoq|^2$ to break $\Tt_{\rhop}$. For $\rhoq$, we assign the vev
\be\label{eq:rhoqvev}
\langle\rhoq\rangle=r_{\!\rhoq}\taub.
\ee 
At $\langle\rhop\rangle$ (\ref{eq:rhopsvevs}) and $\langle\rhoq\rangle$ (\ref{eq:rhoqvev}), the above-mentioned cross term vanishes. The vev (\ref{eq:rhoqvev}) is fully defined up to its norm by its stabilizer, TABLE~\ref{tab:flavonstabilizers}. Hence, it is guaranteed to be a stationary point. 

To explicitly construct the potential of $\rhoq$, we obtain the tensor product expansion,
\be
\rhoq\otimes\rhoq=|\rhoq|^2\oplus{\rhoq}^{\!\!2}.
\ee
The general renormalizable potential of $\rhoq$ is
\begin{align}
\text{\small$\mathcal{V}_{\rhoq}$}&\text{\small$=c_{85}\,m^2|\rhoq|^2+c_{86} |\rhoq|^4$}\label{eq:rhopqot1}\\
&\text{\small$= k_{70} (|\rhoq|^2-r_{\!\rhoq}^2)^2-k_{70}r_{\!\rhoq}^4$}.\label{eq:rhoqpot2}
\end{align}
Its minimum is a circle with $\Tt_{\rhoq}=O(2)_{\rhoq}$ generated by $\rhoq\rightarrow e^{i\theta}\rhoq$ and $\rhoq\rightarrow \rhoq^*$. To break $\Tt_{\rhoq}$, we introduce the flavon $\rhor$. The relevant tensor products are
\begin{align}
\rhor\otimes\rhor&=|\rhor|^2\oplus{\rhor}^{\!\!\!2\alpha}\oplus{\rhor}^{\!\!\!2\beta},\\
\rhoq\otimes\rhor&=\rhoq^*\rhor\oplus\rhoq\rhor,
\end{align}
where
\be
{\rhor}^{\!\!\!2\alpha}=\text{re}({\rhor}^{\!\!2}),\qquad{\rhor}^{\!\!\!2\beta}=\text{im}({\rhor}^{\!\!2}).
\ee
How these objects transform is given in TABLE~\ref{tab:rhosector}. We can see that ${\rhoq}^{\!\!2}$ and $\rhoq^*\rhor$ transform in the same way. Therefore, we can construct the cross term $|{\rhoq}^{\!\!2}-\frac{r_{\!\rhoq}}{r_{\!\rhor}}\rhoq^*\rhor|^2$ to break $\Tt_{\rhoq}$. For $\rhor$, we assign the vev
\be\label{eq:rhorvev}
\langle\rhor\rangle=-r_{\!\rhor}\tau.
\ee
At $\langle\rhoq\rangle$ (\ref{eq:rhoqvev}) and $\langle\rhor\rangle$ (\ref{eq:rhorvev}), the above-mentioned cross term vanishes. 

Using $\rhor$, we construct
\begin{align}
\text{\small$\breve{\mathcal{V}}_{\rhor}$}&\text{\small$=c_{87}\,m^2|\rhor|^2+c_{88}|\rhor|^4$}\label{eq:rhorpot1}\\
&\text{\small$=k_{71}(|\rhor|^2-r_{\!\rhor}^2)^2-k_{71}r_{\!\rhor}^4$},\label{eq:rhorpot2}\\
\text{\small$\mathcal{V}_{\rhor}$}&\text{\small$=\breve{\mathcal{V}}_{\rhor}+c_{89} ({\rhor}^{\!\!\!2\alpha})^2$}\label{eq:rhorpot3}\\
&\text{\small$=\breve{\mathcal{V}}_{\rhor}+k_{72} ({\rhor}^{\!\!\!2\alpha})^2$}.\label{eq:rhorpot4}
\end{align}
$\breve{\mathcal{V}}_{\rhor}$ has a circle as its minimum with $\Tb_{\rhor}=O(2 )_{\rhor}$ generated by $\rhor\rightarrow e^{i\theta}\rhor$ and $\rhor\rightarrow \rhor^*$. The term $({\rhor}^{\!\!\!2\alpha})^2$ (\ref{eq:rhorpot3}) breaks $\Tb_{\rhor}$ to $\Ga_{\rhor}=D_{8\rhor}$ generated by $\rhor\rightarrow i\rhor$ and $\rhor\rightarrow \rhor^*$. The circle gets stratified under $D_{8\rhor}$. The stabilizer of $\langle\rhor\rangle$ (\ref{eq:rhorvev}) is $\Ht_{\rhor}=\Ha_{\rhor}=Z_{2\rhor}$ generated by $\rhor\rightarrow i\rhor^*$. We also have $\Tt_{\rhor}=1$. The stabilizer $Z_{2\rhor}$ fully determines $\langle\rhor\rangle$ in the circle; hence, we are guaranteed to obtain it as a stationary point of $\mathcal{V}_{\rhor}$ (\ref{eq:rhorpot3}). The rewritten expression (\ref{eq:rhorpot4}) shows that the stationary point can be obtained as a minimum.

We combine the potentials of $\rhoq$ and $\rhor$ to obtain 
\be\label{eq:rhoqrpot1}
\text{\small$\breve{\mathcal{V}}_{\rhoqr}=\mathcal{V}_{\rhoq}+\mathcal{V}_{\rhor}$},
\ee
where $\rhoqr=(\rhoq,\rhor)^\T$. It has the minimum $(\langle\rhoq\rangle,\langle\rhor\rangle)^\T$ with the symmetry groups $\Hb_{\rhoqr}=\Ht_{\rhoq}\times \Ht_{\rhor}=Z_{2\rhor}$ and $\Tb_{\rhoqr}=\Tt_{\rhoq}\times \Tt_{\rhor}=O(2)_{\rhoq}$. We couple $\rhoq$ with $\rhor$ to break $\Tb_{\rhoqr}$ as follows,
\begin{align}
\text{\small$\tilde{\mathcal{V}}_{\rhoqr}$}&\text{\small$=\breve{\mathcal{V}}_{\rhoqr}+c_{90}\,\text{re}(\rhoq^3\rhor^*)$}\label{eq:rhoqrpot2}\\
&\text{\small$=\breve{\mathcal{V}}_{\rhoqr}+k_{73} |{\rhoq}^{\!\!2}\!-\!\frac{r_{\!\rhoq}}{r_{\!\rhor}}\rhoq^*\rhor|^2+\frac{k_{73}}{2} (|{\rhoq}|^2-\frac{r_{\!\rhoq}^2}{r_{\!\rhor}^2}|\rhor|^2)^2$}.\label{eq:rhoqrpot3}
\end{align}
The term $\text{re}(\rhoq^3\rhor^*)$ breaks $\Tb_{\rhoqr}$ to $\Ga_{\rhoqr}=D_{6\rhoq}$ generated by $\rhoq\rightarrow\om\rhoq$ and $\rhop\rightarrow-i\rhoq^*$. The circle gets stratified under $\Ga_{\rhoqr}$. The vev (\ref{eq:rhoqvev}) has the stabilizer $\Ha_{\rhoqr}=Z_{2\rhoq}$ generated by $\rhoq\rightarrow-i\rhoq^*$. Since $\Ha_{\rhoqr}$ fully determines $\langle\rhoq\rangle$ in the circle, we are guaranteed to obtain it as a stationary point of $\tilde{\mathcal{V}}_{\rhoqr}$ (\ref{eq:rhoqrpot2}). The rewritten potential (\ref{eq:rhoqrpot3}) shows that we can obtain the stationary point as a minimum. $\tilde{\mathcal{V}}_{\rhoqr}$ (\ref{eq:rhoqrpot3}) contains six arbitrary constants $r_{\!\rhoq}, r_{\!\rhor}, k_{70}, .., k_{73}$, which are functions of the six coefficients $c_{85}, .., c_{90}$ in (\ref{eq:rhoqrpot2}). We have the symmetry groups $\Ht_{\rhoqr}=\Ha_{\rhoqr}\times\Hb_{\rhoqr}=Z_{2\rhoq}\times Z_{2\rhor}$ and $\Tt_{\rhoqr}=1$.

We obtain the general potential involving $\rhoqr$ as 
\begin{align}
\text{\small$\mathcal{V}_{\rhoqr}$}&\text{\small$=\tilde{\mathcal{V}}_{\rhoqr}+c_{91}|{\rhoq}|^2|{\rhor}|^2$}\label{eq:rhoqrpot4}\\
&\text{\small$=\tilde{\mathcal{V}}_{\rhoqr}+k_{74} (|{\rhoq}|^2-\frac{r_{\!\rhoq}^2}{r_{\!\rhor}^2}|\rhor|^2)^2$}.\label{eq:rhoqrpot5}
\end{align}
The cross term $|{\rhoq}|^2|{\rhor}|^2$ is compatible; hence, our vevs $\langle\rhoq\rangle$ (\ref{eq:rhoqvev}) and $\langle\rhor\rangle$ (\ref{eq:rhorvev}) and the symmetry group $\Ht_{\rhoqr}=Z_{2\rhoq}\times Z_{2\rhor}$ remain unspoiled.

We combine $\mathcal{V}_{\rhop}$ and $\mathcal{V}_{\rhoqr}$ to obtain
\be\label{eq:rhopqrpot1}
\text{\small$\breve{\mathcal{V}}_{\rhopqr}=\mathcal{V}_{\rhop}+\mathcal{V}_{\rhoqr}$},
\ee
where $\rhopqr=(\rhop,\rhoqr)^\T=(\rhop,\rhoq,\rhor)^\T$. It has a minimum corresponding to the one-dimensional manifold $\frac{r_{\!\rhop}}{\sqrt{2}}(e^{i\theta},e^{-i\theta})^\T$ for $\rhop$ and the vevs $\langle\rhoq\rangle$ and $\langle\rhor\rangle$ for $\rhoq$ and $\rhor$, respectively. The symmetry groups of this minimum are $\Hb_{\rhopqr}=\Ht_{\rhop}\times \Ht_{\rhoqr}=Z_{2\rhop}\times Z_{2\rhoq}\times Z_{2\rhor}$ and $\Tb_{\rhopqr}=\Tt_{\rhop}\times \Tt_{\rhoqr}=O(2)_{\rhop}$.

We couple $\rhop$ with $\rhoq$ to break the $\Tb_{\rhopqr}$ and obtain
\begin{align}
\text{\small$\tilde{\mathcal{V}}_{\rhopqr}$}&\text{\small$=\breve{\mathcal{V}}_{\rhopqr}+c_{92}\,m\,\text{re}({\rhop}^{\!\!\!\!2\epsilon}\rhoq^*)$}\label{eq:rhopqrpot2}\\
&\text{\small$=\breve{\mathcal{V}}_{\rhopqr}+k_{75} |{\rhop}^{\!\!\!\!2\epsilon}-\frac{r_{\!\rhop}^2}{r_{\!\rhoq}}\rhoq|^2$}.\label{eq:rhopqrpot3}
\end{align}
The term $\text{re}({\rhop}^{\!\!\!2\epsilon}\rhoq^*)$ breaks $\Tb_{\rhopqr}$ to $\Ga_{\rhopqr}=(Z_2'\times Z_2'')_{\rhop}$ generated by $\rhop\rightarrow-\rhop$ and $\rhop\rightarrow\mattau_2^*\rhop^*$. The manifold $\frac{r_{\!\rhop}}{\sqrt{2}}(e^{i\theta},e^{-i\theta})^\T$ gets stratified under $(Z_2'\times Z_2'')_{\rhop}$. In the manifold, $\langle\rhop\rangle$ (\ref{eq:rhopsvevs}) is fully determined by its stabilizer $\Ha_{\rhopqr}=Z''_{2\rhop}$ generated by $\rhop\rightarrow\mattau_2^*\rhop^*$. Hence, $\langle\rhop\rangle$ is guaranteed to be a stationary point of $\tilde{\mathcal{V}}_{\rhopqr}$ (\ref{eq:rhopqrpot2}). The rewritten potential (\ref{eq:rhopqrpot3}) shows that the stationary point can be obtained as a minimum. $\tilde{\mathcal{V}}_{\rhopqr}$ (\ref{eq:rhopqrpot2}) contains $12$ coefficients, $c_{81}, .., c_{92}$, and correspondingly, $\tilde{\mathcal{V}}_{\rhopqr}$ (\ref{eq:rhopqrpot3}) contains $12$ arbitatry constants, $r_{\!\rhop}, r_{\!\rhoq}, r_{\!\rhor}, k_{67}, .., k_{75}$.  The symmetry groups associated with our minimum are $\Ht_{\rhopqr}=\Ha_{\rhopqr}\times\Hb_{\rhopqr}=Z''_{2\rhop}\times Z_{2\rhop}\times Z_{2\rhoq}\times Z_{2\rhor}=K_{4\rhop}\times Z_{2\rhoq}\times Z_{2\rhor}$ and $\Tt_{\rhopqr}=1$.

The general potential constructed with $\rhopqr$ is
\begin{align}
\text{\small$\mathcal{V}_{\rhopqr}$}&\text{\small$=\tilde{\mathcal{V}}_{\rhopqr}+c_{93}|{\rhop}|^2|{\rhoq}|^2+c_{94}|{\rhop}|^2|{\rhor}|^2$}\label{eq:rhopqrpot4}\\
&\text{\small$=\tilde{\mathcal{V}}_{\rhopqr}+k_{76}(|{\rhop}|^2\!-\frac{r_{\!\rhop}^2}{r_{\!\rhoq}^2}|\rhoq|^2)+k_{77}(|{\rhop}|^2\!-\frac{r_{\!\rhop}^2}{r_{\!\rhor}^2}|\rhor|^2)$}.\label{eq:rhopqrpot5}
\end{align}
The newly added cross terms are compatible. Hence, our vevs and the associated symmetry group remain unspoiled. $\mathcal{V}_{\rhopqr}$ contains $14$ coefficients (\ref{eq:rhopqrpot4}) and $14$ arbitrary constants (\ref{eq:rhopqrpot5}).

Using $\rhos$, we construct
\begin{align}
\text{\small$\breve{\mathcal{V}}_{\rhos}$}&\text{\small$=c_{95}\,m^2|\rhos|^2+c_{96}|\rhos|^4$}\label{eq:rhospot1}\\
&\text{\small$=k_{78}(|\rhos|^2-r_{\!\rhos}^2)^2-k_{78}r_{\!\rhos}^4$},\label{eq:rhospot2}\\
\text{\small$\mathcal{V}_{\rhos}$}&\text{\small$=\breve{\mathcal{V}}_{\rhos}+c_{97} ({\rhos}^{\!\!\!2\beta})^2$}\label{eq:rhospot3}\\
&\text{\small$=\breve{\mathcal{V}}_{\rhos}+k_{79} ({\rhos}^{\!\!\!2\beta})^2$}.\label{eq:rhospot4}
\end{align}
$\mathcal{V}_{\rhos}$ has a circle as its minimum with $\Tb=O(2)_{\rhos}$. The term $({\rhos}^{\!\!\!2\beta})^2$ breaks $\Tb$ to $\Ga_{\rhos}=D_{8\rhos}$ generated by $\rhos\rightarrow i\rhos$ and $\rhos\rightarrow\rhos^*$. The circle gets stratified under $\Ga_{\rhos}$. The stabilizer of $\langle\rhos\rangle$ (\ref{eq:rhopsvevs}) is $\Ha_{\rhos}=Z_{2\rhos}$ generated by $\rhos\rightarrow\rhos^*$. $\Ha_{\rhos}$ fully determines $\langle\rhos\rangle$ in the circle. Hence, we are guaranteed to obtain $\langle\rhos\rangle$ as a stationary point of $\mathcal{V}_{\rhos}$ (\ref{eq:rhospot3}). The rewritten form (\ref{eq:rhospot4}) shows that we can obtain it as a minimum. We have $\Ht_{\rhos}=\Ha_{\rhos}=Z_{2\rhos}$ and $\Tt_{\rhos}=1$.
 
We combine $\mathcal{V}_{\rhopqr}$ and $\mathcal{V}_{\rhos}$ to obtain
\be\label{eq:rhopot1}
\text{\small$\tilde{\mathcal{V}}_{\rho}=\mathcal{V}_{\rhopqr}+\mathcal{V}_{\rhos}$},
\ee
where $\rho=(\rhopqr,\rhos)^\T=(\rhop,\rhoq,\rhor,\rhos)^\T$. It has a discrete minimum at $\langle\rhop\rangle$, $\langle\rhoq\rangle$, $\langle\rhor\rangle$ and $\langle\rhos\rangle$ (\ref{eq:rhopsvevs}), (\ref{eq:rhoqvev}), (\ref{eq:rhorvev}). The symmetry groups of this minimum are $\Ht_{\rho}=\Ht_{\rhopqr}\times\Ht_{\rhos}=K_{4\rhop}\times Z_{2\rhoq}\times Z_{2\rhor}\times Z_{2\rhos}$ and $\Tt_{\rho}=\Tt_{\rhopqr}\times\Tt_{\rhos}=1$. The general potential of $\rho$ is given by
\begin{align}
\text{\small$\mathcal{V}_{\rho}$}&\text{\small$=\tilde{\mathcal{V}}_{\rho}+c_{98}|{\rhop}|^2|{\rhos}|^2+c_{99}|{\rhoq}|^2|{\rhos}|^2+c_{100}|{\rhor}|^2|{\rhos}|^2$}\label{eq:rhopot2}\!\!\!\\
\begin{split}\label{eq:rhopot3}
&\text{\small$=\tilde{\mathcal{V}}_{\rho}+k_{80}(|{\rhop}|^2\!-\frac{r_{\!\rhop}^2}{r_{\!\rhos}^2}|\rhos|^2)+k_{81}(|{\rhoq}|^2\!-\frac{r_{\!\rhoq}^2}{r_{\!\rhos}^2}|\rhos|^2)\!\!\!$}\\
&\text{\small$\qquad\qquad+k_{82}(|{\rhor}|^2\!-\frac{r_{\!\rhor}^2}{r_{\!\rhos}^2}|\rhos|^2)$}.
\end{split}\!\!\!
\end{align}
The newly added cross terms are compatible. Therefore, the form of our vevs, as well as our symmetry groups, are conserved. $\mathcal{V}_{\rho}$ (\ref{eq:rhopot2}), (\ref{eq:rhopot3}) contains $20$ coefficients and $20$ arbitrary constants.

Using $\phia$, $\rho$ and $\xic\lambda$, we construct
\be
\text{\small$\breve{\mathcal{V}}_{\phia\rho\xic\lambda}=\mathcal{V}_{\phia}+\mathcal{V}_{\rho}+\mathcal{V}_{\xic\lambda}$}.
\ee
This potential has a minimum corresponding to a $3$-sphere for $\phia$ and the vevs $\langle\rhop\rangle$, $\langle\rhoq\rangle$, $\langle\rhor\rangle$, $\langle\rhos\rangle$, $\langle\xic\rangle$, $\langle\lamp\rangle$, $\langle\lamq\rangle$ and $\langle\lamr\rangle$ for the rest of the flavons. The minimum has the symmetry groups $\Hb_{\phia\rho\xic\lambda}=\Ht_{\phia}\times\Ht_{\rho}\times\Ht_{\xic\lambda}=K_{4\rhop}\times Z_{2\rhoq}\times Z_{2\rhor}\times Z_{2\rhos}\times K_{4\xic}\times(S_4\times Z_2)_{\lamp}\times(S_4\times Z_2)_{\lamq}\times ((Z_{2_1}\times Z_{2_3})\rtimes Z_2)_{\lamr}$ and $\Tb_{\phia\rho\xic\lambda}=\Tt_{\phia}\times\Tt_{\rho}\times\Tt_{\xic\lambda}=O(4)_{\phia}$.

We couple $\phia$ with $\xic$ to obtain
\begin{align}
\text{\small$\tilde{\mathcal{V}}_{\phia\rho\xic\lambda}$}&\text{\small$=\breve{\mathcal{V}}_{\phia\rho\xic\lambda}+c_{101}\,\text{re}({\phiarho}^{\!\!\!\!\!\!2\alpha\dagger}{\xicrho}^{\!\!\!\!\!\!2\alpha})$}\label{eq:phiafullpot1}\\
\begin{split}
&\text{\small$=\breve{\mathcal{V}}_{\phia\rho\xic\lambda}+k_{83}|{\phiarho}^{\!\!\!\!\!\!2\alpha}-\frac{r_{\!\phia}r_{\!\rhop}}{r_{\!\xic}r_{\!\rhos}}{\xicrho}^{\!\!\!\!\!\!2\alpha}|^2$}\\
&\text{\small$\qquad+\frac{k_{83}}{2}\frac{r_{\!\rhop}^2}{r_{\!\phia}^2} (|{\phia}|^2-\frac{r_{\!\phia}^2}{r_{\!\rhop}^2}|\rhop|^2)^2$}\\
&\text{\small$\qquad+\frac{k_{83}}{2}\frac{r_{\!\phia}^2r_{\!\rhop}^2}{r_{\!\xic}^4}(|{\xic}|^2-\frac{r_{\!\xic}^2}{r_{\!\rhos}^2}|\rhos|^2)^2$}.
\end{split}\label{eq:phiafullpot2}
\end{align}
The newly added term $\text{re}({\phiarho}^{\!\!\!\!\!\!2\alpha\dagger}{\xicrho}^{\!\!\!\!\!\!2\alpha})$ (\ref{eq:phiafullpot1}) breaks $\Tb_{\phia\rho\xic\lambda}=O(4)_{\phia}$ to $\Ga_{\phia\rho\xic\lambda}=(K_4\rtimes Z_2)_{\phia}$ generated by ${\phia}_1\rightarrow{\phia}_1^*$, ${\phia}_2\rightarrow i{\phia}_2^*$ and $\phia\rightarrow\mattau_2\sigx\phia$. The $3$-sphere gets stratified under $(K_4\rtimes Z_2)_{\phia}$. The stabilizer of $\langle\phia\rangle$ (\ref{eq:elemvevs}) is $(K_4\rtimes Z_2)_{\phia}$ itself, i.e.,~we have $\Ha_{\phia\rho\xic\lambda}=(K_4\rtimes Z_2)_{\phia}$. It fully determines $\langle\phia\rangle$ in the $3$-sphere. Hence, $\tilde{\mathcal{V}}_{\phia\rho\xic\lambda}$ is guaranteed to have a stationary point corresponding to $\langle\phia\rangle$ and the vevs of other flavons. The rewritten form (\ref{eq:phiafullpot2}) shows that the stationary point can be obtained as a minimum. $\tilde{\mathcal{V}}_{\phia\rho\xic\lambda}$ (\ref{eq:phiafullpot1}), (\ref{eq:phiafullpot2}) contains $49$ coefficients: $c_{53}, .., c_{101}$ and $49$ arbitrary constants: $r_{\!\phia}, r_{\!\rhop}, r_{\!\rhoq}, r_{\!\rhor}, r_{\!\rhos}, r_{\!\xic}, r_{\lamp}, r_{\lamq}, r_{\lamr}, k_{44}, .., k_{83}$. We obtain $\Ht_{\phia\rho\xic\lambda}= \Ha_{\phia\rho\xic\lambda}\times \Hb_{\phia\rho\xic\lambda}=(K_4\rtimes Z_2)_{\phia}\times K_{4\rhop}\times Z_{2\rhoq}\times Z_{2\rhor}\times Z_{2\rhos}\times K_{4\xic}\times(S_4\times Z_2)_{\lamp}\times(S_4\times Z_2)_{\lamq}\times ((Z_{2_1}\times Z_{2_3})\rtimes Z_2)_{\lamr}$ and $\Tt_{\phia\rho\xic\lambda}=1$.

To help us in constructing the general potential involving these flavons, we consider them as three multiplets:  $\phia$,\,\, $\rho\!=\!(\rhop,\rhoq,\rhor,\rhos)^\T$ and $\xic\lambda\!=\!(\xic,\lamp,\lamq,\lamr)^\T$. In $\tilde{\mathcal{V}}_{\phia\rho\xic\lambda}$ (\ref{eq:phiafullpot1}), (\ref{eq:phiafullpot2}), we have considered all cross terms within each of these multiplets. Now, we need to determine the cross terms among these multiplets. The flavons $\{\phia, \rhop\}$, $\{\xic,\rhop\}$ and $\{\rhop,\rhos\}$ change sign under $\gene_{\phia}$, $\gene_{\xic}$ and $\gene_{\rho}$, respectively, while every other flavon remains invariant, TABLE~\ref{tab:eelemflavons}. Therefore, the only allowed combination where $\phia$ or $\xic$ can appear linearly is $\phia\otimes\rhop\otimes\xic\otimes\rhos$. We have already studied this combination, (\ref{eq:xicrhos}), (\ref{eq:phiarhop}). The only flavons that transform non-trivially under $\genxrho$ and $\genxlam$ are $\rho\!=\!(\rhop,\rhoq,\rhor,\rhos)^\T$ and $\lambda\!=\!(\lamp,\lamq,\lamr)^\T$, respectively. This leads us to conclude that, besides $\phia\otimes\rhop\otimes\xic\otimes\rhos$, the only allowed cross combinations among $\phia$, $\rho$ and $\xic\lambda$ are the ones where they contribute quadratically, i.e.,~$\{\phia\otimes\phia\}$, $\{\rho\otimes\rho\}$ and $\{\xic\otimes\xic,\lambda\otimes\lambda\}$. The irreps obtained from these quadratic tensor products can be found in TABLES~\ref{tab:prodtrof}, \ref{tab:lamsector}, \ref{tab:rhosector}. By examining them, we can see that there is only one allowed cross term given by $\rhor^{2\alpha}\lamp^\T\lamr$. According to corollary~B in \cite{2306.07325}, this term is compatible since both $\rhor^{2\alpha}$ and $\lamp^\T\lamr$ vanish at our assigned minimum $\langle\rhor\rangle$ (\ref{eq:rhorvev}), $\langle\lamp\rangle$ (\ref{eq:lampvev}), $\langle\lamr\rangle$ (\ref{eq:lamrvev}). Besides this term, we also have cross terms that are products of norms. Among $\phia$, $\rho$ and $\xic\lambda$, we have a total of $4+4+16=24$ such cross terms: $|\phia|^2|\rhop|^2$, $..$, $|\rhos|^2|\lamr|^2$. They are also compatible. Therefore, adding these cross terms will not spoil the symmetry group $\Ht_{\phia\rho\xic\lambda}$. With this information, we construct the general renormalizable potential
\begin{align}
\begin{split}
\text{\small$\mathcal{V}_{\phia\rho\xic\lambda}$}&\text{\small$=\tilde{\mathcal{V}}_{\phia\rho\xic\lambda}+c_{102}\rhor^{2\alpha}\lamp^\T\lamr$}\\
&\text{\small$\qquad\quad+c_{103}|\phia|^2|\rhop|^2+...+c_{126}|\rhos|^2|\lamr|^2$}
\end{split}\label{eq:phiaxicrest1}\\
\begin{split}
&\text{\small$=\tilde{\mathcal{V}}_{\phia\rho\xic\lambda}+k_{84}(\rhor^{2\alpha}-\frac{r_{\!\rhor}^2}{r_{\lamp}r_{\lamr}}\lamp^\T\lamr)^2+$}\\
&\text{\small$\quad+k_{85}(|{\phia}|^2\!-\!\frac{r_{\!\phia}^2}{r_{\!\rhop}^2}|\rhop|^2)^2\!+...+k_{108}(|{\rhos}|^2\!-\!\frac{r_{\!\rhos}^2}{r_{\!\lamr}^2}|\lamr|^2)^2$}\!.
\end{split}\label{eq:phiaxicrest2}
\end{align}
This potential has a discrete minimum at our assigned vevs: $\langle\phia\rangle$ (\ref{eq:elemvevs}), $\langle\rhop\rangle$ (\ref{eq:rhopsvevs}), $\langle\rhoq\rangle$ (\ref{eq:rhoqvev}), $\langle\rhor\rangle$ (\ref{eq:rhorvev}), $\langle\rhos\rangle$ (\ref{eq:rhopsvevs}), $\langle\xic\rangle$ (\ref{eq:elemvevs}), $\langle\lamp\rangle$ (\ref{eq:lampvev}), $\langle\lamq\rangle$ (\ref{eq:lamqvev}), $\langle\lamr\rangle$ (\ref{eq:lamrvev}) with $\Ht_{\phia\rho\xic\lambda}=(K_4\rtimes Z_2)_{\phia}\times K_{4\rhop}\times Z_{2\rhoq}\times Z_{2\rhor}\times Z_{2\rhos}\times K_{4\xic}\times(S_4\times Z_2)_{\lamp}\times(S_4\times Z_2)_{\lamq}\times ((Z_{2_1}\times Z_{2_3})\rtimes Z_2)_{\lamr}$. (\ref{eq:phiaxicrest2}) contains $74$ arbitrary constants $r_{\!\phia}, r_{\!\rhop}, r_{\!\rhoq}, r_{\!\rhor}, r_{\!\rhos}, r_{\!\xic}, r_{\lamp}, r_{\lamq}, r_{\lamr}, k_{44}, .., k_{108}$, which are functions of the $74$ coefficients $c_{53}, .., c_{126}$ in (\ref{eq:phiaxicrest1}).

\subsection{The potential involving $\phic$}
\label{sec:phicpot}

Construction for the potential of $\phic$ follows steps very similar to that of $\phia$. Using $\phic$, we construct
\begin{align}
\text{\small$\mathcal{V}_{\phic}$}&\text{\small$=c_{127}\,m^2|\phic|^2+c_{128}|\phic|^4$}\\
&\text{\small$=k_{109}(|\phic|^2-r_{\!\phic}^2)^2-k_{109}r_{\!\phic}^4.$}
\end{align}
It has a $3$-sphere as its minimum with $\Tt_{\phic}=O(4)_{\phic}$. To break $\Tt_{\phic}$, we introduce the auxiliary generator $\genxeta$ and driving flavons $\etap=({\etap}_1, {\etap}_2)^\T$, $\etaq$, $\etar$ and $\etas$, which transform as given in TABLE~\ref{tab:etasector}. We assign the following vevs for these flavons,
\begin{align}
\langle\etap\rangle&=\frac{r_{\!\etap}}{\sqrt{2}}(\eib,\eig)^\T,&\langle\etaq\rangle&=r_{\!\etaq}\tau,\label{eq:etapqvevs}\\
\langle\etar\rangle&=-r_{\!\etar}\taub,&\langle\etas\rangle&=-r_{\!\etas} i.\label{eq:etarsvevs}
\end{align}
The stabilizers of these vevs are provided in TABLE~\ref{tab:flavonstabilizers}. These stabilizers fully determine the corresponding vevs up to their norms. Hence, the vevs are guaranteed to be stationary points of the corresponding potentials.

{\renewcommand{\arraystretch}{1.0}
	\setlength{\tabcolsep}{5pt}
	\begin{table}[tbp]
	\begin{center}
	\begin{tabular}{|c|c c c c c c |c|}
\hline
&&&&&&&\\[-1em]
	&\!$\gent$\!&\!$\genr$\!&$\genzn$&$\gencp$&$\genx$&$\genxeta$&$d$\\
\hline
$\etap$&$1$&$1$&$\sigx$&\!\!$-i\sigy$\!\!&$-\mattau_2^*\matw_2^*\sigy$&$-\sigy()^*$&$4$\\
$\etaq$&$1$&$1$&$()^*$&$()^*$&$i\om()^*$&$-1$&$2$\\
$\etar$&$1$&$1$&$()^*$&$()^*$&$-i()^*$&$-1$&$2$\\
$\etas$ &$1$&$1$&$()^*$&$1$&$i()^*$&$i$&$2$\\
\hline
&&&&&&&\\[-1em]
\!\!\!\!${\phiceta}^{\!\!\!\!\!\!2\alpha}$\!\!\!\!&$\matt_2$&$\matr_2$&$i$&$-1$&$i$& $\sigy()^*$&$4$\\
\!\!\!\!${\phiceta}^{\!\!\!\!\!\!2\beta}$\!\!\!\!&$\matt_2$&$\matr_2$&$-i$&$1$&$-i$& $-\sigy()^*$&$4$\\
\!\!\!\!${\phiceta}^{\!\!\!\!\!\!2\gamma}$\!\!\!\!&\!\!\!$\mati_2\!\otimes\!\matt_2$\!\!\!&\!\!$\mati_2\!\otimes\!\matr_2$\!\!&\!\!\!$i\sigx\!\otimes\!\mati_2$\!\!\!&\!\!\!\!$\sigx\!\otimes\!\mati_2$\!\!\!\!&\!\!\!$-\matw_2^*i\sigy\!\otimes\!\mati_2$\!\!&\!\!\!$-\sigx\!\otimes\!\sigy()^*$\!\!\!&$8$\\
\hline
&&&&&&&\\[-1em]
\!\!\!\!${\xiceta}^{\!\!\!\!\!\!2\alpha}$\!\!\!\!&$\matt_2$&$\matr_2$&$i$&\!\!$-1$\!\!&$i$& $\sigy()^*$&$4$\\
\!\!\!\!${\xiceta}^{\!\!\!\!\!\!2\beta}$\!\!\!\!&$\matt_2$&$\matr_2$&$i$&\!\!$-1$\!\!&$-i$& $-\sigy()^*$&$4$\\
\hline
&&&&&&&\\[-1em]
\!${\etap}^{\!\!\!\!2\alpha}$\!&$1$&$1$&$-1$&$-1$&$-1$&$-1$&$1$\\
\!${\etap}^{\!\!\!\!2\beta}$\!&$1$&$1$&$()^*$&$-()^*$&$i\om()^*$&$-1$&$2$\\
\!${\etap}^{\!\!\!\!2\gamma}$\!&$1$&$1$&$1$&$-1$&$1$&$1$&$1$\\
\!${\etap}^{\!\!\!\!2\delta}$\!&$1$&$1$&$1$&$-1$&$1$&$-1$&$1$\\
\!${\etap}^{\!\!\!\!2\epsilon}$\!&$1$&$1$&$()^*$&$()^*$&$i\om()^*$&$-1$&$2$\\
\!${\etap}^{\!\!\!\!2\zeta}$\!&$1$&$1$&$-()^*$&$-()^*$&$-i\om()^*$&$1$&$2$\\
\hline
&&&&&&&\\[-1em]
\!${\etaq}^{\!\!2}$\!&$1$&$1$&$()^*$&$()^*$&$-\ob()^*$&$1$&$2$\\
\!${\etar}^{\!\!\!2\alpha}$\!&$1$&$1$&$1$&$1$&$-1$&$1$&$1$\\
\!${\etar}^{\!\!\!2\beta}$\!&$1$&$1$&$-1$&$-1$&$1$&$1$&$1$\\
\!\!\!\!$\etaq^*\etar$\!\!\!\!&$1$&$1$&$()^*$&$()^*$&$-\ob()^*$&$1$&$2$\\
\!\!\!\!$\etaq\etar$\!\!\!\!&$1$&$1$&$()^*$&$()^*$&$\om()^*$&$1$&$2$\\
\hline
&&&&&&&\\[-1em]
\!${\etas}^{\!\!\!2\alpha}$\! &$1$&$1$&$1$&$1$&$-1$&$-1$&$1$\\
\!${\etas}^{\!\!\!2\beta}$\!&$1$&$1$&$-1$&$1$&$1$&$-1$&$1$\\
\hline
			\end{tabular}
		\end{center}
\caption{The transformation rules for the driving flavons $\etap$, $\etaq$, $\etar$ and $\etas$ and the irreps obtained from them.}
		\label{tab:etasector}
\end{table}} 

To construct the potentials, we list the relevant tensor products involving $\phic$, $\xic$, $\etap$, $\etaq$, $\etar$ and $\etas$ below,
\begin{align}
\phic\!\otimes\etap&={\phiceta}^{\!\!\!\!\!\!2\alpha}\oplus{\phiceta}^{\!\!\!\!\!\!2\beta}\oplus{\phiceta}^{\!\!\!\!\!\!2\gamma},\\
\xic\!\otimes\etas&={\xiceta}^{\!\!\!\!\!\!2\alpha}\oplus{\xiceta}^{\!\!\!\!\!\!2\beta},\\
\!\!\etap\!\otimes\etap\!&=\!|\etap|^2\!\oplus\!{\etap}^{\!\!\!\!2\alpha}\!\oplus\!{\etap}^{\!\!\!\!2\beta}\!\oplus\!{\etap}^{\!\!\!\!2\gamma}\!\oplus\!{\etap}^{\!\!\!\!2\delta}\oplus\!{\etap}^{\!\!\!\!2\epsilon}\oplus\!{\etap}^{\!\!\!\!2\zeta},\!\!\!\!\\
\etaq\!\otimes\etaq\!&=|\etaq|^2\oplus{\etaq}^{\!\!\!2},\\
\etar\!\otimes\etar\!&=|\etar|^2\oplus{\etar}^{\!\!\!2\alpha}\oplus{\etar}^{\!\!\!2\beta}\!,
\end{align}
\begin{align}
\etaq\!\otimes\etar\!&=\etaq^*\etar\oplus\etaq\etar,\\
\etas\!\otimes\etas\!&=\!|\etas|^2\!\oplus{\etas}^{\!\!\!2\alpha}\!\oplus{\etas}^{\!\!\!2\beta},\qquad\qquad\qquad\qquad
\end{align}
where
\begin{align}
{\phiceta}^{\!\!\!\!\!\!2\alpha}&=\llbracket\phic\rrbracket^\T\etap,&{\phiceta}^{\!\!\!\!\!\!2\beta}&=\llbracket \phic\rrbracket^\T\sigz\etap,\\
{\phiceta}^{\!\!\!\!\!\!2\gamma}&=({\phic}_1{\etap}_2,{\phic}_2{\etap}_2,&\!\!\!\!\!\!\!\!{\phic}_2^{\!\!*}{\etap}_1,&\!-{\phic}_1^{\!\!*}{\etap}_1)^\T,\\
{\xiceta}^{\!\!\!\!\!\!2\alpha}&=\llbracket\xic\rrbracket^\T\llbracket\etas\rrbracket,&{\xiceta}^{\!\!\!\!\!\!2\beta}&=\llbracket \xic\rrbracket^\T\llbracket\etas\rrbracket^*,\\
{\etap}^{\!\!\!\!2\alpha}&={\etap}_1^{\!\!*}{\etap}_1-{\etap}_2^{\!\!*}{\etap}_2,&{\etap}^{\!\!\!\!2\beta}&={\etap}_1^{\!\!*}{\etap}_2,\\
{\etap}^{\!\!\!\!2\gamma}&=\text{re}({\etap}_1{\etap}_2),&{\etap}^{\!\!\!\!2\delta}&=\text{im}({\etap}_1{\etap}_2),\\
{\etap}^{\!\!\!\!2\epsilon}&={\etap}_1^{\!\!*2}+{\etap}_2^2,&{\etap}^{\!\!\!\!2\zeta}&={\etap}_1^{\!\!*2}-{\etap}_2^2,\\
{\etar}^{\!\!\!2\alpha}&=\text{re}({\etar}^{\!\!2}),&{\etar}^{\!\!\!2\beta}&=\text{im}({\etar}^{\!\!2}),\\
{\etas}^{\!\!\!2\alpha}&=\text{re}({\etas}^{\!\!2}), &{\etas}^{\!\!\!2\beta}&=\text{im}({\etas}^{\!\!2}),
\end{align}
with $\llbracket\etas\rrbracket=\frac{1}{\sqrt{2}}(\etas,\etas^*)^\T$. How these irreps transform is given in TABLE~\ref{tab:etasector}. 

Using these driving flavons, we construct the potentials,
\begin{align}
\text{\small$\breve{\mathcal{V}}_{\etap}$}&\text{\small$=c_{129}\,m^2|\etap|^2+c_{130} |\etap|^4$}\\
&\text{\small$=k_{110} (|\etap|^2-r_{\!\etap}^2)^2-k_{110}r_{\!\etap}^4$},\\
\text{\small$\mathcal{V}_{\etap}$}&\text{\small$=\breve{\mathcal{V}}_{\etap}+c_{131} ({\etap}^{\!\!\!\!2\alpha})^2+c_{132} ({\etap}^{\!\!\!\!2\delta})^2\!\!\!$}\\
&\text{\small$=\breve{\mathcal{V}}_{\etap}+k_{111} ({\etap}^{\!\!\!\!2\alpha})^2+k_{112} ({\etap}^{\!\!\!\!2\delta})^2$},\\
\text{\small$\mathcal{V}_{\etaq}$}&\text{\small$=c_{133}\,m^2|\etaq|^2+c_{134}|\etaq|^4$}\\
&\text{\small$= k_{113} (|\etaq|^2-r_{\!\etaq}^2)^2-k_{113}r_{\!\etaq}^4$},\\
\text{\small$\breve{\mathcal{V}}_{\etar}$}&\text{\small$=c_{135}\,m^2|\etar|^2+c_{136} |\etar|^4$}\\
&\text{\small$=k_{114} (|\etar|^2-r_{\!\etar}^2)^2+k_{114}r_{\!\etar}^4$},\\
\text{\small$\mathcal{V}_{\etar}$}&\text{\small$=\breve{\mathcal{V}}_{\etar}+c_{137}({\etar}^{\!\!\!2\alpha})^2$}\\
&\text{\small$=\breve{\mathcal{V}}_{\etar}+k_{115}({\etar}^{\!\!\!2\alpha})^2$},\\
\text{\small$\breve{\mathcal{V}}_{\etaqr}$}&\text{\small$=\mathcal{V}_{\etaq}+\mathcal{V}_{\etar}$},\\
\text{\small$\tilde{\mathcal{V}}_{\etaqr}$}&\text{\small$=\breve{\mathcal{V}}_{\etaqr}+c_{138}\,\text{re}(\etaq^3\etar^*)$}\\
&\text{\small$=\breve{\mathcal{V}}_{\etaqr}+k_{116} |{\etaq}^{\!\!2}-\frac{r_{\!\etaq}}{r_{\!\etar}}\etaq^*\etar|^2+\frac{k_{116}}{2} (|{\etaq}|^2-\frac{r_{\!\etaq}^2}{r_{\!\etar}^2}|\etar|^2)^2$}\!,\\
\text{\small$\mathcal{V}_{\etaqr}$}&\text{\small$=\tilde{\mathcal{V}}_{\etaqr}+c_{139}|{\etaq}|^2|{\etar}|^2$}\\
&\text{\small$=\tilde{\mathcal{V}}_{\etaqr}+k_{117} (|{\etaq}|^2-\frac{r_{\!\etaq}^2}{r_{\!\etar}^2}|\etar|^2)^2$},\\
\text{\small$\breve{\mathcal{V}}_{\etapqr}$}&\text{\small$=\mathcal{V}_{\etap}+\mathcal{V}_{\etaqr}$},\\
\text{\small$\tilde{\mathcal{V}}_{\etapqr}$}&\text{\small$=\breve{\mathcal{V}}_{\etapqr}+c_{140}m\, \text{re}({\etap}^{\!\!\!\!2\epsilon}\etaq^*)$}\\
&\text{\small$=\breve{\mathcal{V}}_{\etapqr}+k_{118} |{\etap}^{\!\!\!\!2\epsilon}-\frac{r_{\!\etap}^2}{r_{\!\etaq}}\etaq|^2$},\\
\text{\small$\mathcal{V}_{\etapqr}$}&\text{\small$=\tilde{\mathcal{V}}_{\etapqr}+c_{141}|{\etap}|^2|{\etaq}|^2+c_{142}|{\etap}|^2|{\etar}|^2$}\\
&\text{\small$=\tilde{\mathcal{V}}_{\etapqr}\!+k_{119} (|{\etap}|^2\!-\!\frac{r_{\!\etap}^2}{r_{\!\etaq}^2}|\etaq|^2)^2\!+k_{120} (|{\etap}|^2\!-\!\frac{r_{\!\etap}^2}{r_{\!\etar}^2}|\etar|^2)^2$}\!,\\
\text{\small$\breve{\mathcal{V}}_{\etas}$}&\text{\small$=c_{143}\,m^2|\etas|^2+c_{144} |\etas|^4$}\\
&\text{\small$=k_{121} (|\etas|^2-r_{\!\etap}^2)^2-k_{121}r_{\!\etap}^4$},
\end{align}
\begin{align}
\text{\small$\mathcal{V}_{\etas}$}&\text{\small$=\breve{\mathcal{V}}_{\etas}+c_{145} ({\etas}^{\!\!\!2\beta})^2$}\\
&\text{\small$=\breve{\mathcal{V}}_{\etas}+k_{122} ({\etas}^{\!\!\!2\beta})^2$},\\
\text{\small$\tilde{\mathcal{V}}_{\eta}$}&\text{\small$=\mathcal{V}_{\etapqr}+\mathcal{V}_{\etas}$},\\
\text{\small$\mathcal{V}_{\eta}$}&\text{\small$=\tilde{\mathcal{V}}_{\eta}\!+\!c_{146}|\etap|^2|\etas|^2\!\!+\!c_{147}|\etaq|^2|\etas|^2\!\!+c_{148}|\etar|^2|\etas|^2\!\!\!$}\label{eq:potetafinalc}\\
\begin{split}
&\text{\small$=\tilde{\mathcal{V}}_{\eta}+k_{123} (|{\etap}|^2\!-\frac{r_{\!\etap}^2}{r_{\!\etas}^2}|\etas|^2)^2$}\\
&\text{\small$\quad\,+k_{124} (|{\etaq}|^2\!-\!\frac{r_{\!\etaq}^2}{r_{\!\etas}^2}|\etas|^2)^2\!+\!k_{125} (|{\etar}|^2\!-\!\frac{r_{\!\etar}^2}{r_{\!\etas}^2}|\etas|^2)^2$}.
\end{split}\label{eq:potetafinalk}
\end{align}
Using the same procedure as followed in the previous section, we can show that $\mathcal{V}_{\eta}$ has a discrete minimum at (\ref{eq:etapqvevs}), (\ref{eq:etarsvevs}). The associated symmetry groups are $\Ht_{\eta}=K_{4\etap}\times Z_{2\etaq}\times Z_{2\etar}\times Z_{2\etas}$ and $\Tt_{\eta}=1$. $\mathcal{V}_{\eta}$ (\ref{eq:potetafinalc}), (\ref{eq:potetafinalk}) contains $20$ coefficients, $c_{129}, .., c_{148}$, and correspondingly $20$ arbitrary constants $r_{\!\etap}, r_{\!\etaq}, r_{\!\etar}, r_{\!\etas}, k_{110}, .., k_{125}$.

We combine $\mathcal{V}_{\phic}$, $\mathcal{V}_{\eta}$ and $\mathcal{V}_{\phia\rho\xic\lambda}$ to obtain,
\be
\text{\small$\breve{\mathcal{V}}_{\phic\eta\phia\rho\xic\lambda}=\mathcal{V}_{\phic}+\mathcal{V}_{\eta}+\mathcal{V}_{\phia\rho\xic\lambda}$}.
\ee
This potential has a minimum corresponding to a $3$-sphere for $\phic$ and the vevs $\langle\etap\rangle$, $\langle\etaq\rangle$, $\langle\etar\rangle$, $\langle\etas\rangle$, $\langle\phia\rangle$, $\langle\rhop\rangle$, $\langle\rhoq\rangle$, $\langle\rhor\rangle$, $\langle\rhos\rangle$, $\langle\xic\rangle$, $\langle\lamp\rangle$, $\langle\lamq\rangle$ and $\langle\lamr\rangle$ for the rest of the flavons. The minimum has the symmetry groups $\Hb_{\phic\eta\phia\rho\xic\lambda}=\Ht_{\phic}\times\Ht_{\eta}\times\Ht_{\phia\rho\xic\lambda}=K_{4\etap}\times Z_{2\etaq}\times Z_{2\etar}\times Z_{2\etas}\times (K_4\rtimes Z_2)_{\phia}\times K_{4\rhop}\times Z_{2\rhoq}\times Z_{2\rhor}\times Z_{2\rhos}\times K_{4\xic}\times(S_4\times Z_2)_{\lamp}\times(S_4\times Z_2)_{\lamq}\times ((Z_{2_1}\times Z_{2_3})\rtimes Z_2)_{\lamr}$ and $\Tb_{\phic\eta\phia\rho\xic\lambda}=\Tt_{\phic}\times\Tt_{\eta}\times\Tt_{\phia\rho\xic\lambda}=O(4)_{\phic}$. We use the cross term between $\phic$ and $\xic$ to break $O(4)_{\phic}$, 
\begin{align}\label{eq:phicfullpot1}
\text{\small$\tilde{\mathcal{V}}_{\phic\eta\phia\rho\xic\lambda}$}&\text{\small$=\breve{\mathcal{V}}_{\phic\eta\phia\rho\xic\lambda}+c_{149}\,\text{re}({\phiceta}^{\!\!\!\!\!\!2\alpha\dagger}{\xiceta}^{\!\!\!\!\!\!2\alpha})$}\qquad\quad
\end{align}
\vspace{-5mm}
\begin{align}\label{eq:phicfullpot2}
\begin{split}
\qquad\quad\,\,\,\,\,&\text{\small$=\breve{\mathcal{V}}_{\phic\eta\phia\rho\xic\lambda}+k_{126}|{\phiceta}^{\!\!\!\!\!\!2\alpha}-\frac{r_{\!\phic}r_{\!\etap}}{r_{\!\xic}r_{\!\etas}}{\xiceta}^{\!\!\!\!\!\!2\alpha}|^2$}\\
&\text{\small$\qquad+\frac{k_{126}}{2}\frac{r_{\!\etap}^2}{r_{\!\phic}^2} (|{\phic}|^2-\frac{r_{\!\phic}^2}{r_{\!\etap}^2}|\etap|^2)^2$}\\
&\text{\small$\qquad+\frac{k_{126}}{2}\frac{r_{\!\phic}^2r_{\!\etap}^2}{r_{\!\xic}^4}(|{\xic}|^2-\frac{r_{\!\xic}^2}{r_{\!\etas}^2}|\etas|^2)^2$}.
\end{split}
\end{align}
The newly added term $\text{re}({\phiceta}^{\!\!\!\!\!\!2\alpha\dagger}{\xiceta}^{\!\!\!\!\!\!2\alpha})$ (\ref{eq:phicfullpot1}) breaks $\Tb_{\phic\eta\phia\rho\xic\lambda}=O(4)_{\phic}$ to $\Ga_{\phic\eta\phia\rho\xic\lambda}=(K_4\rtimes Z_2)_{\phic}$ generated by ${\phic}_2\rightarrow{\phic}_2^*$, ${\phic}_1\rightarrow -i{\phic}_1^*$ and $\phic\rightarrow\mattau_2\sigx\phic$. The $3$-sphere gets stratified under $(K_4\rtimes Z_2)_{\phic}$. The stabilizer of $\langle\phic\rangle$ (\ref{eq:elemvevs}) is $(K_4\rtimes Z_2)_{\phic}$ itself, i.e.,~we have $\Ha_{\phic\eta\phia\rho\xic\lambda}=(K_4\rtimes Z_2)_{\phic}$. It fully determines $\langle\phic\rangle$ in the $3$-sphere; hence, $\tilde{\mathcal{V}}_{\phic\eta\phia\rho\xic\lambda}$ is guaranteed to have a stationary point corresponding to $\langle\phic\rangle$ and the vevs of the other flavons. The associated symmetry groups are $\Ht_{\phic\eta\phia\rho\xic\lambda}= \Ha_{\phic\eta\phia\rho\xic\lambda}\times \Hb_{\phic\eta\phia\rho\xic\lambda}=(K_4\rtimes Z_2)_{\phic}\times K_{4\etap}\times Z_{2\etaq}\times Z_{2\etar}\times Z_{2\etas}\times (K_4\rtimes Z_2)_{\phia}\times K_{4\rhop}\times Z_{2\rhoq}\times Z_{2\rhor}\times Z_{2\rhos}\times K_{4\xic}\times(S_4\times Z_2)_{\lamp}\times(S_4\times Z_2)_{\lamq}\times ((Z_{2_1}\times Z_{2_3})\rtimes Z_2)_{\lamr}$ and  $\Tt_{\phic\eta\phia\rho\xic\lambda}=1$. The rewritten form of the potential (\ref{eq:phicfullpot2}) shows that the stationary point can be obtained as a minimum. (\ref{eq:phicfullpot1}) and (\ref{eq:phicfullpot2})  contain $97$ coefficients $c_{53}, .., c_{149}$ and $97$ arbitrary constants $r_{\!\xic}, .., r_{\!\etas}, k_{44},..,k_{126}$, respectively.

To construct the general potential involving these flavons, we need to include various cross terms that we have not yet considered. To this end, we consider them as three multiplets: $\phic$, $\eta=(\etap,\etaq,\etar,\etas)^\T$ and $\phia\rho\xic\lambda=(\phia,\rhop,\rhoq,\rhor,\rhos,\xic,\lamp,\lamq,\lamr)^\T$, and search for the cross terms among these multiplets. Under $\gene_{\phic}$, $\gene_{\eta}$, $\gene_{\phia}$, $\gene_{\rho}$ and $\gene_{\xic}$, TABLE~\ref{tab:eelemflavons}, the flavons $\{\phic,\etap\}$, $\{\etap,\etas\}$, $\{\phia, \rhop\}$, $\{\rhop,\rhos\}$ and $\{\xic, \rhop, \etap\}$, respectively, change sign while every other flavon remains invariant. This leads us to conclude that the only combinations where $\phic$, $\phia$ and/or $\xic$ can appear linearly are $\phic\otimes\etap\otimes\xic\otimes\etas$ and $\phia\otimes\rhop\otimes\xic\otimes\rhos$. We already analyzed these products earlier in this Section and in Section~\ref{sec:phiapot}. Other than these two cases, every possible cross term among our three multiplets can involve quadratic contributions only from every type of flavon, i.e.,~$\{\phic\otimes\phic\}$, $\{\eta\otimes\eta\}$ and $\{\phia\otimes\phia,\rho\otimes\rho,\xic\otimes\xic, \lambda\otimes\lambda\}$. How the irreps obtained from these tensor products transform is given in TABLES~\ref{tab:prodtrof}, \ref{tab:lamsector}, \ref{tab:rhosector}, \ref{tab:etasector}. From examining them, we find that there exist only four cross terms: $\etar^{2\alpha}\lamp^\T\lamr$, $\rhor^{2\alpha}\etar^{2\alpha}$, $\rhor^{2\beta}\etap^{2\gamma}$ and $\rhop^{2\gamma}\etar^{2\beta}$. 

The irreps $\etar^{2\alpha}$, $\lamp^\T\lamr$ and $\rhor^{2\alpha}$ vanish at our assigned vevs $\langle\etar\rangle$ (\ref{eq:etarsvevs}), $\langle\lamp\rangle$ (\ref{eq:lampvev}), $\langle\lamr\rangle$ (\ref{eq:lamrvev}), and $\langle\rhor\rangle$ (\ref{eq:rhorvev}), implying that $\etar^{2\alpha}\lamp^\T\lamr$ and $\rhor^{2\alpha}\etar^{2\alpha}$ are compatible with $\Ht_{\phic\eta\phia\rho\xic\lambda}$. To verify the compatibility of the term $\rhor^{2\beta}\etap^{2\gamma}$, we investigate how $\Ht_{\phic\eta\phia\rho\xic\lambda}$ acts on $\rhor^{2\beta}$ and $\etap^{2\gamma}$. This corresponds to the actions of $Z_{2\rhor}$ and $K_{4\etap}$ on $\rhor^{2\beta}$ and $\etap^{2\gamma}$, respectively, which turn out to be trivial. In other words, the representations of $\Ht_{\phic\eta\phia\rho\xic\lambda}$ acting on both $\rhor^{2\beta}$ and $\etap^{2\gamma}$ are $\{1\}$. According to corollary~A in \cite{2306.07325}, if the representations of the symmetry group acting on two multiplets are equal, then the term obtained as their product is compatible with the group. Therefore, the term $\rhor^{2\beta}\etap^{2\gamma}$ is compatible. Similarly, we can show that the representations of $\Ht_{\phic\eta\phia\rho\xic\lambda}$ acting on both $\rhop^{2\gamma}$ and $\etar^{2\beta}$ are $\{1\}$; hence, the term $\rhop^{2\gamma}\etar^{2\beta}$ is compatible. Besides these cross terms, we have $4+9+36=49$ products of norms, i.e.,~$|\phic|^2|\etap|^2, .., |\etas|^2|\lamr|^2$. They also are compatible. We include all these cross terms to construct the general potential,
\begin{align}\label{eq:phicfullpot3}
\begin{split}
\!\!\!\!\!\!\!\!\!\!\!\text{\small$\mathcal{V}_{\phic\eta\phia\rho\xic\lambda}$}&\text{\small$=\tilde{\mathcal{V}}_{\phic\eta\phia\rho\xic\lambda}+c_{150}\etar^{2\alpha}\lamp^\T\lamr+c_{151}\rhor^{2\alpha} \etar^{2\alpha}$}\\
&\text{\small$\quad+c_{152}\rhor^{2\beta}\etap^{2\gamma}\!+c_{153}\rhop^{2\gamma}\etar^{2\beta}\!+c_{154}|\phic|^2|\etap|^2\,\,\,\,\,\,$}\\
&\text{\small$\quad+ ... +c_{202}|\etas|^2|\lamr|^2$}
\end{split}
\end{align}
\vspace{-8mm}
\begin{align}\label{eq:phicfullpot4}
\begin{split}
\qquad\quad\,\,\,\,\,&\text{\small$=\tilde{\mathcal{V}}_{\phic\eta\phia\rho\xic\lambda}+k_{127}(\etar^{2\alpha}-\frac{r_{\!\etar}^2}{r_{\lamp}r_{\lamr}}\lamp^\T\lamr)^2$}\\
&\text{\small$\quad +k_{128}(\rhor^{2\alpha}-\frac{r_{\!\rhor}^2}{r_{\!\etar}^2}\etar^{2\alpha})^2+k_{129}(\rhor^{2\beta}-\frac{2r_{\!\rhor}^2}{r_{\!\etap}^2}\etap^{2\gamma})^2\!\!\!$}\\
&\text{\small$\quad+k_{130}(\rhop^{2\gamma}-\frac{r_{\!\rhop}^2}{2r_{\!\etar}^2}\etar^{2\beta})^2+k_{131}(|{\phic}|^2\!-\!\frac{r_{\!\phic}^2}{r_{\!\etap}^2}|\etap|^2)^2$}\\
&\text{\small$\quad+ ... +k_{179}(|{\etas}|^2-\frac{r_{\!\etas}^2}{r_{\!\lamr}^2}|\lamr|^2)^2$}.
\end{split}
\end{align}
Since the newly added cross terms are compatible with $\Ht_{\phic\eta\phia\rho\xic\lambda}$, it remains unaffected. Therefore, $\mathcal{V}_{\phic\eta\phia\rho\xic\lambda}$ (\ref{eq:phicfullpot3}) has a minimum that corresponds to our assigned set of vevs. (\ref{eq:phicfullpot3}) and (\ref{eq:phicfullpot4}) contain $150$ coefficients and $150$ arbitrary constants, respectively.

\subsection{The potential involving all the flavons}

In the model, we introduced $7$ flavons, $\chi$, $\xia$, $\xib$, $\xic$, $\phia$, $\phib$, $\phic$, that appear in the fermion mass matrices and $16$ driving flavons, $\mup$, $\muq$, $\nup$, $\nuq$, $\nur$, $\lamp$, $\lamq$, $\lamr$, $\rhop$, $\rhoq$, $\rhor$, $\rhos$, $\etap$, $\etaq$, $\etar$, $\etas$. By involving these $23$ flavons, we construct
\be\label{eq:vtilde}
\text{\small$\tilde{\mathcal{V}}=\mathcal{V}_{\chi}+\mathcal{V}_{\xia}+\mathcal{V}_{\xib\mu}+\mathcal{V}_{\phib\nu}+\mathcal{V}_{\phic\eta\phia\rho\xic\lambda}$}.
\ee
This potential has a discrete minimum that corresponds to the various vevs we previously assigned to the flavons, i.e.,
\begin{equation}\label{eq:fullvev}
\begin{split}
&\langle\chi\rangle, \langle\xia\rangle, \langle\xib\rangle, \langle\xic\rangle, \langle\phia\rangle, \langle\phib\rangle, \langle\phic\rangle,\quad\text{(\ref{eq:elemvevs})}\\
&\langle\mup\rangle, \langle\muq\rangle, \quad\text{(\ref{eq:mupvev}), (\ref{eq:muqvev})}\\
&\langle\nup\rangle, \langle\nuq\rangle, \langle\nur\rangle, \quad\text{(\ref{eq:nuprvevs}), (\ref{eq:nuqvev})}\\
&\langle\lamp\rangle, \langle\lamq\rangle, \langle\lamr\rangle,\quad\text{(\ref{eq:lampvev})-(\ref{eq:lamrvev})}\\
&\langle\rhop\rangle, \langle\rhoq\rangle, \langle\rhor\rangle, \langle\rhos\rangle, \quad\text{(\ref{eq:rhopsvevs}), (\ref{eq:rhoqvev}), (\ref{eq:rhorvev})}\\
&\langle\etap\rangle, \langle\etaq\rangle, \langle\etar\rangle, \langle\etas\rangle. \quad\text{(\ref{eq:etapqvevs}), (\ref{eq:etarsvevs})}
\end{split}
\end{equation}
The symmetry group associated with this minimum is
\begin{equation}\label{eq:htilde}
\begin{split}
\Ht&=\Ht_{\chi}\times\Ht_{\xia}\times\Ht_{\xib\mu}\times\Ht_{\phib\nu}\times\Ht_{\phic\eta\phia\rho\xic\lambda}\\
&=S_{3\chi}\times Z_{2\xia}\times (K_4\rtimes Z_2)_{\xib}\times (K_4\rtimes Z_2)_{\mup}\times Z_{2\muq}\\
&\quad\times (K_4\rtimes Z_2)_{\phib}\times (K_4\rtimes Z_2)_{\nup}\times Z_{2\nuq}\times Z_{2\nur}\\
&\quad\times (K_4\rtimes Z_2)_{\phic}\times K_{4\etap}\times Z_{2\etaq}\times Z_{2\etar}\times Z_{2\etas}\\
&\quad\times (K_4\rtimes Z_2)_{\phia}\times K_{4\rhop}\times Z_{2\rhoq}\times Z_{2\rhor}\times Z_{2\rhos}\times K_{4\xic}\\
&\quad\times(S_4\times Z_2)_{\lamp}\times(S_4\times Z_2)_{\lamq}\times ((Z_{2_1}\times Z_{2_3})\rtimes Z_2)_{\lamr}.
\end{split}
\end{equation}
$\tilde{\mathcal{V}}$ (\ref{eq:vtilde}) contains $202$ coefficients $c_1, .., c_{202}$ and $202$ arbitrary constants $r_{\!\chi}, .., r_{\!\etas}, k_1, .., k_{179}$.

To construct the general potential involving all the flavons, we consider them as five multiplets: $\chi$, $\xia$, $\xib\mu=(\xib,\mup,\muq)^\T$, $\phib\nu=(\phib,\nup,\nuq,\nur)^\T$ and $\phic\eta\phia\rho\xic\lambda=(\phic,\etap,\etaq,\etar,\etas,\phia,\rhop,\rhoq,\rhor,\rhos,\xic,\lamp,\lamq,\lamr)^\T$. Let us search for the cross terms among these five multiplets. Flavons in each multiplet, with the exception of $\xia$, transform under one or more generators whose action is trivial on the flavons in the other multiplets. Therefore, no flavon, with the possible exception of $\xia$, can couple linearly with the flavons in another multiplet. The flavons that transform nontrivially under $\gene_{\xia}$, TABLE~\ref{tab:eelemflavons}, are $\xia$, $\xib$ and $\mup$ only. This implies that $\xia$ can couple linearly with $\xib$ and/or $\mup$ only. On the other hand, $\xib$ is the only flavon that changes sign under $\gene_{\xib}$, and $\mup$ and $\muq$ are the only flavons that change sign under $\gene_\mu$, TABLE~\ref{tab:eelemflavons}. Therefore, the only possible cases of $\xia$ coupling linearly with flavons in another multiplet are $\xia\otimes\xib\otimes\xib$, $\xia\otimes\mup\otimes\mup$ and $\xia\otimes\mup\otimes\muq$. Towards the end of Section~\ref{sec:ssb}, we studied $\xia\otimes\xib\otimes\xib$ and found that no irrep obtained from this tensor product is an invariant. How the various irreps obtained from $\mup\otimes\mup$ and $\mup\otimes\muq$ transform is given in TABLE~\ref{tab:musector}, which rules out obtaining invariants from $\xia\otimes\mup\otimes\mup$ and $\xia\otimes\mup\otimes\muq$ also. Therefore, no flavon, including $\xia$, can couple linearly with flavons in another multiplet. This would imply that in cross terms, the flavons in the various multiplets can contribute quadratically only, i.e.,~ $\{\chi\otimes\chi\}$, $\{\xia\otimes\xia\}$, $\{\xib\otimes\xib,\mu\otimes\mu\}$, $\{\phib\otimes\phib,\nu\otimes\nu\}$ and $\{\phic\otimes\phic,\eta\otimes\eta,\phia\otimes\phia,\rho\otimes\rho,\xic\otimes\xic,\lambda\otimes\lambda\}$. How the irreps obtained from these tensor products transform is given in TABLES~\ref{tab:prodtrof}, \ref{tab:musector}, \ref{tab:nusector}, \ref{tab:lamsector}, \ref{tab:rhosector}, \ref{tab:etasector}. By examining them, we obtain four cross terms: ${\phib}^{\!\!\!\!2\alpha}{\lamp}^{\!\!\!\T}\!\lamq$, ${\nup}^{\!\!\!2\alpha}{\lamp}^{\!\!\!\T}\!\lamq$, ${\nuq}^{\!\!\!2\alpha}{\lamp}^{\!\!\!\T}\!\lamq$, ${\nur}^{\!\!2\alpha}{\lamp}^{\!\!\!\T}\!\lamq$. Since ${\phib}^{\!\!\!\!2\alpha}$, ${\lamp}^{\!\!\!\T}\!\lamq$, ${\nup}^{\!\!\!2\alpha}$, ${\nuq}^{\!\!\!2\alpha}$ and ${\nur}^{\!\!2\alpha}$ vanish at our assigned vev (\ref{eq:fullvev}), these cross terms are compatible with $\Ht$ (\ref{eq:htilde}). We also have cross terms that are products of norms, i.e.,~$|\chi|^2|\xia|^2$, .., $|\nur|^2|\lamr|^2$, which consists of a total of $1+3+4+14+3+4+14+12+42+56=153$ terms. We include all these cross terms to construct the general renormalizable potential containing the $23$ flavons,
\begin{equation}
\begin{aligned}\label{eq:fullpot1}
\!\!\!\!\text{\small$\mathcal{V}$}&\text{\small$=\tilde{\mathcal{V}}+c_{203}{\phib}^{\!\!\!\!2\alpha}{\lamp}^{\!\!\!\T}\!\lamq+c_{204}{\nup}^{\!\!\!2\alpha}{\lamp}^{\!\!\!\T}\!\lamq+c_{205} {\nuq}^{\!\!\!2\alpha}{\lamp}^{\!\!\!\T}\!\lamq$}\\
&\text{\small$\quad+c_{206}{\nur}^{\!\!2\alpha}{\lamp}^{\!\!\!\T}\!\lamq+c_{207}|\chi|^2|\xia|^2+...+c_{359}|\nur|^2|\etas|^2$}\!\!\!\!
\end{aligned}
\end{equation}
\vspace{-5mm}
\begin{equation}
\begin{aligned}\label{eq:fullpot2}
\,\,\,\,\,&\text{\small$=\tilde{\mathcal{V}}+k_{180}({\phib}^{\!\!\!\!2\alpha}-\frac{r_{\!\phib}^2}{r_{\lamp}r_{\lamq}}{\lamp}^{\!\!\!\T}\!\lamq)^2+k_{181}({\nup}^{\!\!\!2\alpha}-\frac{r_{\!\nup}^2}{r_{\lamp}r_{\lamq}}{\lamp}^{\!\!\!\T}\!\lamq)^2$}\\
&\text{\small$\quad+k_{182}({\nuq}^{\!\!\!2\alpha}-\frac{r_{\!\nuq}^2}{r_{\lamp}r_{\lamq}}{\lamp}^{\!\!\!\T}\!\lamq)^2+k_{183}({\nur}^{\!\!2\alpha}-\frac{r_{\!\nur}^2}{r_{\lamp}r_{\lamq}}{\lamp}^{\!\!\!\T}\!\lamq)^2$}\\
&\text{\small$\quad+k_{184}(|\chi|^2\!-\!\frac{v_{\chi}^2}{r_{\!\xia}^2}|\xia|^2)^2\!+...+\!k_{336}(|{\nur}|^2\!-\!\frac{r_{\!\nur}^2}{r_{\lamr}^2}|\lamr|^2)^2.\!\!\!\!\!\!\!\!$}
\end{aligned}
\end{equation}
Since the newly added cross terms are compatible, $\Ht$ (\ref{eq:htilde}) remains unaffected, and our vev (\ref{eq:fullvev}) is obtained as a minimum of $\mathcal{V}$ (\ref{eq:fullpot1}). The rewritten potential (\ref{eq:fullpot2}), which makes the minimum manifest, contains $359$ arbitrary constants, consisting of the norms of the $23$ flavons and $k_1, ..., k_{336}$. They are functions of the $359$ coefficients $c_1, ..., c_{359}$ present in (\ref{eq:fullpot1}).

In this Appendix, we introduced several auxiliary generators beyond $\genx$. This enlarges $G_{\!f}$ and $G_x$ compared to their definitions provided in Section~\ref{sec:aux}. We may find the large number of driving flavons and the corresponding enlargement of the groups aesthetically unappealing. We intend to tackle this issue in future works where we will break the accidental continuous symmetries of the renormalizable flavon potential by utilizing the non-renormalizable terms up to a specific order instead of utilizing the driving flavons. Such an approach would require all the renormalizable cross terms, as well as all the non-renormalizable cross terms up to the specific order, to satisfy the compatibility condition so that they do not spoil our assigned vevs.

\section{Driving flavons and the mass terms} 
\label{sec:appfive}

As we saw in Section~\ref{sec:lagrangian} and TABLE~\ref{tab:eelems}, the combination of flavons that couple with $L^\dagger$ and $N$ to form the Dirac mass term must change sign under $\gene_{\xib}$ and $\gene_{\xic}$ and must remain invariant under $\gene_{\xia}$. It must also remain invariant under $\genzl$, $\gene_{\phia}$, $\gene_{\phib}$, $\gene_{\phic}$, $\gene_\mu$, $\gene_\nu$, $\gene_\lambda$, $\gene_\rho$ and $\gene_\eta$. From TABLE~\ref{tab:eelemflavons}, we can see that no flavon or pair of flavons that transforms in this way exists. At the cubic order, only one such combination exists, which is nothing but $\xia\otimes\xib\otimes\xic$. The combination that couples with $i\sigy N$ and $N$ to form the Majorana mass term must change sign under $\gene_{\phia}$, $\gene_{\phib}$ and $\gene_{\phic}$ and must remain invariant under $\genzl$, $\gene_{\xia}$, $\gene_{\xib}$, $\gene_{\xic}$, $\gene_\mu$, $\gene_\nu$, $\gene_\lambda$, $\gene_\rho$ and $\gene_\eta$. By examining TABLE~\ref{tab:eelemflavons}, we can show that the cubic product $\phia\otimes\phib\otimes\phic$ is the one and only leading order combination that transforms in this way. Thus, we conclude that the driving flavons cannot appear in the mass matrices at the leading order. For a discussion on the tensor product expansions of $\xia\otimes\xib\otimes\xic$ and $\phia\otimes\phib\otimes\phic$, please see Appendix~\ref{sec:apptwo}.

\bibliography{second_paper.bib, noninspire.bib}

\begin{thebibliography}{71}%
\makeatletter
\providecommand \@ifxundefined [1]{%
 \@ifx{#1\undefined}
}%
\providecommand \@ifnum [1]{%
 \ifnum #1\expandafter \@firstoftwo
 \else \expandafter \@secondoftwo
 \fi
}%
\providecommand \@ifx [1]{%
 \ifx #1\expandafter \@firstoftwo
 \else \expandafter \@secondoftwo
 \fi
}%
\providecommand \natexlab [1]{#1}%
\providecommand \enquote  [1]{``#1''}%
\providecommand \bibnamefont  [1]{#1}%
\providecommand \bibfnamefont [1]{#1}%
\providecommand \citenamefont [1]{#1}%
\providecommand \href@noop [0]{\@secondoftwo}%
\providecommand \href [0]{\begingroup \@sanitize@url \@href}%
\providecommand \@href[1]{\@@startlink{#1}\@@href}%
\providecommand \@@href[1]{\endgroup#1\@@endlink}%
\providecommand \@sanitize@url [0]{\catcode `\\12\catcode `\$12\catcode
  `\&12\catcode `\#12\catcode `\^12\catcode `\_12\catcode `\%12\relax}%
\providecommand \@@startlink[1]{}%
\providecommand \@@endlink[0]{}%
\providecommand \url  [0]{\begingroup\@sanitize@url \@url }%
\providecommand \@url [1]{\endgroup\@href {#1}{\urlprefix }}%
\providecommand \urlprefix  [0]{URL }%
\providecommand \Eprint [0]{\href }%
\providecommand \doibase [0]{https://doi.org/}%
\providecommand \selectlanguage [0]{\@gobble}%
\providecommand \bibinfo  [0]{\@secondoftwo}%
\providecommand \bibfield  [0]{\@secondoftwo}%
\providecommand \translation [1]{[#1]}%
\providecommand \BibitemOpen [0]{}%
\providecommand \bibitemStop [0]{}%
\providecommand \bibitemNoStop [0]{.\EOS\space}%
\providecommand \EOS [0]{\spacefactor3000\relax}%
\providecommand \BibitemShut  [1]{\csname bibitem#1\endcsname}%
\let\auto@bib@innerbib\@empty
\bibitem [{\citenamefont {Babu}\ and\ \citenamefont
  {Gabriel}(2010)}]{1006.0203}%
  \BibitemOpen
  \bibfield  {author} {\bibinfo {author} {\bibfnamefont {K.}~\bibnamefont
  {Babu}}\ and\ \bibinfo {author} {\bibfnamefont {S.}~\bibnamefont {Gabriel}},\
  }\bibfield  {title} {\bibinfo {title} {{Semidirect Product Groups, Vacuum
  Alignment and Tribimaximal Neutrino Mixing}},\ }\href
  {https://doi.org/10.1103/PhysRevD.82.073014} {\bibfield  {journal} {\bibinfo
  {journal} {Phys. Rev. D}\ }\textbf {\bibinfo {volume} {82}},\ \bibinfo
  {pages} {073014} (\bibinfo {year} {2010})},\ \Eprint
  {https://arxiv.org/abs/1006.0203} {arXiv:1006.0203 [hep-ph]} \BibitemShut
  {NoStop}%
\bibitem [{\citenamefont {Holthausen}\ and\ \citenamefont
  {Schmidt}(2012)}]{1111.1730}%
  \BibitemOpen
  \bibfield  {author} {\bibinfo {author} {\bibfnamefont {M.}~\bibnamefont
  {Holthausen}}\ and\ \bibinfo {author} {\bibfnamefont {M.~A.}\ \bibnamefont
  {Schmidt}},\ }\bibfield  {title} {\bibinfo {title} {{Natural Vacuum Alignment
  from Group Theory: The Minimal Case}},\ }\href
  {https://doi.org/10.1007/JHEP01(2012)126} {\bibfield  {journal} {\bibinfo
  {journal} {JHEP}\ }\textbf {\bibinfo {volume} {01}},\ \bibinfo {pages}
  {126}},\ \Eprint {https://arxiv.org/abs/1111.1730} {arXiv:1111.1730 [hep-ph]}
  \BibitemShut {NoStop}%
\bibitem [{\citenamefont {Holthausen}\ \emph
  {et~al.}(2013{\natexlab{a}})\citenamefont {Holthausen}, \citenamefont
  {Lindner},\ and\ \citenamefont {Schmidt}}]{1211.5143}%
  \BibitemOpen
  \bibfield  {author} {\bibinfo {author} {\bibfnamefont {M.}~\bibnamefont
  {Holthausen}}, \bibinfo {author} {\bibfnamefont {M.}~\bibnamefont
  {Lindner}},\ and\ \bibinfo {author} {\bibfnamefont {M.~A.}\ \bibnamefont
  {Schmidt}},\ }\bibfield  {title} {\bibinfo {title} {{Lepton flavor at the
  electroweak scale: A complete $A_{4}$ model}},\ }\href
  {https://doi.org/10.1103/PhysRevD.87.033006} {\bibfield  {journal} {\bibinfo
  {journal} {Phys. Rev. D}\ }\textbf {\bibinfo {volume} {87}},\ \bibinfo
  {pages} {033006} (\bibinfo {year} {2013}{\natexlab{a}})},\ \Eprint
  {https://arxiv.org/abs/1211.5143} {arXiv:1211.5143 [hep-ph]} \BibitemShut
  {NoStop}%
\bibitem [{\citenamefont {Altarelli}\ and\ \citenamefont
  {Feruglio}(2005)}]{hep-ph/0504165}%
  \BibitemOpen
  \bibfield  {author} {\bibinfo {author} {\bibfnamefont {G.}~\bibnamefont
  {Altarelli}}\ and\ \bibinfo {author} {\bibfnamefont {F.}~\bibnamefont
  {Feruglio}},\ }\bibfield  {title} {\bibinfo {title} {{Tri-bimaximal neutrino
  mixing from discrete symmetry in extra dimensions}},\ }\href
  {https://doi.org/10.1016/j.nuclphysb.2005.05.005} {\bibfield  {journal}
  {\bibinfo  {journal} {Nucl. Phys. B}\ }\textbf {\bibinfo {volume} {720}},\
  \bibinfo {pages} {64} (\bibinfo {year} {2005})},\ \Eprint
  {https://arxiv.org/abs/hep-ph/0504165} {arXiv:hep-ph/0504165} \BibitemShut
  {NoStop}%
\bibitem [{\citenamefont {Altarelli}\ and\ \citenamefont
  {Feruglio}(2006)}]{hep-ph/0512103}%
  \BibitemOpen
  \bibfield  {author} {\bibinfo {author} {\bibfnamefont {G.}~\bibnamefont
  {Altarelli}}\ and\ \bibinfo {author} {\bibfnamefont {F.}~\bibnamefont
  {Feruglio}},\ }\bibfield  {title} {\bibinfo {title} {{Tri-bimaximal neutrino
  mixing, A(4) and the modular symmetry}},\ }\href
  {https://doi.org/10.1016/j.nuclphysb.2006.02.015} {\bibfield  {journal}
  {\bibinfo  {journal} {Nucl. Phys. B}\ }\textbf {\bibinfo {volume} {741}},\
  \bibinfo {pages} {215} (\bibinfo {year} {2006})},\ \Eprint
  {https://arxiv.org/abs/hep-ph/0512103} {arXiv:hep-ph/0512103} \BibitemShut
  {NoStop}%
\bibitem [{\citenamefont {He}\ \emph {et~al.}(2006)\citenamefont {He},
  \citenamefont {Keum},\ and\ \citenamefont {Volkas}}]{hep-ph/0601001}%
  \BibitemOpen
  \bibfield  {author} {\bibinfo {author} {\bibfnamefont {X.-G.}\ \bibnamefont
  {He}}, \bibinfo {author} {\bibfnamefont {Y.-Y.}\ \bibnamefont {Keum}},\ and\
  \bibinfo {author} {\bibfnamefont {R.~R.}\ \bibnamefont {Volkas}},\ }\bibfield
   {title} {\bibinfo {title} {{A(4) flavor symmetry breaking scheme for
  understanding quark and neutrino mixing angles}},\ }\href
  {https://doi.org/10.1088/1126-6708/2006/04/039} {\bibfield  {journal}
  {\bibinfo  {journal} {JHEP}\ }\textbf {\bibinfo {volume} {04}},\ \bibinfo
  {pages} {039}},\ \Eprint {https://arxiv.org/abs/hep-ph/0601001}
  {arXiv:hep-ph/0601001} \BibitemShut {NoStop}%
\bibitem [{\citenamefont {Kadosh}\ and\ \citenamefont
  {Pallante}(2010)}]{1004.0321}%
  \BibitemOpen
  \bibfield  {author} {\bibinfo {author} {\bibfnamefont {A.}~\bibnamefont
  {Kadosh}}\ and\ \bibinfo {author} {\bibfnamefont {E.}~\bibnamefont
  {Pallante}},\ }\bibfield  {title} {\bibinfo {title} {{An A(4) flavor model
  for quarks and leptons in warped geometry}},\ }\href
  {https://doi.org/10.1007/JHEP08(2010)115} {\bibfield  {journal} {\bibinfo
  {journal} {JHEP}\ }\textbf {\bibinfo {volume} {08}},\ \bibinfo {pages}
  {115}},\ \Eprint {https://arxiv.org/abs/1004.0321} {arXiv:1004.0321 [hep-ph]}
  \BibitemShut {NoStop}%
\bibitem [{\citenamefont {Callen}\ and\ \citenamefont
  {Volkas}(2012)}]{1205.3617}%
  \BibitemOpen
  \bibfield  {author} {\bibinfo {author} {\bibfnamefont {B.~D.}\ \bibnamefont
  {Callen}}\ and\ \bibinfo {author} {\bibfnamefont {R.~R.}\ \bibnamefont
  {Volkas}},\ }\bibfield  {title} {\bibinfo {title} {{Large lepton mixing
  angles from a 4+1-dimensional SU(5) x A(4) domain-wall braneworld model}},\
  }\href {https://doi.org/10.1103/PhysRevD.86.056007} {\bibfield  {journal}
  {\bibinfo  {journal} {Phys. Rev. D}\ }\textbf {\bibinfo {volume} {86}},\
  \bibinfo {pages} {056007} (\bibinfo {year} {2012})},\ \Eprint
  {https://arxiv.org/abs/1205.3617} {arXiv:1205.3617 [hep-ph]} \BibitemShut
  {NoStop}%
\bibitem [{\citenamefont {Krishnan}(2020)}]{1901.01205}%
  \BibitemOpen
  \bibfield  {author} {\bibinfo {author} {\bibfnamefont {R.}~\bibnamefont
  {Krishnan}},\ }\bibfield  {title} {\bibinfo {title} {{Fully Constrained Mass
  Matrix: Can Symmetries alone determine the Flavon Vacuum Alignments?}},\
  }\href {https://doi.org/10.1103/PhysRevD.101.075004} {\bibfield  {journal}
  {\bibinfo  {journal} {Phys. Rev. D}\ }\textbf {\bibinfo {volume} {101}},\
  \bibinfo {pages} {075004} (\bibinfo {year} {2020})},\ \Eprint
  {https://arxiv.org/abs/1901.01205} {arXiv:1901.01205 [hep-ph]} \BibitemShut
  {NoStop}%
\bibitem [{\citenamefont {Krishnan}(2022)}]{1912.02451}%
  \BibitemOpen
  \bibfield  {author} {\bibinfo {author} {\bibfnamefont {R.}~\bibnamefont
  {Krishnan}},\ }\bibfield  {title} {\bibinfo {title} {{$\text {TM}_1$ neutrino
  mixing with $\sin \theta _{13}=\frac{1}{\sqrt{3}}\sin \frac{\pi }{12}$}},\
  }\href {https://doi.org/10.1140/epjp/s13360-022-02706-7} {\bibfield
  {journal} {\bibinfo  {journal} {Eur. Phys. J. Plus}\ }\textbf {\bibinfo
  {volume} {137}},\ \bibinfo {pages} {496} (\bibinfo {year} {2022})},\ \Eprint
  {https://arxiv.org/abs/1912.02451} {arXiv:1912.02451 [hep-ph]} \BibitemShut
  {NoStop}%
\bibitem [{\citenamefont {King}(2015)}]{1510.02091}%
  \BibitemOpen
  \bibfield  {author} {\bibinfo {author} {\bibfnamefont {S.~F.}\ \bibnamefont
  {King}},\ }\bibfield  {title} {\bibinfo {title} {{Models of Neutrino Mass,
  Mixing and CP Violation}},\ }\href
  {https://doi.org/10.1088/0954-3899/42/12/123001} {\bibfield  {journal}
  {\bibinfo  {journal} {J. Phys. G}\ }\textbf {\bibinfo {volume} {42}},\
  \bibinfo {pages} {123001} (\bibinfo {year} {2015})},\ \Eprint
  {https://arxiv.org/abs/1510.02091} {arXiv:1510.02091 [hep-ph]} \BibitemShut
  {NoStop}%
\bibitem [{\citenamefont {Pakvasa}\ and\ \citenamefont
  {Sugawara}(1979)}]{Pakvasa:1978tx}%
  \BibitemOpen
  \bibfield  {author} {\bibinfo {author} {\bibfnamefont {S.}~\bibnamefont
  {Pakvasa}}\ and\ \bibinfo {author} {\bibfnamefont {H.}~\bibnamefont
  {Sugawara}},\ }\bibfield  {title} {\bibinfo {title} {{Mass of the t Quark in
  SU(2) x U(1)}},\ }\href {https://doi.org/10.1016/0370-2693(79)90436-2}
  {\bibfield  {journal} {\bibinfo  {journal} {Phys. Lett. B}\ }\textbf
  {\bibinfo {volume} {82}},\ \bibinfo {pages} {105} (\bibinfo {year}
  {1979})}\BibitemShut {NoStop}%
\bibitem [{\citenamefont {Brown}\ \emph {et~al.}(1984)\citenamefont {Brown},
  \citenamefont {Pakvasa}, \citenamefont {Sugawara},\ and\ \citenamefont
  {Yamanaka}}]{Brown:1984dk}%
  \BibitemOpen
  \bibfield  {author} {\bibinfo {author} {\bibfnamefont {T.}~\bibnamefont
  {Brown}}, \bibinfo {author} {\bibfnamefont {S.}~\bibnamefont {Pakvasa}},
  \bibinfo {author} {\bibfnamefont {H.}~\bibnamefont {Sugawara}},\ and\
  \bibinfo {author} {\bibfnamefont {Y.}~\bibnamefont {Yamanaka}},\ }\bibfield
  {title} {\bibinfo {title} {{Neutrino Masses, Mixing and Oscillations in S(4)
  Model of Permutation Symmetry}},\ }\href
  {https://doi.org/10.1103/PhysRevD.30.255} {\bibfield  {journal} {\bibinfo
  {journal} {Phys. Rev. D}\ }\textbf {\bibinfo {volume} {30}},\ \bibinfo
  {pages} {255} (\bibinfo {year} {1984})}\BibitemShut {NoStop}%
\bibitem [{\citenamefont {Lee}\ and\ \citenamefont
  {Mohapatra}(1994)}]{hep-ph/9403201}%
  \BibitemOpen
  \bibfield  {author} {\bibinfo {author} {\bibfnamefont {D.-G.}\ \bibnamefont
  {Lee}}\ and\ \bibinfo {author} {\bibfnamefont {R.}~\bibnamefont
  {Mohapatra}},\ }\bibfield  {title} {\bibinfo {title} {{An SO(10) x S(4)
  scenario for naturally degenerate neutrinos}},\ }\href
  {https://doi.org/10.1016/0370-2693(94)91091-X} {\bibfield  {journal}
  {\bibinfo  {journal} {Phys. Lett. B}\ }\textbf {\bibinfo {volume} {329}},\
  \bibinfo {pages} {463} (\bibinfo {year} {1994})},\ \Eprint
  {https://arxiv.org/abs/hep-ph/9403201} {arXiv:hep-ph/9403201} \BibitemShut
  {NoStop}%
\bibitem [{\citenamefont {Mohapatra}\ \emph {et~al.}(2004)\citenamefont
  {Mohapatra}, \citenamefont {Parida},\ and\ \citenamefont
  {Rajasekaran}}]{hep-ph/0301234}%
  \BibitemOpen
  \bibfield  {author} {\bibinfo {author} {\bibfnamefont {R.~N.}\ \bibnamefont
  {Mohapatra}}, \bibinfo {author} {\bibfnamefont {M.~K.}\ \bibnamefont
  {Parida}},\ and\ \bibinfo {author} {\bibfnamefont {G.}~\bibnamefont
  {Rajasekaran}},\ }\bibfield  {title} {\bibinfo {title} {{High scale mixing
  unification and large neutrino mixing angles}},\ }\href
  {https://doi.org/10.1103/PhysRevD.69.053007} {\bibfield  {journal} {\bibinfo
  {journal} {Phys. Rev. D}\ }\textbf {\bibinfo {volume} {69}},\ \bibinfo
  {pages} {053007} (\bibinfo {year} {2004})},\ \Eprint
  {https://arxiv.org/abs/hep-ph/0301234} {arXiv:hep-ph/0301234} \BibitemShut
  {NoStop}%
\bibitem [{\citenamefont {Ma}(2006)}]{hep-ph/0508231}%
  \BibitemOpen
  \bibfield  {author} {\bibinfo {author} {\bibfnamefont {E.}~\bibnamefont
  {Ma}},\ }\bibfield  {title} {\bibinfo {title} {{Neutrino mass matrix from
  S(4) symmetry}},\ }\href {https://doi.org/10.1016/j.physletb.2005.10.019}
  {\bibfield  {journal} {\bibinfo  {journal} {Phys. Lett. B}\ }\textbf
  {\bibinfo {volume} {632}},\ \bibinfo {pages} {352} (\bibinfo {year}
  {2006})},\ \Eprint {https://arxiv.org/abs/hep-ph/0508231}
  {arXiv:hep-ph/0508231} \BibitemShut {NoStop}%
\bibitem [{\citenamefont {Hagedorn}\ \emph {et~al.}(2006)\citenamefont
  {Hagedorn}, \citenamefont {Lindner},\ and\ \citenamefont
  {Mohapatra}}]{hep-ph/0602244}%
  \BibitemOpen
  \bibfield  {author} {\bibinfo {author} {\bibfnamefont {C.}~\bibnamefont
  {Hagedorn}}, \bibinfo {author} {\bibfnamefont {M.}~\bibnamefont {Lindner}},\
  and\ \bibinfo {author} {\bibfnamefont {R.~N.}\ \bibnamefont {Mohapatra}},\
  }\bibfield  {title} {\bibinfo {title} {{S(4) flavor symmetry and fermion
  masses: Towards a grand unified theory of flavor}},\ }\href
  {https://doi.org/10.1088/1126-6708/2006/06/042} {\bibfield  {journal}
  {\bibinfo  {journal} {JHEP}\ }\textbf {\bibinfo {volume} {06}},\ \bibinfo
  {pages} {042}},\ \Eprint {https://arxiv.org/abs/hep-ph/0602244}
  {arXiv:hep-ph/0602244} \BibitemShut {NoStop}%
\bibitem [{\citenamefont {Caravaglios}\ and\ \citenamefont
  {Morisi}(2007)}]{hep-ph/0611078}%
  \BibitemOpen
  \bibfield  {author} {\bibinfo {author} {\bibfnamefont {F.}~\bibnamefont
  {Caravaglios}}\ and\ \bibinfo {author} {\bibfnamefont {S.}~\bibnamefont
  {Morisi}},\ }\bibfield  {title} {\bibinfo {title} {{Gauge boson families in
  grand unified theories of fermion masses: $E^4_6 \rtimes S_{4}$}},\ }\href
  {https://doi.org/10.1142/S0217751X07036646} {\bibfield  {journal} {\bibinfo
  {journal} {Int. J. Mod. Phys. A}\ }\textbf {\bibinfo {volume} {22}},\
  \bibinfo {pages} {2469} (\bibinfo {year} {2007})},\ \Eprint
  {https://arxiv.org/abs/hep-ph/0611078} {arXiv:hep-ph/0611078} \BibitemShut
  {NoStop}%
\bibitem [{\citenamefont {Zhang}(2007)}]{hep-ph/0612214}%
  \BibitemOpen
  \bibfield  {author} {\bibinfo {author} {\bibfnamefont {H.}~\bibnamefont
  {Zhang}},\ }\bibfield  {title} {\bibinfo {title} {{Flavor S(4) x Z(2)
  symmetry and neutrino mixing}},\ }\href
  {https://doi.org/10.1016/j.physletb.2007.09.003} {\bibfield  {journal}
  {\bibinfo  {journal} {Phys. Lett. B}\ }\textbf {\bibinfo {volume} {655}},\
  \bibinfo {pages} {132} (\bibinfo {year} {2007})},\ \Eprint
  {https://arxiv.org/abs/hep-ph/0612214} {arXiv:hep-ph/0612214} \BibitemShut
  {NoStop}%
\bibitem [{\citenamefont {Koide}(2007)}]{0705.2275}%
  \BibitemOpen
  \bibfield  {author} {\bibinfo {author} {\bibfnamefont {Y.}~\bibnamefont
  {Koide}},\ }\bibfield  {title} {\bibinfo {title} {{S(4) flavor symmetry
  embedded into SU(3) and lepton masses and mixing}},\ }\href
  {https://doi.org/10.1088/1126-6708/2007/08/086} {\bibfield  {journal}
  {\bibinfo  {journal} {JHEP}\ }\textbf {\bibinfo {volume} {08}},\ \bibinfo
  {pages} {086}},\ \Eprint {https://arxiv.org/abs/0705.2275} {arXiv:0705.2275
  [hep-ph]} \BibitemShut {NoStop}%
\bibitem [{\citenamefont {Bazzocchi}\ and\ \citenamefont
  {Morisi}(2009)}]{0811.0345}%
  \BibitemOpen
  \bibfield  {author} {\bibinfo {author} {\bibfnamefont {F.}~\bibnamefont
  {Bazzocchi}}\ and\ \bibinfo {author} {\bibfnamefont {S.}~\bibnamefont
  {Morisi}},\ }\bibfield  {title} {\bibinfo {title} {{S(4) as a natural flavor
  symmetry for lepton mixing}},\ }\href
  {https://doi.org/10.1103/PhysRevD.80.096005} {\bibfield  {journal} {\bibinfo
  {journal} {Phys. Rev. D}\ }\textbf {\bibinfo {volume} {80}},\ \bibinfo
  {pages} {096005} (\bibinfo {year} {2009})},\ \Eprint
  {https://arxiv.org/abs/0811.0345} {arXiv:0811.0345 [hep-ph]} \BibitemShut
  {NoStop}%
\bibitem [{\citenamefont {Krishnan}\ \emph {et~al.}(2013)\citenamefont
  {Krishnan}, \citenamefont {Harrison},\ and\ \citenamefont
  {Scott}}]{1211.2000}%
  \BibitemOpen
  \bibfield  {author} {\bibinfo {author} {\bibfnamefont {R.}~\bibnamefont
  {Krishnan}}, \bibinfo {author} {\bibfnamefont {P.~F.}\ \bibnamefont
  {Harrison}},\ and\ \bibinfo {author} {\bibfnamefont {W.~G.}\ \bibnamefont
  {Scott}},\ }\bibfield  {title} {\bibinfo {title} {{Simplest Neutrino Mixing
  from S4 Symmetry}},\ }\href {https://doi.org/10.1007/JHEP04(2013)087}
  {\bibfield  {journal} {\bibinfo  {journal} {JHEP}\ }\textbf {\bibinfo
  {volume} {04}},\ \bibinfo {pages} {087}},\ \Eprint
  {https://arxiv.org/abs/1211.2000} {arXiv:1211.2000 [hep-ph]} \BibitemShut
  {NoStop}%
\bibitem [{\citenamefont {Ecker}\ \emph {et~al.}(1987)\citenamefont {Ecker},
  \citenamefont {Grimus},\ and\ \citenamefont {Neufeld}}]{Ecker:1987qp}%
  \BibitemOpen
  \bibfield  {author} {\bibinfo {author} {\bibfnamefont {G.}~\bibnamefont
  {Ecker}}, \bibinfo {author} {\bibfnamefont {W.}~\bibnamefont {Grimus}},\ and\
  \bibinfo {author} {\bibfnamefont {H.}~\bibnamefont {Neufeld}},\ }\bibfield
  {title} {\bibinfo {title} {{A Standard Form for Generalized {CP}
  Transformations}},\ }\href {https://doi.org/10.1088/0305-4470/20/12/010}
  {\bibfield  {journal} {\bibinfo  {journal} {J. Phys. A}\ }\textbf {\bibinfo
  {volume} {20}},\ \bibinfo {pages} {L807} (\bibinfo {year}
  {1987})}\BibitemShut {NoStop}%
\bibitem [{\citenamefont {Neufeld}\ \emph {et~al.}(1988)\citenamefont
  {Neufeld}, \citenamefont {Grimus},\ and\ \citenamefont
  {Ecker}}]{Neufeld:1987wa}%
  \BibitemOpen
  \bibfield  {author} {\bibinfo {author} {\bibfnamefont {H.}~\bibnamefont
  {Neufeld}}, \bibinfo {author} {\bibfnamefont {W.}~\bibnamefont {Grimus}},\
  and\ \bibinfo {author} {\bibfnamefont {G.}~\bibnamefont {Ecker}},\ }\bibfield
   {title} {\bibinfo {title} {{Generalized {CP} Invariance, Neutral Flavor
  Conservation and the Structure of the Mixing Matrix}},\ }\href
  {https://doi.org/10.1142/S0217751X88000254} {\bibfield  {journal} {\bibinfo
  {journal} {Int. J. Mod. Phys. A}\ }\textbf {\bibinfo {volume} {3}},\ \bibinfo
  {pages} {603} (\bibinfo {year} {1988})}\BibitemShut {NoStop}%
\bibitem [{\citenamefont {Feruglio}\ \emph {et~al.}(2013)\citenamefont
  {Feruglio}, \citenamefont {Hagedorn},\ and\ \citenamefont
  {Ziegler}}]{1211.5560}%
  \BibitemOpen
  \bibfield  {author} {\bibinfo {author} {\bibfnamefont {F.}~\bibnamefont
  {Feruglio}}, \bibinfo {author} {\bibfnamefont {C.}~\bibnamefont {Hagedorn}},\
  and\ \bibinfo {author} {\bibfnamefont {R.}~\bibnamefont {Ziegler}},\
  }\bibfield  {title} {\bibinfo {title} {{Lepton Mixing Parameters from
  Discrete and CP Symmetries}},\ }\href
  {https://doi.org/10.1007/JHEP07(2013)027} {\bibfield  {journal} {\bibinfo
  {journal} {JHEP}\ }\textbf {\bibinfo {volume} {07}},\ \bibinfo {pages}
  {027}},\ \Eprint {https://arxiv.org/abs/1211.5560} {arXiv:1211.5560 [hep-ph]}
  \BibitemShut {NoStop}%
\bibitem [{\citenamefont {Holthausen}\ \emph
  {et~al.}(2013{\natexlab{b}})\citenamefont {Holthausen}, \citenamefont
  {Lindner},\ and\ \citenamefont {Schmidt}}]{1211.6953}%
  \BibitemOpen
  \bibfield  {author} {\bibinfo {author} {\bibfnamefont {M.}~\bibnamefont
  {Holthausen}}, \bibinfo {author} {\bibfnamefont {M.}~\bibnamefont
  {Lindner}},\ and\ \bibinfo {author} {\bibfnamefont {M.~A.}\ \bibnamefont
  {Schmidt}},\ }\bibfield  {title} {\bibinfo {title} {{CP and Discrete Flavour
  Symmetries}},\ }\href {https://doi.org/10.1007/JHEP04(2013)122} {\bibfield
  {journal} {\bibinfo  {journal} {JHEP}\ }\textbf {\bibinfo {volume} {04}},\
  \bibinfo {pages} {122}},\ \Eprint {https://arxiv.org/abs/1211.6953}
  {arXiv:1211.6953 [hep-ph]} \BibitemShut {NoStop}%
\bibitem [{\citenamefont {Ding}\ \emph {et~al.}(2013)\citenamefont {Ding},
  \citenamefont {King}, \citenamefont {Luhn},\ and\ \citenamefont
  {Stuart}}]{1303.6180}%
  \BibitemOpen
  \bibfield  {author} {\bibinfo {author} {\bibfnamefont {G.-J.}\ \bibnamefont
  {Ding}}, \bibinfo {author} {\bibfnamefont {S.~F.}\ \bibnamefont {King}},
  \bibinfo {author} {\bibfnamefont {C.}~\bibnamefont {Luhn}},\ and\ \bibinfo
  {author} {\bibfnamefont {A.~J.}\ \bibnamefont {Stuart}},\ }\bibfield  {title}
  {\bibinfo {title} {{Spontaneous CP violation from vacuum alignment in $S_4$
  models of leptons}},\ }\href {https://doi.org/10.1007/JHEP05(2013)084}
  {\bibfield  {journal} {\bibinfo  {journal} {JHEP}\ }\textbf {\bibinfo
  {volume} {05}},\ \bibinfo {pages} {084}},\ \Eprint
  {https://arxiv.org/abs/1303.6180} {arXiv:1303.6180 [hep-ph]} \BibitemShut
  {NoStop}%
\bibitem [{\citenamefont {Chen}\ \emph {et~al.}(2014)\citenamefont {Chen},
  \citenamefont {Fallbacher}, \citenamefont {Mahanthappa}, \citenamefont
  {Ratz},\ and\ \citenamefont {Trautner}}]{1402.0507}%
  \BibitemOpen
  \bibfield  {author} {\bibinfo {author} {\bibfnamefont {M.-C.}\ \bibnamefont
  {Chen}}, \bibinfo {author} {\bibfnamefont {M.}~\bibnamefont {Fallbacher}},
  \bibinfo {author} {\bibfnamefont {K.~T.}\ \bibnamefont {Mahanthappa}},
  \bibinfo {author} {\bibfnamefont {M.}~\bibnamefont {Ratz}},\ and\ \bibinfo
  {author} {\bibfnamefont {A.}~\bibnamefont {Trautner}},\ }\bibfield  {title}
  {\bibinfo {title} {{CP Violation from Finite Groups}},\ }\href
  {https://doi.org/10.1016/j.nuclphysb.2014.03.023} {\bibfield  {journal}
  {\bibinfo  {journal} {Nucl. Phys. B}\ }\textbf {\bibinfo {volume} {883}},\
  \bibinfo {pages} {267} (\bibinfo {year} {2014})},\ \Eprint
  {https://arxiv.org/abs/1402.0507} {arXiv:1402.0507 [hep-ph]} \BibitemShut
  {NoStop}%
\bibitem [{\citenamefont {{The GAP~Group}}(2022)}]{GAP4}%
  \BibitemOpen
  \bibfield  {author} {\bibinfo {author} {\bibnamefont {{The GAP~Group}}},\
  }\href@noop {} {\emph {\bibinfo {title} {{GAP -- Groups, Algorithms, and
  Programming, Version 4.12.2}}}} (\bibinfo {year} {2022}),\ \bibinfo {note}
  {\url{https://www.gap-system.org}}\BibitemShut {NoStop}%
\bibitem [{\citenamefont {Krishnan}(2023)}]{2306.07325}%
  \BibitemOpen
  \bibfield  {author} {\bibinfo {author} {\bibfnamefont {R.}~\bibnamefont
  {Krishnan}},\ }\bibfield  {title} {\bibinfo {title} {{Homogeneous linear
  intrinsic constraints in the stationary manifold of a $G$-invariant
  potential}},\ }\href@noop {} {\  (\bibinfo {year} {2023})},\ \Eprint
  {https://arxiv.org/abs/2306.07325} {arXiv:2306.07325 [hep-ph]} \BibitemShut
  {NoStop}%
\bibitem [{\citenamefont {Michel}(1971)}]{MICHEL1971}%
  \BibitemOpen
  \bibfield  {author} {\bibinfo {author} {\bibfnamefont {L.}~\bibnamefont
  {Michel}},\ }\bibfield  {title} {\bibinfo {title} {Points critiques des
  fonctions invariantes sur une g-variete},\ }\href@noop {} {\bibfield
  {journal} {\bibinfo  {journal} {C. R. Acad. Sci. Paris A}\ }\textbf {\bibinfo
  {volume} {272}},\ \bibinfo {pages} {433} (\bibinfo {year}
  {1971})}\BibitemShut {NoStop}%
\bibitem [{\citenamefont {Michel}\ and\ \citenamefont
  {Zhilinskii}(2001)}]{MICHEL200111}%
  \BibitemOpen
  \bibfield  {author} {\bibinfo {author} {\bibfnamefont {L.}~\bibnamefont
  {Michel}}\ and\ \bibinfo {author} {\bibfnamefont {B.}~\bibnamefont
  {Zhilinskii}},\ }\bibfield  {title} {\bibinfo {title} {Symmetry, invariants,
  topology. basic tools},\ }\href
  {https://doi.org/10.1016/S0370-1573(00)00088-0} {\bibfield  {journal}
  {\bibinfo  {journal} {Physics Reports}\ }\textbf {\bibinfo {volume} {341}},\
  \bibinfo {pages} {11} (\bibinfo {year} {2001})},\ \bibinfo {note} {symmetry,
  invariants, topology}\BibitemShut {NoStop}%
\bibitem [{\citenamefont {Golubitsky}\ \emph {et~al.}(1988)\citenamefont
  {Golubitsky}, \citenamefont {Stewart},\ and\ \citenamefont
  {Schaeffer}}]{GOLUBITSKY1988}%
  \BibitemOpen
  \bibfield  {author} {\bibinfo {author} {\bibfnamefont {M.}~\bibnamefont
  {Golubitsky}}, \bibinfo {author} {\bibfnamefont {I.}~\bibnamefont
  {Stewart}},\ and\ \bibinfo {author} {\bibfnamefont {D.~G.}\ \bibnamefont
  {Schaeffer}},\ }\bibfield  {title} {\bibinfo {title} {Singularities and
  groups in bifurcation theory},\ }in\ \href
  {https://doi.org/10.1007/978-1-4612-5034-0} {\emph {\bibinfo {booktitle}
  {Singularities and groups in bifurcation theory}}}\ (\bibinfo  {publisher}
  {Springer},\ \bibinfo {address} {New York},\ \bibinfo {year}
  {1988})\BibitemShut {NoStop}%
\bibitem [{\citenamefont {Krishnan}\ \emph {et~al.}(2018)\citenamefont
  {Krishnan}, \citenamefont {Harrison},\ and\ \citenamefont
  {Scott}}]{1801.10197}%
  \BibitemOpen
  \bibfield  {author} {\bibinfo {author} {\bibfnamefont {R.}~\bibnamefont
  {Krishnan}}, \bibinfo {author} {\bibfnamefont {P.}~\bibnamefont {Harrison}},\
  and\ \bibinfo {author} {\bibfnamefont {W.}~\bibnamefont {Scott}},\ }\bibfield
   {title} {\bibinfo {title} {{Fully Constrained Majorana Neutrino Mass
  Matrices Using $\Sigma(72\times 3)$}},\ }\href
  {https://doi.org/10.1140/epjc/s10052-018-5516-7} {\bibfield  {journal}
  {\bibinfo  {journal} {Eur. Phys. J. C}\ }\textbf {\bibinfo {volume} {78}},\
  \bibinfo {pages} {74} (\bibinfo {year} {2018})},\ \Eprint
  {https://arxiv.org/abs/1801.10197} {arXiv:1801.10197 [hep-ph]} \BibitemShut
  {NoStop}%
\bibitem [{\citenamefont {Xing}\ and\ \citenamefont
  {Zhou}(2007)}]{hep-ph/0607302}%
  \BibitemOpen
  \bibfield  {author} {\bibinfo {author} {\bibfnamefont {Z.-z.}\ \bibnamefont
  {Xing}}\ and\ \bibinfo {author} {\bibfnamefont {S.}~\bibnamefont {Zhou}},\
  }\bibfield  {title} {\bibinfo {title} {{Tri-bimaximal Neutrino Mixing and
  Flavor-dependent Resonant Leptogenesis}},\ }\href
  {https://doi.org/10.1016/j.physletb.2007.08.009} {\bibfield  {journal}
  {\bibinfo  {journal} {Phys. Lett. B}\ }\textbf {\bibinfo {volume} {653}},\
  \bibinfo {pages} {278} (\bibinfo {year} {2007})},\ \Eprint
  {https://arxiv.org/abs/hep-ph/0607302} {arXiv:hep-ph/0607302} \BibitemShut
  {NoStop}%
\bibitem [{\citenamefont {Albright}\ and\ \citenamefont
  {Rodejohann}(2009)}]{0812.0436}%
  \BibitemOpen
  \bibfield  {author} {\bibinfo {author} {\bibfnamefont {C.~H.}\ \bibnamefont
  {Albright}}\ and\ \bibinfo {author} {\bibfnamefont {W.}~\bibnamefont
  {Rodejohann}},\ }\bibfield  {title} {\bibinfo {title} {{Comparing Trimaximal
  Mixing and Its Variants with Deviations from Tri-bimaximal Mixing}},\ }\href
  {https://doi.org/10.1140/epjc/s10052-009-1074-3} {\bibfield  {journal}
  {\bibinfo  {journal} {Eur. Phys. J. C}\ }\textbf {\bibinfo {volume} {62}},\
  \bibinfo {pages} {599} (\bibinfo {year} {2009})},\ \Eprint
  {https://arxiv.org/abs/0812.0436} {arXiv:0812.0436 [hep-ph]} \BibitemShut
  {NoStop}%
\bibitem [{\citenamefont {Albright}\ \emph {et~al.}(2010)\citenamefont
  {Albright}, \citenamefont {Dueck},\ and\ \citenamefont
  {Rodejohann}}]{1004.2798}%
  \BibitemOpen
  \bibfield  {author} {\bibinfo {author} {\bibfnamefont {C.~H.}\ \bibnamefont
  {Albright}}, \bibinfo {author} {\bibfnamefont {A.}~\bibnamefont {Dueck}},\
  and\ \bibinfo {author} {\bibfnamefont {W.}~\bibnamefont {Rodejohann}},\
  }\bibfield  {title} {\bibinfo {title} {{Possible Alternatives to
  Tri-bimaximal Mixing}},\ }\href
  {https://doi.org/10.1140/epjc/s10052-010-1492-2} {\bibfield  {journal}
  {\bibinfo  {journal} {Eur. Phys. J. C}\ }\textbf {\bibinfo {volume} {70}},\
  \bibinfo {pages} {1099} (\bibinfo {year} {2010})},\ \Eprint
  {https://arxiv.org/abs/1004.2798} {arXiv:1004.2798 [hep-ph]} \BibitemShut
  {NoStop}%
\bibitem [{\citenamefont {Antusch}\ \emph {et~al.}(2012)\citenamefont
  {Antusch}, \citenamefont {King}, \citenamefont {Luhn},\ and\ \citenamefont
  {Spinrath}}]{1108.4278}%
  \BibitemOpen
  \bibfield  {author} {\bibinfo {author} {\bibfnamefont {S.}~\bibnamefont
  {Antusch}}, \bibinfo {author} {\bibfnamefont {S.~F.}\ \bibnamefont {King}},
  \bibinfo {author} {\bibfnamefont {C.}~\bibnamefont {Luhn}},\ and\ \bibinfo
  {author} {\bibfnamefont {M.}~\bibnamefont {Spinrath}},\ }\bibfield  {title}
  {\bibinfo {title} {{Trimaximal mixing with predicted $\theta_{13}$ from a new
  type of constrained sequential dominance}},\ }\href
  {https://doi.org/10.1016/j.nuclphysb.2011.11.009} {\bibfield  {journal}
  {\bibinfo  {journal} {Nucl. Phys. B}\ }\textbf {\bibinfo {volume} {856}},\
  \bibinfo {pages} {328} (\bibinfo {year} {2012})},\ \Eprint
  {https://arxiv.org/abs/1108.4278} {arXiv:1108.4278 [hep-ph]} \BibitemShut
  {NoStop}%
\bibitem [{\citenamefont {de~Medeiros~Varzielas}\ and\ \citenamefont
  {Lavoura}(2013)}]{1212.3247}%
  \BibitemOpen
  \bibfield  {author} {\bibinfo {author} {\bibfnamefont {I.}~\bibnamefont
  {de~Medeiros~Varzielas}}\ and\ \bibinfo {author} {\bibfnamefont
  {L.}~\bibnamefont {Lavoura}},\ }\bibfield  {title} {\bibinfo {title}
  {{Flavour models for $TM_{1}$ lepton mixing}},\ }\href
  {https://doi.org/10.1088/0954-3899/40/8/085002} {\bibfield  {journal}
  {\bibinfo  {journal} {J. Phys. G}\ }\textbf {\bibinfo {volume} {40}},\
  \bibinfo {pages} {085002} (\bibinfo {year} {2013})},\ \Eprint
  {https://arxiv.org/abs/1212.3247} {arXiv:1212.3247 [hep-ph]} \BibitemShut
  {NoStop}%
\bibitem [{\citenamefont {King}(2013{\natexlab{a}})}]{1304.6264}%
  \BibitemOpen
  \bibfield  {author} {\bibinfo {author} {\bibfnamefont {S.~F.}\ \bibnamefont
  {King}},\ }\bibfield  {title} {\bibinfo {title} {{Minimal predictive see-saw
  model with normal neutrino mass hierarchy}},\ }\href
  {https://doi.org/10.1007/JHEP07(2013)137} {\bibfield  {journal} {\bibinfo
  {journal} {JHEP}\ }\textbf {\bibinfo {volume} {07}},\ \bibinfo {pages}
  {137}},\ \Eprint {https://arxiv.org/abs/1304.6264} {arXiv:1304.6264 [hep-ph]}
  \BibitemShut {NoStop}%
\bibitem [{\citenamefont {Luhn}(2013)}]{1306.2358}%
  \BibitemOpen
  \bibfield  {author} {\bibinfo {author} {\bibfnamefont {C.}~\bibnamefont
  {Luhn}},\ }\bibfield  {title} {\bibinfo {title} {{Trimaximal TM$_{1}$
  neutrino mixing in S$_{4}$ with spontaneous CP violation}},\ }\href
  {https://doi.org/10.1016/j.nuclphysb.2013.07.003} {\bibfield  {journal}
  {\bibinfo  {journal} {Nucl. Phys. B}\ }\textbf {\bibinfo {volume} {875}},\
  \bibinfo {pages} {80} (\bibinfo {year} {2013})},\ \Eprint
  {https://arxiv.org/abs/1306.2358} {arXiv:1306.2358 [hep-ph]} \BibitemShut
  {NoStop}%
\bibitem [{\citenamefont {Li}\ and\ \citenamefont {Ding}(2014)}]{1312.4401}%
  \BibitemOpen
  \bibfield  {author} {\bibinfo {author} {\bibfnamefont {C.-C.}\ \bibnamefont
  {Li}}\ and\ \bibinfo {author} {\bibfnamefont {G.-J.}\ \bibnamefont {Ding}},\
  }\bibfield  {title} {\bibinfo {title} {{Generalised CP and trimaximal $TM_1$
  lepton mixing in $S_4$ family symmetry}},\ }\href
  {https://doi.org/10.1016/j.nuclphysb.2014.02.002} {\bibfield  {journal}
  {\bibinfo  {journal} {Nucl. Phys. B}\ }\textbf {\bibinfo {volume} {881}},\
  \bibinfo {pages} {206} (\bibinfo {year} {2014})},\ \Eprint
  {https://arxiv.org/abs/1312.4401} {arXiv:1312.4401 [hep-ph]} \BibitemShut
  {NoStop}%
\bibitem [{\citenamefont {Zhao}(2015)}]{1509.06915}%
  \BibitemOpen
  \bibfield  {author} {\bibinfo {author} {\bibfnamefont {Z.-h.}\ \bibnamefont
  {Zhao}},\ }\bibfield  {title} {\bibinfo {title} {{Modified Friedberg-Lee
  symmetry for neutrino mixing}},\ }\href
  {https://doi.org/10.1103/PhysRevD.92.113001} {\bibfield  {journal} {\bibinfo
  {journal} {Phys. Rev. D}\ }\textbf {\bibinfo {volume} {92}},\ \bibinfo
  {pages} {113001} (\bibinfo {year} {2015})},\ \Eprint
  {https://arxiv.org/abs/1509.06915} {arXiv:1509.06915 [hep-ph]} \BibitemShut
  {NoStop}%
\bibitem [{\citenamefont {Gautam}(2018)}]{1802.00425}%
  \BibitemOpen
  \bibfield  {author} {\bibinfo {author} {\bibfnamefont {R.~R.}\ \bibnamefont
  {Gautam}},\ }\bibfield  {title} {\bibinfo {title} {{Trimaximal mixing with a
  texture zero}},\ }\href {https://doi.org/10.1103/PhysRevD.97.055022}
  {\bibfield  {journal} {\bibinfo  {journal} {Phys. Rev. D}\ }\textbf {\bibinfo
  {volume} {97}},\ \bibinfo {pages} {055022} (\bibinfo {year} {2018})},\
  \Eprint {https://arxiv.org/abs/1802.00425} {arXiv:1802.00425 [hep-ph]}
  \BibitemShut {NoStop}%
\bibitem [{\citenamefont {Shimizu}\ \emph {et~al.}(2017)\citenamefont
  {Shimizu}, \citenamefont {Takagi},\ and\ \citenamefont
  {Tanimoto}}]{1709.02136}%
  \BibitemOpen
  \bibfield  {author} {\bibinfo {author} {\bibfnamefont {Y.}~\bibnamefont
  {Shimizu}}, \bibinfo {author} {\bibfnamefont {K.}~\bibnamefont {Takagi}},\
  and\ \bibinfo {author} {\bibfnamefont {M.}~\bibnamefont {Tanimoto}},\
  }\bibfield  {title} {\bibinfo {title} {{Towards the minimal seesaw model via
  CP violation of neutrinos}},\ }\href
  {https://doi.org/10.1007/JHEP11(2017)201} {\bibfield  {journal} {\bibinfo
  {journal} {JHEP}\ }\textbf {\bibinfo {volume} {11}},\ \bibinfo {pages}
  {201}},\ \Eprint {https://arxiv.org/abs/1709.02136} {arXiv:1709.02136
  [hep-ph]} \BibitemShut {NoStop}%
\bibitem [{\citenamefont {Rodejohann}\ and\ \citenamefont
  {Xu}(2017)}]{1705.02027}%
  \BibitemOpen
  \bibfield  {author} {\bibinfo {author} {\bibfnamefont {W.}~\bibnamefont
  {Rodejohann}}\ and\ \bibinfo {author} {\bibfnamefont {X.-J.}\ \bibnamefont
  {Xu}},\ }\bibfield  {title} {\bibinfo {title} {{Trimaximal $\mu$-$\tau$
  reflection symmetry}},\ }\href {https://doi.org/10.1103/PhysRevD.96.055039}
  {\bibfield  {journal} {\bibinfo  {journal} {Phys. Rev. D}\ }\textbf {\bibinfo
  {volume} {96}},\ \bibinfo {pages} {055039} (\bibinfo {year} {2017})},\
  \Eprint {https://arxiv.org/abs/1705.02027} {arXiv:1705.02027 [hep-ph]}
  \BibitemShut {NoStop}%
\bibitem [{\citenamefont {King}(2013{\natexlab{b}})}]{1305.4846}%
  \BibitemOpen
  \bibfield  {author} {\bibinfo {author} {\bibfnamefont {S.~F.}\ \bibnamefont
  {King}},\ }\bibfield  {title} {\bibinfo {title} {{Minimal see-saw model
  predicting best fit lepton mixing angles}},\ }\href
  {https://doi.org/10.1016/j.physletb.2013.06.013} {\bibfield  {journal}
  {\bibinfo  {journal} {Phys. Lett. B}\ }\textbf {\bibinfo {volume} {724}},\
  \bibinfo {pages} {92} (\bibinfo {year} {2013}{\natexlab{b}})},\ \Eprint
  {https://arxiv.org/abs/1305.4846} {arXiv:1305.4846 [hep-ph]} \BibitemShut
  {NoStop}%
\bibitem [{\citenamefont {King}(2016)}]{1512.07531}%
  \BibitemOpen
  \bibfield  {author} {\bibinfo {author} {\bibfnamefont {S.~F.}\ \bibnamefont
  {King}},\ }\bibfield  {title} {\bibinfo {title} {{Littlest Seesaw}},\ }\href
  {https://doi.org/10.1007/JHEP02(2016)085} {\bibfield  {journal} {\bibinfo
  {journal} {JHEP}\ }\textbf {\bibinfo {volume} {02}},\ \bibinfo {pages}
  {085}},\ \Eprint {https://arxiv.org/abs/1512.07531} {arXiv:1512.07531
  [hep-ph]} \BibitemShut {NoStop}%
\bibitem [{\citenamefont {King}\ and\ \citenamefont {Luhn}(2016)}]{1607.05276}%
  \BibitemOpen
  \bibfield  {author} {\bibinfo {author} {\bibfnamefont {S.~F.}\ \bibnamefont
  {King}}\ and\ \bibinfo {author} {\bibfnamefont {C.}~\bibnamefont {Luhn}},\
  }\bibfield  {title} {\bibinfo {title} {{Littlest Seesaw model from S$_{4}
  \times$ U(1)}},\ }\href {https://doi.org/10.1007/JHEP09(2016)023} {\bibfield
  {journal} {\bibinfo  {journal} {JHEP}\ }\textbf {\bibinfo {volume} {09}},\
  \bibinfo {pages} {023}},\ \Eprint {https://arxiv.org/abs/1607.05276}
  {arXiv:1607.05276 [hep-ph]} \BibitemShut {NoStop}%
\bibitem [{\citenamefont {Chakraborty}\ \emph {et~al.}(2020)\citenamefont
  {Chakraborty}, \citenamefont {Krishnan},\ and\ \citenamefont
  {Ghosal}}]{2003.00506}%
  \BibitemOpen
  \bibfield  {author} {\bibinfo {author} {\bibfnamefont {M.}~\bibnamefont
  {Chakraborty}}, \bibinfo {author} {\bibfnamefont {R.}~\bibnamefont
  {Krishnan}},\ and\ \bibinfo {author} {\bibfnamefont {A.}~\bibnamefont
  {Ghosal}},\ }\bibfield  {title} {\bibinfo {title} {{Predictive $S_4$ flavon
  model with $\text{TM}_1$ mixing and baryogenesis through leptogenesis}},\
  }\href {https://doi.org/10.1007/JHEP09(2020)025} {\bibfield  {journal}
  {\bibinfo  {journal} {JHEP}\ }\textbf {\bibinfo {volume} {09}},\ \bibinfo
  {pages} {025}},\ \Eprint {https://arxiv.org/abs/2003.00506} {arXiv:2003.00506
  [hep-ph]} \BibitemShut {NoStop}%
\bibitem [{\citenamefont {de~Salas}\ \emph {et~al.}(2021)\citenamefont
  {de~Salas}, \citenamefont {Forero}, \citenamefont {Gariazzo}, \citenamefont
  {Mart\'\i{}nez-Mirav\'e}, \citenamefont {Mena}, \citenamefont {Ternes},
  \citenamefont {T\'ortola},\ and\ \citenamefont {Valle}}]{2006.11237}%
  \BibitemOpen
  \bibfield  {author} {\bibinfo {author} {\bibfnamefont {P.~F.}\ \bibnamefont
  {de~Salas}}, \bibinfo {author} {\bibfnamefont {D.~V.}\ \bibnamefont
  {Forero}}, \bibinfo {author} {\bibfnamefont {S.}~\bibnamefont {Gariazzo}},
  \bibinfo {author} {\bibfnamefont {P.}~\bibnamefont {Mart\'\i{}nez-Mirav\'e}},
  \bibinfo {author} {\bibfnamefont {O.}~\bibnamefont {Mena}}, \bibinfo {author}
  {\bibfnamefont {C.~A.}\ \bibnamefont {Ternes}}, \bibinfo {author}
  {\bibfnamefont {M.}~\bibnamefont {T\'ortola}},\ and\ \bibinfo {author}
  {\bibfnamefont {J.~W.~F.}\ \bibnamefont {Valle}},\ }\bibfield  {title}
  {\bibinfo {title} {{2020 global reassessment of the neutrino oscillation
  picture}},\ }\href {https://doi.org/10.1007/JHEP02(2021)071} {\bibfield
  {journal} {\bibinfo  {journal} {JHEP}\ }\textbf {\bibinfo {volume} {02}},\
  \bibinfo {pages} {071}},\ \Eprint {https://arxiv.org/abs/2006.11237}
  {arXiv:2006.11237 [hep-ph]} \BibitemShut {NoStop}%
\bibitem [{\citenamefont {Dolinski}\ \emph {et~al.}(2019)\citenamefont
  {Dolinski}, \citenamefont {Poon},\ and\ \citenamefont
  {Rodejohann}}]{1902.04097}%
  \BibitemOpen
  \bibfield  {author} {\bibinfo {author} {\bibfnamefont {M.~J.}\ \bibnamefont
  {Dolinski}}, \bibinfo {author} {\bibfnamefont {A.~W.~P.}\ \bibnamefont
  {Poon}},\ and\ \bibinfo {author} {\bibfnamefont {W.}~\bibnamefont
  {Rodejohann}},\ }\bibfield  {title} {\bibinfo {title} {{Neutrinoless
  Double-Beta Decay: Status and Prospects}},\ }\href
  {https://doi.org/10.1146/annurev-nucl-101918-023407} {\bibfield  {journal}
  {\bibinfo  {journal} {Ann. Rev. Nucl. Part. Sci.}\ }\textbf {\bibinfo
  {volume} {69}},\ \bibinfo {pages} {219} (\bibinfo {year} {2019})},\ \Eprint
  {https://arxiv.org/abs/1902.04097} {arXiv:1902.04097 [nucl-ex]} \BibitemShut
  {NoStop}%
\bibitem [{\citenamefont {Gando}\ \emph {et~al.}(2016)\citenamefont {Gando}
  \emph {et~al.}}]{1605.02889}%
  \BibitemOpen
  \bibfield  {author} {\bibinfo {author} {\bibfnamefont {A.}~\bibnamefont
  {Gando}} \emph {et~al.} (\bibinfo {collaboration} {KamLAND-Zen}),\ }\bibfield
   {title} {\bibinfo {title} {{Search for Majorana Neutrinos near the Inverted
  Mass Hierarchy Region with KamLAND-Zen}},\ }\href
  {https://doi.org/10.1103/PhysRevLett.117.082503} {\bibfield  {journal}
  {\bibinfo  {journal} {Phys. Rev. Lett.}\ }\textbf {\bibinfo {volume} {117}},\
  \bibinfo {pages} {082503} (\bibinfo {year} {2016})},\ \bibinfo {note}
  {[Addendum: Phys.Rev.Lett. 117, 109903 (2016)]},\ \Eprint
  {https://arxiv.org/abs/1605.02889} {arXiv:1605.02889 [hep-ex]} \BibitemShut
  {NoStop}%
\bibitem [{\citenamefont {Anton}\ \emph {et~al.}(2019)\citenamefont {Anton}
  \emph {et~al.}}]{1906.02723}%
  \BibitemOpen
  \bibfield  {author} {\bibinfo {author} {\bibfnamefont {G.}~\bibnamefont
  {Anton}} \emph {et~al.} (\bibinfo {collaboration} {EXO-200}),\ }\bibfield
  {title} {\bibinfo {title} {{Search for Neutrinoless Double-$\beta$ Decay with
  the Complete EXO-200 Dataset}},\ }\href
  {https://doi.org/10.1103/PhysRevLett.123.161802} {\bibfield  {journal}
  {\bibinfo  {journal} {Phys. Rev. Lett.}\ }\textbf {\bibinfo {volume} {123}},\
  \bibinfo {pages} {161802} (\bibinfo {year} {2019})},\ \Eprint
  {https://arxiv.org/abs/1906.02723} {arXiv:1906.02723 [hep-ex]} \BibitemShut
  {NoStop}%
\bibitem [{\citenamefont {Adams}\ \emph {et~al.}(2022)\citenamefont {Adams}
  \emph {et~al.}}]{2104.06906}%
  \BibitemOpen
  \bibfield  {author} {\bibinfo {author} {\bibfnamefont {D.~Q.}\ \bibnamefont
  {Adams}} \emph {et~al.} (\bibinfo {collaboration} {CUORE}),\ }\bibfield
  {title} {\bibinfo {title} {{Search for Majorana neutrinos exploiting
  millikelvin cryogenics with CUORE}},\ }\href
  {https://doi.org/10.1038/s41586-022-04497-4} {\bibfield  {journal} {\bibinfo
  {journal} {Nature}\ }\textbf {\bibinfo {volume} {604}},\ \bibinfo {pages}
  {53} (\bibinfo {year} {2022})},\ \Eprint {https://arxiv.org/abs/2104.06906}
  {arXiv:2104.06906 [nucl-ex]} \BibitemShut {NoStop}%
\bibitem [{\citenamefont {Di~Valentino}\ \emph {et~al.}(2021)\citenamefont
  {Di~Valentino}, \citenamefont {Gariazzo},\ and\ \citenamefont
  {Mena}}]{2106.15267}%
  \BibitemOpen
  \bibfield  {author} {\bibinfo {author} {\bibfnamefont {E.}~\bibnamefont
  {Di~Valentino}}, \bibinfo {author} {\bibfnamefont {S.}~\bibnamefont
  {Gariazzo}},\ and\ \bibinfo {author} {\bibfnamefont {O.}~\bibnamefont
  {Mena}},\ }\bibfield  {title} {\bibinfo {title} {{Most constraining
  cosmological neutrino mass bounds}},\ }\href
  {https://doi.org/10.1103/PhysRevD.104.083504} {\bibfield  {journal} {\bibinfo
   {journal} {Phys. Rev. D}\ }\textbf {\bibinfo {volume} {104}},\ \bibinfo
  {pages} {083504} (\bibinfo {year} {2021})},\ \Eprint
  {https://arxiv.org/abs/2106.15267} {arXiv:2106.15267 [astro-ph.CO]}
  \BibitemShut {NoStop}%
\bibitem [{\citenamefont {Loureiro}\ \emph {et~al.}(2019)\citenamefont
  {Loureiro} \emph {et~al.}}]{1811.02578}%
  \BibitemOpen
  \bibfield  {author} {\bibinfo {author} {\bibfnamefont {A.}~\bibnamefont
  {Loureiro}} \emph {et~al.},\ }\bibfield  {title} {\bibinfo {title} {{On The
  Upper Bound of Neutrino Masses from Combined Cosmological Observations and
  Particle Physics Experiments}},\ }\href
  {https://doi.org/10.1103/PhysRevLett.123.081301} {\bibfield  {journal}
  {\bibinfo  {journal} {Phys. Rev. Lett.}\ }\textbf {\bibinfo {volume} {123}},\
  \bibinfo {pages} {081301} (\bibinfo {year} {2019})},\ \Eprint
  {https://arxiv.org/abs/1811.02578} {arXiv:1811.02578 [astro-ph.CO]}
  \BibitemShut {NoStop}%
\bibitem [{\citenamefont {Abdullahi}\ \emph {et~al.}(2023)\citenamefont
  {Abdullahi} \emph {et~al.}}]{2203.08039}%
  \BibitemOpen
  \bibfield  {author} {\bibinfo {author} {\bibfnamefont {A.~M.}\ \bibnamefont
  {Abdullahi}} \emph {et~al.},\ }\bibfield  {title} {\bibinfo {title} {{The
  present and future status of heavy neutral leptons}},\ }\href
  {https://doi.org/10.1088/1361-6471/ac98f9} {\bibfield  {journal} {\bibinfo
  {journal} {J. Phys. G}\ }\textbf {\bibinfo {volume} {50}},\ \bibinfo {pages}
  {020501} (\bibinfo {year} {2023})},\ \Eprint
  {https://arxiv.org/abs/2203.08039} {arXiv:2203.08039 [hep-ph]} \BibitemShut
  {NoStop}%
\bibitem [{\citenamefont {Novichkov}\ \emph {et~al.}(2021)\citenamefont
  {Novichkov}, \citenamefont {Penedo},\ and\ \citenamefont
  {Petcov}}]{2006.03058}%
  \BibitemOpen
  \bibfield  {author} {\bibinfo {author} {\bibfnamefont {P.~P.}\ \bibnamefont
  {Novichkov}}, \bibinfo {author} {\bibfnamefont {J.~T.}\ \bibnamefont
  {Penedo}},\ and\ \bibinfo {author} {\bibfnamefont {S.~T.}\ \bibnamefont
  {Petcov}},\ }\bibfield  {title} {\bibinfo {title} {{Double cover of modular
  $S_4$ for flavour model building}},\ }\href
  {https://doi.org/10.1016/j.nuclphysb.2020.115301} {\bibfield  {journal}
  {\bibinfo  {journal} {Nucl. Phys. B}\ }\textbf {\bibinfo {volume} {963}},\
  \bibinfo {pages} {115301} (\bibinfo {year} {2021})},\ \Eprint
  {https://arxiv.org/abs/2006.03058} {arXiv:2006.03058 [hep-ph]} \BibitemShut
  {NoStop}%
\bibitem [{\citenamefont {Liu}\ \emph {et~al.}(2021)\citenamefont {Liu},
  \citenamefont {Yao},\ and\ \citenamefont {Ding}}]{2006.10722}%
  \BibitemOpen
  \bibfield  {author} {\bibinfo {author} {\bibfnamefont {X.-G.}\ \bibnamefont
  {Liu}}, \bibinfo {author} {\bibfnamefont {C.-Y.}\ \bibnamefont {Yao}},\ and\
  \bibinfo {author} {\bibfnamefont {G.-J.}\ \bibnamefont {Ding}},\ }\bibfield
  {title} {\bibinfo {title} {{Modular invariant quark and lepton models in
  double covering of $S_4$ modular group}},\ }\href
  {https://doi.org/10.1103/PhysRevD.103.056013} {\bibfield  {journal} {\bibinfo
   {journal} {Phys. Rev. D}\ }\textbf {\bibinfo {volume} {103}},\ \bibinfo
  {pages} {056013} (\bibinfo {year} {2021})},\ \Eprint
  {https://arxiv.org/abs/2006.10722} {arXiv:2006.10722 [hep-ph]} \BibitemShut
  {NoStop}%
\bibitem [{\citenamefont {Ding}\ \emph {et~al.}(2023)\citenamefont {Ding},
  \citenamefont {Liu},\ and\ \citenamefont {Yao}}]{2211.04546}%
  \BibitemOpen
  \bibfield  {author} {\bibinfo {author} {\bibfnamefont {G.-J.}\ \bibnamefont
  {Ding}}, \bibinfo {author} {\bibfnamefont {X.-G.}\ \bibnamefont {Liu}},\ and\
  \bibinfo {author} {\bibfnamefont {C.-Y.}\ \bibnamefont {Yao}},\ }\bibfield
  {title} {\bibinfo {title} {{A minimal modular invariant neutrino model}},\
  }\href {https://doi.org/10.1007/JHEP01(2023)125} {\bibfield  {journal}
  {\bibinfo  {journal} {JHEP}\ }\textbf {\bibinfo {volume} {01}},\ \bibinfo
  {pages} {125}},\ \Eprint {https://arxiv.org/abs/2211.04546} {arXiv:2211.04546
  [hep-ph]} \BibitemShut {NoStop}%
\bibitem [{\citenamefont {Harrison}\ \emph {et~al.}(2002)\citenamefont
  {Harrison}, \citenamefont {Perkins},\ and\ \citenamefont
  {Scott}}]{hep-ph/0202074}%
  \BibitemOpen
  \bibfield  {author} {\bibinfo {author} {\bibfnamefont {P.~F.}\ \bibnamefont
  {Harrison}}, \bibinfo {author} {\bibfnamefont {D.~H.}\ \bibnamefont
  {Perkins}},\ and\ \bibinfo {author} {\bibfnamefont {W.~G.}\ \bibnamefont
  {Scott}},\ }\bibfield  {title} {\bibinfo {title} {{Tri-bimaximal mixing and
  the neutrino oscillation data}},\ }\href
  {https://doi.org/10.1016/S0370-2693(02)01336-9} {\bibfield  {journal}
  {\bibinfo  {journal} {Phys. Lett. B}\ }\textbf {\bibinfo {volume} {530}},\
  \bibinfo {pages} {167} (\bibinfo {year} {2002})},\ \Eprint
  {https://arxiv.org/abs/hep-ph/0202074} {arXiv:hep-ph/0202074} \BibitemShut
  {NoStop}%
\bibitem [{\citenamefont {Jarlskog}(1985{\natexlab{a}})}]{Jarlskog:1985ht}%
  \BibitemOpen
  \bibfield  {author} {\bibinfo {author} {\bibfnamefont {C.}~\bibnamefont
  {Jarlskog}},\ }\bibfield  {title} {\bibinfo {title} {{Commutator of the Quark
  Mass Matrices in the Standard Electroweak Model and a Measure of Maximal CP
  Violation}},\ }\href {https://doi.org/10.1103/PhysRevLett.55.1039} {\bibfield
   {journal} {\bibinfo  {journal} {Phys. Rev. Lett.}\ }\textbf {\bibinfo
  {volume} {55}},\ \bibinfo {pages} {1039} (\bibinfo {year}
  {1985}{\natexlab{a}})}\BibitemShut {NoStop}%
\bibitem [{\citenamefont {Jarlskog}(1985{\natexlab{b}})}]{Jarlskog:1985cw}%
  \BibitemOpen
  \bibfield  {author} {\bibinfo {author} {\bibfnamefont {C.}~\bibnamefont
  {Jarlskog}},\ }\bibfield  {title} {\bibinfo {title} {{A Basis Independent
  Formulation of the Connection Between Quark Mass Matrices, CP Violation and
  Experiment}},\ }\href {https://doi.org/10.1007/BF01565198} {\bibfield
  {journal} {\bibinfo  {journal} {Z. Phys. C}\ }\textbf {\bibinfo {volume}
  {29}},\ \bibinfo {pages} {491} (\bibinfo {year}
  {1985}{\natexlab{b}})}\BibitemShut {NoStop}%
\bibitem [{\citenamefont {Zhou}(2012)}]{1205.0761}%
  \BibitemOpen
  \bibfield  {author} {\bibinfo {author} {\bibfnamefont {S.}~\bibnamefont
  {Zhou}},\ }\bibfield  {title} {\bibinfo {title} {{Lepton Flavor Mixing
  Pattern and Neutrino Mass Matrix after the Daya Bay Experiment}},\
  }\href@noop {} {\  (\bibinfo {year} {2012})},\ \Eprint
  {https://arxiv.org/abs/1205.0761} {arXiv:1205.0761 [hep-ph]} \BibitemShut
  {NoStop}%
\bibitem [{\citenamefont {Harrison}\ and\ \citenamefont
  {Scott}(2002{\natexlab{a}})}]{hep-ph/0203209}%
  \BibitemOpen
  \bibfield  {author} {\bibinfo {author} {\bibfnamefont {P.~F.}\ \bibnamefont
  {Harrison}}\ and\ \bibinfo {author} {\bibfnamefont {W.~G.}\ \bibnamefont
  {Scott}},\ }\bibfield  {title} {\bibinfo {title} {{Symmetries and
  generalizations of tri - bimaximal neutrino mixing}},\ }\href
  {https://doi.org/10.1016/S0370-2693(02)01753-7} {\bibfield  {journal}
  {\bibinfo  {journal} {Phys. Lett. B}\ }\textbf {\bibinfo {volume} {535}},\
  \bibinfo {pages} {163} (\bibinfo {year} {2002}{\natexlab{a}})},\ \Eprint
  {https://arxiv.org/abs/hep-ph/0203209} {arXiv:hep-ph/0203209} \BibitemShut
  {NoStop}%
\bibitem [{\citenamefont {Babu}\ \emph {et~al.}(2003)\citenamefont {Babu},
  \citenamefont {Ma},\ and\ \citenamefont {Valle}}]{hep-ph/0206292}%
  \BibitemOpen
  \bibfield  {author} {\bibinfo {author} {\bibfnamefont {K.~S.}\ \bibnamefont
  {Babu}}, \bibinfo {author} {\bibfnamefont {E.}~\bibnamefont {Ma}},\ and\
  \bibinfo {author} {\bibfnamefont {J.~W.~F.}\ \bibnamefont {Valle}},\
  }\bibfield  {title} {\bibinfo {title} {{Underlying A(4) symmetry for the
  neutrino mass matrix and the quark mixing matrix}},\ }\href
  {https://doi.org/10.1016/S0370-2693(02)03153-2} {\bibfield  {journal}
  {\bibinfo  {journal} {Phys. Lett. B}\ }\textbf {\bibinfo {volume} {552}},\
  \bibinfo {pages} {207} (\bibinfo {year} {2003})},\ \Eprint
  {https://arxiv.org/abs/hep-ph/0206292} {arXiv:hep-ph/0206292} \BibitemShut
  {NoStop}%
\bibitem [{\citenamefont {Ma}(2002)}]{hep-ph/0207352}%
  \BibitemOpen
  \bibfield  {author} {\bibinfo {author} {\bibfnamefont {E.}~\bibnamefont
  {Ma}},\ }\bibfield  {title} {\bibinfo {title} {{The All purpose neutrino mass
  matrix}},\ }\href {https://doi.org/10.1103/PhysRevD.66.117301} {\bibfield
  {journal} {\bibinfo  {journal} {Phys. Rev. D}\ }\textbf {\bibinfo {volume}
  {66}},\ \bibinfo {pages} {117301} (\bibinfo {year} {2002})},\ \Eprint
  {https://arxiv.org/abs/hep-ph/0207352} {arXiv:hep-ph/0207352} \BibitemShut
  {NoStop}%
\bibitem [{\citenamefont {Harrison}\ and\ \citenamefont
  {Scott}(2002{\natexlab{b}})}]{hep-ph/0210197}%
  \BibitemOpen
  \bibfield  {author} {\bibinfo {author} {\bibfnamefont {P.~F.}\ \bibnamefont
  {Harrison}}\ and\ \bibinfo {author} {\bibfnamefont {W.~G.}\ \bibnamefont
  {Scott}},\ }\bibfield  {title} {\bibinfo {title} {{mu - tau reflection
  symmetry in lepton mixing and neutrino oscillations}},\ }\href
  {https://doi.org/10.1016/S0370-2693(02)02772-7} {\bibfield  {journal}
  {\bibinfo  {journal} {Phys. Lett. B}\ }\textbf {\bibinfo {volume} {547}},\
  \bibinfo {pages} {219} (\bibinfo {year} {2002}{\natexlab{b}})},\ \Eprint
  {https://arxiv.org/abs/hep-ph/0210197} {arXiv:hep-ph/0210197} \BibitemShut
  {NoStop}%
\bibitem [{\citenamefont {Grimus}\ and\ \citenamefont
  {Lavoura}(2004)}]{hep-ph/0305309}%
  \BibitemOpen
  \bibfield  {author} {\bibinfo {author} {\bibfnamefont {W.}~\bibnamefont
  {Grimus}}\ and\ \bibinfo {author} {\bibfnamefont {L.}~\bibnamefont
  {Lavoura}},\ }\bibfield  {title} {\bibinfo {title} {{A Nonstandard CP
  transformation leading to maximal atmospheric neutrino mixing}},\ }\href
  {https://doi.org/10.1016/j.physletb.2003.10.075} {\bibfield  {journal}
  {\bibinfo  {journal} {Phys. Lett. B}\ }\textbf {\bibinfo {volume} {579}},\
  \bibinfo {pages} {113} (\bibinfo {year} {2004})},\ \Eprint
  {https://arxiv.org/abs/hep-ph/0305309} {arXiv:hep-ph/0305309} \BibitemShut
  {NoStop}%
\bibitem [{\citenamefont {Krishnan}(2021)}]{2011.11653}%
  \BibitemOpen
  \bibfield  {author} {\bibinfo {author} {\bibfnamefont {R.}~\bibnamefont
  {Krishnan}},\ }\bibfield  {title} {\bibinfo {title} {{Symmetries of
  stationary points of the $G$-invariant potential and the framework of the
  auxiliary group}},\ }\href {https://doi.org/10.1103/PhysRevD.103.L051701}
  {\bibfield  {journal} {\bibinfo  {journal} {Phys. Rev. D}\ }\textbf {\bibinfo
  {volume} {103}},\ \bibinfo {pages} {051701} (\bibinfo {year} {2021})},\
  \Eprint {https://arxiv.org/abs/2011.11653} {arXiv:2011.11653 [hep-ph]}
  \BibitemShut {NoStop}%
\end{thebibliography}%

\end{document}